\RequirePackage{fix-cm}
\documentclass[11pt]{article}
\newcounter{ourcount}
\usepackage{amsmath,amsfonts,amssymb,graphicx,epsfig,pdflscape,multirow,theorem,gensymb}
\usepackage{lscape,arydshln,float}
\usepackage[matrix,arrow,curve]{xy} 
\numberwithin{equation}{section}
\graphicspath{{./eps/}}
\usepackage[nosort]{cite}
\usepackage[usenames]{color}
\usepackage{pstricks}
\usepackage{pst-plot}
\usepackage[backref=false]{hyperref}
\usepackage[capitalise,noabbrev]{cleveref} 
\usepackage{float}
\usepackage[11pt]{moresize}
\hypersetup{
colorlinks=true,
citecolor=red,
linkcolor=darkblue,
urlcolor=darkblue
}
\definecolor{darkblue}{rgb}{0,0,.8}
\definecolor{red}{rgb}{1,0,0}

\theorembodyfont{\itshape} 
\theoremheaderfont{\scshape}
\theoremstyle{plain}

\newtheorem{Conjecture}{Conjecture}

\numberwithin{equation}{section}

\crefname{Conjecture}{Conjecture}{Conjectures}

\newcommand{\nc}{\newcommand}

\def\arxiv#1#2{\href{http://arxiv.org/abs/#1}{\textsf{arXiv:#1 #2}}}
\nc{\ir}{\mathrm{i}}
\nc{\dd}{\mathrm{d}}   
\nc{\eE}{\mathsf{e}}
\nc{\bib}{\bibitem}
\nc{\be}{\begin{equation}}
\nc{\ee}{\end{equation}}
\nc{\chit}{\raisebox{0.25ex}{$\chi$}}
\nc{\dtl}{\mathsf{dTL}}
\nc{\pdtl}{\mathsf{pdTL}}
\nc{\Dbh}{\mbox{\boldmath $\widehat D$}}
\nc{\Dh}{\mbox{$\hat D$}}
\nc{\Dbb}{\mbox{\boldmath $\bar D$}}
\nc{\Dbm}{\mbox{\boldmath $\mathcal D$}}
\nc{\Dbt}{\mbox{\boldmath $\tilde{D}$}}
\nc{\Tbt}{\mbox{\boldmath $\tilde{T}$}}
\nc{\Tbh}{\mbox{\boldmath $\widehat{T}$}}

\nc{\setS}{\mathcal S}

\nc{\db}{\mbox{\boldmath $d$}}
\nc{\Ab}{\mbox{\boldmath $A$}}
\nc{\Bb}{\mbox{\boldmath $B$}}
\nc{\Cb}{\mbox{\boldmath $C$}}
\nc{\Db}{\mbox{\boldmath $D$}}
\nc{\eb}{\mbox{\boldmath $e$}}
\nc{\Fb}{\mbox{\boldmath $F$}}
\nc{\Fbt}{\mbox{\boldmath $\tilde{F}$}}
\nc{\fb}{\mbox{\boldmath $f$}}
\nc{\fbt}{\mbox{\boldmath $\tilde{f}$}}
\nc{\Gb}{\mbox{\boldmath $G$}}
\nc{\Hb}{\mbox{\boldmath $H$}}
\nc{\Ib}{\mbox{\boldmath $I$}}
\nc{\Jb}{\mbox{\boldmath $J$}}
\nc{\Kb}{\mbox{\boldmath $K$}}
\nc{\Lb}{\mbox{\boldmath $L$}}
\nc{\Mb}{\mbox{\boldmath $M$}}
\nc{\Pb}{\mbox{\boldmath $P$}}
\nc{\Qb}{\mbox{\boldmath $Q$}}
\nc{\Rb}{\mbox{\boldmath $R$}}
\nc{\Tb}{\mbox{\boldmath $T$}}
\nc{\Tbb}{\mbox{\boldmath $\bar T$}}
\nc{\Tbm}{\mbox{\boldmath $\mathcal T$}}
\nc{\tb}{\mbox{\boldmath $t$}}
\nc{\Ub}{\mbox{\boldmath $U$}}
\nc{\Vb}{\mbox{\boldmath $V$}}
\nc{\Wb}{\mbox{\boldmath $W$}}
\nc{\xb}{\mbox{\boldmath $x$}}
\nc{\yb}{\mbox{\boldmath $y$}}
\nc{\Zb}{\mbox{\boldmath $Z$}}
\nc{\Lambdab}{\boldsymbol{\Lambda}}

\def\mbar{\overline m}
\def\deltabar{\bar \Delta}
\def\Ebar{\bar E}
\def\qbar{\bar q}

\nc{\Atwotwo}{\mbox{$A_2^{\textrm{\fontsize{7pt}{7pt}\selectfont $(2)$}}$}}
\nc{\Aoneone}{\mbox{$A_1^{\textrm{\fontsize{7pt}{7pt}\selectfont $(1)$}}$}}
\nc{\amf}{\mbox{$\mathfrak a$}}
\nc{\bmf}{\mbox{$\mathfrak b$}}
\nc{\cmf}{\mbox{$\mathfrak c$}}
\nc{\dmf}{\mbox{$\mathfrak d$}}
\nc{\fmf}{\mbox{$\mathfrak f$}}
\nc{\gmf}{\mbox{$\mathfrak g$}}
\nc{\Amf}{\mbox{$\mathfrak A$}}
\nc{\Bmf}{\mbox{$\mathfrak B$}}
\nc{\Cmf}{\mbox{$\mathfrak C$}}
\nc{\Dmf}{\mbox{$\mathfrak D$}}
\nc{\Fmf}{\mbox{$\mathfrak F$}}
\nc{\asf}{\mbox{$\mathsf a$}}
\nc{\bsf}{\mbox{$\mathsf b$}}
\nc{\csf}{\mbox{$\mathsf c$}}
\nc{\dsf}{\mbox{$\mathsf d$}}
\nc{\fsf}{\mbox{$\mathsf f$}}
\nc{\gsf}{\mbox{$\mathsf g$}}
\nc{\Asf}{\mbox{$\mathsf A$}}
\nc{\Bsf}{\mbox{$\mathsf B$}}
\nc{\Csf}{\mbox{$\mathsf C$}}
\nc{\Dsf}{\mbox{$\mathsf D$}}

\nc{\repV}{\mathsf{V}}
\nc{\repW}{\mathsf{W}}
\newrgbcolor{darkgreen}{0., 0.733333, 0.0621042}
\definecolor{lightblue}{rgb}{.7,.7,1}
\definecolor{lightestblue}{rgb}{.95,.95,1}
\definecolor{lightlightblue}{rgb}{.85,.85,1}
\definecolor{midblue}{rgb}{.7,.7,1}

\nc{\elegant}{1.5pt}
\nc{\moyen}{1.0pt}
\nc{\mince}{0.5pt}

\def\vvdots{\mathinner{\mkern1mu\raise1pt\vbox{\kern7pt\hbox{.}}\mkern2mu
  \raise4pt\hbox{.}\mkern2mu\raise7pt\hbox{.}\mkern1mu}}

\def\loopa{
\psframe[linewidth=.25pt](0,0)(1,1)
}
\def\loopb{
\psframe[linewidth=.25pt](0,0)(1,1)
\psarc[linewidth=1.5pt,linecolor=blue](0,1){.5}{-90}{0}
}
\def\loopc{
\psframe[linewidth=.25pt](0,0)(1,1)
\psarc[linewidth=1.5pt,linecolor=blue](1,0){.5}{90}{180}
}
\def\loopd{
\psframe[linewidth=.25pt](0,0)(1,1)
\psarc[linewidth=1.5pt,linecolor=blue](0,0){.5}{0}{90}
}
\def\loope{
\psframe[linewidth=.25pt](0,0)(1,1)
\psarc[linewidth=1.5pt,linecolor=blue](1,1){.5}{180}{270}
}
\def\loopf{
\psframe[linewidth=.25pt](0,0)(1,1)
\psline[linewidth=1.5pt,linecolor=blue](0.5,0)(0.5,1)
}
\def\loopg{
\psframe[linewidth=.25pt](0,0)(1,1)
\psline[linewidth=1.5pt,linecolor=blue](0,0.5)(1,0.5)
}
\def\looph{
\psframe[linewidth=.25pt](0,0)(1,1)
\psarc[linewidth=1.5pt,linecolor=blue](1,0){.5}{90}{180}
\psarc[linewidth=1.5pt,linecolor=blue](0,1){.5}{-90}{0}
}
\def\loopi{
\psframe[linewidth=.25pt](0,0)(1,1)
\psarc[linewidth=1.5pt,linecolor=blue](0,0){.5}{0}{90}
\psarc[linewidth=1.5pt,linecolor=blue](1,1){.5}{180}{270}
}

\def\facegrid#1#2{
\psframe[fillstyle=solid,fillcolor=lightlightblue,linewidth=0pt]#1#2
\psgrid[gridlabels=0pt,subgriddiv=1]#1#2}

\def\specialcircle#1{
\pspolygon[fillstyle=solid,fillcolor=white,linewidth=0.5pt](#1,#1)(#1,-#1)(-#1,-#1)(-#1,#1)
}

\def\dloopone{
\psframe[linewidth=.25pt](0,0)(1,1)
\psline[linewidth=.35pt]{-}(0,0)(1,1)
}
\def\dlooptwo{
\psframe[linewidth=.25pt](0,0)(1,1)
\psline[linewidth=.35pt]{-}(0,0)(1,1)
\psarc[linewidth=1.5pt,linecolor=blue](0,1){.5}{-90}{0}
}
\def\dloopthr{
\psframe[linewidth=.25pt](0,0)(1,1)
\psline[linewidth=.35pt]{-}(0,0)(1,1)
\psarc[linewidth=1.5pt,linecolor=blue](1,0){.5}{90}{180}
}
\def\dloopfou{
\psframe[linewidth=.25pt](0,0)(1,1)
\psline[linewidth=.35pt]{-}(0,0)(1,1)
\psarc[linewidth=1.5pt,linecolor=blue](0,0){.5}{0}{90}
}
\def\dloopfiv{
\psframe[linewidth=.25pt](0,0)(1,1)
\psline[linewidth=.35pt]{-}(0,0)(1,1)
\psarc[linewidth=1.5pt,linecolor=blue](1,1){.5}{180}{270}
}
\def\dloopsix{
\psframe[linewidth=.25pt](0,0)(1,1)
\psline[linewidth=.35pt]{-}(0,0)(1,1)
\psline[linewidth=1.5pt,linecolor=blue](0,0.5)(1,0.5)
}
\def\dloopsev{
\psframe[linewidth=.25pt](0,0)(1,1)
\psline[linewidth=.35pt]{-}(0,0)(1,1)
\psline[linewidth=1.5pt,linecolor=blue](0.5,0)(0.5,1)
}
\def\dloopeig{
\psframe[linewidth=.25pt](0,0)(1,1)
\psline[linewidth=.35pt]{-}(0,0)(1,1)
\psarc[linewidth=1.5pt,linecolor=blue](1,0){.5}{90}{180}
\psarc[linewidth=1.5pt,linecolor=blue](0,1){.5}{-90}{0}
}

\def\oloopone{
\psframe[linewidth=.25pt](0,0)(1,1)
\psline[linewidth=.35pt]{-}(0,1)(1,0)
}
\def\olooptwo{
\psframe[linewidth=.25pt](0,0)(1,1)
\psline[linewidth=.35pt]{-}(0,1)(1,0)
\psarc[linewidth=1.5pt,linecolor=blue](0,1){.5}{-90}{0}
}
\def\oloopthr{
\psframe[linewidth=.25pt](0,0)(1,1)
\psline[linewidth=.35pt]{-}(0,1)(1,0)
\psarc[linewidth=1.5pt,linecolor=blue](1,0){.5}{90}{180}
}
\def\oloopfou{
\psframe[linewidth=.25pt](0,0)(1,1)
\psline[linewidth=.35pt]{-}(0,1)(1,0)
\psarc[linewidth=1.5pt,linecolor=blue](0,0){.5}{0}{90}
}
\def\oloopfiv{
\psframe[linewidth=.25pt](0,0)(1,1)
\psline[linewidth=.35pt]{-}(0,1)(1,0)
\psarc[linewidth=1.5pt,linecolor=blue](1,1){.5}{180}{270}
}
\def\oloopsix{
\psframe[linewidth=.25pt](0,0)(1,1)
\psline[linewidth=.35pt]{-}(0,1)(1,0)
\psline[linewidth=1.5pt,linecolor=blue](0,0.5)(1,0.5)
}
\def\oloopsev{
\psframe[linewidth=.25pt](0,0)(1,1)
\psline[linewidth=.35pt]{-}(0,1)(1,0)
\psline[linewidth=1.5pt,linecolor=blue](0.5,0)(0.5,1)
}

\def\oloopnin{
\psframe[linewidth=.25pt](0,0)(1,1)
\psline[linewidth=.35pt]{-}(0,1)(1,0)
\psarc[linewidth=1.5pt,linecolor=blue](0,0){.5}{0}{90}
\psarc[linewidth=1.5pt,linecolor=blue](1,1){.5}{180}{270}
}

\def\loopab{
\psframe[linewidth=.25pt](0,0)(1,1)
\psarc[linewidth=1.5pt,linecolor=blue](1,0){.5}{90}{180}
\psarc[linewidth=1.5pt,linecolor=blue](0,1){.5}{-90}{0}
\psline[linewidth=3pt,linecolor=blue](0,0)(1,1)
}
\def\loopar{
\psframe[linewidth=.25pt](0,0)(1,1)
\psarc[linewidth=1.5pt,linecolor=blue](1,0){.5}{90}{180}
\psarc[linewidth=1.5pt,linecolor=blue](0,1){.5}{-90}{0}
\psline[linewidth=3pt,linecolor=red](0,0)(1,1)
}
\def\loopbb{
\psframe[linewidth=.25pt](0,0)(1,1)
\psarc[linewidth=1.5pt,linecolor=blue](0,0){.5}{0}{90}
\psarc[linewidth=1.5pt,linecolor=blue](1,1){.5}{180}{270}
\psline[linewidth=3pt,linecolor=blue](1,0)(0,1)
}
\def\loopbr{
\psframe[linewidth=.25pt](0,0)(1,1)
\psarc[linewidth=1.5pt,linecolor=blue](0,0){.5}{0}{90}
\psarc[linewidth=1.5pt,linecolor=blue](1,1){.5}{180}{270}
\psline[linewidth=3pt,linecolor=red](1,0)(0,1)
}

\def\hexab{
\pspolygon[fillstyle=solid,linewidth=0.5pt,linecolor=black,fillcolor=lightblue](0,0)(0,1)(0.866,1.5)(1.73205,1)(1.73205,0)(0.866,-0.5)
}

\def\hexaw{
\pspolygon[fillstyle=solid,linewidth=0.5pt,linecolor=black,fillcolor=white](0,0)(0,1)(0.866,1.5)(1.73205,1)(1.73205,0)(0.866,-0.5)
}

\def\hc{
\pspolygon[fillstyle=solid,linewidth=0.5pt,linecolor=black,fillcolor=lightblue](0,0)(0,1)(0.866,1.5)(1.73205,1)(1.73205,0)(0.866,-0.5)
}

\def\hu{
\pspolygon[fillstyle=solid,linewidth=0.5pt,linecolor=black,fillcolor=white](0,0)(0,1)(0.866,1.5)(1.73205,1)(1.73205,0)(0.866,-0.5)
}

\def\trida{
\pspolygon[fillstyle=solid,linewidth=0.5pt,linecolor=black,fillcolor=lightlightblue](0.866,0.5)(-0.866,0.5)(0,-1)
}
\def\tridb{
\pspolygon[fillstyle=solid,linewidth=0.5pt,linecolor=black,fillcolor=lightlightblue](0.866,0.5)(-0.866,0.5)(0,-1)
\psarc[linewidth=1.5pt,linecolor=blue](0,-1){0.866}{60}{120}
}
\def\tridc{
\pspolygon[fillstyle=solid,linewidth=0.5pt,linecolor=black,fillcolor=lightlightblue](0.866,0.5)(-0.866,0.5)(0,-1)
\psarc[linewidth=1.5pt,linecolor=blue](-0.866,0.5){0.866}{-60}{0}
}

\def\tridd{
\pspolygon[fillstyle=solid,linewidth=0.5pt,linecolor=black,fillcolor=lightlightblue](0.866,0.5)(-0.866,0.5)(0,-1)
\psarc[linewidth=1.5pt,linecolor=blue](0.866,0.5){0.866}{180}{240}
}

\def\triua{
\pspolygon[fillstyle=solid,linewidth=0.5pt,linecolor=black,fillcolor=lightlightblue](0.866,-1)(-0.866,-1)(0,0.5)
}
\def\triub{
\pspolygon[fillstyle=solid,linewidth=0.5pt,linecolor=black,fillcolor=lightlightblue](0.866,-1)(-0.866,-1)(0,0.5)
\psarc[linewidth=1.5pt,linecolor=blue](0,0.5){0.866}{240}{300}
}

\def\triuc{
\pspolygon[fillstyle=solid,linewidth=0.5pt,linecolor=black,fillcolor=lightlightblue](0.866,-1)(-0.866,-1)(0,0.5)
\psarc[linewidth=1.5pt,linecolor=blue](-0.866,-1){0.866}{0}{60}
}

\def\triud{
\pspolygon[fillstyle=solid,linewidth=0.5pt,linecolor=black,fillcolor=lightlightblue](0.866,-1)(-0.866,-1)(0,0.5)
\psarc[linewidth=1.5pt,linecolor=blue](0.866,-1){0.866}{120}{180}
}

\renewcommand{\ge}{\geqslant}
\renewcommand{\le}{\leqslant}

\nc{\proof}{{\scshape Proof.\ }} 				
\nc{\eproof}{{\hfill \rule{0.5em}{0.5em}\medskip}}

\begin{document}

\topmargin -5mm
\oddsidemargin 5mm
\vspace*{-2cm}

\makeatletter 
\newcommand\Larger{\@setfontsize\semiHuge{20.00}{23.78}}
\makeatother 

\setcounter{page}{1}
\mbox{}\vspace{1cm}
\begin{center}

{\Larger \bf \mbox{Critical site percolation on the triangular lattice:}
\\[0.3cm] 
\mbox{From integrability to conformal partition functions}
}

\end{center}

\vspace{1cm}
\begin{center}
{\vspace{-5mm}\Large Alexi Morin-Duchesne$^{a}$, Andreas Kl\"umper$^b$, Paul A.~Pearce$^{c,d}$}
\\[.5cm]
{\em { }$^a$ Department of Applied Mathematics, Computer Science and Statistics \\ Ghent University, 9000 Ghent, Belgium}
\\[.2cm]
  {\em { }$^b$Fakult\"at f\"ur Mathematik und Naturwissenschaften \\ Bergische
 Universit\"at Wuppertal, 42097 Wuppertal, Germany}
\\[.2cm]
{\em { }$^c$School of Mathematics and Statistics, University of Melbourne\\
Parkville, Victoria 3010, Australia}
\\[.2cm]
{\em { }$^d$School of Mathematics and Physics, University of Queensland}\\
{\em St Lucia, Brisbane, Queensland 4072, Australia}
\\[.2cm] 
{\tt alexi.morin.duchesne\,@\,gmail.com}
\qquad
{\tt kluemper\,@\,uni-wuppertal.de}
\qquad
{\tt papearce\,@unimelb.edu.au}
\end{center}

\vspace{6mm}
\centerline{{\bf{Abstract}}}
\vskip.3cm
\noindent 
Critical site percolation on the triangular lattice is described by the Yang-Baxter solvable dilute $\Atwotwo$ loop model with crossing parameter specialized to $\lambda=\tfrac{\pi}{3}$, corresponding to the contractible loop fugacity $\beta=-2\cos 4\lambda=1$. We study the functional relations satisfied by the commuting transfer matrices of this model and the associated Bethe ansatz equations. The single and double row transfer matrices are respectively endowed with strip and periodic boundary conditions, and are elements of the ordinary and periodic dilute Temperley-Lieb algebras. The standard modules for these algebras are labeled by the number of defects $d$ and, in the latter case, also by the twist~$\eE^{\ir\gamma}$. Nonlinear integral equation techniques are used to analytically solve the Bethe ansatz functional equations in the scaling limit for the central charge $c=0$ and conformal weights $\Delta, \bar \Delta$. For the ground states, we find $\Delta=\Delta_{1,d+1}$ for strip boundary conditions and $(\Delta,\bar\Delta)=(\Delta_{\gamma/\pi,d/2},\Delta_{\gamma/\pi,-d/2})$ for periodic boundary conditions, where $\Delta_{r,s}=\frac1{24}\big((3r-2s)^2-1\big)$. We give explicit conjectures for the scaling limit of the trace of the transfer matrix in each standard module. For $d\le 8$, these conjectures are supported by numerical solutions of the logarithmic form of the Bethe ansatz equations for the leading $20$ or more conformal eigenenergies. With these conjectures, we apply the Markov traces to obtain the conformal partition functions on the cylinder and torus. These precisely coincide with our previous results for critical bond percolation on the square lattice, described by the dense $\Aoneone$ loop model with $\lambda=\tfrac{\pi}{3}$. The concurrence of all this conformal data provides compelling evidence supporting a strong form of universality between these two stochastic models as logarithmic conformal field theories.

\newpage
\tableofcontents

\newpage
\hyphenpenalty=30000

\setcounter{footnote}{0}

%
\section{Introduction}
%

Percolation theory~\cite{BroadHamm57,KestonPerc82,Stauffer92,Grimmet97} is a central pillar within statistical physics exemplifying the underlying principles of phase transitions and critical phenomena~\cite{PhaseTransitions} with a myriad of applications in a diverse range of fields including physics, biology, ecology and social sciences. In percolation, the bonds or sites of a regular lattice are occupied independently and randomly with a probability $p$ and a geometric phase transition occurs with the appearance of an infinite connected cluster as $p$ is increased beyond a critical threshold $p_c$.

In this paper, we are interested in percolation in two dimensions and study critical site percolation on the triangular lattice. This work forms a companion to a previous paper~\cite{MDKP2017} in which we studied critical bond percolation on the square lattice. By a fortunate happenstance, bond percolation on the square lattice and site percolation on the triangular lattice are both Yang-Baxter solvable~\cite{BaxBook} at the common percolation threshold $p_c=\tfrac{1}{2}$~\cite{SykesEssam64}. In hindsight, it turns out to be advantageous to study the associated loop models where the domain wall boundaries surrounding the percolation clusters constitute nonlocal degrees of freedom in the form of loop segments. Bond percolation on the square lattice is described by the dense $\Aoneone$ loop model with crossing parameter $\lambda=\tfrac{\pi}{3}$ and is underpinned by the dense Temperley-Lieb algebra~\cite{TL71,Jones}. Site percolation on the triangular lattice is described by the dilute $\Atwotwo$ loop model with crossing parameter $\lambda=\tfrac{\pi}{3}$ and is underpinned by the dilute Temperley-Lieb algebra~\cite{GP93,P94,Grimm96,BSA14}. 

Despite their commonalities, these percolating systems display some differences. In particular, the loop and vertex models describing bond and site percolation have underlying $s\ell(2)$ and $s\ell(3)$ structures, for bond and site percolation respectively. In practice, Yang-Baxter integrability means that both of these models can be solved exactly in intricate detail including in the continuum scaling limit, where the scaling behaviour is manifest and precisely described by conformal field theory (CFT)~\cite{FMS}. The study of the conformal description of critical percolation in two dimensions was initiated by Saleur and Duplantier~\cite{Saluer87,SaleurDup87,Saleur92} in the late eighties and subsequently received a significant impetus arising from the study of crossing probabilities~\cite{LanglandsEtAl92,Cardy92,Watts96}. These models were later studied by analysing the fractal dimensions~\cite{AAMRH,JankeSchakel05d,JankeSchakel06b,SAPR2009,DengEtAl,SAPR2012} of the boundaries of various types of clusters. Notably, the rigorous proof of conformal invariance in the continuum scaling limit was provided~\cite{Smirnov,KSmirnov} for critical percolation on the triangular lattice and its associated dilute $\Atwotwo$ loop model (namely the model we study in this paper), with deep connections to $SLE_6$~\cite{SmirnovW01,KNienhuis,CNewman}. 

According to the {\it universality hypothesis}, the critical behaviour of statistical systems without long-ranged interactions should depend only on (i) the lattice dimension and (ii) the symmetry of the local spin/order parameter. It should moreover be insensitive to other physical details such as the crystalline lattice structure and its sublattice symmetry. Fundamentally, in two dimensions, this asserts that the critical exponents should only depend on the local symmetry group for the order parameter. For the Ising model this symmetry group is ${\Bbb Z}_2$ and for the $Q$-state Potts model it is the symmetric group~${\cal S}_Q$. The local degrees of freedom for percolation, namely the site or bond occupation numbers, are independent and identically distributed. They exhibit a $\mathbb Z_2$ symmetry and have trivial correlations. In contrast, seen as the $Q \to 1$ limit of the Potts model \cite{DC98}, the local symmetry group of percolation ${\cal S}_1$ is trivial.

For percolation in two dimensions, the known values of the standard critical exponents related to (i) the number of clusters per site, (ii) the percolation probability, (iii) the truncated mean cluster size, (iv) the cluster volume, (v) the correlation length and (vi) the decay with separation of the probability that two sites belong to the same cluster are respectively
\begin{subequations}
\begin{alignat}{3}
\alpha&=\frac{2\Delta_t-1}{\Delta_t-1}=-\frac{2}{3},\qquad&&  
\beta=\frac{\Delta_h}{1-\Delta_t}=\frac{5}{36},\qquad  
&&\gamma=\frac{2\Delta_h-1}{\Delta_t-1}=\frac{43}{18},\\[0.2cm]
\delta&=\frac{1-\Delta_h}{\Delta_h}=\frac{91}{5},\qquad&&
\nu=\frac{1}{2(1-\Delta_t)}=\frac{4}{3},\qquad && \eta=4\Delta_h=\frac{5}{24}.
\end{alignat}
\end{subequations}
For example,  as $p\to p_c^+$, 
the percolation probability behaves as $P(p)\sim (p-p_c)^\beta$. 
Only two of these six exponents are independent since, by scaling relations~\cite{CardyRGScaling}, they are all expressible in terms of the thermal and magnetic conformal weights
\begin{equation}
\Delta_t=\Delta_{2,1}=\Delta_{0,2}=\frac{5}{8},\qquad \Delta_h=\Delta_{\frac{1}{2},0}=\frac{5}{96}.
\end{equation}

Conventionally, the set of critical exponents is often said to determine the {\em universality class\/} of critical behaviour, but this is somewhat simplistic. 
Nonetheless, for statistical systems described by {\em rational} CFTs~\cite{MooreSeiberg}, a complete delineation of universality classes is possible. A CFT is {\em rational\/} if it admits a {\em finite\/} number of irreducible representations of the conformal Virasoro chiral algebra (or its extension) corresponding to the scaling operators of the theory and these representations close under fusion. The Ising model, 3-state Potts model and the more general family of $A$-$D$-$E$ minimal models~\cite{BPZ,ABF,FB,Pasquier87,CIZ87} are examples of rational CFTs. These theories are completely classified by a finite set of Moore-Seiberg conformal data~\cite{MooreSeiberg} which includes  (i)~the central charge $c$, (ii)~the operator content, (iii)~the associated conformal weights $\Delta$ and conformal characters $\chit_\Delta(q)$ and (iv)~the conformal partition functions. For example, in this more precise sense, the hard square model (which appears not to be Yang-Baxter solvable) is asserted~\cite{GuoBlote} to be in the same universality class as the Ising model. Likewise, hard hexagons is asserted~\cite{KP91,Pearce93} to lie in the same universality class as the 3-state Potts model.

Folklore dictates that, in any given lattice dimension, bond and site percolation lie in the same universality class independent of the lattice structure. More specifically, it is expected that bond percolation on the square lattice and site percolation on the triangular lattice lie in the same universality class. But, as CFTs,  these theories are not {\em rational\/} --- instead they are {\em logarithmic} CFTs~\cite{Gurarie,PRZ2006,RS07,SpecialIssue}. This implies that the Virasoro algebra admits reducible yet indecomposable representations entailing non-trivial Jordan blocks and the existence of an infinite number of scaling operators. This raises the central question of this paper which is --- to what extent can {\em universality} be demonstrated between these two logarithmic CFTs?

To answer this question, we study the dilute $\Atwotwo$ loop model with crossing parameter $\lambda=\tfrac{\pi}{3}$, focussing on the use of transfer matrix functional relations and the Bethe ansatz equations to obtain the conformal eigenenergies and conformal partition functions. We study single and double row transfer matrices, $\Tbh(u)$ and $\Dbh(u)$, with periodic and strip boundary conditions respectively, and the corresponding partition functions on the torus and cylinder. These geometries allow for loops with non-trivial windings around any periods. Loops are said to be contractible if they can be deformed to a point and non-contractible otherwise. It is then customary to assign different fugacities according to the winding: $\alpha$ for non-contractible loops and $\beta$ for contractible loops. For loop models associated with percolation, the fugacity of the contractible loops is set to $\beta=1$, whereas there is a freedom in the choice of~$\alpha$.

For rational CFTs, the torus partition function is fixed by conformal invariance and modular invariance \cite{FMS}. It is a sum of terms of the form $C_{\Delta,\bar\Delta}q^\Delta \bar q^{\bar \Delta}$, where $q$ and $\bar q$ are the modular nomes and $C_{\Delta,\bar\Delta}$ is an integer coefficient that reveals the operator content and the degeneracy for the pair $(\Delta,\bar\Delta)$ of conformal weights. Similarly, the cylinder partition functions are sums of terms of the form $C_{\Delta}q^\Delta$ with integer coefficients $C_{\Delta}$ that reveals the field content of the boundary CFT. Things are a bit different for loop models and their underlying logarithmic CFTs. For generic real values of the fugacity~$\alpha$ of the non-contractible loops, the torus and cylinder partition functions are still sums of terms of the same form as above, however the coefficients $C_{\Delta,\bar\Delta}$ and $C_{\Delta}$ are in general not integers. For percolation, the values $\alpha = 1$ and $\alpha=2$ are two exceptions. For $\alpha = 1$, the classical partition function per site/bond is trivially $Z = 1$ from which little can be learned. Much of our efforts will thus be directed at the value $\alpha=2$. Nonetheless, we will also compute the partition functions for the other values of $\alpha$.

We point out that there exists an extensive literature on the three types of $\Atwotwo$ lattice models based on the vertex, RSOS and loop representations of the underlying dilute Temperley-Lieb algebras. Indeed, for general $\lambda\in (0,\tfrac{\pi}{3})$, much is known about the $\Atwotwo$ lattice models for the 19-vertex Izergin-Korepin model~\cite{IK1981,AMN1995,Fan97,Utiel2003}, dilute RSOS models~\cite{Kuniba1991,WNS1992,Roche92,W93,WPSN1994,BNW94,ZPG1995,Suzuki1998} as well as the dilute $\Atwotwo$  loop model~\cite{Nienhuis90,DJS2010,Garboli12,FeherNien2015,GarbNien2017a,GarbNien2017b,MDP19,BMDSA21} including the application of nonlinear integral equations (NLIEs)~\cite{BNW89,WBN92,ZB97}. The case $\lambda=\tfrac{\pi}{3}$ of interest here is special because the formula for the bulk free energy \cite{WBN92} valid generically for $\lambda\in (0,\tfrac{\pi}{3})$ is divergent in the limit $\lambda \to \frac \pi 3$, a problem that we will circumvent by choosing different normalisations for the face operators and transfer matrices. We will here focus on loop representations, namely the standard modules over the ordinary and periodic dilute Temperley-Lieb algebras $\dtl_N(\beta)$ and $\pdtl_N(\beta)$. These are respectively denoted as $\repV_{N,d}$ and $\repW_{N,d,\omega}$, where $d$ is the number of defects $d$ and $\omega=\eE^{\ir\gamma}$ is a twist, relevant only for periodic boundary conditions. 

Let us summarize the main ideas and new results of this paper. First, we push further the investigation of the integrability of the model of critical site percolation on the triangular lattice. That the model is integrable in the bulk was previously established by F\'eher and Nienhuis \cite{FeherNien2015}, who described the local map from configurations of site percolation to those of the dilute $\Atwotwo$ loop model. Here, we establish that the model is also integrable when defined on a lattice with a boundary, namely we show that a boundary decorated with percolation sites of a given fixed colour can be described in terms of a solution of the boundary Yang-Baxter equation in the $\Atwotwo$ model, first obtained by Yung and Batchelor \cite{YB95} and later studied by Dubail, Jacobsen and Saleur \cite{DJS10}. We then use the Markov traces of the dilute Temperley-Lieb algebras to express the partition functions in terms of the traces of the commuting transfer matrices over the standard modules. We in fact consider two partition functions $\mathcal Z_{\textrm{cyl}}^{\textrm{\tiny$(i)$}}$ on the cylinder and four partition functions $\mathcal Z_{\textrm{tor}}^{\textrm{\tiny$(h,v)$}}$ on the torus, with $i,h,v \in \{0,1\}$. These correspond to different boundary conditions assigned to the percolation clusters on these geometries.

Next, we study the scaling limit of the model from the large-$N$ behaviour of the transfer matrix eigenvalues. We compute analytically the leading finite-size correction for the ground states eigenvalues in each standard module using the methods of functional relations and NLIEs. This allows us to extract the conformal weights
\be
\label{eq:Delta.periodic}
\Delta = \Delta_{1,d+1} = \frac{d(d-6)}2
\ee
for strip boundary conditions and
\be
\label{eq:Delta.strip}
\Delta = \Delta_{\gamma/\pi,d/2} = \frac1{24} \bigg[\displaystyle\Big(\frac{3\gamma}\pi-d\Big)^2-1\bigg], \qquad \bar \Delta = \Delta_{\gamma/\pi,-d/2} = \frac1{24} \bigg[\displaystyle\Big(\frac{3\gamma}\pi+d\Big)^2-1\bigg]
\ee 
for periodic boundary conditions, where $\Delta_{r,s}=\frac{(3r-2s)^2-1}{24}$ is the usual Kac formula for $c = 0$. For periodic boundary conditions, the groundstate conformal weights in $\repW_{N,0,\omega}$ reproduce the values previously obtained for the ground state of the Izergin-Korepin model by Warnaar, Batchelor and Nienhuis \cite{WBN92}, and Zhou and Batchelor \cite{ZB97}.
Going beyond the ground states, we conjecture character formulas for the trace of the rescaled transfer matrix (to the power $M$) in each standard module:
\begin{subequations}
\label{eq:trace.conjectures}
\begin{alignat}{2}
\mathrm{tr}_{\raisebox{-0.05cm}{\tiny$\repW_{N,d,\omega}$}} \widehat \Tb(u)^M &\xrightarrow{N \to \infty}
\frac{1}{(q)_\infty (\bar q_\infty)}
\sum_{\ell=-\infty}^{\infty}  q^{\Delta_{\gamma/\pi, 3\ell+d/2}} \bar q^{\Delta_{\gamma/\pi, 3\ell-d/2}},
\\[0.15cm]
\mathrm{tr}_{\raisebox{-0.05cm}{\tiny$\repV_{N,d}$}} \widehat \Db(u) &\xrightarrow{N \to \infty}
 q^{\Delta_{1,d+1}} \frac{1-q^{d+1}}{(q)_\infty},
\end{alignat}
\end{subequations}
where $\omega = \eE^{\ir \gamma}$.
This conjecture thus gives the conformal dimensions of all the scaling states arising in each standard module. Lastly, we combine the Markov trace~\cite{Jones83,RJ06,DJS09,MDPR13} with the above conjectures to obtain the partition functions for $\alpha = 2$, namely
\begin{subequations}
\begin{alignat}{2}
\mathcal Z_{\textrm{cyl}}^{\textrm{\tiny$(0)$}} &= \varkappa_1(q) + 3\varkappa_3(q) - \varkappa_5(q) -6\frac{\dd}{\dd z} \Big(\varkappa_1(q,z) - \varkappa_3(q,z) + \varkappa_5(q,z) \Big)\Big|_{z=1}\ ,
\\[0.15cm]
\mathcal Z_{\textrm{cyl}}^{\textrm{\tiny$(1)$}} &= 2\varkappa_1(q)  + 4 \varkappa_5(q) +6\frac{\dd}{\dd z} \Big(\varkappa_1(q,z) - \varkappa_3(q,z) + \varkappa_5(q,z) \Big)\Big|_{z=1}\ ,
\end{alignat}
on the cylinder, and
\be
\begin{array}{l}
\displaystyle \mathcal Z_{\textrm{tor}}^{\textrm{\tiny$(0,v)$}} = \sum_{j=0}^6 \Big((-1)^{v j} d_j |\varkappa_j(q)|^2\Big) - 2(-1)^v,
\\[0.3cm]
\displaystyle \mathcal Z_{\textrm{tor}}^{\textrm{\tiny$(1,v)$}} = \sum_{j=0}^6 \Big((-1)^{v (j+1)} d_j \varkappa_j(q)\varkappa_{6-j}(\bar q)\Big) + 2,
\end{array}
\qquad
d_j=
\left\{\begin{array}{ll}
1&j=0, 6,\\[0.15cm]
2&j=1, \dots, 5,
\end{array}\right.
\quad v \in \{0,1\},
\ee
\end{subequations}
on the torus, each one expressed in terms of affine $u(1)$ characters. The precise concurrence with the analytic results for critical bond percolation on the square lattice~\cite{MDKP2017} reveals a strong form of universality between the two models. Further evidence for this strong form of universality is revealed by computing the partition functions of the two models for $\alpha\in \mathbb C$. We find that these twisted partition functions again coincide and that they reproduce known functions obtained previously from Coulomb gas arguments \cite{FSZ87,RS01}.

The outline of the paper is as follows. In \cref{sec:CFT.data}, we recall the conformal description of critical percolation in two dimensions. In \cref{sec:perco.triangular}, we define site percolation on the triangular lattice and the mapping onto a dilute loop model with the relevant boundary conditions on the cylinder and torus. In \cref{sec:A22.def}, we define the $\Atwotwo$ loop model on the cylinder and torus, and give their description in terms of the ordinary and periodic dilute Temperley-Lieb algebras, their standard representations and their commuting transfer matrices. We also review the result of F\'eher and Nienhuis \cite{FeherNien2015}, whereby the model of site percolation corresponds to the $\Atwotwo$ loop model with the crossing parameter specialized to $\lambda=\tfrac{\pi}{3}$. In \cref{sec:partition.functions}, we first study the large-$N$ expansion of the eigenvalues of the transfer matrices in the standard modules. Based on numerics, we conjecture the formulas for the scaling limit of these traces in each of these standard modules. With these conjectures and the Markov traces, we derive expressions for the four conformal partition functions on the torus and the two partition functions on the cylinder.

In \cref{sec:fun.rel,sec:NLIE.FSS}, we present an exact derivation of the groundstate finite size corrections, which confirms the leading term of the trace conjectures \eqref{eq:trace.conjectures}. First, the functional relations and Bethe ansatz equations for site percolation are derived in \cref{sec:fun.rel}, for both periodic and strip boundary conditions. We use the fusion hierarchy and its closure relation~\cite{MDP19} to show that $\Tbh(u)$ satisfies a cubic functional equation. Next we recall the form of Baxter's functional $\widehat T$-$Q$ equations and the resulting Bethe ansatz equations. We rewrite these in terms of two linear functional equations satisfied by an eigenvalue $\widehat T(u)$ of the single row transfer matrix, its corresponding Baxter $Q(u)$ function and a suitably chosen additional auxiliary function $P(u)$. These $s\ell(3)$ linear equations are the counterpart of the known $s\ell(2)$ linear equation of Fabricius and McCoy~\cite{FabriciusMcCoy01}. Analogs of all of these functional equations are also presented for $\Dbh(u)$. In \cref{sec:NLIE.FSS}, the methods of \cite{KBP91,KP92,FK99,J08} are used to convert the Bethe ansatz functional equations for $\widehat T(u)$ to nonlinear integral equations (NLIEs) and to analytically solve them for the ground state conformal weights in \eqref{eq:Delta.periodic}. Due to the similar form of the functional equations, the same analytic calculations yield the groundstate conformal weights in \eqref{eq:Delta.strip} for the eigenvalues $\widehat D(u)$ of the double row transfer matrix. The proof of the needed dilogarithm identities is presented in \cref{app:dilog.ints}. Finally, \cref{sec:Numerics} presents our numerical evidence supporting the trace conjectures \eqref{eq:trace.conjectures}. These numerical calculatons were carried out using the logarithmic form of the Bethe ansatz equations, which are more stable for iterative solutions. We solve these equations for the leading conformal eigenenergies for both $\widehat T(u)$ and $\widehat D(u)$, in the standard modules with $0\le d\le 8$. The numerical results are tabulated in \cref{sec:Tabulated}. We make some concluding remarks in \cref{sec:conclusion} and comment on the relation with the $Y$-system of Gliozzi and Tateo in \cref{app:Y.system}.

\subsection{Conformal description of critical percolation in two dimensions}\label{sec:CFT.data}

In this section, before specializing to the case $\lambda=\tfrac{\pi}{3}$ relevant to percolation, we recall the basic conformal data for the general dense $\Aoneone$ and dilute $\Atwotwo$ loop models with crossing parameter
\be
\lambda=\left\{\begin{array}{cl} \frac{(p'-p)\pi}{p'}&\Aoneone,\\[0.15cm]
\frac{(2p'-p)\pi}{4p'}&\Atwotwo,
\end{array}\right.\qquad p<p',\qquad \textrm{gcd}(p,p')=1,
\ee
with $\eE^{\ir \lambda}$ a root of unity. We refer to the dense $\Aoneone$ and dilute $\Atwotwo$ models, at these roots of unity, as the ${\cal LM}(p,p')$ and ${\cal DLM}(p,p')$ models respectively, with regimes
\be
\begin{array}{rll}
\Aoneone: &{\cal LM}(p,p'):   &\hspace{0.86cm}0<\lambda<\pi,\\[0.25cm]
\Atwotwo: &{\cal DLM}(p,p'):&\left\{\begin{array}{cl}
\hspace{0.21cm}0<\lambda<\tfrac{\pi}{2} & \textrm{Regime I,} \\[0.1cm]
\tfrac{2\pi}{3}<\lambda<\pi & \textrm{Regime II,} \\[0.1cm]
\hspace{0.26cm}\tfrac{\pi}{2}<\lambda<\tfrac{2\pi}{3} & \textrm{Regime III.} \\
\end{array}\right.
\end{array}\label{genericRegimes}
\ee
These regimes are valid for the isotropic models, namely for the spectral parameter set to $u = \frac \lambda 2$ for the $\Aoneone$ model and $u = \frac {3\lambda} 2$ for the $\Atwotwo$ model. Critical bond percolation on the square lattice coincides with the logarithmic minimal model ${\cal LM}(2,3)$ 
and critical site percolation on the triangular lattice is the model ${\cal DLM}(2,3)$ belonging to Regime I. 

In the continuum scaling limit, these models are described by logarithmic conformal field theories~\cite{Gurarie,PRZ2006,RS07,SpecialIssue}.
For the ${\cal LM}(p,p')$ and ${\cal DLM}(p,p')$ models, the common central charges and conformal weights in the infinitely extended Kac tables are
\begin{subequations}
\begin{alignat}{3}
c^{p,p'}&=1-\frac{6(p-p')^2}{pp'}, \qquad p<p',\qquad &&p,p'=1,2,3,\ldots\label{central}
\\
\Delta_{r,s}^{p,p'}&=\frac{(p'r-ps)^2-(p-p')^2}{4pp'}, \quad\qquad &&r,s=1,2,3,\ldots\label{Delta}
\end{alignat}
\end{subequations}

\begin{figure}[t]
\psset{unit=0.7cm}
\begin{equation*}
\begin{pspicture}(0,-.3)(7,11)
\psframe[linewidth=0pt,fillstyle=solid,fillcolor=lightestblue](0,0)(7,11)
\psframe[linewidth=0pt,fillstyle=solid,fillcolor=lightlightblue](1,0)(2,11)
\psframe[linewidth=0pt,fillstyle=solid,fillcolor=lightlightblue](3,0)(4,11)
\psframe[linewidth=0pt,fillstyle=solid,fillcolor=lightlightblue](5,0)(6,11)
\psframe[linewidth=0pt,fillstyle=solid,fillcolor=lightlightblue](0,2)(7,3)
\psframe[linewidth=0pt,fillstyle=solid,fillcolor=lightlightblue](0,5)(7,6)
\psframe[linewidth=0pt,fillstyle=solid,fillcolor=lightlightblue](0,8)(7,9)
\multiput(0,0)(0,3){3}{\multiput(0,0)(2,0){3}{\psframe[linewidth=0pt,fillstyle=solid,fillcolor=midblue](1,2)(2,3)}}
\psgrid[gridlabels=0pt,subgriddiv=1]
\rput(.5,10.6){$\vdots$}\rput(1.5,10.6){$\vdots$}\rput(2.5,10.6){$\vdots$}
\rput(3.5,10.6){$\vdots$}\rput(4.5,10.6){$\vdots$}\rput(5.5,10.6){$\vdots$}
\rput(6.5,10.5){$\vvdots$}
\rput(.5,9.5){$12$}\rput(1.5,9.5){$\frac{65}8$}\rput(2.5,9.5){$5$}
\rput(3.5,9.5){$\frac{21}8$}\rput(4.5,9.5){$1$}\rput(5.5,9.5){$\frac{1}8$}\rput(6.525,9.5){$\cdots$}
\rput(.5,8.5){$\frac{28}3$}\rput(1.5,8.5){$\frac{143}{24}$}\rput(2.5,8.5){$\frac{10}3$}
\rput(3.5,8.5){$\frac{35}{24}$}\rput(4.5,8.5){$\frac 13$}\rput(5.5,8.5){$-\frac{1}{24}$}
\rput(6.525,8.5){$\cdots$}
\rput(.5,7.5){$7$}\rput(1.5,7.5){$\frac {33}8$}\rput(2.5,7.5){$2$}
\rput(3.5,7.5){$\frac{5}8$}\rput(4.5,7.5){$0$}\rput(5.5,7.5){$\frac{1}8$}
\rput(6.525,7.5){$\cdots$}
\rput(.5,6.5){$5$}\rput(1.5,6.5){$\frac {21}8$}\rput(2.5,6.5){$1$}
\rput(3.5,6.5){$\frac{1}8$}\rput(4.5,6.5){$0$}\rput(5.5,6.5){$\frac{5}8$}\rput(6.525,6.5){$\cdots$}
\rput(.5,5.5){$\frac{10}3$}\rput(1.5,5.5){$\frac {35}{24}$}\rput(2.5,5.5){$\frac 13$}
\rput(3.5,5.5){$-\frac{1}{24}$}\rput(4.5,5.5){$\frac 13$}\rput(5.5,5.5){$\frac{35}{24}$}
\rput(6.525,5.5){$\cdots$}
\rput(.5,4.5){$2$}\rput(1.5,4.5){$\frac 58$}\rput(2.5,4.5){$0$}\rput(3.5,4.5){$\frac{1}8$}
\rput(4.5,4.5){$1$}\rput(5.5,4.5){$\frac{21}8$}\rput(6.525,4.5){$\cdots$}
\rput(.5,3.5){$1$}\rput(1.5,3.5){$\frac 18$}\rput(2.5,3.5){$0$}\rput(3.5,3.5){$\frac{5}8$}
\rput(4.5,3.5){$2$}\rput(5.5,3.5){$\frac{33}8$}\rput(6.525,3.5){$\cdots$}
\rput(.5,2.5){$\frac 13$}\rput(1.5,2.5){$-\frac 1{24}$}\rput(2.5,2.5){$\frac 13$}
\rput(3.5,2.5){$\frac{35}{24}$}\rput(4.5,2.5){$\frac{10}3$}\rput(5.5,2.5){$\frac{143}{24}$}
\rput(6.525,2.5){$\cdots$}
\rput(.5,1.5){$0$}\rput(1.5,1.5){$\frac 18$}\rput(2.5,1.5){$1$}\rput(3.5,1.5){$\frac{21}8$}
\rput(4.5,1.5){$5$}\rput(5.5,1.5){$\frac{65}8$}\rput(6.525,1.5){$\cdots$}
\rput(.5,.5){$0$}\rput(1.5,.5){$\frac 58$}\rput(2.5,.5){$2$}\rput(3.5,.5){$\frac{33}8$}\rput(4.5,.5){$7$}
\rput(5.5,.5){$\frac{85}8$}\rput(6.525,.5){$\cdots$}
{\color{blue}
\rput(.5,-.5){$1$}
\rput(1.5,-.5){$2$}
\rput(2.5,-.5){$3$}
\rput(3.5,-.5){$4$}
\rput(4.5,-.5){$5$}
\rput(5.5,-.5){$6$}
\rput(6.5,-.5){$r$}
\rput(-.5,.5){$1$}
\rput(-.5,1.5){$2$}
\rput(-.5,2.5){$3$}
\rput(-.5,3.5){$4$}
\rput(-.5,4.5){$5$}
\rput(-.5,5.5){$6$}
\rput(-.5,6.5){$7$}
\rput(-.5,7.5){$8$}
\rput(-.5,8.5){$9$}
\rput(-.5,9.5){$10$}
\rput(-.5,10.5){$s$}}
\end{pspicture}
\hspace{3cm}
\begin{pspicture}(-0.5,-0.3)(5,15)
\psframe[fillstyle=solid,fillcolor=white,linewidth=0pt](0,0)(5,15)
\psframe[linewidth=0pt,fillstyle=solid,fillcolor=lightestblue](0,0)(5,15)
\psgrid[gridlabels=0pt,subgriddiv=1](0,0)(5,15)
{\color{blue}
\rput(0.5,-0.5){$0$}
\rput(1.5,-0.5){$1$}
\rput(2.5,-0.5){$2$}
\rput(3.5,-0.5){$3$}
\rput(4.5,-0.5){$r$}
\rput(-0.5,0.5){$0$}
\rput(-0.5,1.5){$\frac12$}
\rput(-0.5,2.5){$1$}
\rput(-0.5,3.5){$\frac32$}
\rput(-0.5,4.5){$2$}
\rput(-0.5,5.5){$\frac52$}
\rput(-0.5,6.5){$3$}
\rput(-0.5,7.5){$\frac72$}
\rput(-0.5,8.5){$4$}
\rput(-0.5,9.5){$\frac92$}
\rput(-0.5,10.5){$5$}
\rput(-0.5,11.5){$\frac{11}2$}
\rput(-0.5,12.5){$6$}
\rput(-0.5,13.5){$\frac{13}2$}
\rput(-0.5,14.5){$s$}
}
\rput(0.5,0.5){$-\frac1{24}$}
\rput(0.5,1.5){$0$}
\rput(0.5,2.5){$\frac18$}
\rput(0.5,3.5){$\frac13$}
\rput(0.5,4.5){$\frac58$}
\rput(0.5,5.5){$1$}
\rput(0.5,6.5){$\frac{35}{24}$}
\rput(0.5,7.5){$2$}
\rput(0.5,8.5){$\frac{21}{8}$}
\rput(0.5,9.5){$\frac{10}{3}$}
\rput(0.5,10.5){$\frac{33}{8}$}
\rput(0.5,11.5){$5$}
\rput(0.5,12.5){$\frac{143}{24}$}
\rput(0.5,13.5){$7$}
\rput(0.5,14.60){$\vdots$}
\rput(1.5,0.5){$\frac13$}
\rput(1.5,1.5){$\frac18$}
\rput(1.5,2.5){$0$}
\rput(1.5,3.5){$-\frac1{24}$}
\rput(1.5,4.5){$0$}
\rput(1.5,5.5){$\frac18$}
\rput(1.5,6.5){$\frac13$}
\rput(1.5,7.5){$\frac58$}
\rput(1.5,8.5){$1$}
\rput(1.5,9.5){$\frac{35}{24}$}
\rput(1.5,10.5){$2$}
\rput(1.5,11.5){$\frac{21}{8}$}
\rput(1.5,12.5){$\frac{10}{3}$}
\rput(1.5,13.5){$\frac{33}{8}$}
\rput(1.5,14.60){$\vdots$}
\rput(2.5,0.5){$\frac{35}{24}$}
\rput(2.5,1.5){$1$}
\rput(2.5,2.5){$\frac{5}{8}$}
\rput(2.5,3.5){$\frac{1}{3}$}
\rput(2.5,4.5){$\frac{1}{8}$}
\rput(2.5,5.5){$0$}
\rput(2.5,6.5){$-\frac{1}{24}$}
\rput(2.5,7.5){$0$}
\rput(2.5,8.5){$\frac{1}{8}$}
\rput(2.5,9.5){$\frac{1}{3}$}
\rput(2.5,10.5){$\frac{5}{8}$}
\rput(2.5,11.5){$1$}
\rput(2.5,12.5){$\frac{35}{24}$}
\rput(2.5,13.5){$2$}
\rput(2.5,14.60){$\vdots$}
\rput(3.5,0.5){$\frac{10}{3}$}
\rput(3.5,1.5){$\frac{21}{8}$}
\rput(3.5,2.5){$2$}
\rput(3.5,3.5){$\frac{35}{24}$}
\rput(3.5,4.5){$1$}
\rput(3.5,5.5){$\frac{5}{8}$}
\rput(3.5,6.5){$\frac{1}{3}$}
\rput(3.5,7.5){$\frac{1}{8}$}
\rput(3.5,8.5){$0$}
\rput(3.5,9.5){$-\frac{1}{24}$}
\rput(3.5,10.5){$0$}
\rput(3.5,11.5){$\frac{1}{8}$}
\rput(3.5,12.5){$\frac{1}{3}$}
\rput(3.5,13.5){$\frac{5}{8}$}
\rput(3.5,14.60){$\vdots$}
\multiput(0,0)(0,1){14}{\rput(4.5,0.525){$\cdots$}}
\rput(4.5,14.50){$\vvdots$}
\end{pspicture}
\end{equation*}
\caption{{\it Left}: Infinitely extended Virasoro Kac table of conformal weights $\Delta_{r,s}$, with $r,s \in \mathbb Z_{>0}$ of critical percolation with strip boundary conditions. The entries are given different colours depending on whether $r$ is a multiple of $p=2$, $s$ is a multiple of $p'=3$, or both. {\it Right}: The conformal weights $\Delta_{r,s}$ for critical percolation with $r \in \mathbb Z_{\ge 0}$ and $s \in \frac12\mathbb Z_{\ge 0}$ for critical percolation with twisted periodic boundary conditions, for $\omega = \pm 1$.}
\label{fig:VirKac}
\end{figure}
Specializing to $(p,p')=(2,3)$ with $\lambda=\tfrac{\pi}{3}$ for critical percolation gives
\be
c=0,\quad\qquad \Delta_{r,s}=\frac{(3r-2s)^2-1}{24}. \label{Delta}
\ee
The extended integer Kac table of critical percolation for strip boundary conditions is shown in \cref{fig:VirKac}. The extended Kac table with integer $r$ and half-integer $s\in\tfrac{1}{2}{\Bbb Z}$ values, relevant to the periodic boundary conditions when the twist is set to $\omega = \pm 1$, is also shown. The associated characters for the strip boundary conditions are the Kac characters~\cite{PRZ2006}
\be
\chit_{r,s}(q)=q^{\Delta_{r,s}}\frac{1-q^{rs}}{(q)_\infty},\qquad r,s=1,2,3,\ldots\label{eq:rsChars}
\ee
The relevant characters~\cite{PR2011} for the cylinder and torus partition functions are the affine $u(1)$ characters
\begin{equation}
\varkappa_j(q) = \varkappa_j^{n}(q,1), \qquad \varkappa_j^{n}(q,z)=\frac{\Theta_{j,n}(q,z)}{q^{1/24}(q)_\infty}=\frac{q^{-1/24}}{(q)_\infty} \sum_{k\in{\mathbb Z}} z^kq^{(j+2kn)^2/4n},
\label{eq:u1chars}
\end{equation}
with $n = pp' = 6$ for percolation. These satisfy the relations
\be
\varkappa_{2n+j}^n(z,q)=z^{-1}\varkappa_{j}^n(z,q), \qquad \varkappa_{2n-j}^n(z,q)=z^{-1}\varkappa_{j}^n(z^{-1},q).
\ee

%
\section{Site percolation on the triangular lattice}\label{sec:perco.triangular}
%

\subsection{Definition of the model on the torus and cylinder}\label{sec:def.model.torus.cylinder}

We are interested in site percolation on a triangular lattice with the geometry of the cylinder and torus. In a configuration of this model, each site is either coloured with probability $p$ or uncoloured with probability $1-p$. Here we focus on the critical value $p_c = \frac12$ where the sites are coloured and uncoloured with equal probabilities. In this presentation, the sites are at the centers of the hexagonal cells, and the role of coloured and uncoloured cells are interchangeable. \cref{fig:perco.configurations.torus,fig:perco.configurations.cylinder} give examples of configurations of site percolation on the torus and cylinder. 

\begin{samepage}
\begin{figure}[t!]
\centering
\begin{pspicture}[shift=-1.7](-3.3,-1.75)(3,2)
\rput(0,0){\includegraphics[width=.45\textwidth]{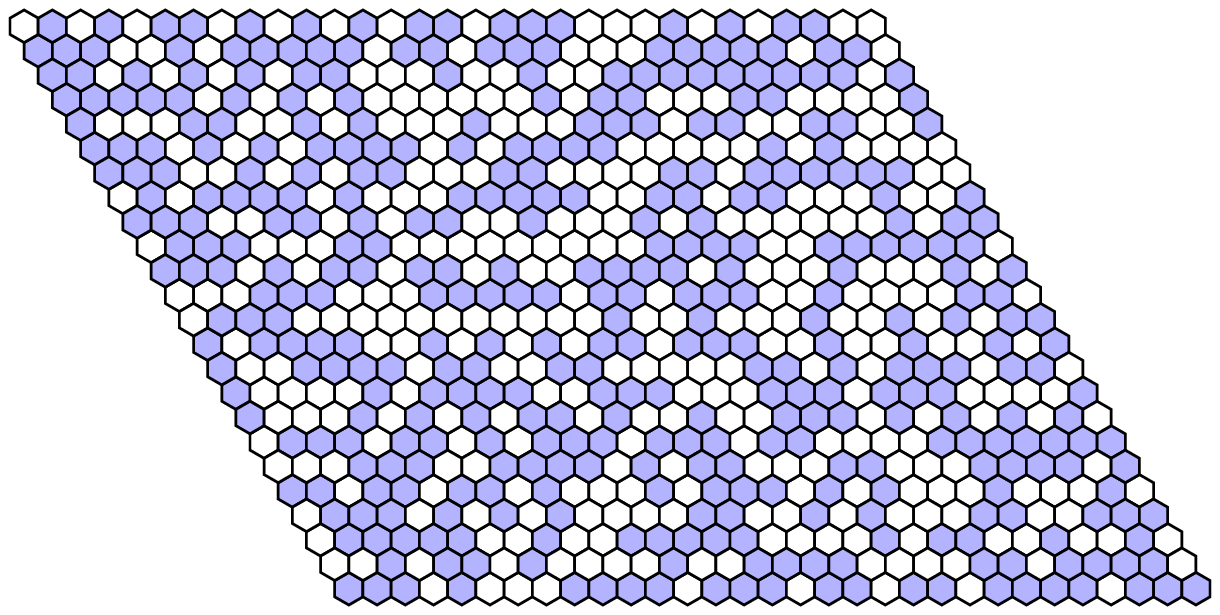}}
\end{pspicture}
\qquad \qquad \qquad \qquad
\begin{pspicture}[shift=-1.7](-3.3,-1.5)(3,1.5)
\rput(0,0){\includegraphics[width=.65\textwidth]{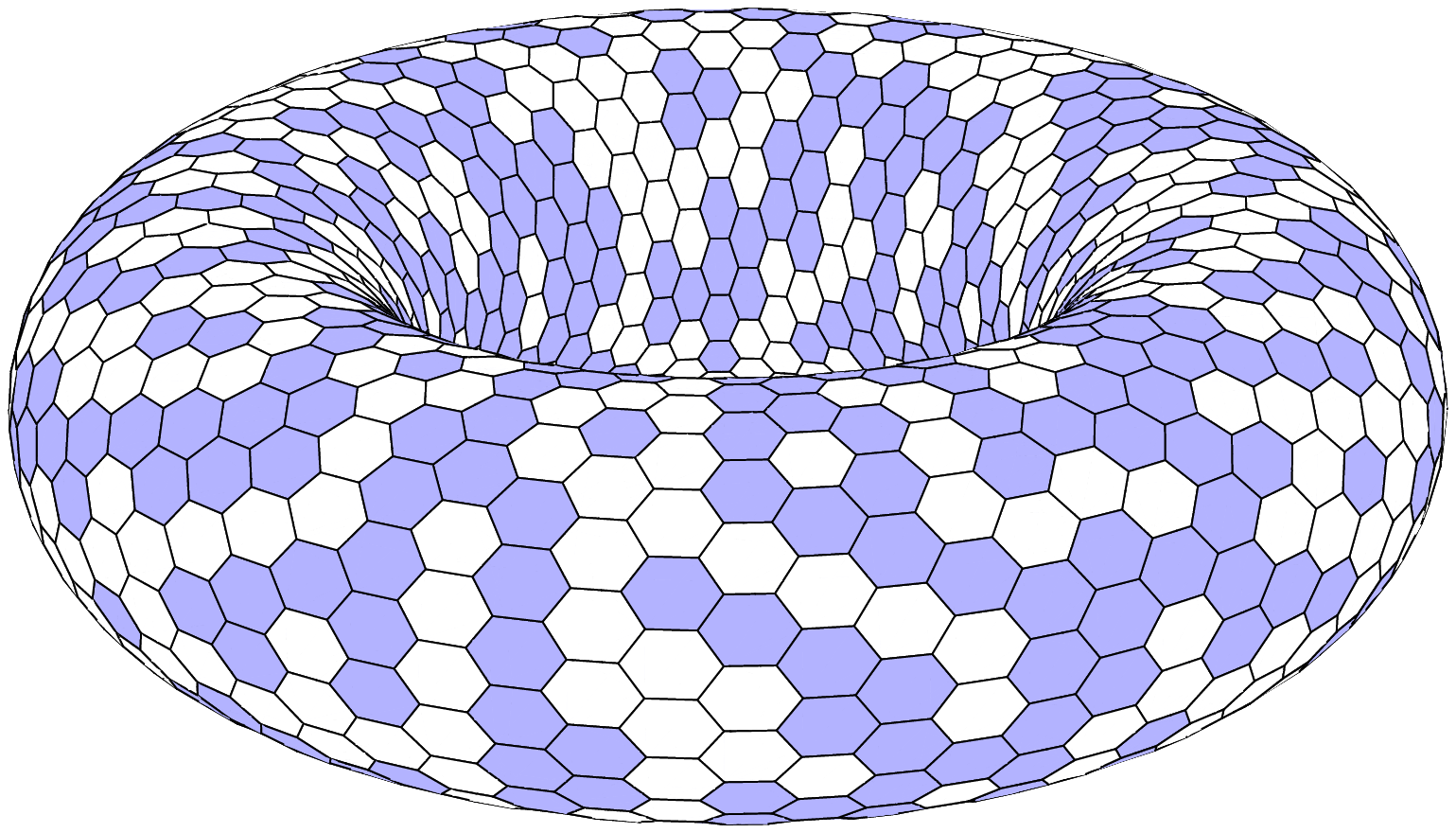}}
\end{pspicture}
\caption{Configurations of critical site percolation with doubly-periodic boundary conditions, drawn in the plane and on the torus.}
\label{fig:perco.configurations.torus}
\end{figure}
\nopagebreak
\begin{figure}[t!] 
\centering
\begin{pspicture}[shift=-1.7](-3.3,-2.2)(3,2.2)
\rput(0,0){\includegraphics[width=.35\textwidth]{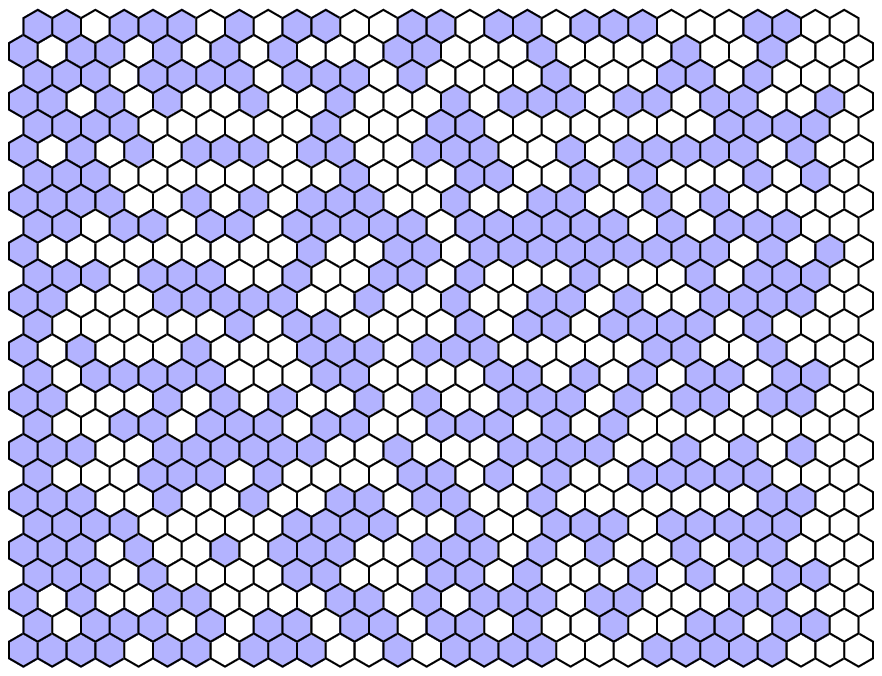}}
\end{pspicture}
\qquad \qquad \qquad \qquad
\begin{pspicture}[shift=-1.7](-3.3,-2.2)(3,2.2)
\rput(0,0){\includegraphics[width=.50\textwidth]{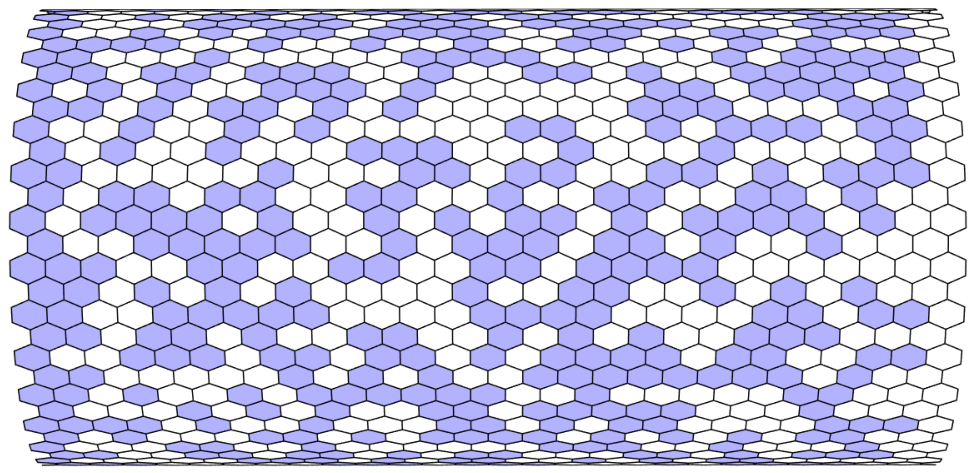}}
\end{pspicture}
\caption{Configurations of critical site percolation with a periodic boundary condition along one direction and fixed boundary conditions along the second direction, drawn in the plane and on the cylinder. The fixed colour is different on the left and right boundaries, and the corresponding partition function is $Z_{\textrm{cyl}}^{\textrm{\tiny$(1)$}}(\alpha)$.}
\label{fig:perco.configurations.cylinder}
\end{figure}
\end{samepage}

Configurations of this model on the cylinder or torus can be drawn in the plane. For toroidal boundary conditions, we draw the lattice in the plane as a parallelogram with horizontal length $N$ and diagonal length $M$. For instance, these lengths are $M=24$ and $N=31$ in the left panel of \cref{fig:perco.configurations.torus}. Domain walls between the purple and white hexagons trace out curves on the lattice that divide the cells into connected clusters. These curves can be non-contractible, meaning that they wrap the torus non-trivially, $i \in \mathbb Z$ times around the horizontal period and $j \in \mathbb Z_{\ge 0}$ times around the vertical period, with $i$ and $j$ coprime. We write this condition in terms of their greatest common divisor as $i \wedge j = 1$, with the convention $i \wedge 0 = 0 \wedge i = i$. Moreover, we restrict $j$ to be greater than or equal to zero, whereas $i$ can be either positive or negative. Following a curve moving upwards on the torus, we assign it a positive value of $i$ if it winds around the torus horizontally by moving towards the right, and a negative integer if it winds by moving towards the left. A curve that winds around the torus $i$ times along one axis and $j$ times around the second axis always draws the contour of a cluster that has the same property. There are two more types of clusters on the torus. A cluster can have {\it cross-topology}, namely, it can wrap around both periods of the torus. Finally, a cluster need not wrap around either of the two periods. In these last two cases, the curves delimiting the boundary of the clusters are contractible loops. We assign to each configuration $\sigma$ the weight 
\be
\label{eq:n.sigma}
w(\sigma) = \alpha^{n(\sigma)}, \qquad n(\sigma) = \sum_{i,j|i\wedge j = 1} \!\!n_{i,j}(\sigma),
\ee
where $n_{i,j}(\sigma)$ is the number of non-contractible loops in $\sigma$ that encircle the torus $i$ times in one direction and $j$ times in the other. Thus the contractible loops are assigned a unit weight. The partition function on the torus is then given by
\be
\label{eq:Z.perco.tor}
Z_{\textrm{tor}}(\alpha) = \sum_\sigma \alpha^{n(\sigma)}.
\ee
Each non-contractible coloured cluster has two non-contractible loops on its boundary. As a result, $n(\sigma)$ is always an even number and $Z(\alpha)$ is an even polynomial in $\alpha$. Also, it is clear that 
\be
\label{eq:power.of.2}
Z_{\textrm{tor}}(\alpha=1) = 2^{MN}.
\ee

For cylindrical boundary conditions, we draw the lattice in the plane as a rectangle of $2M$ rows that alternate between $N+1$ and $N+2$ hexagonal faces. The example on the left panel of \cref{fig:perco.configurations.cylinder} has $M = 13$ and $N=28$. On each of the boundaries of the cylinder, we assign a fixed colour, either purple or white, to all the sites belonging to the first two columns of sites. As is the case on the torus, the domain walls of purple and white hexagons trace out curves on the lattice that divide the faces into connected clusters. These curves are either non-contractible, meaning that they encircle the cylinder, or they are contractible, in which case they can be continuously deformed to a point. We again denote the number of non-contractible curves by $n(\sigma)$ and define the cylinder partition function as
\be
\label{eq:Z.perco.cyl}
Z_{\textrm{cyl}}(\alpha) = \sum_\sigma \alpha^{n(\sigma)}.
\ee
This partition function is an even and odd polynomial in $\alpha$ if the two boundaries of the cylinder are assigned identical or different colours. We denote the corresponding partition functions as $Z_{\textrm{cyl}}^{\textrm{\tiny$(0)$}}(\alpha)$ and $Z_{\textrm{cyl}}^{\textrm{\tiny$(1)$}}(\alpha)$, respectively. Clearly, we have 
\be
\label{eq:power.of.2.v2}
Z_{\textrm{cyl}}^{\textrm{\tiny$(0)$}}(\alpha=1) = Z_{\textrm{cyl}}^{\textrm{\tiny$(1)$}}(\alpha=1) = 2^{M(2N-1)}.
\ee

\subsection{Map to dilute loop configurations on the square lattice}\label{sec:PercToLoops}

To study the statistical and conformal properties of critical site percolation on the triangular lattice, it is convenient to use the idea of F\'eher and Nienhuis \cite{FeherNien2015} and view it as a dilute loop model defined on a square lattice. The transformation that maps percolation configurations into configurations of the loop model is defined in two steps. The first step consists in drawing triangular tiles centered on each site of the hexagonal lattice and whose corners lie in the centers of the three adjacent hexagonal cells. If the three cells are either all purple or all white, we draw no loop segment inside the triangle. Otherwise, we draw a loop segment that follows the domain wall separating the purple and white clusters. These local rules are summarised as
\be
\psset{unit=0.35cm}
\begin{pspicture}[shift=-1.9](0,-0.7)(4,3.5)
\rput(0,0){\hexab}\rput(1.732,0){\hexab}\rput(0.866,1.5){\hexab}
\rput(1.732,1.5){\triua}
\end{pspicture}
\qquad
\begin{pspicture}[shift=-1.9](0,-0.7)(4,3.5)
\rput(0,0){\hexab}\rput(1.732,0){\hexab}\rput(0.866,1.5){\hexaw}
\rput(1.732,1.5){\triub}
\end{pspicture}
\qquad
\begin{pspicture}[shift=-1.9](0,-0.7)(4,3.5)
\rput(0,0){\hexaw}\rput(1.732,0){\hexab}\rput(0.866,1.5){\hexab}
\rput(1.732,1.5){\triuc}
\end{pspicture}
\qquad
\begin{pspicture}[shift=-1.9](0,-0.7)(4,3.5)
\rput(0,0){\hexab}\rput(1.732,0){\hexaw}\rput(0.866,1.5){\hexab}
\rput(1.732,1.5){\triud}
\end{pspicture}
\ee
with similar diagrams with the purple and white colours interchanged, or with the diagrams flipped vertically. This first step thus produces a configuration of a dilute loop model living on triangular tiles.

\begin{figure}
\begin{alignat*}{2}
\psset{unit=0.30cm}
\begin{pspicture}[shift=-5.4](-7,-1)(10.7,10.5)
\rput(-6.928,9.0){\hc}\rput(-5.196,9.0){\hc}\rput(-3.464,9.0){\hu}\rput(-1.732,9.0){\hc}\rput(0.000,9.0){\hu}\rput(1.732,9.0){\hc}\rput(3.464,9.0){\hu}
\rput(-6.062,7.5){\hu}\rput(-4.330,7.5){\hc}\rput(-2.598,7.5){\hu}\rput(-0.866,7.5){\hc}\rput(0.866,7.5){\hc}\rput(2.598,7.5){\hu}\rput(4.33,7.5){\hc}
\rput(-5.196,6.0){\hu}\rput(-3.464,6.0){\hc}\rput(-1.732,6.0){\hu}\rput(0.000,6.0){\hu}\rput(1.732,6.0){\hc}\rput(3.464,6.0){\hc}\rput(5.196,6.0){\hu}
\rput(-4.330,4.5){\hc}\rput(-2.598,4.5){\hu}\rput(-0.866,4.5){\hc}\rput(0.866,4.5){\hu}\rput(2.598,4.5){\hu}\rput(4.330,4.5){\hc}\rput(6.062,4.5){\hu}
\rput(-3.464,3.0){\hc}\rput(-1.732,3.0){\hc}\rput(0.000,3.0){\hc}\rput(1.732,3.0){\hc}\rput(3.464,3.0){\hc}\rput(5.196,3.0){\hu}\rput(6.928,3.0){\hc}
\rput(-2.598,1.5){\hc}\rput(-0.866,1.5){\hc}\rput(0.866,1.5){\hu}\rput(2.598,1.5){\hc}\rput(4.33,1.5){\hu}\rput(6.062,1.5){\hc}\rput(7.794,1.5){\hc}
\rput(-1.732,0.0){\hu}\rput(0.000,0.0){\hc}\rput(1.732,0.0){\hc}\rput(3.464,0.0){\hc}\rput(5.196,0.0){\hu}\rput(6.928,0.0){\hu}\rput(8.66,0.0){\hc}
\rput(-0.866,-1.5){\hc}\rput(0.866,-1.5){\hu}\rput(2.598,-1.5){\hc}\rput(4.33,-1.5){\hc}\rput(6.062,-1.5){\hu}\rput(7.794,-1.5){\hc}\rput(9.526,-1.5){\hu}
\end{pspicture}
\longrightarrow
\begin{pspicture}[shift=-5.4](-6.9,-1)(11,10.5)
\rput(-6.928,9.0){\hc}\rput(-5.196,9.0){\hc}\rput(-3.464,9.0){\hu}\rput(-1.732,9.0){\hc}\rput(0.000,9.0){\hu}\rput(1.732,9.0){\hc}\rput(3.464,9.0){\hu}
\rput(-6.062,7.5){\hu}\rput(-4.330,7.5){\hc}\rput(-2.598,7.5){\hu}\rput(-0.866,7.5){\hc}\rput(0.866,7.5){\hc}\rput(2.598,7.5){\hu}\rput(4.330,7.5){\hc}
\rput(-5.196,6.0){\hu}\rput(-3.464,6.0){\hc}\rput(-1.732,6.0){\hu}\rput(0.000,6.0){\hu}\rput(1.732,6.0){\hc}\rput(3.464,6.0){\hc}\rput(5.196,6.0){\hu}
\rput(-4.330,4.5){\hc}\rput(-2.598,4.5){\hu}\rput(-0.866,4.5){\hc}\rput(0.866,4.5){\hu}\rput(2.598,4.5){\hu}\rput(4.330,4.5){\hc}\rput(6.062,4.5){\hu}
\rput(-3.464,3.0){\hc}\rput(-1.732,3.0){\hc}\rput(0.000,3.0){\hc}\rput(1.732,3.0){\hc}\rput(3.464,3.0){\hc}\rput(5.196,3.0){\hu}\rput(6.928,3.0){\hc}
\rput(-2.598,1.5){\hc}\rput(-0.866,1.5){\hc}\rput(0.866,1.5){\hu}\rput(2.598,1.5){\hc}\rput(4.330,1.5){\hu}\rput(6.062,1.5){\hc}\rput(7.794,1.5){\hc}
\rput(-1.732,0.0){\hu}\rput(0.000,0,0){\hc}\rput(1.732,0.0){\hc}\rput(3.464,0.0){\hc}\rput(5.196,0.0){\hu}\rput(6.928,0.0){\hu}\rput(8.660,0.0){\hc}
\rput(-0.866,-1.5){\hc}\rput(0.866,-1.5){\hu}\rput(2.598,-1.5){\hc}\rput(4.33,-1.5){\hc}\rput(6.062,-1.5){\hu}\rput(7.794,-1.5){\hc}\rput(9.526,-1.5){\hu}
\rput(-7.794,10.5){\tridd}\rput(-6.062,10.5){\tridd}\rput(-4.33013,10.5){\tridc}\rput(-2.598,10.5){\tridb}\rput(-0.866,10.5){\tridd}\rput(0.866,10.5){\tridd}\rput(2.598,10.5){\tridd}
\rput(-6.928,10.5){\triuc}\rput(-5.19615,10.5){\triub}\rput(-3.4641,10.5){\triud}\rput(-1.732,10.5){\triuc}\rput(0,10.5){\triuc}\rput(1.732,10.5){\triuc}\rput(3.464,10.5){\triuc}
\rput(-6.928,9){\tridc}\rput(-5.196,9){\tridb}\rput(-3.464,9){\tridd}\rput(-1.732,9){\tridd}\rput(0,9){\tridd}\rput(1.732,9){\tridc}\rput(3.464,9){\tridc}
\rput(-6.062,9){\triud}\rput(-4.330,9){\triuc}\rput(-2.599,9){\triuc}\rput(-0.866,9){\triuc}\rput(0.866,9){\triub}\rput(2.598,9){\triud}\rput(4.33,9){\triud}
\rput(-6.06218,7.5){\tridc}\rput(-4.33013,7.5){\tridd}\rput(-2.598,7.5){\tridd}\rput(-0.866,7.5){\tridd}\rput(0.866,7.5){\tridb}\rput(2.598,7.5){\tridd}\rput(4.33013,7.5){\tridc}
\rput(-5.19615,7.5){\triua}\rput(-3.4641,7.5){\triuc}\rput(-1.732,7.5){\triuc}\rput(0,7.5){\triub}\rput(1.732,7.5){\triuc}\rput(3.464,7.5){\triub}\rput(5.19615,7.5){\triud}
\rput(-5.196,6){\trida}\rput(-3.464,6){\tridc}\rput(-1.732,6){\tridc}\rput(0,6){\tridb}\rput(1.732,6){\tridd}\rput(3.464,6){\tridb}\rput(5.196,6){\tridd}
\rput(-4.330,6){\triud}\rput(-2.599,6){\triud}\rput(-0.866,6){\triud}\rput(0.866,6){\triuc}\rput(2.598,6){\triub}\rput(4.33,6){\triuc}\rput(6.062,6){\triuc}
\rput(-4.330,4.5){\tridc}\rput(-2.599,4.5){\tridd}\rput(-0.866,4.5){\tridc}\rput(0.866,4.5){\tridd}\rput(2.598,4.5){\tridb}\rput(4.33,4.5){\tridc}\rput(6.062,4.5){\tridc}
\rput(-3.464,4.5){\triua}\rput(-1.732,4.5){\triub}\rput(0,4.5){\triua}\rput(1.732,4.5){\triub}\rput(3.464,4.5){\triub}\rput(5.196,4.5){\triud}\rput(6.928,4.5){\triud}
\rput(-3.464,3){\trida}\rput(-1.732,3){\trida}\rput(0,3){\trida}\rput(1.732,3){\tridb}\rput(3.464,3){\trida}\rput(5.196,3){\tridc}\rput(6.928,3){\tridc}
\rput(-2.598,3){\triua}\rput(-0.866,3){\triua}\rput(0.866,3){\triud}\rput(2.598,3){\triuc}\rput(4.33,3){\triud}\rput(6.062,3){\triud}\rput(7.794,3){\triua}
\rput(-2.598,1.5){\trida}\rput(-0.866,1.5){\tridb}\rput(0.866,1.5){\tridd}\rput(2.598,1.5){\tridc}\rput(4.33,1.5){\tridd}\rput(6.062,1.5){\tridd}\rput(7.794,1.5){\tridb}
\rput(-1.732,1.5){\triud}\rput(0,1.5){\triuc}\rput(1.732,1.5){\triub}\rput(3.464,1.5){\triua}\rput(5.196,1.5){\triuc}\rput(6.928,1.5){\triub}\rput(8.660,1.5){\triuc}
\rput(-1.732,0){\tridc}\rput(0,0){\tridc}\rput(1.732,0){\tridb}\rput(3.464,0){\trida}\rput(5.196,0){\tridd}\rput(6.928,0){\trida}\rput(8.660,0){\tridc}
\rput(-0.866,0){\triud}\rput(0.866,0){\triud}\rput(2.598,0){\triuc}\rput(4.33,0){\triua}\rput(6.062,0){\triuc}\rput(7.794,0){\triud}\rput(9.526,0){\triud}
\end{pspicture}
&\longrightarrow\qquad
\psset{unit=0.5cm}
\begin{pspicture}[shift=-3.9](-2,-1)(5,7)
\facegrid{(-2,-1)}{(5,7)}
\rput(-2,6){\dloopsev}\rput(-1,6){\dloopfiv}\rput(0,6){\dloopeig}\rput(1,6){\dloopfou}\rput(2,6){\dloopsev}\rput(3,6){\dloopsev}\rput(4,6){\dloopsev}
\rput(-2,5){\dloopeig}\rput(-1,5){\dloopfou}\rput(0,5){\dloopsev}\rput(1,5){\dloopsev}\rput(2,5){\dloopfiv}\rput(3,5){\dloopeig}\rput(4,5){\dloopeig}
\rput(-2,4){\dlooptwo}\rput(-1,4){\dloopsev}\rput(0,4){\dloopsev}\rput(1,4){\dloopfiv}\rput(2,4){\dloopfou}\rput(3,4){\dloopfiv}\rput(4,4){\dloopeig}
\rput(-2,3){\dloopthr}\rput(-1,3){\dloopeig}\rput(0,3){\dloopeig}\rput(1,3){\dloopfou}\rput(2,3){\dloopfiv}\rput(3,3){\dloopfou}\rput(4,3){\dloopsev}
\rput(-2,2){\dlooptwo}\rput(-1,2){\dloopfiv}\rput(0,2){\dlooptwo}\rput(1,2){\dloopfiv}\rput(2,2){\dloopsix}\rput(3,2){\dloopeig}\rput(4,2){\dloopeig}
\rput(-2,1){\dloopone}\rput(-1,1){\dloopone}\rput(0,1){\dloopthr}\rput(1,1){\dloopfou}\rput(2,1){\dloopthr}\rput(3,1){\dloopeig}\rput(4,1){\dlooptwo}
\rput(-2,0){\dloopthr}\rput(-1,0){\dloopfou}\rput(0,0){\dloopfiv}\rput(1,0){\dlooptwo}\rput(2,0){\dloopsev}\rput(3,0){\dloopfiv}\rput(4,0){\dloopfou}
\rput(-2,-1){\dloopeig}\rput(-1,-1){\dloopeig}\rput(0,-1){\dloopfou}\rput(1,-1){\dloopone}\rput(2,-1){\dloopsev}\rput(3,-1){\dloopthr}\rput(4,-1){\dloopeig}
\end{pspicture}
\\[0.7cm]
\psset{unit=0.30cm}
\begin{pspicture}[shift=-5.4](-2,-1)(11.5,10.5)
\rput(-1.732,9.0){\hc}\rput(0.000,9.0){\hu}\rput(1.732,9.0){\hu}\rput(3.464,9.0){\hc}\rput(5.196,9.0){\hu}\rput(6.928,9.0){\hu}\rput(8.660,9.0){\hc}\rput(10.392,9.0){\hu}
\rput(-0.866,7.5){\hc}\rput(0.866,7.5){\hc}\rput(2.598,7.5){\hu}\rput(4.330,7.5){\hc}\rput(6.062,7.5){\hc}\rput(7.794,7.5){\hc}\rput(9.526,7.5){\hu}
\rput(-1.732,6.0){\hc}\rput(0.000,6.0){\hc}\rput(1.732,6.0){\hc}\rput(3.464,6.0){\hc}\rput(5.196,6.0){\hu}\rput(6.928,6.0){\hc}\rput(8.660,6.0){\hu}\rput(10.392,6.0){\hu}
\rput(-0.866,4.5){\hc}\rput(0.866,4.5){\hu}\rput(2.598,4.5){\hu}\rput(4.330,4.5){\hu}\rput(6.062,4.5){\hc}\rput(7.794,4.5){\hc}\rput(9.526,4.5){\hu}
\rput(-1.732,3.0){\hc}\rput(0.000,3.0){\hu}\rput(1.732,3.0){\hc}\rput(3.464,3.0){\hu}\rput(5.196,3.0){\hu}\rput(6.928,3.0){\hc}\rput(8.660,3.0){\hc}\rput(10.392,3.0){\hu}
\rput(-0.866,1.5){\hc}\rput(0.866,1.5){\hu}\rput(2.598,1.5){\hu}\rput(4.330,1.5){\hc}\rput(6.062,1.5){\hu}\rput(7.794,1.5){\hu}\rput(9.526,1.5){\hu}
\rput(-1.732,0.0){\hc}\rput(0.000,0.0){\hc}\rput(1.732,0.0){\hu}\rput(3.464,0.0){\hc}\rput(5.196,0.0){\hc}\rput(6.928,0.0){\hu}\rput(8.660,0.0){\hc}\rput(10.392,0.0){\hu}
\rput(-0.866,-1.5){\hc}\rput(0.866,-1.5){\hc}\rput(2.598,-1.5){\hc}\rput(4.33,-1.5){\hc}\rput(6.062,-1.5){\hu}\rput(7.794,-1.5){\hc}\rput(9.526,-1.5){\hu}
\end{pspicture}
\qquad \longrightarrow \qquad
\begin{pspicture}[shift=-5.4](-1.4,-1)(11.5,10.5)
\rput(-1.732,9.0){\hc}\rput(0.000,9.0){\hu}\rput(1.732,9.0){\hu}\rput(3.464,9.0){\hc}\rput(5.196,9.0){\hu}\rput(6.928,9.0){\hu}\rput(8.660,9.0){\hc}\rput(10.392,9.0){\hu}
\rput(-0.866,7.5){\hc}\rput(0.866,7.5){\hc}\rput(2.598,7.5){\hu}\rput(4.330,7.5){\hc}\rput(6.062,7.5){\hc}\rput(7.794,7.5){\hc}\rput(9.526,7.5){\hu}
\rput(-1.732,6.0){\hc}\rput(0.000,6.0){\hc}\rput(1.732,6.0){\hc}\rput(3.464,6.0){\hc}\rput(5.196,6.0){\hu}\rput(6.928,6.0){\hc}\rput(8.660,6.0){\hu}\rput(10.392,6.0){\hu}
\rput(-0.866,4.5){\hc}\rput(0.866,4.5){\hu}\rput(2.598,4.5){\hu}\rput(4.330,4.5){\hu}\rput(6.062,4.5){\hc}\rput(7.794,4.5){\hc}\rput(9.526,4.5){\hu}
\rput(-1.732,3.0){\hc}\rput(0.000,3.0){\hu}\rput(1.732,3.0){\hc}\rput(3.464,3.0){\hu}\rput(5.196,3.0){\hu}\rput(6.928,3.0){\hc}\rput(8.660,3.0){\hc}\rput(10.392,3.0){\hu}
\rput(-0.866,1.5){\hc}\rput(0.866,1.5){\hu}\rput(2.598,1.5){\hu}\rput(4.330,1.5){\hc}\rput(6.062,1.5){\hu}\rput(7.794,1.5){\hu}\rput(9.526,1.5){\hu}
\rput(-1.732,0.0){\hc}\rput(0.000,0.0){\hc}\rput(1.732,0.0){\hu}\rput(3.464,0.0){\hc}\rput(5.196,0.0){\hc}\rput(6.928,0.0){\hu}\rput(8.660,0.0){\hc}\rput(10.392,0.0){\hu}
\rput(-0.866,-1.5){\hc}\rput(0.866,-1.5){\hc}\rput(2.598,-1.5){\hc}\rput(4.33,-1.5){\hc}\rput(6.062,-1.5){\hu}\rput(7.794,-1.5){\hc}\rput(9.526,-1.5){\hu}
\rput(0.866,10.5){\tridb}\rput(2.598,10.5){\tridb}\rput(4.330,10.5){\trida}\rput(6.062,10.5){\tridc}\rput(7.794,10.5){\tridd}\rput(9.526,10.5){\tridd}
\rput(0.000,10.5){\triud}\rput(1.732,10.5){\triub}\rput(3.464,10.5){\triuc}\rput(5.196,10.5){\triud}\rput(6.928,10.5){\triua}\rput(8.660,10.5){\triuc}\rput(10.392,10.5){\triuc}
\rput(0.000,9.0){\tridd}\rput(1.732,9.0){\tridb}\rput(3.464,9.0){\tridd}\rput(5.196,9.0){\tridd}\rput(6.928,9.0){\tridb}\rput(8.660,9.0){\tridc}\rput(10.392,9.0){\tridc}
\rput(0.866,9.0){\triub}\rput(2.598,9.0){\triuc}\rput(4.330,9.0){\triuc}\rput(6.062,9.0){\triub}\rput(7.794,9.0){\triub}\rput(9.526,9.0){\triud}
\rput(0.866,7.5){\trida}\rput(2.598,7.5){\tridd}\rput(4.330,7.5){\tridc}\rput(6.062,7.5){\tridb}\rput(7.794,7.5){\trida}\rput(9.526,7.5){\tridc}
\rput(0.000,7.5){\triua}\rput(1.732,7.5){\triua}\rput(3.464,7.5){\triub}\rput(5.196,7.5){\triud}\rput(6.928,7.5){\triuc}\rput(8.660,7.5){\triud}\rput(10.392,7.5){\triua}
\rput(0.000,6.0){\trida}\rput(1.732,6.0){\tridb}\rput(3.464,6.0){\tridb}\rput(5.196,6.0){\tridc}\rput(6.928,6.0){\tridc}\rput(8.660,6.0){\tridd}\rput(10.392,6.0){\trida}
\rput(0.866,6.0){\triud}\rput(2.598,6.0){\triub}\rput(4.330,6.0){\triub}\rput(6.062,6.0){\triud}\rput(7.794,6.0){\triua}\rput(9.526,6.0){\triuc}
\rput(0.866,4.5){\tridc}\rput(2.598,4.5){\tridb}\rput(4.330,4.5){\trida}\rput(6.062,4.5){\tridd}\rput(7.794,4.5){\trida}\rput(9.526,4.5){\tridd}
\rput(0.000,4.5){\triud}\rput(1.732,4.5){\triud}\rput(3.464,4.5){\triuc}\rput(5.196,4.5){\triua}\rput(6.928,4.5){\triuc}\rput(8.660,4.5){\triua}\rput(10.392,4.5){\triuc}
\rput(0.000,3.0){\tridd}\rput(1.732,3.0){\tridd}\rput(3.464,3.0){\tridc}\rput(5.196,3.0){\tridb}\rput(6.928,3.0){\tridd}\rput(8.660,3.0){\tridb}\rput(10.392,3.0){\tridc}
\rput(0.866,3.0){\triuc}\rput(2.598,3.0){\triub}\rput(4.330,3.0){\triud}\rput(6.062,3.0){\triuc}\rput(7.794,3.0){\triub}\rput(9.526,3.0){\triub}
\rput(0.866,1.5){\tridd}\rput(2.598,1.5){\trida}\rput(4.330,1.5){\tridc}\rput(6.062,1.5){\tridd}\rput(7.794,1.5){\trida}\rput(9.526,1.5){\tridb}
\rput(0.000,1.5){\triua}\rput(1.732,1.5){\triuc}\rput(3.464,1.5){\triud}\rput(5.196,1.5){\triua}\rput(6.928,1.5){\triuc}\rput(8.660,1.5){\triud}\rput(10.392,1.5){\triuc}
\rput(0.000,0.0){\trida}\rput(1.732,0.0){\tridd}\rput(3.464,0.0){\tridc}\rput(5.196,0.0){\trida}\rput(6.928,0.0){\tridc}\rput(8.660,0.0){\tridc}\rput(10.392,0.0){\tridc}
\rput(0.866,0.0){\triua}\rput(2.598,0.0){\triub}\rput(4.330,0.0){\triua}\rput(6.062,0.0){\triud}\rput(7.794,0.0){\triud}\rput(9.526,0.0){\triud}
\end{pspicture}
\qquad &\longrightarrow\qquad
\psset{unit=0.50cm}
\begin{pspicture}[shift=-3.9](0,0)(7,8)
\facegrid{(0,0)}{(6,8)}
\rput(0,7){\oloopthr}\rput(1,7){\oloopsix}\rput(2,7){\oloopfou}\rput(3,7){\oloopsev}\rput(4,7){\oloopfiv}\rput(5,7){\oloopnin}
\rput(0,6){\dloopfiv}\rput(1,6){\dloopfou}\rput(2,6){\dloopsev}\rput(3,6){\dloopfiv}\rput(4,6){\dloopsix}\rput(5,6){\dloopeig}
\rput(0,5){\oloopone}\rput(1,5){\oloopfiv}\rput(2,5){\olooptwo}\rput(3,5){\oloopthr}\rput(4,5){\oloopfou}\rput(5,5){\oloopsev}
\rput(0,4){\dloopthr}\rput(1,4){\dloopsix}\rput(2,4){\dloopsix}\rput(3,4){\dloopeig}\rput(4,4){\dlooptwo}\rput(5,4){\dloopsev}
\rput(0,3){\oloopsev}\rput(1,3){\oloopthr}\rput(2,3){\oloopfou}\rput(3,3){\oloopfiv}\rput(4,3){\oloopfou}\rput(5,3){\oloopfiv}
\rput(0,2){\dloopsev}\rput(1,2){\dloopfiv}\rput(2,2){\dloopeig}\rput(3,2){\dloopfou}\rput(4,2){\dloopfiv}\rput(5,2){\dloopsix}
\rput(0,1){\oloopfiv}\rput(1,1){\oloopfou}\rput(2,1){\oloopsev}\rput(3,1){\oloopfiv}\rput(4,1){\oloopfou}\rput(5,1){\oloopthr}
\rput(0,0){\dloopone}\rput(1,0){\dloopfiv}\rput(2,0){\dlooptwo}\rput(3,0){\dloopthr}\rput(4,0){\dloopeig}\rput(5,0){\dloopeig}
\multiput(6,0)(0,2){4}{\pspolygon[fillstyle=solid,fillcolor=lightlightblue](0,0)(0,1)(1,1)
\pspolygon[fillstyle=solid,fillcolor=lightlightblue](0,1)(0,2)(1,1)}
\psarc[linecolor=blue,linewidth=1.5pt](6,1){0.5}{-90}{90}
\psarc[linecolor=blue,linewidth=1.5pt](6,3){0.5}{-90}{90}
\psarc[linecolor=blue,linewidth=1.5pt](6,7){0.5}{-90}{90}
\end{pspicture}
\end{alignat*}
\caption{The maps from percolation configurations to loop configurations, with toroidal and cylindrical boundary conditions.}
\label{fig:the.map}
\end{figure}
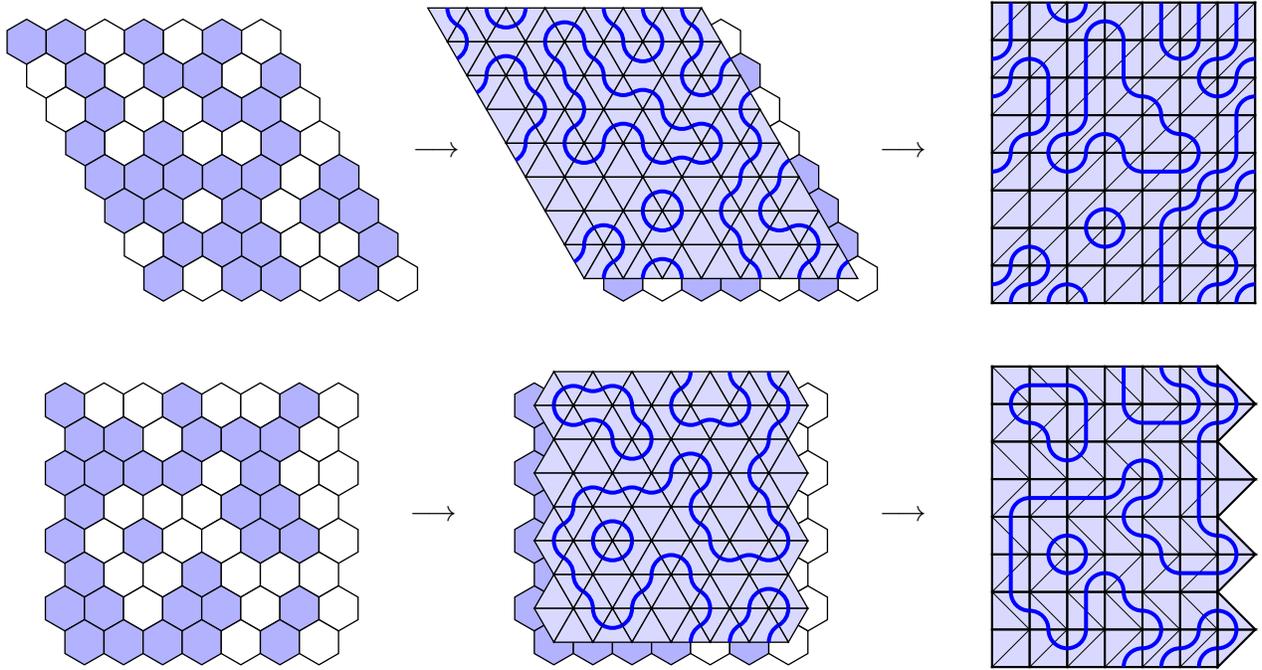

On the torus, the second step is to apply a shear that transforms the parallelogram into a rectangle of height $M$ and width $N$. An example illustrating the map applied to a percolation configuration on the torus is given in the upper panel of \cref{fig:the.map}. Combining together pairs of triangles into square tiles, we see that the square tiles formed are given by the following eight diagrams:
\be
\label{eq:8Tiles1}
\psset{unit=0.8}
\begin{pspicture}[shift=-.50](0,-.2)(1,1.2)
\facegrid{(0,0)}{(1,1)}
\rput[bl](0,0){\loopa}
\end{pspicture} \ ,
\quad \
\begin{pspicture}[shift=-.50](0,-.2)(1,1.2)
\facegrid{(0,0)}{(1,1)}
\rput[bl](0,0){\loopb}
\end{pspicture} \ ,
\quad \
\begin{pspicture}[shift=-.50](0,-.2)(1,1.2)
\facegrid{(0,0)}{(1,1)}
\rput[bl](0,0){\loopc}
\end{pspicture} \ ,
\quad \
\begin{pspicture}[shift=-.50](0,-.2)(1,1.2)
\facegrid{(0,0)}{(1,1)}
\rput[bl](0,0){\loopd}
\end{pspicture} \ ,
\quad \
\begin{pspicture}[shift=-.50](0,-.2)(1,1.2)
\facegrid{(0,0)}{(1,1)}
\rput[bl](0,0){\loope}
\end{pspicture} \ ,
\quad \
\begin{pspicture}[shift=-.50](0,-.2)(1,1.2)
\facegrid{(0,0)}{(1,1)}
\rput[bl](0,0){\loopf}
\end{pspicture} \ ,
\quad \
\begin{pspicture}[shift=-.50](0,-.2)(1,1.2)
\facegrid{(0,0)}{(1,1)}
\rput[bl](0,0){\loopg}
\end{pspicture}\ ,
\quad \
\begin{pspicture}[shift=-.50](0,-.2)(1,1.2)
\facegrid{(0,0)}{(1,1)}
\rput[bl](0,0){\looph}
\end{pspicture}\ .
\ee
This second step thus produces a configuration of a dilute loop model on the square lattice with these eight admissible tiles and with periodic boundary conditions in both directions.

On the cylinder, the sheer is applied in different directions for the even and odd rows of the lattice. This transformation is illustrated in the lower panel of \cref{fig:the.map}. The result is a configuration of the dilute loop model on a square lattice with $2M\times N$ square tiles, $M$ pairs of triangular tiles on the right boundary, and a straight left boundary. The loop segments can never touch the boundary. In this case, the square tiles of the odd rows (counted from the bottom) are given by the eight tiles \eqref{eq:8Tiles1}, whereas those of the even rows are given by the eight tiles
\be
\label{eq:8Tiles2}
\psset{unit=0.8}
\begin{pspicture}[shift=-.50](0,-.2)(1,1.2)
\facegrid{(0,0)}{(1,1)}
\rput[bl](0,0){\loopa}
\end{pspicture} \ ,
\quad \
\begin{pspicture}[shift=-.50](0,-.2)(1,1.2)
\facegrid{(0,0)}{(1,1)}
\rput[bl](0,0){\loopb}
\end{pspicture} \ ,
\quad \
\begin{pspicture}[shift=-.50](0,-.2)(1,1.2)
\facegrid{(0,0)}{(1,1)}
\rput[bl](0,0){\loopc}
\end{pspicture} \ ,
\quad \
\begin{pspicture}[shift=-.50](0,-.2)(1,1.2)
\facegrid{(0,0)}{(1,1)}
\rput[bl](0,0){\loopd}
\end{pspicture} \ ,
\quad \
\begin{pspicture}[shift=-.50](0,-.2)(1,1.2)
\facegrid{(0,0)}{(1,1)}
\rput[bl](0,0){\loope}
\end{pspicture} \ ,
\quad \
\begin{pspicture}[shift=-.50](0,-.2)(1,1.2)
\facegrid{(0,0)}{(1,1)}
\rput[bl](0,0){\loopf}
\end{pspicture} \ ,
\quad \
\begin{pspicture}[shift=-.50](0,-.2)(1,1.2)
\facegrid{(0,0)}{(1,1)}
\rput[bl](0,0){\loopg}
\end{pspicture}\ ,
\quad \
\begin{pspicture}[shift=-.50](0,-.2)(1,1.2)
\facegrid{(0,0)}{(1,1)}
\rput[bl](0,0){\loopi}
\end{pspicture}\ .
\ee
Only the last one is different. Moreover, the boundary triangles on the right boundary are occupied by one of the two diagrams
\be
\psset{unit=0.8}
\begin{pspicture}[shift=-.9](0,0)(1,2)
\pspolygon[fillstyle=solid,fillcolor=lightlightblue](0,0)(0,2)(1,1)
\psarc[linecolor=blue,linewidth=1.5pt](0,1){0.5}{-90}{90}
\end{pspicture}\ , \qquad
\begin{pspicture}[shift=-.9](0,0)(1,2)
\pspolygon[fillstyle=solid,fillcolor=lightlightblue](0,0)(0,2)(1,1)
\end{pspicture}\ .
\ee

This transformation produces configurations of loop segments where each loop is part of a closed loop, namely there are no free ends. This map is not one-to-one. On the torus, loop configurations with odd numbers of non-contractible loops are never produced by this map. Moreover, each configuration with an even number of non-contractible loops is produced twice, as interchanging the role of purple and white hexagonal cells leads to the same loop configuration. The map is not bijective on the cylinder either. Indeed, the loop configurations produced by the map have only even and odd numbers of non-contractible loops, if the boundary sites have identical or different colours, respectively. In this case however, each loop configuration in the image has a unique pre-image. This is due to the fixed boundary conditions, which break the invariance under the interchange of purple and white hexagons.

\subsection{Antiperiodic boundary conditions on the torus}\label{sec:antiperiodic}

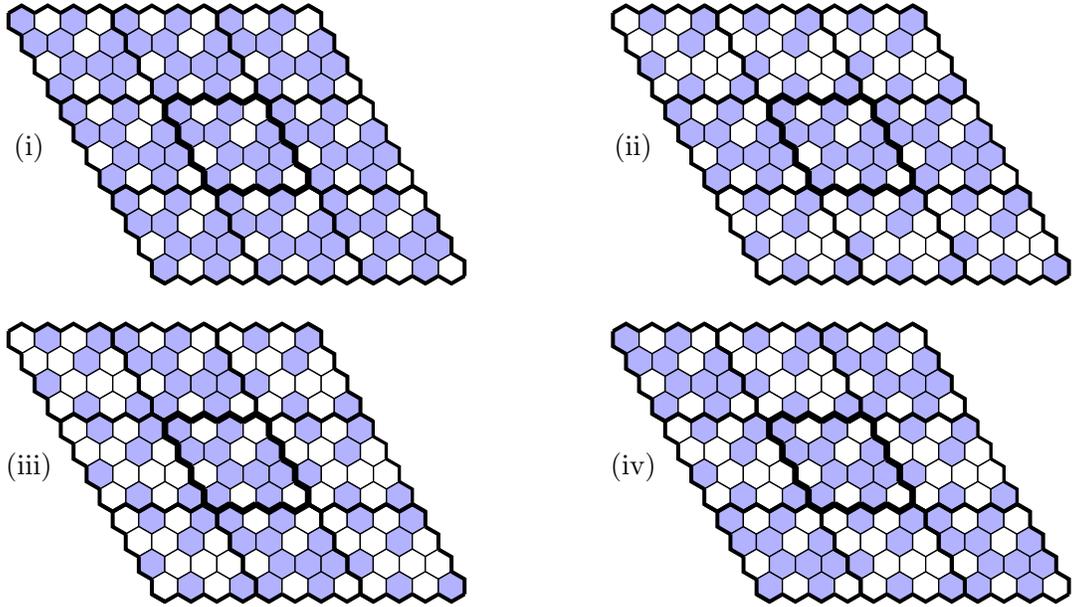
\begin{figure}
\begin{center}
\psset{unit=0.20cm}
\begin{pspicture}(-13,-7)(15,10)
\multiput(-6.9282,0)(6.9282,0){3}{
\rput(-3.464,3){\hexab}\rput(-1.732,3){\hexaw}\rput(0,3){\hexab}\rput(1.732,3){\hexaw}
\rput(-2.598,1.5){\hexab}\rput(-0.866,1.5){\hexab}\rput(0.866,1.5){\hexaw}\rput(2.598,1.5){\hexab}
\rput(-1.732,0){\hexaw}\rput(0,0){\hexab}\rput(1.732,0){\hexab}\rput(3.464,0){\hexab}
\rput(-0.866,-1.5){\hexab}\rput(0.866,-1.5){\hexaw}\rput(2.598,-1.5){\hexab}\rput(4.33,-1.5){\hexaw}
\rput(-3.4641,6){
\rput(-3.464,3){\hexab}\rput(-1.732,3){\hexaw}\rput(0,3){\hexab}\rput(1.732,3){\hexaw}
\rput(-2.598,1.5){\hexab}\rput(-0.866,1.5){\hexab}\rput(0.866,1.5){\hexaw}\rput(2.598,1.5){\hexab}
\rput(-1.732,0){\hexaw}\rput(0,0){\hexab}\rput(1.732,0){\hexab}\rput(3.464,0){\hexab}
\rput(-0.866,-1.5){\hexab}\rput(0.866,-1.5){\hexaw}\rput(2.598,-1.5){\hexab}\rput(4.33,-1.5){\hexaw}
}
\rput(3.4641,-6){
\rput(-3.464,3){\hexab}\rput(-1.732,3){\hexaw}\rput(0,3){\hexab}\rput(1.732,3){\hexaw}
\rput(-2.598,1.5){\hexab}\rput(-0.866,1.5){\hexab}\rput(0.866,1.5){\hexaw}\rput(2.598,1.5){\hexab}
\rput(-1.732,0){\hexaw}\rput(0,0){\hexab}\rput(1.732,0){\hexab}\rput(3.464,0){\hexab}
\rput(-0.866,-1.5){\hexab}\rput(0.866,-1.5){\hexaw}\rput(2.598,-1.5){\hexab}\rput(4.33,-1.5){\hexaw}
}
}
\multiput(-6.9282,0)(6.9282,0){3}{
\multiput(0,0)(1.73,0){4}{\psline[linewidth=1.5pt]{-}(-0.866,-1.5)(0,-2)(0.866,-1.5)}
\multiput(-1.73,6)(1.73,0){4}{\psline[linewidth=1.5pt]{-}(-1.73,-2)(-0.866,-1.5)(0,-2)}
\multiput(0,0)(-0.866,1.5){4}{\psline[linewidth=1.5pt]{-}(0,-2)(-0.866,-1.5)(-0.866,-0.5)}
\multiput(6.9282,0)(-0.866,1.5){4}{\psline[linewidth=1.25pt]{-}(-0.866,-1.5)(-0.866,-0.5)(-1.73,0)}
}
\multiput(-10.3923,6)(6.9282,0){3}{
\multiput(0,0)(1.73,0){4}{\psline[linewidth=1.5pt]{-}(-0.866,-1.5)(0,-2)(0.866,-1.5)}
\multiput(-1.73,6)(1.73,0){4}{\psline[linewidth=1.5pt]{-}(-1.73,-2)(-0.866,-1.5)(0,-2)}
\multiput(0,0)(-0.866,1.5){4}{\psline[linewidth=1.5pt]{-}(0,-2)(-0.866,-1.5)(-0.866,-0.5)}
\multiput(6.9282,0)(-0.866,1.5){4}{\psline[linewidth=1.5pt]{-}(-0.866,-1.5)(-0.866,-0.5)(-1.73,0)}
}
\multiput(-3.4641,-6)(6.9282,0){3}{
\multiput(0,0)(1.73,0){4}{\psline[linewidth=1.5pt]{-}(-0.866,-1.5)(0,-2)(0.866,-1.5)}
\multiput(-1.73,6)(1.73,0){4}{\psline[linewidth=1.5pt]{-}(-1.73,-2)(-0.866,-1.5)(0,-2)}
\multiput(0,0)(-0.866,1.5){4}{\psline[linewidth=1.5pt]{-}(0,-2)(-0.866,-1.5)(-0.866,-0.5)}
\multiput(6.9282,0)(-0.866,1.5){4}{\psline[linewidth=1.5pt]{-}(-0.866,-1.5)(-0.866,-0.5)(-1.73,0)}
}
\multiput(0,0)(1.73,0){4}{\psline[linewidth=2.5pt]{-}(-0.866,-1.5)(0,-2)(0.866,-1.5)}
\multiput(-1.73,6)(1.73,0){4}{\psline[linewidth=2.5pt]{-}(-1.73,-2)(-0.866,-1.5)(0,-2)}
\multiput(0,0)(-0.866,1.5){4}{\psline[linewidth=2.5pt]{-}(0,-2)(-0.866,-1.5)(-0.866,-0.5)}
\multiput(6.9282,0)(-0.866,1.5){4}{\psline[linewidth=2.5pt]{-}(-0.866,-1.5)(-0.866,-0.5)(-1.73,0)}
\rput(-12.5,1){(i)}
\end{pspicture}
\qquad \qquad \qquad
\begin{pspicture}(-13,-7)(15,10)
\multiput(-6.9282,0)(6.9282,0){3}{
\rput(-3.464,3){\hexab}\rput(-1.732,3){\hexaw}\rput(0,3){\hexab}\rput(1.732,3){\hexaw}
\rput(-2.598,1.5){\hexab}\rput(-0.866,1.5){\hexab}\rput(0.866,1.5){\hexaw}\rput(2.598,1.5){\hexab}
\rput(-1.732,0){\hexaw}\rput(0,0){\hexab}\rput(1.732,0){\hexab}\rput(3.464,0){\hexab}
\rput(-0.866,-1.5){\hexab}\rput(0.866,-1.5){\hexaw}\rput(2.598,-1.5){\hexab}\rput(4.33,-1.5){\hexaw}
\rput(-3.4641,6){
\rput(-3.464,3){\hexaw}\rput(-1.732,3){\hexab}\rput(0,3){\hexaw}\rput(1.732,3){\hexab}
\rput(-2.598,1.5){\hexaw}\rput(-0.866,1.5){\hexaw}\rput(0.866,1.5){\hexab}\rput(2.598,1.5){\hexaw}
\rput(-1.732,0){\hexab}\rput(0,0){\hexaw}\rput(1.732,0){\hexaw}\rput(3.464,0){\hexaw}
\rput(-0.866,-1.5){\hexaw}\rput(0.866,-1.5){\hexab}\rput(2.598,-1.5){\hexaw}\rput(4.33,-1.5){\hexab}
}
\rput(3.4641,-6){
\rput(-3.464,3){\hexaw}\rput(-1.732,3){\hexab}\rput(0,3){\hexaw}\rput(1.732,3){\hexab}
\rput(-2.598,1.5){\hexaw}\rput(-0.866,1.5){\hexaw}\rput(0.866,1.5){\hexab}\rput(2.598,1.5){\hexaw}
\rput(-1.732,0){\hexab}\rput(0,0){\hexaw}\rput(1.732,0){\hexaw}\rput(3.464,0){\hexaw}
\rput(-0.866,-1.5){\hexaw}\rput(0.866,-1.5){\hexab}\rput(2.598,-1.5){\hexaw}\rput(4.33,-1.5){\hexab}
}
}
\multiput(-6.9282,0)(6.9282,0){3}{
\multiput(0,0)(1.73,0){4}{\psline[linewidth=1.5pt]{-}(-0.866,-1.5)(0,-2)(0.866,-1.5)}
\multiput(-1.73,6)(1.73,0){4}{\psline[linewidth=1.5pt]{-}(-1.73,-2)(-0.866,-1.5)(0,-2)}
\multiput(0,0)(-0.866,1.5){4}{\psline[linewidth=1.5pt]{-}(0,-2)(-0.866,-1.5)(-0.866,-0.5)}
\multiput(6.9282,0)(-0.866,1.5){4}{\psline[linewidth=1.25pt]{-}(-0.866,-1.5)(-0.866,-0.5)(-1.73,0)}
}
\multiput(-10.3923,6)(6.9282,0){3}{
\multiput(0,0)(1.73,0){4}{\psline[linewidth=1.5pt]{-}(-0.866,-1.5)(0,-2)(0.866,-1.5)}
\multiput(-1.73,6)(1.73,0){4}{\psline[linewidth=1.5pt]{-}(-1.73,-2)(-0.866,-1.5)(0,-2)}
\multiput(0,0)(-0.866,1.5){4}{\psline[linewidth=1.5pt]{-}(0,-2)(-0.866,-1.5)(-0.866,-0.5)}
\multiput(6.9282,0)(-0.866,1.5){4}{\psline[linewidth=1.5pt]{-}(-0.866,-1.5)(-0.866,-0.5)(-1.73,0)}
}
\multiput(-3.4641,-6)(6.9282,0){3}{
\multiput(0,0)(1.73,0){4}{\psline[linewidth=1.5pt]{-}(-0.866,-1.5)(0,-2)(0.866,-1.5)}
\multiput(-1.73,6)(1.73,0){4}{\psline[linewidth=1.5pt]{-}(-1.73,-2)(-0.866,-1.5)(0,-2)}
\multiput(0,0)(-0.866,1.5){4}{\psline[linewidth=1.5pt]{-}(0,-2)(-0.866,-1.5)(-0.866,-0.5)}
\multiput(6.9282,0)(-0.866,1.5){4}{\psline[linewidth=1.5pt]{-}(-0.866,-1.5)(-0.866,-0.5)(-1.73,0)}
}
\multiput(0,0)(1.73,0){4}{\psline[linewidth=2.5pt]{-}(-0.866,-1.5)(0,-2)(0.866,-1.5)}
\multiput(-1.73,6)(1.73,0){4}{\psline[linewidth=2.5pt]{-}(-1.73,-2)(-0.866,-1.5)(0,-2)}
\multiput(0,0)(-0.866,1.5){4}{\psline[linewidth=2.5pt]{-}(0,-2)(-0.866,-1.5)(-0.866,-0.5)}
\multiput(6.9282,0)(-0.866,1.5){4}{\psline[linewidth=2.5pt]{-}(-0.866,-1.5)(-0.866,-0.5)(-1.73,0)}
\rput(-12.5,1){(ii)}
\end{pspicture}
\\[0.6cm]
\psset{unit=0.20cm}
\begin{pspicture}(-13,-5)(15,11)
\multiput(3.464,-6)(-3.464,6){3}{
\rput(-3.464,3){\hexab}\rput(-1.732,3){\hexaw}\rput(0,3){\hexab}\rput(1.732,3){\hexaw}
\rput(-2.598,1.5){\hexab}\rput(-0.866,1.5){\hexab}\rput(0.866,1.5){\hexaw}\rput(2.598,1.5){\hexab}
\rput(-1.732,0){\hexaw}\rput(0,0){\hexab}\rput(1.732,0){\hexab}\rput(3.464,0){\hexab}
\rput(-0.866,-1.5){\hexab}\rput(0.866,-1.5){\hexaw}\rput(2.598,-1.5){\hexab}\rput(4.33,-1.5){\hexaw}
\rput(6.9282,0){
\rput(-3.464,3){\hexaw}\rput(-1.732,3){\hexab}\rput(0,3){\hexaw}\rput(1.732,3){\hexab}
\rput(-2.598,1.5){\hexaw}\rput(-0.866,1.5){\hexaw}\rput(0.866,1.5){\hexab}\rput(2.598,1.5){\hexaw}
\rput(-1.732,0){\hexab}\rput(0,0){\hexaw}\rput(1.732,0){\hexaw}\rput(3.464,0){\hexaw}
\rput(-0.866,-1.5){\hexaw}\rput(0.866,-1.5){\hexab}\rput(2.598,-1.5){\hexaw}\rput(4.33,-1.5){\hexab}
}
\rput(-6.9282,0){
\rput(-3.464,3){\hexaw}\rput(-1.732,3){\hexab}\rput(0,3){\hexaw}\rput(1.732,3){\hexab}
\rput(-2.598,1.5){\hexaw}\rput(-0.866,1.5){\hexaw}\rput(0.866,1.5){\hexab}\rput(2.598,1.5){\hexaw}
\rput(-1.732,0){\hexab}\rput(0,0){\hexaw}\rput(1.732,0){\hexaw}\rput(3.464,0){\hexaw}
\rput(-0.866,-1.5){\hexaw}\rput(0.866,-1.5){\hexab}\rput(2.598,-1.5){\hexaw}\rput(4.33,-1.5){\hexab}
}
}
\multiput(-6.9282,0)(6.9282,0){3}{
\multiput(0,0)(1.73,0){4}{\psline[linewidth=1.5pt]{-}(-0.866,-1.5)(0,-2)(0.866,-1.5)}
\multiput(-1.73,6)(1.73,0){4}{\psline[linewidth=1.5pt]{-}(-1.73,-2)(-0.866,-1.5)(0,-2)}
\multiput(0,0)(-0.866,1.5){4}{\psline[linewidth=1.5pt]{-}(0,-2)(-0.866,-1.5)(-0.866,-0.5)}
\multiput(6.9282,0)(-0.866,1.5){4}{\psline[linewidth=1.25pt]{-}(-0.866,-1.5)(-0.866,-0.5)(-1.73,0)}
}
\multiput(-10.3923,6)(6.9282,0){3}{
\multiput(0,0)(1.73,0){4}{\psline[linewidth=1.5pt]{-}(-0.866,-1.5)(0,-2)(0.866,-1.5)}
\multiput(-1.73,6)(1.73,0){4}{\psline[linewidth=1.5pt]{-}(-1.73,-2)(-0.866,-1.5)(0,-2)}
\multiput(0,0)(-0.866,1.5){4}{\psline[linewidth=1.5pt]{-}(0,-2)(-0.866,-1.5)(-0.866,-0.5)}
\multiput(6.9282,0)(-0.866,1.5){4}{\psline[linewidth=1.5pt]{-}(-0.866,-1.5)(-0.866,-0.5)(-1.73,0)}
}
\multiput(-3.4641,-6)(6.9282,0){3}{
\multiput(0,0)(1.73,0){4}{\psline[linewidth=1.5pt]{-}(-0.866,-1.5)(0,-2)(0.866,-1.5)}
\multiput(-1.73,6)(1.73,0){4}{\psline[linewidth=1.5pt]{-}(-1.73,-2)(-0.866,-1.5)(0,-2)}
\multiput(0,0)(-0.866,1.5){4}{\psline[linewidth=1.5pt]{-}(0,-2)(-0.866,-1.5)(-0.866,-0.5)}
\multiput(6.9282,0)(-0.866,1.5){4}{\psline[linewidth=1.5pt]{-}(-0.866,-1.5)(-0.866,-0.5)(-1.73,0)}
}
\multiput(0,0)(1.73,0){4}{\psline[linewidth=2.5pt]{-}(-0.866,-1.5)(0,-2)(0.866,-1.5)}
\multiput(-1.73,6)(1.73,0){4}{\psline[linewidth=2.5pt]{-}(-1.73,-2)(-0.866,-1.5)(0,-2)}
\multiput(0,0)(-0.866,1.5){4}{\psline[linewidth=2.5pt]{-}(0,-2)(-0.866,-1.5)(-0.866,-0.5)}
\multiput(6.9282,0)(-0.866,1.5){4}{\psline[linewidth=2.5pt]{-}(-0.866,-1.5)(-0.866,-0.5)(-1.73,0)}
\rput(-12.5,1){(iii)}
\end{pspicture}
\qquad\qquad\qquad
\begin{pspicture}(-13,-5)(15,11)
\rput(-3.464,3){\hexab}\rput(-1.732,3){\hexaw}\rput(0,3){\hexab}\rput(1.732,3){\hexaw}
\rput(-2.598,1.5){\hexab}\rput(-0.866,1.5){\hexab}\rput(0.866,1.5){\hexaw}\rput(2.598,1.5){\hexab}
\rput(-1.732,0){\hexaw}\rput(0,0){\hexab}\rput(1.732,0){\hexab}\rput(3.464,0){\hexab}
\rput(-0.866,-1.5){\hexab}\rput(0.866,-1.5){\hexaw}\rput(2.598,-1.5){\hexab}\rput(4.33,-1.5){\hexaw}
\rput(6.9282,0){
\rput(-3.464,3){\hexaw}\rput(-1.732,3){\hexab}\rput(0,3){\hexaw}\rput(1.732,3){\hexab}
\rput(-2.598,1.5){\hexaw}\rput(-0.866,1.5){\hexaw}\rput(0.866,1.5){\hexab}\rput(2.598,1.5){\hexaw}
\rput(-1.732,0){\hexab}\rput(0,0){\hexaw}\rput(1.732,0){\hexaw}\rput(3.464,0){\hexaw}
\rput(-0.866,-1.5){\hexaw}\rput(0.866,-1.5){\hexab}\rput(2.598,-1.5){\hexaw}\rput(4.33,-1.5){\hexab}
}
\rput(-3.4641,6){
\rput(-3.464,3){\hexaw}\rput(-1.732,3){\hexab}\rput(0,3){\hexaw}\rput(1.732,3){\hexab}
\rput(-2.598,1.5){\hexaw}\rput(-0.866,1.5){\hexaw}\rput(0.866,1.5){\hexab}\rput(2.598,1.5){\hexaw}
\rput(-1.732,0){\hexab}\rput(0,0){\hexaw}\rput(1.732,0){\hexaw}\rput(3.464,0){\hexaw}
\rput(-0.866,-1.5){\hexaw}\rput(0.866,-1.5){\hexab}\rput(2.598,-1.5){\hexaw}\rput(4.33,-1.5){\hexab}
}
\rput(-6.9282,0){
\rput(-3.464,3){\hexaw}\rput(-1.732,3){\hexab}\rput(0,3){\hexaw}\rput(1.732,3){\hexab}
\rput(-2.598,1.5){\hexaw}\rput(-0.866,1.5){\hexaw}\rput(0.866,1.5){\hexab}\rput(2.598,1.5){\hexaw}
\rput(-1.732,0){\hexab}\rput(0,0){\hexaw}\rput(1.732,0){\hexaw}\rput(3.464,0){\hexaw}
\rput(-0.866,-1.5){\hexaw}\rput(0.866,-1.5){\hexab}\rput(2.598,-1.5){\hexaw}\rput(4.33,-1.5){\hexab}
}
\rput(3.4641,-6){
\rput(-3.464,3){\hexaw}\rput(-1.732,3){\hexab}\rput(0,3){\hexaw}\rput(1.732,3){\hexab}
\rput(-2.598,1.5){\hexaw}\rput(-0.866,1.5){\hexaw}\rput(0.866,1.5){\hexab}\rput(2.598,1.5){\hexaw}
\rput(-1.732,0){\hexab}\rput(0,0){\hexaw}\rput(1.732,0){\hexaw}\rput(3.464,0){\hexaw}
\rput(-0.866,-1.5){\hexaw}\rput(0.866,-1.5){\hexab}\rput(2.598,-1.5){\hexaw}\rput(4.33,-1.5){\hexab}
}
\rput(-3.4641,-6){
\rput(-3.464,3){\hexab}\rput(-1.732,3){\hexaw}\rput(0,3){\hexab}\rput(1.732,3){\hexaw}
\rput(-2.598,1.5){\hexab}\rput(-0.866,1.5){\hexab}\rput(0.866,1.5){\hexaw}\rput(2.598,1.5){\hexab}
\rput(-1.732,0){\hexaw}\rput(0,0){\hexab}\rput(1.732,0){\hexab}\rput(3.464,0){\hexab}
\rput(-0.866,-1.5){\hexab}\rput(0.866,-1.5){\hexaw}\rput(2.598,-1.5){\hexab}\rput(4.33,-1.5){\hexaw}
}
\rput(10.3923,-6){
\rput(-3.464,3){\hexab}\rput(-1.732,3){\hexaw}\rput(0,3){\hexab}\rput(1.732,3){\hexaw}
\rput(-2.598,1.5){\hexab}\rput(-0.866,1.5){\hexab}\rput(0.866,1.5){\hexaw}\rput(2.598,1.5){\hexab}
\rput(-1.732,0){\hexaw}\rput(0,0){\hexab}\rput(1.732,0){\hexab}\rput(3.464,0){\hexab}
\rput(-0.866,-1.5){\hexab}\rput(0.866,-1.5){\hexaw}\rput(2.598,-1.5){\hexab}\rput(4.33,-1.5){\hexaw}
}
\rput(3.4641,6){
\rput(-3.464,3){\hexab}\rput(-1.732,3){\hexaw}\rput(0,3){\hexab}\rput(1.732,3){\hexaw}
\rput(-2.598,1.5){\hexab}\rput(-0.866,1.5){\hexab}\rput(0.866,1.5){\hexaw}\rput(2.598,1.5){\hexab}
\rput(-1.732,0){\hexaw}\rput(0,0){\hexab}\rput(1.732,0){\hexab}\rput(3.464,0){\hexab}
\rput(-0.866,-1.5){\hexab}\rput(0.866,-1.5){\hexaw}\rput(2.598,-1.5){\hexab}\rput(4.33,-1.5){\hexaw}
}
\rput(-10.3923,6){
\rput(-3.464,3){\hexab}\rput(-1.732,3){\hexaw}\rput(0,3){\hexab}\rput(1.732,3){\hexaw}
\rput(-2.598,1.5){\hexab}\rput(-0.866,1.5){\hexab}\rput(0.866,1.5){\hexaw}\rput(2.598,1.5){\hexab}
\rput(-1.732,0){\hexaw}\rput(0,0){\hexab}\rput(1.732,0){\hexab}\rput(3.464,0){\hexab}
\rput(-0.866,-1.5){\hexab}\rput(0.866,-1.5){\hexaw}\rput(2.598,-1.5){\hexab}\rput(4.33,-1.5){\hexaw}
}
\multiput(-6.9282,0)(6.9282,0){3}{
\multiput(0,0)(1.73,0){4}{\psline[linewidth=1.5pt]{-}(-0.866,-1.5)(0,-2)(0.866,-1.5)}
\multiput(-1.73,6)(1.73,0){4}{\psline[linewidth=1.5pt]{-}(-1.73,-2)(-0.866,-1.5)(0,-2)}
\multiput(0,0)(-0.866,1.5){4}{\psline[linewidth=1.5pt]{-}(0,-2)(-0.866,-1.5)(-0.866,-0.5)}
\multiput(6.9282,0)(-0.866,1.5){4}{\psline[linewidth=1.25pt]{-}(-0.866,-1.5)(-0.866,-0.5)(-1.73,0)}
}
\multiput(-10.3923,6)(6.9282,0){3}{
\multiput(0,0)(1.73,0){4}{\psline[linewidth=1.5pt]{-}(-0.866,-1.5)(0,-2)(0.866,-1.5)}
\multiput(-1.73,6)(1.73,0){4}{\psline[linewidth=1.5pt]{-}(-1.73,-2)(-0.866,-1.5)(0,-2)}
\multiput(0,0)(-0.866,1.5){4}{\psline[linewidth=1.5pt]{-}(0,-2)(-0.866,-1.5)(-0.866,-0.5)}
\multiput(6.9282,0)(-0.866,1.5){4}{\psline[linewidth=1.5pt]{-}(-0.866,-1.5)(-0.866,-0.5)(-1.73,0)}
}
\multiput(-3.4641,-6)(6.9282,0){3}{
\multiput(0,0)(1.73,0){4}{\psline[linewidth=1.5pt]{-}(-0.866,-1.5)(0,-2)(0.866,-1.5)}
\multiput(-1.73,6)(1.73,0){4}{\psline[linewidth=1.5pt]{-}(-1.73,-2)(-0.866,-1.5)(0,-2)}
\multiput(0,0)(-0.866,1.5){4}{\psline[linewidth=1.5pt]{-}(0,-2)(-0.866,-1.5)(-0.866,-0.5)}
\multiput(6.9282,0)(-0.866,1.5){4}{\psline[linewidth=1.5pt]{-}(-0.866,-1.5)(-0.866,-0.5)(-1.73,0)}
}
\multiput(0,0)(1.73,0){4}{\psline[linewidth=2.5pt]{-}(-0.866,-1.5)(0,-2)(0.866,-1.5)}
\multiput(-1.73,6)(1.73,0){4}{\psline[linewidth=2.5pt]{-}(-1.73,-2)(-0.866,-1.5)(0,-2)}
\multiput(0,0)(-0.866,1.5){4}{\psline[linewidth=2.5pt]{-}(0,-2)(-0.866,-1.5)(-0.866,-0.5)}
\multiput(6.9282,0)(-0.866,1.5){4}{\psline[linewidth=2.5pt]{-}(-0.866,-1.5)(-0.866,-0.5)(-1.73,0)}
\rput(-12.5,1){(iv)}
\end{pspicture}
\end{center}
\caption{Configurations of site percolation on the torus assigned with boundary conditions corresponding to (i) $Z_{\textrm{tor}}^{\textrm{\tiny$(0,0)$}}(\alpha)$, (ii) $Z_{\textrm{tor}}^{\textrm{\tiny$(0,1)$}}(\alpha)$, (iii) $Z_{\textrm{tor}}^{\textrm{\tiny$(1,0)$}}(\alpha)$, and (iv) $Z_{\textrm{tor}}^{\textrm{\tiny$(1,1)$}}(\alpha)$. The fundamental domain is delimited by a heavy zigzag boundary.}
\label{fig:anti.periodic.configs}
\end{figure}

In this subsection, we describe the model of site percolation on a torus with antiperiodic boundary conditions. On the lattice, each pair of opposite edges of the parallelogram can be assigned periodic or antiperiodic boundary conditions. In this context, an antiperiodic boundary condition means that, as we move through this edge of the parallelogram from one elementary domain to the next, all the coloured sites are changed to uncoloured sites, and vice versa. There are therefore four possible toroidal boundary conditions. Examples of configurations for these four boundary conditions are given in \cref{fig:anti.periodic.configs}. We denote the four resulting partition functions by $Z_{\textrm{tor}}^{\textrm{\tiny$(h,v)$}}(\alpha)$, with
\be
h,v=\left\{\begin{array}{cl}
0&\mbox{periodic},\\[0.1cm]
1&\mbox{antiperiodic},
\end{array}\right.
\ee
where the first and second index label the boundary condition along the horizontal and vertical axes of the parallelogram, respectively.

Let us discuss the properties of the non-contractible clusters and curves for the four cases. First, $Z_{\textrm{tor}}^{\textrm{\tiny$(0,0)$}}(\alpha)$ corresponds to the partition function already defined in \eqref{eq:Z.perco.tor} with purely periodic boundary conditions. The curves that act as contours for its non-contractible clusters can have any winding $(i,j)$ with $i\wedge j =1$. There are always an even number of such curves. Second, for $Z_{\textrm{tor}}^{\textrm{\tiny$(0,1)$}}(\alpha)$, the contributing configurations may only have clusters with windings $(i,j)$ with $i\wedge j = 1$, $i$ odd and $j$ even, as otherwise the antiperiodicity of the boundary condition in the vertical direction is violated. Likewise for $Z_{\textrm{tor}}^{\textrm{\tiny$(1,0)$}}(\alpha)$, the only possible windings $(i,j)$ have $i$ even and $j$ odd. Finally for $Z_{\textrm{tor}}^{\textrm{\tiny$(1,1)$}}(\alpha)$, the only possible windings have both $i$ and $j$ odd. For the last three cases, the number of non-contractible curves is always odd.

%
\section{The dilute $\boldsymbol{\Atwotwo}$ loop model}\label{sec:A22.def}
%

\subsection{Definition of face operators}

In this section, we describe the local operators of the general $\Atwotwo$ loop model. The elementary face operator is defined as the linear combination of nine elementary tiles
\begin{alignat}{2}
\psset{unit=0.5}
\label{eq:face.op}
\begin{pspicture}[shift=-.90](-1,-1)(1,1)
\pspolygon[fillstyle=solid,fillcolor=lightlightblue,linewidth=1pt](1,0)(0,1)(-1,0)(0,-1)
\psarc[linewidth=0.025]{-}(0,-1){0.25}{45}{135}
\rput(0,0){$u$}
\end{pspicture}
\ = \ &\psset{unit=0.5}
\rho_1(u)\ 
\begin{pspicture}[shift=-.90](-1,-1)(1,1)
\pspolygon[fillstyle=solid,fillcolor=lightlightblue,linewidth=1pt](1,0)(0,1)(-1,0)(0,-1)
\pscircle[fillstyle=solid,fillcolor=black](-0.5,-0.5){0.09}
\pscircle[fillstyle=solid,fillcolor=black](0.5,-0.5){0.09}
\pscircle[fillstyle=solid,fillcolor=black](0.5,0.5){0.09}
\pscircle[fillstyle=solid,fillcolor=black](-0.5,0.5){0.09}
\end{pspicture}
\ + \rho_2(u)\ 
\begin{pspicture}[shift=-.90](-1,-1)(1,1)
\pspolygon[fillstyle=solid,fillcolor=lightlightblue,linewidth=1pt](1,0)(0,1)(-1,0)(0,-1)
\psarc[linecolor=blue,linewidth=\elegant](-1,0){0.71}{-45}{45}
\pscircle[fillstyle=solid,fillcolor=black](0.5,-0.5){0.09}
\pscircle[fillstyle=solid,fillcolor=black](0.5,0.5){0.09}
\end{pspicture}
\ + \rho_3(u)\ 
\begin{pspicture}[shift=-.90](-1,-1)(1,1)
\pspolygon[fillstyle=solid,fillcolor=lightlightblue,linewidth=1pt](1,0)(0,1)(-1,0)(0,-1)
\psarc[linecolor=blue,linewidth=\elegant](1,0){0.71}{135}{225}
\pscircle[fillstyle=solid,fillcolor=black](-0.5,-0.5){0.09}
\pscircle[fillstyle=solid,fillcolor=black](-0.5,0.5){0.09}
\end{pspicture}
\ + \rho_4(u)\ 
\begin{pspicture}[shift=-.90](-1,-1)(1,1)
\pspolygon[fillstyle=solid,fillcolor=lightlightblue,linewidth=1pt](1,0)(0,1)(-1,0)(0,-1)
\psarc[linecolor=blue,linewidth=\elegant](0,-1){0.71}{45}{135}
\pscircle[fillstyle=solid,fillcolor=black](0.5,0.5){0.09}
\pscircle[fillstyle=solid,fillcolor=black](-0.5,0.5){0.09}
\end{pspicture}
\ + \rho_5(u)\ 
\begin{pspicture}[shift=-.90](-1,-1)(1,1)
\pspolygon[fillstyle=solid,fillcolor=lightlightblue,linewidth=1pt](1,0)(0,1)(-1,0)(0,-1)
\psarc[linecolor=blue,linewidth=\elegant](0,1){0.71}{-135}{-45}
\pscircle[fillstyle=solid,fillcolor=black](-0.5,-0.5){0.09}
\pscircle[fillstyle=solid,fillcolor=black](0.5,-0.5){0.09}
\end{pspicture}
\nonumber\\[0.2cm] \ &\hspace{0.2cm}+ \rho_6(u)\ 
\psset{unit=0.5}
\begin{pspicture}[shift=-.90](-1,-1)(1,1)
\pspolygon[fillstyle=solid,fillcolor=lightlightblue,linewidth=1pt](1,0)(0,1)(-1,0)(0,-1)
\psline[linecolor=blue,linewidth=\elegant](-0.5,0.5)(0.5,-0.5)
\pscircle[fillstyle=solid,fillcolor=black](-0.5,-0.5){0.09}
\pscircle[fillstyle=solid,fillcolor=black](0.5,0.5){0.09}
\end{pspicture}
\ + \rho_7(u)\ 
\begin{pspicture}[shift=-.90](-1,-1)(1,1)
\pspolygon[fillstyle=solid,fillcolor=lightlightblue,linewidth=1pt](1,0)(0,1)(-1,0)(0,-1)
\psline[linecolor=blue,linewidth=\elegant](-0.5,-0.5)(0.5,0.5)
\pscircle[fillstyle=solid,fillcolor=black](0.5,-0.5){0.09}
\pscircle[fillstyle=solid,fillcolor=black](-0.5,0.5){0.09}
\end{pspicture}
\ + \rho_8(u)\ 
\begin{pspicture}[shift=-.90](-1,-1)(1,1)
\pspolygon[fillstyle=solid,fillcolor=lightlightblue,linewidth=1pt](1,0)(0,1)(-1,0)(0,-1)
\psarc[linecolor=blue,linewidth=\elegant](1,0){0.71}{135}{225}
\psarc[linecolor=blue,linewidth=\elegant](-1,0){0.71}{-45}{45}
\end{pspicture}
\ + \rho_9(u)\ 
\begin{pspicture}[shift=-.90](-1,-1)(1,1)
\pspolygon[fillstyle=solid,fillcolor=lightlightblue,linewidth=1pt](1,0)(0,1)(-1,0)(0,-1)
\psarc[linecolor=blue,linewidth=\elegant](0,1){0.71}{-135}{-45}
\psarc[linecolor=blue,linewidth=\elegant](0,-1){0.71}{45}{135}
\end{pspicture}
\ \ .
\end{alignat}
The local Boltzmann weights are
\begin{subequations}
\label{eq:weights}
\begin{alignat}{3}
&\rho_1(u)= s(2\lambda) s(3\lambda) + s(u) s(3\lambda-u), \qquad
&&\rho_{6}(u)=\rho_{7}(u)= s(u) s(3\lambda-u), \\[0.15cm]
&\rho_{2}(u) = \rho_{3}(u) = s(2\lambda) s(3\lambda-u), \qquad
&&\rho_8(u)= s(2\lambda-u)s(3\lambda-u),\\[0.15cm]
&\rho_{4}(u)=\rho_{5}(u)=s(2\lambda) s(u), \qquad
&&\rho_9(u)= -s(u)s(\lambda-u),
\end{alignat}
\end{subequations}
where 
\be
s(u) = \frac{\sin u}{\sin \lambda}.
\ee
The parameters $u$ and $\lambda$ are the spectral and crossing parameters, respectively. The weight $\beta$ of non-contractible loops is parameterised in terms of the crossing parameter as
\be
\beta = -2 \cos 4\lambda.
\ee
The boundary face operator is
\be
\label{eq:bdy.face.op}
\psset{unit=0.7cm}
\begin{pspicture}[shift=-.90](0,-1)(1,1)
\pspolygon[fillstyle=solid,fillcolor=lightlightblue,linewidth=0.75pt](0,0)(1,1)(1,-1)
\rput(0.65,0){$u$}
\end{pspicture}
\ = 
\kappa_1(u) \ \ 
\begin{pspicture}[shift=-.90](0,-1)(1,1)
\pspolygon[fillstyle=solid,fillcolor=lightlightblue,linewidth=0.75pt](0,0)(1,1)(1,-1)
\psarc[linecolor=blue,linewidth=1.5pt]{-}(0,0){0.61}{-90}{90}
\end{pspicture}
\ + 
\kappa_2(u)\ \ 
\begin{pspicture}[shift=-.90](0,-1)(1,1)
\pspolygon[fillstyle=solid,fillcolor=lightlightblue,linewidth=0.75pt](0,0)(1,1)(1,-1)
\pscircle[fillstyle=solid,fillcolor=black](0.5,0.5){0.07}
\pscircle[fillstyle=solid,fillcolor=black](0.5,-0.5){0.07}
\end{pspicture}\ \ .
\ee
The weights $\kappa_1$ and $\kappa_2$ are chosen from one of two choices that solve the boundary Yang-Baxter equation \cite{YB95,DJS10}:
\begin{subequations}
\begin{alignat}{3}
\textrm{choice 1:} \qquad &\kappa^+_1(u) = \sin(\tfrac{3\lambda}2-u), \qquad &&\kappa^+_2(u) = \sin(\tfrac{3\lambda}2+u),
\\[0.15cm]
\textrm{choice 2:} \qquad &\kappa^-_1(u) = \cos(\tfrac{3\lambda}2-u), \qquad &&\kappa^-_2(u) = \cos(\tfrac{3\lambda}2+u).
\end{alignat}
\end{subequations}

%
\subsection{Partition functions on the torus and cylinder}\label{sec:lattice.partition.functions}
%

On the torus, a configuration of the $\Atwotwo$ loop model is a decoration with one of the nine tiles of each of the $M\times N$ square faces, in such a way that every loop segment is part of a closed loop, namely there are no free ends. This lattice is drawn in the left panel of \cref{fig:lattices.loops}. As stated earlier, a loop can wind non-trivially around the torus $j \in \mathbb Z_{\ge 0}$ times vertically and $i \in \mathbb Z$ times horizontally, with $i$ and $j$ coprime. For instance, embedded on the $1 \times 1$ torus, the tiles
$\,
\psset{unit=0.35}
\begin{pspicture}[shift=-.20](0,0)(1,1)
\facegrid{(0,0)}{(1,1)}
\rput[bl](0,0){\looph}
\end{pspicture}
\, $
and
$\,
\psset{unit=0.35}
\begin{pspicture}[shift=-.20](0,0)(1,1)
\facegrid{(0,0)}{(1,1)}
\rput[bl](0,0){\loopi}
\end{pspicture}
\, $
have windings $(i,j) = (1,1)$ and $(-1,1)$, respectively. We assign a weight $\alpha_{i,j}$ to a loop that has the winding $(i,j)$. Any given configuration $c$ has at most one type of non-trivial windings, but it can have more than one loop with this winding. We denote their number in $c$ by $n_{i,j}(c)$. Likewise, we denote by $n_\beta(c)$ the number of contractible loops of $c$. The weights of the configurations and the partition function are 
\be
\label{eq:w.and.Z}
w(c) = \beta^{n_{\beta}(c)}\prod_{i\wedge j = 1}\alpha_{i,j}^{n_{i,j}(c)}  \prod_{k=1}^9 \rho_k(u)^{n_k(c)}, \qquad
\widetilde Z_{\textrm{tor}} = \sum_c w(c),
\ee
where $n_k(c)$ is the number of occurences of the $k$-th tile in \eqref{eq:face.op}.

We also define the $\Atwotwo$ loop model on the cylinder with $2M\times N$ bulk tiles, $M$ triangular tiles on the left boundary and $M$ triangular tiles on the right boundary. A configuration is obtained by replacing each square face with one of the nine tiles of \eqref{eq:face.op}, and each boundary triangle with one of the two possible tiles in \eqref{eq:bdy.face.op}. This lattice is drawn in the right panel of \cref{fig:lattices.loops}. We note that the orientation of the faces, indicated by the quarter arc in a corner, is different on odd and even rows of the lattice, which then modifies the counting of the numbers $n_k(c)$ accordingly. Moreover the weights of the triangular tiles are $\kappa_1(u)$ and $\kappa_2(u)$ on the right boundary, and $\kappa_1(3\lambda - u)$ and $\kappa_2(3\lambda -u)$ on the left boundary. On a given boundary, we make the same choice of weights, either $\kappa_{\ell}^+$ or $\kappa_{\ell}^-$, for all of its boundary face operators. However, one is free to make the same or a different choice for the left and right boundaries, leading to four possible boundary conditions for the full system. Non-contractible loops are assigned a fugacity $\alpha$, whereas contractible ones are given a fugacity $\beta$. We denote their respective numbers in $c$ by $n_\alpha(c)$ and $n_\beta(c)$. The weights of the configurations and the partition function are 
\be
\label{eq:w.and.Z}
w(c) = \alpha^{n_{\alpha}(c)}\beta^{n_{\beta}(c)} \prod_{k=1}^9 \rho_k(u)^{n_k(c)} \prod_{\ell=1}^2 \kappa_\ell(3\lambda - u)^{m_\ell(c)}\kappa_\ell(u)^{m'_\ell(c)},\qquad
\widetilde Z_{\textrm{cyl}} = \sum_c w(c),
\ee
where $m_\ell(c)$ and $m'_\ell(c)$ count the number of boundary triangular tiles of type $\ell \in \{1,2\}$ on the left and right boundaries, respectively.

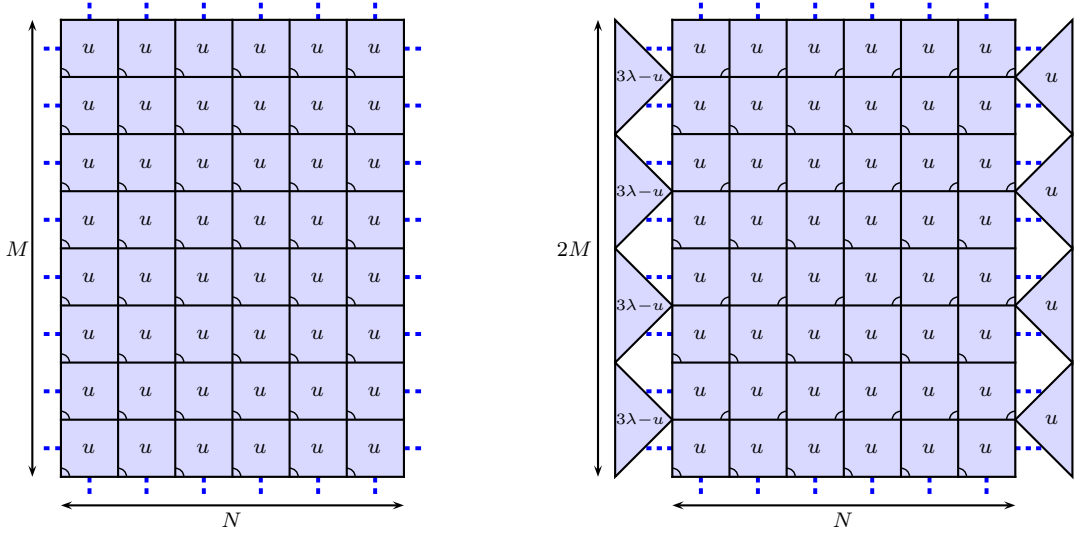
\begin{figure}[t!]
\centering
\psset{unit=0.76cm}
\begin{pspicture}[shift=-3.2](-0.5,-0.6)(7,8)
\multiput(0,0)(0,1){8}{\psline[linewidth=1.5pt,linecolor=blue,linestyle=dashed,dash=2pt 2pt]{-}(0.0,0.5)(-0.3,0.5)}
\multiput(0,0)(0,1){8}{\psline[linewidth=1.5pt,linecolor=blue,linestyle=dashed,dash=2pt 2pt]{-}(6.0,0.5)(6.3,0.5)}
\multiput(0,0)(1,0){6}{\psline[linewidth=1.5pt,linecolor=blue,linestyle=dashed,dash=2pt 2pt]{-}(0.5,0.0)(0.5,-0.3)}
\multiput(0,0)(1,0){6}{\psline[linewidth=1.5pt,linecolor=blue,linestyle=dashed,dash=2pt 2pt]{-}(0.5,8.0)(0.5,8.3)}
\facegrid{(0,0)}{(6,8)}
\multiput(0,0)(0,1){8}{\multiput(0,0)(1,0){6}{\rput(0.5,0.5){$_u$}\psarc[linewidth=0.025]{-}(0,0){0.15}{0}{90}}} 
\psline{<->}(0,-0.5)(6,-0.5)\rput(3,-0.75){$_N$}
\psline{<->}(-0.5,0)(-0.5,8)\rput(-0.75,4){$_M$}
\end{pspicture}
\qquad \qquad
\begin{pspicture}[shift=-3.2](-1.5,-0.6)(7,8)
\multiput(0,0)(0,1){8}{\psline[linewidth=1.5pt,linecolor=blue,linestyle=dashed,dash=2pt 2pt]{-}(0,0.5)(-1,0.5)}
\multiput(0,0)(0,1){8}{\psline[linewidth=1.5pt,linecolor=blue,linestyle=dashed,dash=2pt 2pt]{-}(6,0.5)(7,0.5)}
\multiput(0,0)(1,0){6}{\psline[linewidth=1.5pt,linecolor=blue,linestyle=dashed,dash=2pt 2pt]{-}(0.5,0)(0.5,-0.3)}
\multiput(0,0)(1,0){6}{\psline[linewidth=1.5pt,linecolor=blue,linestyle=dashed,dash=2pt 2pt]{-}(0.5,8)(0.5,8.3)}
\facegrid{(0,0)}{(6,8)}
\multiput(0,0)(0,2){4}{\pspolygon[fillstyle=solid,fillcolor=lightlightblue](0,1)(-1,0)(-1,2)\rput(-0.55,1){\scriptsize$_{3\lambda-u}$}}
\multiput(6,0)(0,2){4}{\pspolygon[fillstyle=solid,fillcolor=lightlightblue](0,1)(1,0)(1,2)\rput(0.65,1){$_u$}}
\multiput(0,0)(0,2){4}{\multiput(0,0)(1,0){6}{\rput(0.5,0.5){$_u$}\psarc[linewidth=0.025]{-}(0,0){0.15}{0}{90}}}
\multiput(0,1)(0,2){4}{\multiput(0,0)(1,0){6}{\rput(0.5,0.5){$_u$}\psarc[linewidth=0.025]{-}(1,0){0.15}{90}{180}}} 
\psline{<->}(0,-0.5)(6,-0.5)\rput(3,-0.75){$_N$}
\psline{<->}(-1.3,0)(-1.3,8)\rput(-1.7,4){$_{2M}$}
\end{pspicture}
\caption{The lattices with toroidal and cylindrical boundary conditions on which the $\Atwotwo$ model is defined, with $(M,N) = (8,6)$ and $(4,6)$, respectively.}
\label{fig:lattices.loops}
\end{figure}

\subsection{Dilute Temperley-Lieb algebras}

\paragraph{Site percolation and dilute connectivities.}

The dilute $\Atwotwo$ loop model is described using diagrammatic algebras: the dilute Temperley-Lieb algebra $\dtl_N(\beta)$ and its periodic incarnation $\pdtl_N(\beta)$. The vector spaces on which these are defined are spanned by objects called {\it connectivities} or {\it connectivity diagrams}. 

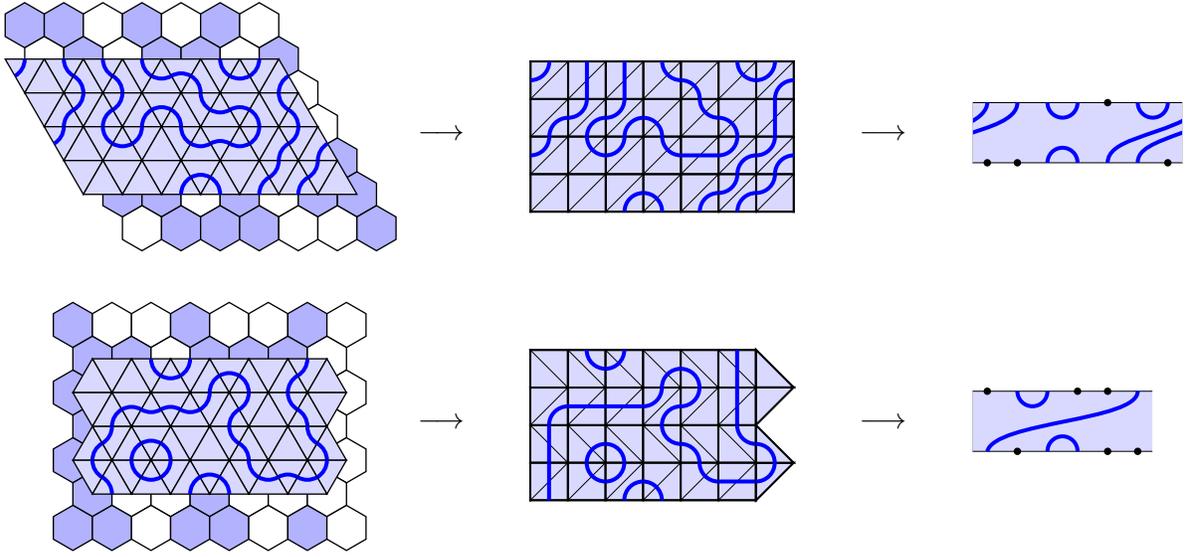
\begin{figure}
\begin{alignat*}{3}
\psset{unit=0.30cm}
\begin{pspicture}[shift=-5.4](-6.9,-1)(11,10.5)
\rput(-6.928,9.0){\hc}\rput(-5.196,9.0){\hc}\rput(-3.464,9.0){\hu}\rput(-1.732,9.0){\hc}\rput(0.000,9.0){\hu}\rput(1.732,9.0){\hc}\rput(3.464,9.0){\hu}
\rput(-6.062,7.5){\hu}\rput(-4.330,7.5){\hc}\rput(-2.598,7.5){\hu}\rput(-0.866,7.5){\hc}\rput(0.866,7.5){\hc}\rput(2.598,7.5){\hu}\rput(4.330,7.5){\hc}
\rput(-5.196,6.0){\hu}\rput(-3.464,6.0){\hc}\rput(-1.732,6.0){\hu}\rput(0.000,6.0){\hu}\rput(1.732,6.0){\hc}\rput(3.464,6.0){\hc}\rput(5.196,6.0){\hu}
\rput(-4.330,4.5){\hc}\rput(-2.598,4.5){\hu}\rput(-0.866,4.5){\hc}\rput(0.866,4.5){\hu}\rput(2.598,4.5){\hu}\rput(4.330,4.5){\hc}\rput(6.062,4.5){\hu}
\rput(-3.464,3.0){\hc}\rput(-1.732,3.0){\hc}\rput(0.000,3.0){\hc}\rput(1.732,3.0){\hc}\rput(3.464,3.0){\hc}\rput(5.196,3.0){\hu}\rput(6.928,3.0){\hc}
\rput(-2.598,1.5){\hc}\rput(-0.866,1.5){\hc}\rput(0.866,1.5){\hu}\rput(2.598,1.5){\hc}\rput(4.330,1.5){\hu}\rput(6.062,1.5){\hc}\rput(7.794,1.5){\hc}
\rput(-1.732,0.0){\hu}\rput(0.000,0,0){\hc}\rput(1.732,0.0){\hc}\rput(3.464,0.0){\hc}\rput(5.196,0.0){\hu}\rput(6.928,0.0){\hu}\rput(8.660,0.0){\hc}
\rput(-6.06218,7.5){\tridc}\rput(-4.33013,7.5){\tridd}\rput(-2.598,7.5){\tridd}\rput(-0.866,7.5){\tridd}\rput(0.866,7.5){\tridb}\rput(2.598,7.5){\tridd}\rput(4.33013,7.5){\tridc}
\rput(-5.19615,7.5){\triua}\rput(-3.4641,7.5){\triuc}\rput(-1.732,7.5){\triuc}\rput(0,7.5){\triub}\rput(1.732,7.5){\triuc}\rput(3.464,7.5){\triub}\rput(5.19615,7.5){\triud}
\rput(-5.196,6){\trida}\rput(-3.464,6){\tridc}\rput(-1.732,6){\tridc}\rput(0,6){\tridb}\rput(1.732,6){\tridd}\rput(3.464,6){\tridb}\rput(5.196,6){\tridd}
\rput(-4.330,6){\triud}\rput(-2.599,6){\triud}\rput(-0.866,6){\triud}\rput(0.866,6){\triuc}\rput(2.598,6){\triub}\rput(4.33,6){\triuc}\rput(6.062,6){\triuc}
\rput(-4.330,4.5){\tridc}\rput(-2.599,4.5){\tridd}\rput(-0.866,4.5){\tridc}\rput(0.866,4.5){\tridd}\rput(2.598,4.5){\tridb}\rput(4.33,4.5){\tridc}\rput(6.062,4.5){\tridc}
\rput(-3.464,4.5){\triua}\rput(-1.732,4.5){\triub}\rput(0,4.5){\triua}\rput(1.732,4.5){\triub}\rput(3.464,4.5){\triub}\rput(5.196,4.5){\triud}\rput(6.928,4.5){\triud}
\rput(-3.464,3){\trida}\rput(-1.732,3){\trida}\rput(0,3){\trida}\rput(1.732,3){\tridb}\rput(3.464,3){\trida}\rput(5.196,3){\tridc}\rput(6.928,3){\tridc}
\rput(-2.598,3){\triua}\rput(-0.866,3){\triua}\rput(0.866,3){\triud}\rput(2.598,3){\triuc}\rput(4.33,3){\triud}\rput(6.062,3){\triud}\rput(7.794,3){\triua}
\end{pspicture}
&\longrightarrow\qquad
\psset{unit=0.5cm}
\begin{pspicture}[shift=-1.9](-2,1)(5,5)
\facegrid{(-2,1)}{(5,5)}
\rput(-2,4){\dlooptwo}\rput(-1,4){\dloopsev}\rput(0,4){\dloopsev}\rput(1,4){\dloopfiv}\rput(2,4){\dloopfou}\rput(3,4){\dloopfiv}\rput(4,4){\dloopeig}
\rput(-2,3){\dloopthr}\rput(-1,3){\dloopeig}\rput(0,3){\dloopeig}\rput(1,3){\dloopfou}\rput(2,3){\dloopfiv}\rput(3,3){\dloopfou}\rput(4,3){\dloopsev}
\rput(-2,2){\dlooptwo}\rput(-1,2){\dloopfiv}\rput(0,2){\dlooptwo}\rput(1,2){\dloopfiv}\rput(2,2){\dloopsix}\rput(3,2){\dloopeig}\rput(4,2){\dloopeig}
\rput(-2,1){\dloopone}\rput(-1,1){\dloopone}\rput(0,1){\dloopthr}\rput(1,1){\dloopfou}\rput(2,1){\dloopthr}\rput(3,1){\dloopeig}\rput(4,1){\dlooptwo}
\end{pspicture}
\qquad&&\longrightarrow\qquad
\psset{unit=1cm}
\begin{pspicture}[shift=-0.3](-0.0,0)(2.8,0.8)
\pspolygon[fillstyle=solid,fillcolor=lightlightblue,linecolor=black,linewidth=0pt](0,0)(0,0.8)(2.8,0.8)(2.8,0)(0,0)
\psarc[linecolor=blue,linewidth=\elegant]{-}(1.2,0){0.2}{0}{180}
\psarc[linecolor=blue,linewidth=\elegant]{-}(1.2,0.8){0.2}{180}{0}
\psarc[linecolor=blue,linewidth=\elegant]{-}(2.4,0.8){0.2}{180}{0}
\psbezier[linecolor=blue,linewidth=\elegant]{-}(1.8,0.0)(1.8,0.3)(2.4,0.35)(2.82,0.57)
\psbezier[linecolor=blue,linewidth=\elegant]{-}(2.2,0.0)(2.2,0.18)(2.4,0.26)(2.82,0.41)
\psbezier[linecolor=blue,linewidth=\elegant]{-}(0.2,0.8)(0.2,0.7)(0.1,0.6)(-0.03,0.54)
\psbezier[linecolor=blue,linewidth=\elegant]{-}(0.6,0.8)(0.6,0.57)(0.2,0.47)(-0.03,0.39)
\pscircle[fillstyle=solid,fillcolor=black](1.8,0.8){0.035}
\pscircle[fillstyle=solid,fillcolor=black](0.2,0){0.035}
\pscircle[fillstyle=solid,fillcolor=black](0.6,0){0.035}
\pscircle[fillstyle=solid,fillcolor=black](2.6,0){0.035}
\psframe[fillstyle=solid,linecolor=white,linewidth=0pt](-0.04,0)(0,0.8)
\psframe[fillstyle=solid,linecolor=white,linewidth=0pt](2.8,0)(2.84,0.8)
\end{pspicture}
\\[0.4cm]
\psset{unit=0.30cm}
\begin{pspicture}[shift=-4.9](-1.4,0)(11.5,10.5)
\rput(-1.732,9.0){\hc}\rput(0.000,9.0){\hu}\rput(1.732,9.0){\hu}\rput(3.464,9.0){\hc}\rput(5.196,9.0){\hu}\rput(6.928,9.0){\hu}\rput(8.660,9.0){\hc}\rput(10.392,9.0){\hu}
\rput(-0.866,7.5){\hc}\rput(0.866,7.5){\hc}\rput(2.598,7.5){\hu}\rput(4.330,7.5){\hc}\rput(6.062,7.5){\hc}\rput(7.794,7.5){\hc}\rput(9.526,7.5){\hu}
\rput(-1.732,6.0){\hc}\rput(0.000,6.0){\hc}\rput(1.732,6.0){\hc}\rput(3.464,6.0){\hc}\rput(5.196,6.0){\hu}\rput(6.928,6.0){\hc}\rput(8.660,6.0){\hu}\rput(10.392,6.0){\hu}
\rput(-0.866,4.5){\hc}\rput(0.866,4.5){\hu}\rput(2.598,4.5){\hu}\rput(4.330,4.5){\hu}\rput(6.062,4.5){\hc}\rput(7.794,4.5){\hc}\rput(9.526,4.5){\hu}
\rput(-1.732,3.0){\hc}\rput(0.000,3.0){\hu}\rput(1.732,3.0){\hc}\rput(3.464,3.0){\hu}\rput(5.196,3.0){\hu}\rput(6.928,3.0){\hc}\rput(8.660,3.0){\hc}\rput(10.392,3.0){\hu}
\rput(-0.866,1.5){\hc}\rput(0.866,1.5){\hu}\rput(2.598,1.5){\hu}\rput(4.330,1.5){\hc}\rput(6.062,1.5){\hu}\rput(7.794,1.5){\hu}\rput(9.526,1.5){\hu}
\rput(-1.732,0.0){\hc}\rput(0.000,0.0){\hc}\rput(1.732,0.0){\hu}\rput(3.464,0.0){\hc}\rput(5.196,0.0){\hc}\rput(6.928,0.0){\hu}\rput(8.660,0.0){\hc}\rput(10.392,0.0){\hu}
\rput(0.866,7.5){\trida}\rput(2.598,7.5){\tridd}\rput(4.330,7.5){\tridc}\rput(6.062,7.5){\tridb}\rput(7.794,7.5){\trida}\rput(9.526,7.5){\tridc}
\rput(0.000,7.5){\triua}\rput(1.732,7.5){\triua}\rput(3.464,7.5){\triub}\rput(5.196,7.5){\triud}\rput(6.928,7.5){\triuc}\rput(8.660,7.5){\triud}\rput(10.392,7.5){\triua}
\rput(0.000,6.0){\trida}\rput(1.732,6.0){\tridb}\rput(3.464,6.0){\tridb}\rput(5.196,6.0){\tridc}\rput(6.928,6.0){\tridc}\rput(8.660,6.0){\tridd}\rput(10.392,6.0){\trida}
\rput(0.866,6.0){\triud}\rput(2.598,6.0){\triub}\rput(4.330,6.0){\triub}\rput(6.062,6.0){\triud}\rput(7.794,6.0){\triua}\rput(9.526,6.0){\triuc}
\rput(0.866,4.5){\tridc}\rput(2.598,4.5){\tridb}\rput(4.330,4.5){\trida}\rput(6.062,4.5){\tridd}\rput(7.794,4.5){\trida}\rput(9.526,4.5){\tridd}
\rput(0.000,4.5){\triud}\rput(1.732,4.5){\triud}\rput(3.464,4.5){\triuc}\rput(5.196,4.5){\triua}\rput(6.928,4.5){\triuc}\rput(8.660,4.5){\triua}\rput(10.392,4.5){\triuc}
\rput(0.000,3.0){\tridd}\rput(1.732,3.0){\tridd}\rput(3.464,3.0){\tridc}\rput(5.196,3.0){\tridb}\rput(6.928,3.0){\tridd}\rput(8.660,3.0){\tridb}\rput(10.392,3.0){\tridc}
\rput(0.866,3.0){\triuc}\rput(2.598,3.0){\triub}\rput(4.330,3.0){\triud}\rput(6.062,3.0){\triuc}\rput(7.794,3.0){\triub}\rput(9.526,3.0){\triub}
\end{pspicture}
\qquad &\longrightarrow\qquad
\psset{unit=0.50cm}
\begin{pspicture}[shift=-1.9](0,2)(7,6)
\facegrid{(0,2)}{(6,6)}
\rput(0,5){\oloopone}\rput(1,5){\oloopfiv}\rput(2,5){\olooptwo}\rput(3,5){\oloopthr}\rput(4,5){\oloopfou}\rput(5,5){\oloopsev}
\rput(0,4){\dloopthr}\rput(1,4){\dloopsix}\rput(2,4){\dloopsix}\rput(3,4){\dloopeig}\rput(4,4){\dlooptwo}\rput(5,4){\dloopsev}
\rput(0,3){\oloopsev}\rput(1,3){\oloopthr}\rput(2,3){\oloopfou}\rput(3,3){\oloopfiv}\rput(4,3){\oloopfou}\rput(5,3){\oloopfiv}
\rput(0,2){\dloopsev}\rput(1,2){\dloopfiv}\rput(2,2){\dloopeig}\rput(3,2){\dloopfou}\rput(4,2){\dloopfiv}\rput(5,2){\dloopsix}
\multiput(6,2)(0,2){2}{\pspolygon[fillstyle=solid,fillcolor=lightlightblue](0,0)(0,1)(1,1)
\pspolygon[fillstyle=solid,fillcolor=lightlightblue](0,1)(0,2)(1,1)}
\psarc[linecolor=blue,linewidth=1.5pt](6,3){0.5}{-90}{90}
\end{pspicture}
\qquad&&\longrightarrow\qquad
\psset{unit=1cm}
\begin{pspicture}[shift=-0.3](-0.0,0)(2.4,0.8)
\pspolygon[fillstyle=solid,fillcolor=lightlightblue,linecolor=black,linewidth=0pt](0,0)(0,0.8)(2.4,0.8)(2.4,0)(0,0)
\psarc[linecolor=blue,linewidth=\elegant]{-}(1.2,0){0.2}{0}{180}
\psarc[linecolor=blue,linewidth=\elegant]{-}(0.8,0.8){0.2}{180}{0}
\psbezier[linecolor=blue,linewidth=\elegant]{-}(0.2,0.0)(0.3,0.4)(2.2,0.4)(2.2,0.8)
\pscircle[fillstyle=solid,fillcolor=black](0.2,0.8){0.035}
\pscircle[fillstyle=solid,fillcolor=black](1.4,0.8){0.035}
\pscircle[fillstyle=solid,fillcolor=black](1.8,0.8){0.035}
\pscircle[fillstyle=solid,fillcolor=black](0.6,0){0.035}
\pscircle[fillstyle=solid,fillcolor=black](1.8,0){0.035}
\pscircle[fillstyle=solid,fillcolor=black](2.2,0){0.035}
\psframe[fillstyle=solid,linecolor=white,linewidth=0pt](-0.04,0)(0,0.8)
\psframe[fillstyle=solid,linecolor=white,linewidth=0pt](2.4,0)(2.44,0.8)
\end{pspicture}
\end{alignat*}
\caption{The map from horizontal sections of percolation configurations to connectivity diagrams, with periodic and strip boundary conditions. The resulting connectivities respectively belong to $\pdtl_7(\alpha,\beta)$ and $\dtl_6(\beta)$.}
\label{fig:strips.and.connectivities}
\end{figure}

For the model of site percolation, these objects arise when considering horizontal sections of the cylinder or torus. Applied to only the hexagons overlapping the section, the map discussed in \cref{sec:PercToLoops} produces configurations of loop segments that either close into loops or are attached to the top and bottom segments of the studied section. For each segment, the $N$ possible locations where the loops may be attached are called {\it nodes}. These nodes may also be left vacant. A connectivity segment is then obtained by straightening the loop segments that connect the top and bottom segments, and by removing the loops. Examples are given in \cref{fig:strips.and.connectivities}.

\paragraph{Definition of the algebras.}

The algebra $\dtl_N(\beta)$ was first studied in \cite{GP93,P94,Grimm96} and its representation theory was investigated in \cite{BSA14}. Here we introduce $\pdtl_N(\alpha,\beta)$ following the conventions of \cite{MDP19}, and $\dtl_N(\beta)$ as a subalgebra of $\pdtl_N(\alpha,\beta)$. The algebra $\pdtl_N(\alpha,\beta)$ is the linear span of connectivity diagrams drawn in a rectangular box with periodic boundary conditions in the horizontal direction. The top and bottom segments of the box each have $N$ marked nodes labelled from left to right with the integers $1, \dots, N$. In a connectivity diagram, these are either connected pairwise by a loop segment or are left vacant. Here are examples of connectivity diagrams for $N=6$:
\be
\label{eq:connect.examples}
a_1 = \ 
\begin{pspicture}[shift=-0.3](-0.0,0)(2.4,0.8)
\pspolygon[fillstyle=solid,fillcolor=lightlightblue,linecolor=black,linewidth=0pt](0,0)(0,0.8)(2.4,0.8)(2.4,0)(0,0)
\psarc[linecolor=blue,linewidth=\elegant]{-}(1.6,0){0.2}{0}{180}
\psarc[linecolor=blue,linewidth=\elegant]{-}(0.8,0.8){0.2}{180}{0}
\psbezier[linecolor=blue,linewidth=\elegant]{-}(0.2,0)(0.2,0.4)(1.0,0.4)(1.0,0)
\psbezier[linecolor=blue,linewidth=\elegant]{-}(0.2,0.8)(0.3,0.2)(2.2,0.3)(2.2,0.8)
\pscircle[fillstyle=solid,fillcolor=black](1.4,0.8){0.035}
\pscircle[fillstyle=solid,fillcolor=black](1.8,0.8){0.035}
\pscircle[fillstyle=solid,fillcolor=black](2.2,0){0.035}
\pscircle[fillstyle=solid,fillcolor=black](0.6,0){0.035}
\psframe[fillstyle=solid,linecolor=white,linewidth=0pt](-0.04,0)(0,0.8)
\psframe[fillstyle=solid,linecolor=white,linewidth=0pt](2.4,0)(2.44,0.8)
\end{pspicture}\ ,
\qquad
a_2 = \ 
\begin{pspicture}[shift=-0.3](-0.0,0)(2.4,0.8)
\pspolygon[fillstyle=solid,fillcolor=lightlightblue,linecolor=black,linewidth=0pt](0,0)(0,0.8)(2.4,0.8)(2.4,0)(0,0)
\psarc[linecolor=blue,linewidth=\elegant]{-}(0.8,0){0.2}{0}{180}
\psbezier[linecolor=blue,linewidth=\elegant]{-}(1.4,0.8)(1.4,0.4)(2.2,0.4)(2.2,0)
\psbezier[linecolor=blue,linewidth=\elegant]{-}(0.2,0)(0.2,0.28)(0,0.39)(-0.02,0.43)
\psbezier[linecolor=blue,linewidth=\elegant]{-}(2.2,0.8)(2.2,0.52)(2.4,0.41)(2.42,0.37)
\pscircle[fillstyle=solid,fillcolor=black](1.4,0){0.035}
\pscircle[fillstyle=solid,fillcolor=black](1.8,0){0.035}
\pscircle[fillstyle=solid,fillcolor=black](0.2,0.8){0.035}
\pscircle[fillstyle=solid,fillcolor=black](0.6,0.8){0.035}
\pscircle[fillstyle=solid,fillcolor=black](1.0,0.8){0.035}
\pscircle[fillstyle=solid,fillcolor=black](1.8,0.8){0.035}
\psframe[fillstyle=solid,linecolor=white,linewidth=0pt](-0.04,0)(0,0.8)
\psframe[fillstyle=solid,linecolor=white,linewidth=0pt](2.4,0)(2.44,0.8)
\end{pspicture}\ ,
\qquad
a_3 = \ 
\begin{pspicture}[shift=-0.3](-0.0,0)(2.4,0.8)
\pspolygon[fillstyle=solid,fillcolor=lightlightblue,linecolor=black,linewidth=0pt](0,0)(0,0.8)(2.4,0.8)(2.4,0)(0,0)
\psarc[linecolor=blue,linewidth=\elegant]{-}(2.4,0){0.2}{90}{180}
\psarc[linecolor=blue,linewidth=\elegant]{-}(0,0){0.2}{0}{90}
\psbezier[linecolor=blue,linewidth=\elegant]{-}(0.2,0.8)(0.2,0.4)(1.0,0.4)(1.0,0.8)
\psbezier[linecolor=blue,linewidth=\elegant]{-}(1.0,0)(1.0,0.4)(1.4,0.4)(1.4,0.8)
\psbezier[linecolor=blue,linewidth=\elegant]{-}(0.6,0)(0.6,0.4)(0,0.4)(-0.04,0.4)
\psbezier[linecolor=blue,linewidth=\elegant]{-}(1.8,0.8)(1.8,0.4)(2.4,0.4)(2.44,0.4)
\psframe[fillstyle=solid,linecolor=white,linewidth=0pt](-0.04,0)(0,0.8)
\psframe[fillstyle=solid,linecolor=white,linewidth=0pt](2.4,0)(2.44,0.8)
\pscircle[fillstyle=solid,fillcolor=black](1.4,0){0.035}
\pscircle[fillstyle=solid,fillcolor=black](1.8,0){0.035}
\pscircle[fillstyle=solid,fillcolor=black](0.6,0.8){0.035}
\pscircle[fillstyle=solid,fillcolor=black](2.2,0.8){0.035}
\end{pspicture}\ .
\ee
Alternatively, these diagrams can be drawn in a planar representation, with the loop segments living inside an annulus, see for instance \cite{IMD21}. 

The product $a_1a_2$ of two connectivities of $\pdtl_N(\alpha,\beta)$ is obtained by stacking $a_2$ above $a_1$. If one or more loop segments are connected to a vacant site, the result is set to zero. Otherwise, the new connectivity is obtained by reading the connections of the nodes on the top and bottom segments of the new, larger box. The result of the product $a_1a_2$ is this new connectivity diagram multiplied by factors of $\beta$ and $\alpha$ for each closed contractible and non-contractible loop formed in the process, respectively. This product is then linearly extended to all elements of $\pdtl_N(\alpha,\beta)$. Here are examples illustrating this product for $N=6$:
\begin{subequations}
\begin{alignat}{2}
a_1 a_2 &= \ 
\begin{pspicture}[shift=-0.7](-0.0,0)(2.4,1.6)
\pspolygon[fillstyle=solid,fillcolor=lightlightblue,linecolor=black,linewidth=0pt](0,0)(0,0.8)(2.4,0.8)(2.4,0)(0,0)
\psarc[linecolor=blue,linewidth=\elegant]{-}(1.6,0){0.2}{0}{180}
\psarc[linecolor=blue,linewidth=\elegant]{-}(0.8,0.8){0.2}{180}{0}
\psbezier[linecolor=blue,linewidth=\elegant]{-}(0.2,0)(0.2,0.4)(1.0,0.4)(1.0,0)
\psbezier[linecolor=blue,linewidth=\elegant]{-}(0.2,0.8)(0.3,0.2)(2.2,0.3)(2.2,0.8)
\pscircle[fillstyle=solid,fillcolor=black](1.4,0.8){0.035}
\pscircle[fillstyle=solid,fillcolor=black](1.8,0.8){0.035}
\pscircle[fillstyle=solid,fillcolor=black](2.2,0){0.035}
\pscircle[fillstyle=solid,fillcolor=black](0.6,0){0.035}
\psframe[fillstyle=solid,linecolor=white,linewidth=0pt](-0.04,0)(0,0.8)
\psframe[fillstyle=solid,linecolor=white,linewidth=0pt](2.4,0)(2.44,0.8)
\rput(0,0.8){
\pspolygon[fillstyle=solid,fillcolor=lightlightblue,linecolor=black,linewidth=0pt](0,0)(0,0.8)(2.4,0.8)(2.4,0)(0,0)
\psarc[linecolor=blue,linewidth=\elegant]{-}(0.8,0){0.2}{0}{180}
\psbezier[linecolor=blue,linewidth=\elegant]{-}(1.4,0.8)(1.4,0.4)(2.2,0.4)(2.2,0)
\psbezier[linecolor=blue,linewidth=\elegant]{-}(0.2,0)(0.2,0.28)(0,0.39)(-0.02,0.43)
\psbezier[linecolor=blue,linewidth=\elegant]{-}(2.2,0.8)(2.2,0.52)(2.4,0.41)(2.42,0.37)
\pscircle[fillstyle=solid,fillcolor=black](1.4,0){0.035}
\pscircle[fillstyle=solid,fillcolor=black](1.8,0){0.035}
\pscircle[fillstyle=solid,fillcolor=black](0.2,0.8){0.035}
\pscircle[fillstyle=solid,fillcolor=black](0.6,0.8){0.035}
\pscircle[fillstyle=solid,fillcolor=black](1.0,0.8){0.035}
\pscircle[fillstyle=solid,fillcolor=black](1.8,0.8){0.035}
\psframe[fillstyle=solid,linecolor=white,linewidth=0pt](-0.04,0)(0,0.8)
\psframe[fillstyle=solid,linecolor=white,linewidth=0pt](2.4,0)(2.44,0.8)
}
\end{pspicture}
\ = \beta \ \,
\begin{pspicture}[shift=-0.3](-0.0,0)(2.4,0.8)
\pspolygon[fillstyle=solid,fillcolor=lightlightblue,linecolor=black,linewidth=0pt](0,0)(0,0.8)(2.4,0.8)(2.4,0)(0,0)
\psarc[linecolor=blue,linewidth=\elegant]{-}(1.6,0){0.2}{0}{180}
\psbezier[linecolor=blue,linewidth=\elegant]{-}(0.2,0)(0.2,0.4)(1.0,0.4)(1.0,0)
\psbezier[linecolor=blue,linewidth=\elegant]{-}(1.4,0.8)(1.4,0.3)(0.1,0.4)(-0.02,0.56)
\psbezier[linecolor=blue,linewidth=\elegant]{-}(2.2,0.8)(2.2,0.6)(2.4,0.55)(2.42,0.53)
\pscircle[fillstyle=solid,fillcolor=black](0.6,0){0.035}
\pscircle[fillstyle=solid,fillcolor=black](2.2,0){0.035}
\pscircle[fillstyle=solid,fillcolor=black](0.2,0.8){0.035}
\pscircle[fillstyle=solid,fillcolor=black](0.6,0.8){0.035}
\pscircle[fillstyle=solid,fillcolor=black](1.0,0.8){0.035}
\pscircle[fillstyle=solid,fillcolor=black](1.8,0.8){0.035}
\psframe[fillstyle=solid,linecolor=white,linewidth=0pt](-0.04,0)(0,0.8)
\psframe[fillstyle=solid,linecolor=white,linewidth=0pt](2.4,0)(2.44,0.8)
\end{pspicture}\ \ ,
\end{alignat}
\begin{alignat}{2}
a_1 a_3 &= \ 
\begin{pspicture}[shift=-0.7](-0.0,0)(2.4,1.6)
\pspolygon[fillstyle=solid,fillcolor=lightlightblue,linecolor=black,linewidth=0pt](0,0)(0,0.8)(2.4,0.8)(2.4,0)(0,0)
\psarc[linecolor=blue,linewidth=\elegant]{-}(1.6,0){0.2}{0}{180}
\psarc[linecolor=blue,linewidth=\elegant]{-}(0.8,0.8){0.2}{180}{0}
\psbezier[linecolor=blue,linewidth=\elegant]{-}(0.2,0)(0.2,0.4)(1.0,0.4)(1.0,0)
\psbezier[linecolor=blue,linewidth=\elegant]{-}(0.2,0.8)(0.3,0.2)(2.2,0.3)(2.2,0.8)
\pscircle[fillstyle=solid,fillcolor=black](1.4,0.8){0.035}
\pscircle[fillstyle=solid,fillcolor=black](1.8,0.8){0.035}
\pscircle[fillstyle=solid,fillcolor=black](2.2,0){0.035}
\pscircle[fillstyle=solid,fillcolor=black](0.6,0){0.035}
\psframe[fillstyle=solid,linecolor=white,linewidth=0pt](-0.04,0)(0,0.8)
\psframe[fillstyle=solid,linecolor=white,linewidth=0pt](2.4,0)(2.44,0.8)
\rput(0,0.8){
\pspolygon[fillstyle=solid,fillcolor=lightlightblue,linecolor=black,linewidth=0pt](0,0)(0,0.8)(2.4,0.8)(2.4,0)(0,0)
\psarc[linecolor=blue,linewidth=\elegant]{-}(2.4,0){0.2}{90}{180}
\psarc[linecolor=blue,linewidth=\elegant]{-}(0,0){0.2}{0}{90}
\psbezier[linecolor=blue,linewidth=\elegant]{-}(0.2,0.8)(0.2,0.4)(1.0,0.4)(1.0,0.8)
\psbezier[linecolor=blue,linewidth=\elegant]{-}(1.0,0)(1.0,0.4)(1.4,0.4)(1.4,0.8)
\psbezier[linecolor=blue,linewidth=\elegant]{-}(0.6,0)(0.6,0.4)(0,0.4)(-0.04,0.4)
\psbezier[linecolor=blue,linewidth=\elegant]{-}(1.8,0.8)(1.8,0.4)(2.4,0.4)(2.44,0.4)
\psframe[fillstyle=solid,linecolor=white,linewidth=0pt](-0.04,0)(0,0.8)
\psframe[fillstyle=solid,linecolor=white,linewidth=0pt](2.4,0)(2.44,0.8)
\pscircle[fillstyle=solid,fillcolor=black](1.4,0){0.035}
\pscircle[fillstyle=solid,fillcolor=black](1.8,0){0.035}
\pscircle[fillstyle=solid,fillcolor=black](0.6,0.8){0.035}
\pscircle[fillstyle=solid,fillcolor=black](2.2,0.8){0.035}
}
\end{pspicture}
\ = \alpha \ \,
\begin{pspicture}[shift=-0.3](-0.0,0)(2.4,0.8)
\pspolygon[fillstyle=solid,fillcolor=lightlightblue,linecolor=black,linewidth=0pt](0,0)(0,0.8)(2.4,0.8)(2.4,0)(0,0)
\psarc[linecolor=blue,linewidth=\elegant]{-}(1.6,0){0.2}{0}{180}
\psbezier[linecolor=blue,linewidth=\elegant]{-}(0.2,0)(0.2,0.4)(1.0,0.4)(1.0,0)
\psbezier[linecolor=blue,linewidth=\elegant]{-}(0.2,0.8)(0.2,0.4)(1.0,0.4)(1.0,0.8)
\psbezier[linecolor=blue,linewidth=\elegant]{-}(1.4,0.8)(1.4,0.35)(0.1,0.34)(-0.02,0.44)
\psbezier[linecolor=blue,linewidth=\elegant]{-}(1.8,0.8)(1.8,0.5)(2.4,0.43)(2.42,0.41)
\pscircle[fillstyle=solid,fillcolor=black](0.6,0){0.035}
\pscircle[fillstyle=solid,fillcolor=black](2.2,0){0.035}
\pscircle[fillstyle=solid,fillcolor=black](0.6,0.8){0.035}
\pscircle[fillstyle=solid,fillcolor=black](2.2,0.8){0.035}
\psframe[fillstyle=solid,linecolor=white,linewidth=0pt](-0.04,0)(0,0.8)
\psframe[fillstyle=solid,linecolor=white,linewidth=0pt](2.4,0)(2.44,0.8)
\end{pspicture}\ \ ,
\end{alignat}
\begin{alignat}{2}
a_2 a_3 &= \ 
\begin{pspicture}[shift=-0.7](-0.0,0)(2.4,1.6)
\pspolygon[fillstyle=solid,fillcolor=lightlightblue,linecolor=black,linewidth=0pt](0,0)(0,0.8)(2.4,0.8)(2.4,0)(0,0)
\psarc[linecolor=blue,linewidth=\elegant]{-}(0.8,0){0.2}{0}{180}
\psbezier[linecolor=blue,linewidth=\elegant]{-}(1.4,0.8)(1.4,0.4)(2.2,0.4)(2.2,0)
\psbezier[linecolor=blue,linewidth=\elegant]{-}(0.2,0)(0.2,0.28)(0,0.39)(-0.02,0.43)
\psbezier[linecolor=blue,linewidth=\elegant]{-}(2.2,0.8)(2.2,0.52)(2.4,0.41)(2.42,0.37)
\pscircle[fillstyle=solid,fillcolor=black](1.4,0){0.035}
\pscircle[fillstyle=solid,fillcolor=black](1.8,0){0.035}
\pscircle[fillstyle=solid,fillcolor=black](0.2,0.8){0.035}
\pscircle[fillstyle=solid,fillcolor=black](0.6,0.8){0.035}
\pscircle[fillstyle=solid,fillcolor=black](1.0,0.8){0.035}
\pscircle[fillstyle=solid,fillcolor=black](1.8,0.8){0.035}
\psframe[fillstyle=solid,linecolor=white,linewidth=0pt](-0.04,0)(0,0.8)
\psframe[fillstyle=solid,linecolor=white,linewidth=0pt](2.4,0)(2.44,0.8)
\rput(0,0.8){
\pspolygon[fillstyle=solid,fillcolor=lightlightblue,linecolor=black,linewidth=0pt](0,0)(0,0.8)(2.4,0.8)(2.4,0)(0,0)
\psarc[linecolor=blue,linewidth=\elegant]{-}(2.4,0){0.2}{90}{180}
\psarc[linecolor=blue,linewidth=\elegant]{-}(0,0){0.2}{0}{90}
\psbezier[linecolor=blue,linewidth=\elegant]{-}(0.2,0.8)(0.2,0.4)(1.0,0.4)(1.0,0.8)
\psbezier[linecolor=blue,linewidth=\elegant]{-}(1.0,0)(1.0,0.4)(1.4,0.4)(1.4,0.8)
\psbezier[linecolor=blue,linewidth=\elegant]{-}(0.6,0)(0.6,0.4)(0,0.4)(-0.04,0.4)
\psbezier[linecolor=blue,linewidth=\elegant]{-}(1.8,0.8)(1.8,0.4)(2.4,0.4)(2.44,0.4)
\psframe[fillstyle=solid,linecolor=white,linewidth=0pt](-0.04,0)(0,0.8)
\psframe[fillstyle=solid,linecolor=white,linewidth=0pt](2.4,0)(2.44,0.8)
\pscircle[fillstyle=solid,fillcolor=black](1.4,0){0.035}
\pscircle[fillstyle=solid,fillcolor=black](1.8,0){0.035}
\pscircle[fillstyle=solid,fillcolor=black](0.6,0.8){0.035}
\pscircle[fillstyle=solid,fillcolor=black](2.2,0.8){0.035}
}
\end{pspicture}
\ = 0.
\end{alignat}
\end{subequations}
Because loop segments can wind arbitrarily many times around the periodic boundary conditions, $\pdtl_N(\alpha,\beta)$ is an infinite-dimensional algebra. 

The algebra $\dtl_N(\beta)$ is the finite-dimensional subalgebra of $\pdtl_N(\alpha,\beta)$, generated by connectivity diagrams with loop segments that do not cross the vertical segment that marks the periodic boundary conditions between the sites $1$ and~$N$. For instance, only the connectivity $a_1$ in \eqref{eq:connect.examples} is an element of $\dtl_6(\beta)$.

\paragraph{Standard modules.}
We denote by $\repW_{N,d,\omega}$ and $\repV_{N,d}$ the standard modules over $\pdtl_N(\alpha,\beta)$ and $\dtl_N(\beta)$, respectively. Their vector spaces are spanned by link states on $N$ nodes with $d$ defects, with $0 \le d \le N$. These are diagrams drawn above a segment with $N$ marked nodes that are either connected pairwise, occupied by a vertical segment called a defect, or left vacant. The link states of $\repW_{N,d,\omega}$ may have loop segments that cross the periodic boundary conditions, whereas the link states of $\repV_{N,d}$ may not. 
For example, the link states that span the standard modules for $N = 3$ are
\begin{subequations}
\begin{alignat}{2}
&\begin{array}{l}
\repW_{3,3,\omega}:\qquad 
\begin{pspicture}[shift=0](0.0,0)(1.2,0.5)
\psline[linewidth=\mince](0,0)(1.2,0)
\psline[linecolor=darkgreen,linewidth=\elegant]{-}(0.2,0)(0.2,0.4)
\psline[linecolor=darkgreen,linewidth=\elegant]{-}(0.6,0)(0.6,0.4)
\psline[linecolor=darkgreen,linewidth=\elegant]{-}(1.0,0)(1.0,0.4)
\end{pspicture}\ \ ,
\qquad \qquad
\repW_{3,2,\omega}:\qquad 
\begin{pspicture}[shift=0](0.0,0)(1.2,0.5)
\psline[linewidth=\mince](0,0)(1.2,0)
\psline[linecolor=darkgreen,linewidth=\elegant]{-}(0.2,0)(0.2,0.4)
\psline[linecolor=darkgreen,linewidth=\elegant]{-}(0.6,0)(0.6,0.4)
\pscircle[fillstyle=solid,fillcolor=black](1.0,0){0.04}
\end{pspicture}
\quad
\begin{pspicture}[shift=0](0.0,0)(1.2,0.5)
\psline[linewidth=\mince](0,0)(1.2,0)
\psline[linecolor=darkgreen,linewidth=\elegant]{-}(0.2,0)(0.2,0.4)
\psline[linecolor=darkgreen,linewidth=\elegant]{-}(1.0,0)(1.0,0.4)
\pscircle[fillstyle=solid,fillcolor=black](0.6,0){0.04}
\end{pspicture}
\quad
\begin{pspicture}[shift=0](0.0,0)(1.2,0.5)
\psline[linewidth=\mince](0,0)(1.2,0)
\psline[linecolor=darkgreen,linewidth=\elegant]{-}(0.6,0)(0.6,0.4)
\psline[linecolor=darkgreen,linewidth=\elegant]{-}(1.0,0)(1.0,0.4)
\pscircle[fillstyle=solid,fillcolor=black](0.2,0){0.04}
\end{pspicture}\ \ ,
\\[0.2cm]
\repW_{3,1,\omega}:\qquad 
\begin{pspicture}[shift=0](0.0,0)(1.2,0.5)
\psline[linewidth=\mince](0,0)(1.2,0)
\psarc[linecolor=darkgreen,linewidth=\elegant]{-}(0.8,0){0.2}{0}{180}
\psline[linecolor=darkgreen,linewidth=\elegant]{-}(0.2,0)(0.2,0.4)
\end{pspicture}
\quad
\begin{pspicture}[shift=0](0.0,0)(1.2,0.5)
\psline[linewidth=\mince](0,0)(1.2,0)
\psarc[linecolor=darkgreen,linewidth=\elegant]{-}(0.4,0){0.2}{0}{180}
\psline[linecolor=darkgreen,linewidth=\elegant]{-}(1.0,0)(1.0,0.4)
\end{pspicture}
\quad
\begin{pspicture}[shift=0](0.0,0)(1.2,0.5)
\psline[linewidth=\mince](0,0)(1.2,0)
\psarc[linecolor=darkgreen,linewidth=\elegant]{-}(0,0){0.2}{0}{90}
\psarc[linecolor=darkgreen,linewidth=\elegant]{-}(1.2,0){0.2}{90}{180}
\psline[linecolor=darkgreen,linewidth=\elegant]{-}(0.6,0)(0.6,0.4)
\end{pspicture}
\quad
\begin{pspicture}[shift=0](0.0,0)(1.2,0.5)
\psline[linewidth=\mince](0,0)(1.2,0)
\pscircle[fillstyle=solid,fillcolor=black](0.6,0){0.04}
\pscircle[fillstyle=solid,fillcolor=black](1,0){0.04}
\psline[linecolor=darkgreen,linewidth=\elegant]{-}(0.2,0)(0.2,0.4)
\end{pspicture}
\quad
\begin{pspicture}[shift=0](0.0,0)(1.2,0.5)
\psline[linewidth=\mince](0,0)(1.2,0)
\pscircle[fillstyle=solid,fillcolor=black](0.2,0){0.04}
\pscircle[fillstyle=solid,fillcolor=black](1.0,0){0.04}
\psline[linecolor=darkgreen,linewidth=\elegant]{-}(0.6,0)(0.6,0.4)
\end{pspicture}
\quad
\begin{pspicture}[shift=0](0.0,0)(1.2,0.5)
\psline[linewidth=\mince](0,0)(1.2,0)
\pscircle[fillstyle=solid,fillcolor=black](0.2,0){0.04}
\pscircle[fillstyle=solid,fillcolor=black](0.6,0){0.04}
\psline[linecolor=darkgreen,linewidth=\elegant]{-}(1.0,0)(1.0,0.4)
\end{pspicture}
\ \ ,
\\[0.2cm]%
\repW_{3,0,\omega}:\qquad
\begin{pspicture}[shift=0](0.0,0)(1.2,0.5)
\psline[linewidth=\mince](0,0)(1.2,0)
\pscircle[fillstyle=solid,fillcolor=black](0.2,0){0.04}
\psarc[linecolor=darkgreen,linewidth=\elegant]{-}(0.8,0){0.2}{0}{180}
\end{pspicture}
\quad
\begin{pspicture}[shift=0](0.0,0)(1.2,0.5)
\psline[linewidth=\mince](0,0)(1.2,0)
\pscircle[fillstyle=solid,fillcolor=black](1.0,0){0.04}
\psarc[linecolor=darkgreen,linewidth=\elegant]{-}(0.4,0){0.2}{0}{180}
\end{pspicture}
\quad
\begin{pspicture}[shift=0](0.0,0)(1.2,0.5)
\psline[linewidth=\mince](0,0)(1.2,0)
\pscircle[fillstyle=solid,fillcolor=black](0.6,0){0.04}
\psbezier[linecolor=darkgreen,linewidth=\elegant](0.2,0)(0.2,0.4)(1.0,0.4)(1.0,0)
\end{pspicture}
\quad
\begin{pspicture}[shift=0](0.0,0)(1.2,0.5)
\psline[linewidth=\mince](0,0)(1.2,0)
\pscircle[fillstyle=solid,fillcolor=black](0.2,0){0.04}
\psbezier[linecolor=darkgreen,linewidth=\elegant](1.0,0)(1.0,0.4)(1.8,0.4)(1.8,0)\rput(-1.2,0){\psbezier[linecolor=darkgreen,linewidth=\elegant](1.0,0)(1.0,0.4)(1.8,0.4)(1.8,0)}
\psframe[fillstyle=solid,linecolor=white,linewidth=0pt](1.2,0)(2.5,0.4)
\psframe[fillstyle=solid,linecolor=white,linewidth=0pt](0,0)(-0.5,0.4)
\end{pspicture}
\quad
\begin{pspicture}[shift=0](0.0,0)(1.2,0.5)
\psline[linewidth=\mince](0,0)(1.2,0)
\pscircle[fillstyle=solid,fillcolor=black](1.0,0){0.04}
\psbezier[linecolor=darkgreen,linewidth=\elegant](0.6,0)(0.6,0.4)(1.4,0.4)(1.4,0)\psbezier[linecolor=darkgreen,linewidth=\elegant](-0.1,0.30)(0.1,0.23)(0.2,0.16)(0.2,0)
\psframe[fillstyle=solid,linecolor=white,linewidth=0pt](1.2,0)(2.5,0.4)
\psframe[fillstyle=solid,linecolor=white,linewidth=0pt](0,0)(-0.2,0.4)
\end{pspicture}
\quad
\begin{pspicture}[shift=0](0.0,0)(1.2,0.5)
\psline[linewidth=\mince](0,0)(1.2,0)
\pscircle[fillstyle=solid,fillcolor=black](0.6,0){0.04}
\psarc[linecolor=darkgreen,linewidth=\elegant]{-}(0,0){0.2}{0}{90}
\psarc[linecolor=darkgreen,linewidth=\elegant]{-}(1.2,0){0.2}{90}{180}
\end{pspicture}
\quad
\begin{pspicture}[shift=0](0.0,0)(1.2,0.5)
\psline[linewidth=\mince](0,0)(1.2,0)
\pscircle[fillstyle=solid,fillcolor=black](0.2,0){0.04}
\pscircle[fillstyle=solid,fillcolor=black](0.6,0){0.04}
\pscircle[fillstyle=solid,fillcolor=black](1.0,0){0.04}
\end{pspicture}\ \ ,
\end{array}
\\\intertext{and}
&\begin{array}{l}
\repV_{3,3}:\qquad 
\begin{pspicture}[shift=0](0.0,0)(1.2,0.5)
\psline[linewidth=\mince](0,0)(1.2,0)
\psline[linecolor=darkgreen,linewidth=\elegant]{-}(0.2,0)(0.2,0.4)
\psline[linecolor=darkgreen,linewidth=\elegant]{-}(0.6,0)(0.6,0.4)
\psline[linecolor=darkgreen,linewidth=\elegant]{-}(1.0,0)(1.0,0.4)
\end{pspicture}\ \ ,
\qquad \qquad
\repV_{3,2}:\qquad 
\begin{pspicture}[shift=0](0.0,0)(1.2,0.5)
\psline[linewidth=\mince](0,0)(1.2,0)
\psline[linecolor=darkgreen,linewidth=\elegant]{-}(0.2,0)(0.2,0.4)
\psline[linecolor=darkgreen,linewidth=\elegant]{-}(0.6,0)(0.6,0.4)
\pscircle[fillstyle=solid,fillcolor=black](1.0,0){0.04}
\end{pspicture}
\quad
\begin{pspicture}[shift=0](0.0,0)(1.2,0.5)
\psline[linewidth=\mince](0,0)(1.2,0)
\psline[linecolor=darkgreen,linewidth=\elegant]{-}(0.2,0)(0.2,0.4)
\psline[linecolor=darkgreen,linewidth=\elegant]{-}(1.0,0)(1.0,0.4)
\pscircle[fillstyle=solid,fillcolor=black](0.6,0){0.04}
\end{pspicture}
\quad
\begin{pspicture}[shift=0](0.0,0)(1.2,0.5)
\psline[linewidth=\mince](0,0)(1.2,0)
\psline[linecolor=darkgreen,linewidth=\elegant]{-}(0.6,0)(0.6,0.4)
\psline[linecolor=darkgreen,linewidth=\elegant]{-}(1.0,0)(1.0,0.4)
\pscircle[fillstyle=solid,fillcolor=black](0.2,0){0.04}
\end{pspicture}\ \ ,
\\[0.2cm]
\repV_{3,1}:\qquad 
\begin{pspicture}[shift=0](0.0,0)(1.2,0.5)
\psline[linewidth=\mince](0,0)(1.2,0)
\psarc[linecolor=darkgreen,linewidth=\elegant]{-}(0.8,0){0.2}{0}{180}
\psline[linecolor=darkgreen,linewidth=\elegant]{-}(0.2,0)(0.2,0.4)
\end{pspicture}
\quad
\begin{pspicture}[shift=0](0.0,0)(1.2,0.5)
\psline[linewidth=\mince](0,0)(1.2,0)
\psarc[linecolor=darkgreen,linewidth=\elegant]{-}(0.4,0){0.2}{0}{180}
\psline[linecolor=darkgreen,linewidth=\elegant]{-}(1.0,0)(1.0,0.4)
\end{pspicture}
\quad
\begin{pspicture}[shift=0](0.0,0)(1.2,0.5)
\psline[linewidth=\mince](0,0)(1.2,0)
\pscircle[fillstyle=solid,fillcolor=black](0.6,0){0.04}
\pscircle[fillstyle=solid,fillcolor=black](1,0){0.04}
\psline[linecolor=darkgreen,linewidth=\elegant]{-}(0.2,0)(0.2,0.4)
\end{pspicture}
\quad
\begin{pspicture}[shift=0](0.0,0)(1.2,0.5)
\psline[linewidth=\mince](0,0)(1.2,0)
\pscircle[fillstyle=solid,fillcolor=black](0.2,0){0.04}
\pscircle[fillstyle=solid,fillcolor=black](1.0,0){0.04}
\psline[linecolor=darkgreen,linewidth=\elegant]{-}(0.6,0)(0.6,0.4)
\end{pspicture}
\quad
\begin{pspicture}[shift=0](0.0,0)(1.2,0.5)
\psline[linewidth=\mince](0,0)(1.2,0)
\pscircle[fillstyle=solid,fillcolor=black](0.2,0){0.04}
\pscircle[fillstyle=solid,fillcolor=black](0.6,0){0.04}
\psline[linecolor=darkgreen,linewidth=\elegant]{-}(1.0,0)(1.0,0.4)
\end{pspicture}
\ \ ,
\\[0.2cm]%
\repV_{3,0}:\qquad
\begin{pspicture}[shift=0](0.0,0)(1.2,0.5)
\psline[linewidth=\mince](0,0)(1.2,0)
\pscircle[fillstyle=solid,fillcolor=black](0.2,0){0.04}
\psarc[linecolor=darkgreen,linewidth=\elegant]{-}(0.8,0){0.2}{0}{180}
\end{pspicture}
\quad
\begin{pspicture}[shift=0](0.0,0)(1.2,0.5)
\psline[linewidth=\mince](0,0)(1.2,0)
\pscircle[fillstyle=solid,fillcolor=black](1.0,0){0.04}
\psarc[linecolor=darkgreen,linewidth=\elegant]{-}(0.4,0){0.2}{0}{180}
\end{pspicture}
\quad
\begin{pspicture}[shift=0](0.0,0)(1.2,0.5)
\psline[linewidth=\mince](0,0)(1.2,0)
\pscircle[fillstyle=solid,fillcolor=black](0.6,0){0.04}
\psbezier[linecolor=darkgreen,linewidth=\elegant](0.2,0)(0.2,0.4)(1.0,0.4)(1.0,0)
\end{pspicture}
\quad
\begin{pspicture}[shift=0](0.0,0)(1.2,0.5)
\psline[linewidth=\mince](0,0)(1.2,0)
\pscircle[fillstyle=solid,fillcolor=black](0.2,0){0.04}
\pscircle[fillstyle=solid,fillcolor=black](0.6,0){0.04}
\pscircle[fillstyle=solid,fillcolor=black](1.0,0){0.04}
\end{pspicture}\ \ .
\end{array}
\end{alignat}
\end{subequations}

The action of a connectivity $c \in \pdtl_N(\alpha,\beta)$ on a link state $w$ is obtained by drawing $w$ above~$c$ and reading the new link state. If two defects are connected or if a loop segment is connected to a vacant site, the result is set to zero. Otherwise, for $d = 0$, the product $cw$ is this new link state multiplied by $\alpha^{n_\alpha}\beta^{n_\beta}$ where $n_\alpha$ and $n_\beta$ are the numbers of non-contractible and contractible loops closed in the process. For $d>0$, any non-contractible loop results in a zero weight, as this implies connecting defects. 
Moreover, for $d>0$, the action depends on a twist factor $\omega$ that couples to the winding of the defects around the cylinder. 
Indeed, a defect moving from the top to the bottom that crosses the periodic boundary conditions between the nodes $1$ and $N$ produces a factor of $\omega$ if it crosses it while traveling to the left, and $\omega^{-1}$ if it crosses while travelling to the right.
Then $cw$ is equal to the new link state formed on the bottom segment of the box, times $\beta^{n_\beta}\omega^{n_\omega}$, where $n_\omega$ is an integer that measures the total winding of the defects. Here are examples of this action:
\begin{subequations}
\begin{alignat}{3}
&
\begin{pspicture}[shift=-0.3](-0.0,0)(2.4,1.1)
\pspolygon[fillstyle=solid,fillcolor=lightlightblue,linecolor=black,linewidth=0pt](0,0)(0,0.8)(2.4,0.8)(2.4,0)(0,0)
\psarc[linecolor=blue,linewidth=\elegant]{-}(1.6,0){0.2}{0}{180}
\psarc[linecolor=blue,linewidth=\elegant]{-}(0.8,0.8){0.2}{180}{0}
\psbezier[linecolor=blue,linewidth=\elegant]{-}(0.2,0)(0.2,0.4)(1.0,0.4)(1.0,0)
\psbezier[linecolor=blue,linewidth=\elegant]{-}(0.2,0.8)(0.3,0.2)(2.2,0.3)(2.2,0.8)
\pscircle[fillstyle=solid,fillcolor=black](1.4,0.8){0.035}
\pscircle[fillstyle=solid,fillcolor=black](1.8,0.8){0.035}
\pscircle[fillstyle=solid,fillcolor=black](2.2,0){0.035}
\pscircle[fillstyle=solid,fillcolor=black](0.6,0){0.035}
\psframe[fillstyle=solid,linecolor=white,linewidth=0pt](-0.04,0)(0,0.8)
\psframe[fillstyle=solid,linecolor=white,linewidth=0pt](2.4,0)(2.44,0.8)
\rput(0,0.8){
\psarc[linecolor=blue,linewidth=\elegant]{-}(0.8,0){0.2}{0}{180}
\psarc[linecolor=blue,linewidth=\elegant]{-}(0,0){0.2}{0}{90}
\psarc[linecolor=blue,linewidth=\elegant]{-}(2.4,0){0.2}{90}{180}
}
\end{pspicture} \ = \alpha\beta \ 
\begin{pspicture}[shift=0](0.0,0)(2.4,0.5)
\psline[linewidth=\mince](0,0)(2.4,0)
\pscircle[fillstyle=solid,fillcolor=black](2.2,0){0.035}
\pscircle[fillstyle=solid,fillcolor=black](0.6,0){0.035}
\psarc[linecolor=blue,linewidth=\elegant]{-}(1.6,0){0.2}{0}{180}
\psbezier[linecolor=blue,linewidth=\elegant]{-}(0.2,0)(0.2,0.4)(1.0,0.4)(1.0,0)
\end{pspicture}\ ,
\qquad\quad&
\begin{pspicture}[shift=-0.3](-0.0,0)(2.4,1.1)
\pspolygon[fillstyle=solid,fillcolor=lightlightblue,linecolor=black,linewidth=0pt](0,0)(0,0.8)(2.4,0.8)(2.4,0)(0,0)
\psarc[linecolor=blue,linewidth=\elegant]{-}(0.8,0){0.2}{0}{180}
\psbezier[linecolor=blue,linewidth=\elegant]{-}(1.4,0.8)(1.4,0.4)(2.2,0.4)(2.2,0)
\psbezier[linecolor=blue,linewidth=\elegant]{-}(0.2,0)(0.2,0.28)(0,0.39)(-0.02,0.43)
\psbezier[linecolor=blue,linewidth=\elegant]{-}(2.2,0.8)(2.2,0.52)(2.4,0.41)(2.42,0.37)
\pscircle[fillstyle=solid,fillcolor=black](1.4,0){0.035}
\pscircle[fillstyle=solid,fillcolor=black](1.8,0){0.035}
\pscircle[fillstyle=solid,fillcolor=black](0.2,0.8){0.035}
\pscircle[fillstyle=solid,fillcolor=black](0.6,0.8){0.035}
\pscircle[fillstyle=solid,fillcolor=black](1.0,0.8){0.035}
\pscircle[fillstyle=solid,fillcolor=black](1.8,0.8){0.035}
\psframe[fillstyle=solid,linecolor=white,linewidth=0pt](-0.04,0)(0,0.8)
\psframe[fillstyle=solid,linecolor=white,linewidth=0pt](2.4,0)(2.44,0.8)
\rput(0,0.8){
\psarc[linecolor=blue,linewidth=\elegant]{-}(1.2,0){0.2}{0}{180}
\pscircle[fillstyle=solid,fillcolor=black](0.6,0){0.04}
\psline[linecolor=blue,linewidth=\elegant]{-}(2.2,0)(2.2,0.4)
}
\end{pspicture}\ = 0,
\\[0.5cm]
&
\begin{pspicture}[shift=-0.3](-0.0,0)(2.4,1.1)
\pspolygon[fillstyle=solid,fillcolor=lightlightblue,linecolor=black,linewidth=0pt](0,0)(0,0.8)(2.4,0.8)(2.4,0)(0,0)
\psarc[linecolor=blue,linewidth=\elegant]{-}(2.4,0){0.2}{90}{180}
\psarc[linecolor=blue,linewidth=\elegant]{-}(0,0){0.2}{0}{90}
\psbezier[linecolor=blue,linewidth=\elegant]{-}(0.2,0.8)(0.2,0.4)(1.0,0.4)(1.0,0.8)
\psbezier[linecolor=blue,linewidth=\elegant]{-}(1.0,0)(1.0,0.4)(1.4,0.4)(1.4,0.8)
\psbezier[linecolor=blue,linewidth=\elegant]{-}(0.6,0)(0.6,0.4)(0,0.4)(-0.04,0.4)
\psbezier[linecolor=blue,linewidth=\elegant]{-}(1.8,0.8)(1.8,0.4)(2.4,0.4)(2.44,0.4)
\psframe[fillstyle=solid,linecolor=white,linewidth=0pt](-0.04,0)(0,0.8)
\psframe[fillstyle=solid,linecolor=white,linewidth=0pt](2.4,0)(2.44,0.8)
\pscircle[fillstyle=solid,fillcolor=black](1.4,0){0.035}
\pscircle[fillstyle=solid,fillcolor=black](1.8,0){0.035}
\pscircle[fillstyle=solid,fillcolor=black](0.6,0.8){0.035}
\pscircle[fillstyle=solid,fillcolor=black](2.2,0.8){0.035}
\rput(0,0.8){
\psarc[linecolor=blue,linewidth=\elegant]{-}(1.2,0){0.2}{0}{180}
\psline[linecolor=blue,linewidth=\elegant]{-}(0.2,0)(0.2,0.4)
\psline[linecolor=blue,linewidth=\elegant]{-}(1.8,0)(1.8,0.4)
}
\end{pspicture} \ = \omega^{-1} \ 
\begin{pspicture}[shift=0](0.0,0)(2.4,0.5)
\psline[linewidth=\mince](0,0)(2.4,0)
\psarc[linecolor=blue,linewidth=\elegant]{-}(2.4,0){0.2}{90}{180}
\psarc[linecolor=blue,linewidth=\elegant]{-}(0,0){0.2}{0}{90}
\psline[linecolor=blue,linewidth=\elegant]{-}(0.6,0)(0.6,0.4)
\psline[linecolor=blue,linewidth=\elegant]{-}(1.0,0)(1.0,0.4)
\pscircle[fillstyle=solid,fillcolor=black](1.4,0){0.035}
\pscircle[fillstyle=solid,fillcolor=black](1.8,0){0.035}
\end{pspicture}\ ,
\qquad\quad&
\begin{pspicture}[shift=-0.3](-0.0,0)(2.4,1.1)
\pspolygon[fillstyle=solid,fillcolor=lightlightblue,linecolor=black,linewidth=0pt](0,0)(0,0.8)(2.4,0.8)(2.4,0)(0,0)
\psarc[linecolor=blue,linewidth=\elegant]{-}(1.6,0){0.2}{0}{180}
\psbezier[linecolor=blue,linewidth=\elegant]{-}(0.2,0)(0.2,0.4)(1.0,0.4)(1.0,0)
\psbezier[linecolor=blue,linewidth=\elegant]{-}(0.2,0.8)(0.2,0.4)(1.0,0.4)(1.0,0.8)
\psbezier[linecolor=blue,linewidth=\elegant]{-}(1.4,0.8)(1.4,0.35)(0.1,0.34)(-0.02,0.44)
\psbezier[linecolor=blue,linewidth=\elegant]{-}(1.8,0.8)(1.8,0.5)(2.4,0.43)(2.42,0.41)
\pscircle[fillstyle=solid,fillcolor=black](0.6,0){0.035}
\pscircle[fillstyle=solid,fillcolor=black](2.2,0){0.035}
\pscircle[fillstyle=solid,fillcolor=black](0.6,0.8){0.035}
\pscircle[fillstyle=solid,fillcolor=black](2.2,0.8){0.035}
\psframe[fillstyle=solid,linecolor=white,linewidth=0pt](-0.04,0)(0,0.8)
\psframe[fillstyle=solid,linecolor=white,linewidth=0pt](2.4,0)(2.44,0.8)
\rput(0,0.8){
\psarc[linecolor=blue,linewidth=\elegant]{-}(1.6,0){0.2}{0}{180}
\psline[linecolor=blue,linewidth=\elegant]{-}(0.2,0)(0.2,0.4)
\psline[linecolor=blue,linewidth=\elegant]{-}(1.0,0)(1.0,0.4)
}
\end{pspicture}\ = 0.
\end{alignat}
\end{subequations}
This action on the basis states is linearly extended to all linear combinations of link states. We also note that the standard module with zero defects depends on $\alpha$ but not on $\omega$, so a more natural way to denote it would be $\repW_{N,0,\alpha}$. However, it is usual and useful for practical purposes to parameterize $\alpha$ as
\be
\label{eq:alpha.omega}
\alpha = \omega+\omega^{-1}
\ee 
and denote by $\repW_{N,0,\omega}$ the corresponding module.

The same rules defining the action of the algebra apply for the standard modules $\repV_{N,d}$ over $\dtl_N(\beta)$. In this case however, the representations are independent of $\alpha$ and $\omega$. These standard modules have dimensions given respectively by trinomial coefficients and Motzkin numbers~\cite{TrinomMotzkin} 
\be
\dim \repW_{N,d,\omega} = \binom{N}{d}_{\!\!2}, \qquad \dim \repV_{N,d} = \binom{N}{d}_{\!\!2} -  \binom{N}{d+2}_{\!\!2},
\ee
where the trinomial coefficients are defined by
\be
(x+1+x^{-1})^N = \sum_{k=-N}^N \binom{N}{k}_{\!\!2}\, x^k.
\ee
These dimensions are tabulated in \cref{tab:VW.dims} for $N=1,2, \dots, 7$.
\begin{table}
\begin{center}
$
\begin{array}{ccc}
\dim \repW_{N,d,\omega} & & \dim \repV_{N,d}
\\[0.2cm]
\begin{array}{c | cccccccccccc}
N\backslash d & 0 & 1 & 2 & 3 & 4 & 5 & 6 & 7
\\
\hline\\[-0.3cm]
1& 1 & 1 & & & & & & \\
2& 3 & 2 & 1 & & & & & \\
3& 7 & 6 & 3 & 1 & & & & \\
4& 19 & 16 & 10 & 4 & 1 & & & \\
5& 51 & 45 & 30 & 15 & 5 & 1 & & \\
6& 141 & 126 & 90 & 50 & 21 & 6 & 1 & \\
7& 393 & 357 & 266 & 161 & 77 & 28 & 7 & 1
\end{array}
&\quad &
\begin{array}{c | cccccccccccc}
N\backslash d & 0 & 1 & 2 & 3 & 4 & 5 & 6 & 7
\\
\hline\\[-0.3cm]
1&1 & 1 & & & & & & \\
2& 2 & 2 & 1 & & & & & \\
3& 4 & 5 & 3 & 1 & & & & \\
4& 9 & 12 & 9 & 4 & 1 & & & \\
5& 21 & 30 & 25 & 14 & 5 & 1 & & \\
6& 51 & 76 & 69 & 44 & 20 & 6 & 1 & \\
7& 127 & 196 & 189 & 133 & 70 & 27 & 7 & 1
\end{array}
\end{array}
$
\end{center}
\caption{The dimensions of the standard modules $\repW_{N,d,\omega}$ and $\repV_{N,d}$ for $N = 1,2, \dots, 7$.}
\label{tab:VW.dims}
\end{table}

\subsection{Markov traces}\label{sec:Markov}

The Markov traces~\cite{Jones83,RJ06,DJS09,MDPR13} for the dilute Temperley-Lieb algebras are linear maps 
\be
\mathcal F: \dtl_N(\beta) \to \mathbb C, \qquad
\mathcal F:\pdtl_N(\alpha,\beta) \to \mathbb C.
\ee
Roughly speaking, their purpose is to connect the top and bottom of the connectivity diagrams, and thus embed them in geometries with periodic boundary conditions in the vertical direction. In this way, the connectivity diagrams in $\dtl_N(\beta)$ and $\pdtl_N(\alpha,\beta)$ are embedded on a cylinder and a torus, respectively. Then $\mathcal F(c)$ outputs the product of the weights of the loops created by this embedding:
\be
\mathcal F(c) = \beta^{n_\beta}\times \left\{\begin{array}{cl}
\alpha^{n_{\alpha}} & c\in \dtl_N(\beta), \\[0.15cm]
\displaystyle\prod_{i \wedge j=1} \alpha_{i,j}^{n_{i,j}} & c \in \pdtl_N(\alpha,\beta), 
\end{array}\right.
\ee
where $n_\alpha$, $n_\beta$ and $n_{i,j}$ count the numbers of loops of each possible winding in $c$. 

Remarkably, the Markov traces can be realised in terms of weighted sums of traces over the standard modules. For $\dtl_N(\beta)$, generalizing the arguments used for the usual Temperley-Lieb algebra \cite{RJ06,DJS09}, we find
\be
\label{eq:cylinder.Markov.trace}
\mathcal F(c) = \sum_{d=0}^N U_d (\tfrac \alpha 2) \textrm{tr}_{\raisebox{-0.05cm}{\tiny$\repV_{N,d}$}} (c),
\ee
where $U_n(x)$ is the $n$-th Chebyshev polynomial of the second kind: $U_n(\cos \theta) = \sin\big((n\!+\!1) \theta\big)/\sin \theta$. Similarly for $\pdtl_N(\beta)$, supposing that $c$ is an arbitrary element of the algebra, its trace over the standard modules decomposes as
\be
\label{eq:trace.Cdj}
\textrm{tr}_{\raisebox{-0.05cm}{\tiny$\repW_{N,d,\omega}$}} (c) = \sum_{j = -\infty}^\infty  \omega^{-j} C_{d,j},
\ee
where the coefficients $C_{d,j}$ are independent of $\omega$. Any element of the algebra has only finitely many non-zero coefficients $C_{d,j}$, implying that the sum in \eqref{eq:trace.Cdj} is in fact finite. The Markov trace is then given by
\be
\label{eq:trace.torus}
\mathcal F(c) = \textrm{tr}_{\raisebox{-0.05cm}{\tiny$\repW_{N,0,\omega}$}} (c) \big|_{\alpha \to \alpha_{1,0}} + \sum_{d=1}^N \sum_{j=-\infty}^\infty 2\, T_{d \wedge j} (\tfrac12 \alpha_{\frac j{d \wedge j},\frac{d}{d \wedge j}})\, C_{d,j},
\ee
where $T_n(x)$ is the $n$-th Chebyshev polynomial of the first kind: $T_n(\cos \theta) = \cos n \theta$. Such trace formulas were previously obtained for the dense loop model with periodic boundary conditions in \cite{RJ07,AK10}, and \eqref{eq:trace.torus} extends this to the dilute case.

For both geometries, the only difference with the dense loop model is that the sums over $d$ run over all integers, not only those with the same parity as $N$. In \cref{sec:partition.functions}, we will apply the Markov traces to powers of transfer matrices to compute the partition function on the torus and cylinder. In particular, for the torus, we will consider the Markov trace applied to the single row transfer matrix to the power $M$, in which case the sums over $j$ in \eqref{eq:trace.Cdj} and \eqref{eq:trace.torus} can be restricted to run between $-M$ and $M$.

\subsection{Local relations and transfer matrices}

\paragraph{Local relations.}
The face operator \eqref{eq:face.op} is an element of the dilute Temperley-Lieb algebra $\dtl_N(\beta)$ and of its periodic incarnation $\pdtl_N(\alpha,\beta)$. It satisfies the Yang-Baxter equation and the inversion relation
\be
\psset{unit=0.6}
\begin{pspicture}[shift=-1.9](-1,-2)(2,2)
\pspolygon[fillstyle=solid,fillcolor=lightlightblue,linewidth=1pt](1,0)(0,1)(-1,0)(0,-1)\psarc[linewidth=0.025]{-}(0,-1){0.25}{45}{135}\rput(0,0){\small$u+v$}
\rput(1,1){\pspolygon[fillstyle=solid,fillcolor=lightlightblue,linewidth=1pt](1,0)(0,1)(-1,0)(0,-1)\psarc[linewidth=0.025]{-}(0,-1){0.25}{45}{135}\rput(0,0){\small$u$}}
\rput(1,-1){\pspolygon[fillstyle=solid,fillcolor=lightlightblue,linewidth=1pt](1,0)(0,1)(-1,0)(0,-1)\psarc[linewidth=0.025]{-}(0,-1){0.25}{45}{135}\rput(0,0){\small$v$}}
\psline[linewidth=1.5pt,linecolor=blue,linestyle=dashed,dash=2pt 2pt]{-}(1.5,-0.5)(1.5,0.5)
\end{pspicture}
\ = \
\begin{pspicture}[shift=-1.9](-1,-2)(2,2)
\rput(1,0){\pspolygon[fillstyle=solid,fillcolor=lightlightblue,linewidth=1pt](1,0)(0,1)(-1,0)(0,-1)\psarc[linewidth=0.025]{-}(0,-1){0.25}{45}{135}\rput(0,0){\small$u+v$}}
\rput(0,-1){\pspolygon[fillstyle=solid,fillcolor=lightlightblue,linewidth=1pt](1,0)(0,1)(-1,0)(0,-1)\psarc[linewidth=0.025]{-}(0,-1){0.25}{45}{135}\rput(0,0){\small$u$}}
\rput(0,1){\pspolygon[fillstyle=solid,fillcolor=lightlightblue,linewidth=1pt](1,0)(0,1)(-1,0)(0,-1)\psarc[linewidth=0.025]{-}(0,-1){0.25}{45}{135}\rput(0,0){\small$v$}}
\psline[linewidth=1.5pt,linecolor=blue,linestyle=dashed,dash=2pt 2pt]{-}(-0.5,-0.5)(-0.5,0.5)
\end{pspicture}\ \ ,
\qquad \qquad
\begin{pspicture}[shift=-1.9](-1,-2)(1,2)
\rput(0,-1){\pspolygon[fillstyle=solid,fillcolor=lightlightblue,linewidth=1pt](1,0)(0,1)(-1,0)(0,-1)\psarc[linewidth=0.025]{-}(0,-1){0.25}{45}{135}\rput(0,0){\small$u$}}
\rput(0,1){\pspolygon[fillstyle=solid,fillcolor=lightlightblue,linewidth=1pt](1,0)(0,1)(-1,0)(0,-1)\psarc[linewidth=0.025]{-}(0,-1){0.25}{45}{135}\rput(0,0){\small$-u$}}
\psline[linewidth=1.5pt,linecolor=blue,linestyle=dashed,dash=2pt 2pt]{-}(-0.5,-0.5)(-0.5,0.5)
\psline[linewidth=1.5pt,linecolor=blue,linestyle=dashed,dash=2pt 2pt]{-}(0.5,-0.5)(0.5,0.5)
\end{pspicture}
\ = \rho_8(u)\rho_8(-u)\ \ 
\begin{pspicture}[shift=-.90](-1,-1)(1,1)
\pspolygon[fillstyle=solid,fillcolor=lightlightblue,linewidth=1pt](1,0)(0,1)(-1,0)(0,-1)
\psarc[linecolor=blue,linewidth=\elegant,linestyle=dashed,dash=2pt 2pt](1,0){0.71}{135}{225}
\psarc[linecolor=blue,linewidth=\elegant,linestyle=dashed,dash=2pt 2pt](-1,0){0.71}{-45}{45}
\end{pspicture}\ \ ,
\ee
where the dashed loop segment is the identity strand of the dilute Temperley-Lieb algebra
\be
\label{eq:id1}
\begin{pspicture}[shift=-0.4](-0.2,-0.5)(0.2,0.5)
\pspolygon[fillstyle=solid,fillcolor=lightlightblue,linewidth=1pt](-0.2,-0.5)(0.2,-0.5)(0.2,0.5)(-0.2,0.5)
\psline[linewidth=1.5pt,linecolor=blue,linestyle=dashed,dash=2pt 2pt]{-}(0,-0.5)(0,0.5)
\end{pspicture} 
\ = \
\begin{pspicture}[shift=-0.4](-0.2,-0.5)(0.2,0.5)
\pspolygon[fillstyle=solid,fillcolor=lightlightblue,linewidth=1pt](-0.2,-0.5)(0.2,-0.5)(0.2,0.5)(-0.2,0.5)
\psline[linewidth=1.5pt,linecolor=blue]{-}(0,-0.5)(0,0.5)
\end{pspicture} 
\ + \
\begin{pspicture}[shift=-0.4](-0.2,-0.5)(0.2,0.5)
\pspolygon[fillstyle=solid,fillcolor=lightlightblue,linewidth=1pt](-0.2,-0.5)(0.2,-0.5)(0.2,0.5)(-0.2,0.5)
\pscircle[fillstyle=solid,fillcolor=black](0,-0.5){0.05}
\pscircle[fillstyle=solid,fillcolor=black](0,0.5){0.05}
\end{pspicture}  \ \ .
\ee
Likewise, the boundary face operators \eqref{eq:bdy.face.op} are elements of $\dtl_N(\beta)$ and satisfy the boundary Yang-Baxter equation
\be
\psset{unit=0.7cm}
\begin{pspicture}[shift=-2.4](0,-2)(2,3)
\rput(2,0){\pspolygon[fillstyle=solid,fillcolor=lightlightblue,linewidth=1pt](0,1)(-1,0)(0,-1)\rput(-0.35,0){\small$u$}}
\rput(2,2){\pspolygon[fillstyle=solid,fillcolor=lightlightblue,linewidth=1pt](0,1)(-1,0)(0,-1)\rput(-0.35,0){\small$v$}}
\rput(1,1){\pspolygon[fillstyle=solid,fillcolor=lightlightblue,linewidth=1pt](1,0)(0,1)(-1,0)(0,-1)\psarc[linewidth=0.025]{-}(0,-1){0.25}{45}{135}\rput(0,0){\small$u+v$}}
\rput(1,-1){\pspolygon[fillstyle=solid,fillcolor=lightlightblue,linewidth=1pt](1,0)(0,1)(-1,0)(0,-1)\psarc[linewidth=0.025]{-}(0,-1){0.25}{45}{135}\rput(0,0){\small$u-v$}}
\psline[linewidth=1.5pt,linecolor=blue,linestyle=dashed,dash=2pt 2pt]{-}(0.5,-0.5)(0.5,0.5)
\end{pspicture} 
\ \ = \ \ 
\begin{pspicture}[shift=-2.4](0,-1)(2,4)
\rput(2,0){\pspolygon[fillstyle=solid,fillcolor=lightlightblue,linewidth=1pt](0,1)(-1,0)(0,-1)\rput(-0.35,0){\small$v$}}
\rput(2,2){\pspolygon[fillstyle=solid,fillcolor=lightlightblue,linewidth=1pt](0,1)(-1,0)(0,-1)\rput(-0.35,0){\small$u$}}
\rput(1,1){\pspolygon[fillstyle=solid,fillcolor=lightlightblue,linewidth=1pt](1,0)(0,1)(-1,0)(0,-1)\psarc[linewidth=0.025]{-}(0,-1){0.25}{45}{135}\rput(0,0){\small$u+v$}}
\rput(1,3){\pspolygon[fillstyle=solid,fillcolor=lightlightblue,linewidth=1pt](1,0)(0,1)(-1,0)(0,-1)\psarc[linewidth=0.025]{-}(0,-1){0.25}{45}{135}\rput(0,0){\small$u-v$}}
\psline[linewidth=1.5pt,linecolor=blue,linestyle=dashed,dash=2pt 2pt]{-}(0.5,1.5)(0.5,2.5)
\end{pspicture}\ \ .
\ee

\paragraph{Transfer matrices.}
The single row transfer matrix is an element of $\pdtl_N(\alpha,\beta)$ defined by
\be
\psset{unit=0.9cm}
\Tb(u)
= \ 
\begin{pspicture}[shift=-0.9](-0.3,-0.5)(5.3,1.2)
\facegrid{(0,0)}{(5,1)}
\psarc[linewidth=0.025]{-}(0,0){0.16}{0}{90}
\psarc[linewidth=0.025]{-}(1,0){0.16}{0}{90}
\psarc[linewidth=0.025]{-}(2,0){0.16}{0}{90}
\psarc[linewidth=0.025]{-}(4,0){0.16}{0}{90}
\psline[linewidth=1.5pt,linecolor=blue,linestyle=dashed,dash=2pt 2pt]{-}(0,0.5)(-0.3,0.5)
\psline[linewidth=1.5pt,linecolor=blue,linestyle=dashed,dash=2pt 2pt]{-}(5,0.5)(5.3,0.5)
\rput(0.5,.5){$u$}
\rput(1.5,.5){$u$}
\rput(2.5,0.5){$u$}
\rput(3.5,0.5){$\ldots$}
\rput(4.5,.5){$u$}
\psline{<->}(0,-0.2)(5,-0.2)\rput(2.5,-0.45){$_N$}
\end{pspicture}\ \ .
\ee
The periodic boundary condition is reflected by the identification of the left and right edges with the dashed strand. Likewise, the double row transfer matrix is the element of $\dtl_N(\beta)$ defined as
\be
\psset{unit=0.9cm}
\Db(u)
= \ 
\begin{pspicture}[shift=-1.4](-1,-0.5)(6,2.0)
\psline[linewidth=1.5pt,linecolor=blue,linestyle=dashed,dash=2pt 2pt]{-}(-0.7,1.5)(5.7,1.5)
\psline[linewidth=1.5pt,linecolor=blue,linestyle=dashed,dash=2pt 2pt]{-}(-0.7,0.5)(5.7,0.5)
\facegrid{(0,0)}{(5,2)}
\multirput(0,0)(1,0){5}{\psarc[linewidth=0.025]{-}(0,0){0.16}{0}{90}}
\multirput(0,0)(1,0){5}{\psarc[linewidth=0.025]{-}(1,1){0.16}{90}{180}}
\rput(0.5,.5){$u$}
\rput(1.5,.5){$u$}
\rput(2.5,0.5){$u$}
\rput(3.5,0.5){$\ldots$}
\rput(4.5,.5){$u$}
\rput(0.5,1.5){$u$}
\rput(1.5,1.5){$u$}
\rput(2.5,1.5){$u$}
\rput(3.5,1.5){$\ldots$}
\rput(4.5,1.5){$u$}
\rput(5,1){\pspolygon[fillstyle=solid,fillcolor=lightlightblue,linewidth=0.75pt](0,0)(1,1)(1,-1)\rput(0.65,0){$u$}}
\rput(0,1){\pspolygon[fillstyle=solid,fillcolor=lightlightblue,linewidth=0.75pt](0,0)(-1,1)(-1,-1)\rput(-0.54,0){\scriptsize$3\lambda\!-\!u$}}
\psline{<->}(0,-0.2)(5,-0.2)\rput(2.5,-0.45){$_N$}
\end{pspicture}\ \ .
\ee
As a result of the local relations given above, the transfer matrices obey the commutation relations 
\be
[\Tb(u),\Tb(v)]=0, \qquad [\Db(u),\Db(v)]=0.
\ee
The diagrammatic proof of these relations is done in the planar dilute Temperley-Lieb algebra following the usual arguments \cite{BPO96,PRZ2006}. It can also be performed without resorting to the planar algebra, but at the cost of writing down three independent Yang-Baxter equations \cite{PRVO20}.

\subsection[Site percolation as the dilute $\Atwotwo$ model at $\lambda = \frac\pi3$]{Site percolation as the dilute $\boldsymbol{\Atwotwo}$ model at $\boldsymbol{\lambda = \frac\pi3}$}\label{sec:perco.and.A22}

\paragraph{Local operators.}
In \cref{sec:PercToLoops}, we found that the configurations of site percolation could be mapped to dilute loop configurations on the square lattice. To compute the partition function of site percolation, one must also relate the weights of the percolation configurations with those of the $\Atwotwo$ loop model. In this respect, each of the occurring eight tiles in \eqref{eq:8Tiles1} should be assigned a unit weight, whereas the ninth tile should have a vanishing weight. Moreover, each contractible loop should have the fugacity $\beta = 1$.  This corresponds to the special point $\lambda = u = \frac \pi 3$ of the $\Atwotwo$ loop model.

We stress that the face operator \eqref{eq:face.op} is normalised differently compared to \cite{MDP19}, wherein overall factors of $\sin2\lambda \sin 3\lambda$ were included in the denominators of the Boltzmann weights. With these factors removed, the limit $\lambda \to \frac \pi 3$ of the face operator is well-defined. Moreover, in this limit, there is an overall factor of $s(u)$ in the numerator which we remove to define the suitably normalised face operator
\begin{alignat}{2}
\label{eq:face.op.2}
\psset{unit=0.8}
\begin{pspicture}[shift=-.40](0,0)(1,1)
\facegrid{(0,0)}{(1,1)}
\psline[linewidth=0.025]{-}(0,0.15)(0.15,0.15)(0.15,0)
\rput(.5,.5){$u$}
\end{pspicture}
\ = 
\lim_{\lambda \to \frac \pi 3} \frac1{s(u)} \ 
\begin{pspicture}[shift=-.40](0,0)(1,1)
\facegrid{(0,0)}{(1,1)}
\psarc[linewidth=0.025]{-}(0,0){0.16}{0}{90}
\rput(.5,.5){$u$}
\end{pspicture} 
\
&\psset{unit=0.8}
= s(u) \ 
\begin{pspicture}[shift=-.40](0,0)(1,1)
\facegrid{(0,0)}{(1,1)}
\rput[bl](0,0){\loopa}
\pscircle[fillstyle=solid,fillcolor=black](0.5,0){0.06}
\pscircle[fillstyle=solid,fillcolor=black](1,0.5){0.06}
\pscircle[fillstyle=solid,fillcolor=black](0.5,1){0.06}
\pscircle[fillstyle=solid,fillcolor=black](0,0.5){0.06}
\end{pspicture}
\ + \  
\begin{pspicture}[shift=-.40](0,0)(1,1)
\facegrid{(0,0)}{(1,1)}
\rput[bl](0,0){\loopb}
\pscircle[fillstyle=solid,fillcolor=black](0.5,0){0.06}
\pscircle[fillstyle=solid,fillcolor=black](1,0.5){0.06}
\end{pspicture}
\ + \  
\begin{pspicture}[shift=-.40](0,0)(1,1)
\facegrid{(0,0)}{(1,1)}
\rput[bl](0,0){\loopc}
\pscircle[fillstyle=solid,fillcolor=black](0.5,1){0.06}
\pscircle[fillstyle=solid,fillcolor=black](0,0.5){0.06}
\end{pspicture}
\ + \ 
\begin{pspicture}[shift=-.40](0,0)(1,1)
\facegrid{(0,0)}{(1,1)}
\rput[bl](0,0){\loopd}
\pscircle[fillstyle=solid,fillcolor=black](1,0.5){0.06}
\pscircle[fillstyle=solid,fillcolor=black](0.5,1){0.06}
\end{pspicture}
\ + \ 
\begin{pspicture}[shift=-.40](0,0)(1,1)
\facegrid{(0,0)}{(1,1)}
\rput[bl](0,0){\loope}
\pscircle[fillstyle=solid,fillcolor=black](0.5,0){0.06}
\pscircle[fillstyle=solid,fillcolor=black](0,0.5){0.06}
\end{pspicture}
\nonumber \\[0.2cm]&\psset{unit=0.8}
+ s(u)\  
\begin{pspicture}[shift=-.40](0,0)(1,1)
\facegrid{(0,0)}{(1,1)}
\rput[bl](0,0){\loopf}
\pscircle[fillstyle=solid,fillcolor=black](1,0.5){0.06}
\pscircle[fillstyle=solid,fillcolor=black](0,0.5){0.06}
\end{pspicture}
+ s(u)\
\begin{pspicture}[shift=-.40](0,0)(1,1)
\facegrid{(0,0)}{(1,1)}
\rput[bl](0,0){\loopg}
\pscircle[fillstyle=solid,fillcolor=black](0.5,0){0.06}
\pscircle[fillstyle=solid,fillcolor=black](0.5,1){0.06}
\end{pspicture}
\ + s(u+\tfrac \pi 3)\ 
\begin{pspicture}[shift=-.40](0,0)(1,1)
\facegrid{(0,0)}{(1,1)}
\rput[bl](0,0){\looph}
\end{pspicture}
\ + s(u-\tfrac \pi 3)\ 
\begin{pspicture}[shift=-.40](0,0)(1,1)
\facegrid{(0,0)}{(1,1)}
\rput[bl](0,0){\loopi}
\end{pspicture}\ .
\end{alignat}
At $u=\tfrac\pi 3$, the weight of the last tile vanishes, and the weights of the remaining eight tiles are all equal to $1$. In addition, the face operator at $u= \lambda = \frac \pi 3$ factorises as a product of two triangle operators:
\be
\label{eq:face.op.pi/3}
\psset{unit=0.8}
\begin{pspicture}[shift=-.60](0,-.2)(1,1.2)
\facegrid{(0,0)}{(1,1)}
\psline[linewidth=0.025]{-}(0,0.15)(0.15,0.15)(0.15,0)
\rput(.5,.5){$\frac{\pi}3$}
\end{pspicture}
\ = \ 
\begin{pspicture}[shift=-.60](0,-.2)(1,1.2)
\facegrid{(0,0)}{(1,1)}
\psline{-}(0,0)(1,1)
\rput(0.7,0.3){\specialcircle{0.05}}
\rput(0.3,0.7){\specialcircle{0.05}}
\end{pspicture} 
\ = \ 
\begin{pspicture}[shift=-.60](0,-.2)(1,1.2)
\facegrid{(0,0)}{(1,1)}
\rput[bl](0,0){\loopa}
\pscircle[fillstyle=solid,fillcolor=black](0.5,0){0.06}
\pscircle[fillstyle=solid,fillcolor=black](1,0.5){0.06}
\pscircle[fillstyle=solid,fillcolor=black](0.5,1){0.06}
\pscircle[fillstyle=solid,fillcolor=black](0,0.5){0.06}
\end{pspicture} 
\ + \ 
\begin{pspicture}[shift=-.60](0,-.2)(1,1.2)
\facegrid{(0,0)}{(1,1)}
\rput[bl](0,0){\loopb}
\pscircle[fillstyle=solid,fillcolor=black](0.5,0){0.06}
\pscircle[fillstyle=solid,fillcolor=black](1,0.5){0.06}
\end{pspicture} 
\ + \ 
\begin{pspicture}[shift=-.60](0,-.2)(1,1.2)
\facegrid{(0,0)}{(1,1)}
\rput[bl](0,0){\loopc}
\pscircle[fillstyle=solid,fillcolor=black](0.5,1){0.06}
\pscircle[fillstyle=solid,fillcolor=black](0,0.5){0.06}
\end{pspicture} 
\ + \ 
\begin{pspicture}[shift=-.60](0,-.2)(1,1.2)
\facegrid{(0,0)}{(1,1)}
\rput[bl](0,0){\loopd}
\pscircle[fillstyle=solid,fillcolor=black](1,0.5){0.06}
\pscircle[fillstyle=solid,fillcolor=black](0.5,1){0.06}
\end{pspicture} 
\ + \ 
\begin{pspicture}[shift=-.60](0,-.2)(1,1.2)
\facegrid{(0,0)}{(1,1)}
\rput[bl](0,0){\loope}
\pscircle[fillstyle=solid,fillcolor=black](0.5,0){0.06}
\pscircle[fillstyle=solid,fillcolor=black](0,0.5){0.06}
\end{pspicture} 
\ + \ 
\begin{pspicture}[shift=-.60](0,-.2)(1,1.2)
\facegrid{(0,0)}{(1,1)}
\rput[bl](0,0){\loopf}
\pscircle[fillstyle=solid,fillcolor=black](1,0.5){0.06}
\pscircle[fillstyle=solid,fillcolor=black](0,0.5){0.06}
\end{pspicture} 
\ + \ 
\begin{pspicture}[shift=-.60](0,-.2)(1,1.2)
\facegrid{(0,0)}{(1,1)}
\rput[bl](0,0){\loopg}
\pscircle[fillstyle=solid,fillcolor=black](0.5,0){0.06}
\pscircle[fillstyle=solid,fillcolor=black](0.5,1){0.06}
\end{pspicture} 
\ + \ 
\begin{pspicture}[shift=-.60](0,-.2)(1,1.2)
\facegrid{(0,0)}{(1,1)}
\rput[bl](0,0){\looph}
\end{pspicture}\ \ ,
\ee
where
\be
\begin{pspicture}[shift=-.40](0,-.2)(1,1)
\pspolygon[fillstyle=solid,fillcolor=lightlightblue](0,0)(0.5,0.866)(1,0)
\rput(0.5,0.3){\specialcircle{0.05}}
\end{pspicture} 
\ = \
\begin{pspicture}[shift=-.40](0,-.2)(1,1)
\pspolygon[fillstyle=solid,fillcolor=lightlightblue](0,0)(0.5,0.866)(1,0)
\pscircle[fillstyle=solid,fillcolor=black](0.5,0){0.05}
\pscircle[fillstyle=solid,fillcolor=black](0.25,0.433){0.05}
\pscircle[fillstyle=solid,fillcolor=black](0.75,0.433){0.05}
\end{pspicture} 
\ + \ 
\begin{pspicture}[shift=-.40](0,-.2)(1,1)
\pspolygon[fillstyle=solid,fillcolor=lightlightblue](0,0)(0.5,0.866)(1,0)
\psarc[linewidth=1.5pt,linecolor=blue](0.5,0.866){.5}{-120}{-60}
\pscircle[fillstyle=solid,fillcolor=black](0.5,0){0.05}
\end{pspicture} 
\ + \
\begin{pspicture}[shift=-.40](0,-.2)(1,1)
\pspolygon[fillstyle=solid,fillcolor=lightlightblue](0,0)(0.5,0.866)(1,0)
\psarc[linewidth=1.5pt,linecolor=blue](1,0){.5}{120}{180}
\pscircle[fillstyle=solid,fillcolor=black](0.25,0.433){0.05}
\end{pspicture} 
\ + \
\begin{pspicture}[shift=-.40](0,-.2)(1,1)
\pspolygon[fillstyle=solid,fillcolor=lightlightblue](0,0)(0.5,0.866)(1,0)
\psarc[linewidth=1.5pt,linecolor=blue](0,0){.5}{0}{60}
\pscircle[fillstyle=solid,fillcolor=black](0.75,0.433){0.05}
\end{pspicture} \ .
\ee
These are precisely the weights needed for site percolation. Likewise, the boundary face operator at $\lambda = \frac \pi 3$ simplifies to 
\be
\label{eq:bdy.face.op}
\psset{unit=0.7cm}
\begin{pspicture}[shift=-.90](0,-1)(1,1)
\pspolygon[fillstyle=solid,fillcolor=lightlightblue,linewidth=0.75pt](0,0)(1,1)(1,-1)
\rput(0.65,0){$u$}
\end{pspicture}
\ = \left\{
\begin{array}{ll}
\cos u 
\left( \ \ 
\begin{pspicture}[shift=-.90](0,-1)(1,1)
\pspolygon[fillstyle=solid,fillcolor=lightlightblue,linewidth=0.75pt](0,0)(1,1)(1,-1)
\psarc[linecolor=blue,linewidth=1.5pt]{-}(0,0){0.61}{-90}{90}
\end{pspicture}
\ \ + \ \ 
\begin{pspicture}[shift=-.90](0,-1)(1,1)
\pspolygon[fillstyle=solid,fillcolor=lightlightblue,linewidth=0.75pt](0,0)(1,1)(1,-1)
\pscircle[fillstyle=solid,fillcolor=black](0.5,0.5){0.07}
\pscircle[fillstyle=solid,fillcolor=black](0.5,-0.5){0.07}
\end{pspicture} \ \ \right) \quad& \textrm{choice 1,}
\\[0.8cm]
\sin u 
\left( \ \ 
\begin{pspicture}[shift=-.90](0,-1)(1,1)
\pspolygon[fillstyle=solid,fillcolor=lightlightblue,linewidth=0.75pt](0,0)(1,1)(1,-1)
\psarc[linecolor=blue,linewidth=1.5pt]{-}(0,0){0.61}{-90}{90}
\end{pspicture}
\ \ - \ \ 
\begin{pspicture}[shift=-.90](0,-1)(1,1)
\pspolygon[fillstyle=solid,fillcolor=lightlightblue,linewidth=0.75pt](0,0)(1,1)(1,-1)
\pscircle[fillstyle=solid,fillcolor=black](0.5,0.5){0.07}
\pscircle[fillstyle=solid,fillcolor=black](0.5,-0.5){0.07}
\end{pspicture} \ \ \right) \quad& \textrm{choice 2.}
\end{array}\right.
\ee
Hereafter, we select choice 1 with equal Boltzmann weights, for which the boundary operator is proportional the identity strand \eqref{eq:id1}. 

\paragraph{Transfer matrices.}
With a suitable change of normalisation, the commuting single and double row transfer matrices are
\begin{subequations}
\label{eq:DThat}
\begin{alignat}{3}
\Tbh(u) &= \ \
\psset{unit=0.9cm}
\begin{pspicture}[shift=-0.9](-0.3,-0.5)(5.3,1.0)
\facegrid{(0,0)}{(5,1)}
\psline[linewidth=0.025]{-}(0,0.15)(0.15,0.15)(0.15,0)
\rput(1,0){\psline[linewidth=0.025]{-}(0,0.15)(0.15,0.15)(0.15,0)}
\rput(2,0){\psline[linewidth=0.025]{-}(0,0.15)(0.15,0.15)(0.15,0)}
\rput(4,0){\psline[linewidth=0.025]{-}(0,0.15)(0.15,0.15)(0.15,0)}
\psline[linewidth=1.5pt,linecolor=blue,linestyle=dashed,dash=2pt 2pt]{-}(0,0.5)(-0.3,0.5)
\psline[linewidth=1.5pt,linecolor=blue,linestyle=dashed,dash=2pt 2pt]{-}(5,0.5)(5.3,0.5)
\rput(0.5,.5){$u$}
\rput(1.5,.5){$u$}
\rput(2.5,0.5){$u$}
\rput(3.5,0.5){$\ldots$}
\rput(4.5,.5){$u$}
\psline{<->}(0,-0.2)(5,-0.2)\rput(2.5,-0.45){$_N$}
\end{pspicture}\ 
&&=\lim_{\lambda \to \frac \pi 3} \frac{1}{f(u)}\, \Tb(u), 
\\[0.4cm]
\Dbh(u) &= \ \
\psset{unit=0.9cm}
\begin{pspicture}[shift=-1.4](-0.3,-0.5)(5.3,2.0)
\facegrid{(0,0)}{(5,2)}
\psline[linewidth=0.025]{-}(0,0.15)(0.15,0.15)(0.15,0)
\rput(1,0){\psline[linewidth=0.025]{-}(0,0.15)(0.15,0.15)(0.15,0)}
\rput(2,0){\psline[linewidth=0.025]{-}(0,0.15)(0.15,0.15)(0.15,0)}
\rput(4,0){\psline[linewidth=0.025]{-}(0,0.15)(0.15,0.15)(0.15,0)}
\multirput(1,1)(1,0){3}{\psline[linewidth=0.025]{-}(0,0.15)(-0.15,0.15)(-0.15,0)}
\rput(5,1){\psline[linewidth=0.025]{-}(0,0.15)(-0.15,0.15)(-0.15,0)}
\psarc[linewidth=1.5pt,linecolor=blue,linestyle=dashed,dash=2pt 2pt]{-}(0,1){.5}{90}{270}
\psarc[linewidth=1.5pt,linecolor=blue,linestyle=dashed,dash=2pt 2pt]{-}(5,1){.5}{270}{90}
\rput(0.5,.5){$u$}
\rput(1.5,.5){$u$}
\rput(2.5,0.5){$u$}
\rput(3.5,0.5){$\ldots$}
\rput(4.5,.5){$u$}
\rput(0.5,1.5){$u$}
\rput(1.5,1.5){$u$}
\rput(2.5,1.5){$u$}
\rput(3.5,1.5){$\ldots$}
\rput(4.5,1.5){$u$}
\psline{<->}(0,-0.2)(5,-0.2)\rput(2.5,-0.45){$_N$}
\end{pspicture}\ \ 
&&= -\lim_{\lambda \to \frac \pi 3} \frac{1}{f(u) \cos^2u}\, \Db(u), \label{redDTM}
\end{alignat}
\end{subequations}
where
\be
\label{eq:fk}
f(u) = \left\{\begin{array}{ll}
s(u)^N&\mbox{single row,}\\[0.1cm]
s(u)^{2N}&\mbox{double row.}
\end{array}\right.
\ee
Specializing to $u = \frac \pi 3$ and using \eqref{eq:face.op.pi/3}, we obtain 
\be
\Tbh(\tfrac \pi 3) = \ 
\psset{unit=0.9cm}
\begin{pspicture}[shift=-0.9](-0.2,-0.5)(5.2,1)
\facegrid{(0,0)}{(5,1)}
\multiput(0,0)(1,0){3}{\psline{-}(0,0)(1,1)\rput(0.3,0.7){\specialcircle{0.05}}\rput(0.7,0.3){\specialcircle{0.05}}}
\rput(4,0){\psline{-}(0,0)(1,1)\rput(0.3,0.7){\specialcircle{0.05}}\rput(0.7,0.3){\specialcircle{0.05}}}
\rput(3.5,0.5){$\ldots$}
\psline{<->}(0,-0.2)(5,-0.2)\rput(2.5,-0.45){$_N$}
\psline[linewidth=1.5pt,linecolor=blue,linestyle=dashed,dash=2pt 2pt]{-}(0,0.5)(-0.3,0.5)
\psline[linewidth=1.5pt,linecolor=blue,linestyle=dashed,dash=2pt 2pt]{-}(5,0.5)(5.3,0.5)
\end{pspicture} \ \ ,
\qquad
\Dbh(\tfrac \pi 3) = \ 
\begin{pspicture}[shift=-1.4](-0.5,-0.5)(5.5,2.0)
\facegrid{(0,0)}{(5,2)}
\psarc[linewidth=1.5pt,linecolor=blue,linestyle=dashed,dash=2pt 2pt]{-}(0,1){.5}{90}{270}
\psarc[linewidth=1.5pt,linecolor=blue,linestyle=dashed,dash=2pt 2pt]{-}(5,1){.5}{270}{90}\multiput(0,0)(1,0){3}{\psline{-}(0,0)(1,1)\rput(0.3,0.7){\specialcircle{0.05}}\rput(0.7,0.3){\specialcircle{0.05}}}
\rput(4,0){\psline{-}(0,0)(1,1)\rput(0.3,0.7){\specialcircle{0.05}}\rput(0.7,0.3){\specialcircle{0.05}}}
\multiput(0,1)(1,0){3}{\psline{-}(0,1)(1,0)\rput(0.3,0.3){\specialcircle{0.05}}\rput(0.7,0.7){\specialcircle{0.05}}}
\rput(4,1){\psline{-}(0,1)(1,0)\rput(0.3,0.3){\specialcircle{0.05}}\rput(0.7,0.7){\specialcircle{0.05}}}
\rput(3.5,0.5){$\ldots$}
\rput(3.5,1.5){$\ldots$}
\psline{<->}(0,-0.2)(5,-0.2)\rput(2.5,-0.45){$_N$}
\end{pspicture}\ \ .
\ee
As can be seen from \cref{fig:the.map}, the transfer matrix relevant for the torus partition function is indeed $\Tbh(\tfrac \pi 3)$. However, the transfer matrix relevant for the cylinder partition function is not exactly $\Dbh(\tfrac \pi 3)$, since the boundary conditions are identity strands. They should instead consist of vacancies on the left end and of triangular tiles with vacancies on the right end. To resolve this issue, we note the two identities 
\be
\psset{unit=0.8cm}
\begin{pspicture}[shift=-0.9](0,0)(1,2.0)
\pspolygon[fillstyle=solid,fillcolor=lightlightblue](0,0)(0,1)(1,1)
\pspolygon[fillstyle=solid,fillcolor=lightlightblue](0,2)(0,1)(1,1)
\pscircle[fillstyle=solid,fillcolor=black](0,0.5){0.07}
\pscircle[fillstyle=solid,fillcolor=black](0,1.5){0.07}
\rput(0.3,0.7){\specialcircle{0.05}}
\rput(0.3,1.3){\specialcircle{0.05}}
\end{pspicture}
 \ \ = \frac12 \ \ 
 \begin{pspicture}[shift=-0.9](-0.5,0)(1,2.0)
\pspolygon[fillstyle=solid,fillcolor=lightlightblue](0,0)(0,1)(1,1)
\pspolygon[fillstyle=solid,fillcolor=lightlightblue](0,2)(0,1)(1,1)
\psarc[linecolor=blue,linewidth=1.5pt,linestyle=dashed,dash=2pt 2pt]{-}(0,1){0.5}{90}{-90}
\rput(0.3,0.7){\specialcircle{0.05}}
\rput(0.3,1.3){\specialcircle{0.05}}
\end{pspicture}
\ \ ,
\qquad\quad
\ \ 
\begin{pspicture}[shift=-0.9](0,0)(1,2.0)
\pspolygon[fillstyle=solid,fillcolor=lightlightblue](0,0)(0,1)(1,1)
\pspolygon[fillstyle=solid,fillcolor=lightlightblue](0,2)(0,1)(1,1)
\pscircle[fillstyle=solid,fillcolor=black](0.5,0.5){0.07}
\pscircle[fillstyle=solid,fillcolor=black](0.5,1.5){0.07}
\rput(0.3,0.7){\specialcircle{0.05}}
\rput(0.3,1.3){\specialcircle{0.05}}
\end{pspicture}
 \ \ = \ \ 
\begin{pspicture}[shift=-0.9](0,0)(0.5,2.0)
\psline(0,0)(0,2)
\psarc[linecolor=blue,linewidth=1.5pt,linestyle=dashed,dash=2pt 2pt]{-}(0,1){0.5}{-90}{90}
\end{pspicture}\ \ .
\ee
The correct transfer matrix for the cylinder partition function is then
\be
\label{eq:DbmDbh}
\Dbm = \ \
\psset{unit=0.9cm}
\begin{pspicture}[shift=-1.4](0,-0.5)(6,2.0)
\facegrid{(0,0)}{(5,2)}
\pspolygon[fillstyle=solid,fillcolor=lightlightblue](5,0)(5,1)(6,1)
\pspolygon[fillstyle=solid,fillcolor=lightlightblue](5,2)(5,1)(6,1)
\pscircle[fillstyle=solid,fillcolor=black](0,0.5){0.07}
\pscircle[fillstyle=solid,fillcolor=black](0,1.5){0.07}
\pscircle[fillstyle=solid,fillcolor=black](5.5,0.5){0.07}
\pscircle[fillstyle=solid,fillcolor=black](5.5,1.5){0.07}
\multiput(0,0)(1,0){3}{\psline{-}(0,0)(1,1)\rput(0.3,0.7){\specialcircle{0.05}}\rput(0.7,0.3){\specialcircle{0.05}}}
\rput(4,0){\psline{-}(0,0)(1,1)\rput(0.3,0.7){\specialcircle{0.05}}\rput(0.7,0.3){\specialcircle{0.05}}}
\multiput(0,1)(1,0){3}{\psline{-}(0,1)(1,0)\rput(0.3,0.3){\specialcircle{0.05}}\rput(0.7,0.7){\specialcircle{0.05}}}
\rput(4,1){\psline{-}(0,1)(1,0)\rput(0.3,0.3){\specialcircle{0.05}}\rput(0.7,0.7){\specialcircle{0.05}}}
\rput(5.3,0.7){\specialcircle{0.05}}
\rput(5.3,1.3){\specialcircle{0.05}}
\rput(3.5,0.5){$\ldots$}
\rput(3.5,1.5){$\ldots$}
\psline{<->}(0,-0.2)(5,-0.2)\rput(2.5,-0.45){$_N$}
\end{pspicture} 
\ = \tfrac12 \widehat \Db(\tfrac \pi 3).
\ee
This confirms that the transfer matrices $\Tbm$ and $\Dbm$ needed to compute the partition functions for site percolation are respectively elements of the commuting families $\Tbh(u)$ and $\Dbh(u)$, and thus that the model is Yang-Baxter integrable on both geometries.

\paragraph{Stochasticity and simple eigenvalues.} 
The transfer matrices of critical bond percolation have some eigenvalues that are simple. For the single row transfer matrix, this occurs for even system sizes in the standard module with $d=0$ and $\alpha=1$, and for odd system sizes for $d=1$ and $\omega=1$. Likewise, for the double row transfer matrix, there are simple eigenvalues in the standard modules with $d=0$ and $d=1$. After suitable normalization, these transfer matrices are stochastic matrices and the associated quantum Hamiltonians are intensity matrices~\cite{PRGN2002}. This simple eigenvalue turns out to be the eigenvalue of largest modulus for $0<\textrm{Re}(u)<\frac \pi 3$. The corresponding left eigenstate is the trivial vector with all equal entries. The right eigenvector is non-trivial and has integer entries related to the counting of alternating sign matrices~\cite{BGN2001}. These properties are closely related to the similar features \cite{RS01a,RS01b} for the groundstate of the XXZ Hamiltonian at the combinatorial point $\Delta = -\frac 12$.

Similar ties with combinatorics were found by Garbali and Nienhuis \cite{GarbNien2017a,GarbNien2017b} for the dilute $\Atwotwo$ model with strip boundary conditions at $\lambda = \frac \pi 3$. The boundary condition that they study is however different from the ones considered here, as they focus on a solution of the boundary Yang-Baxter equation with five boundary tiles, including some where the loops are attached to the boundary in the context of the boundary dilute Temperley-Lieb algebra.

It turns out that the transfer matrices \eqref{eq:DThat} also have simple eigenvalues, with special right eigenvectors expected to be related to combinatorial problems. For $\dtl_N(\beta)$, this occurs in the representations $\repV_{N,0}$ and $\repV_{N,1}$, for all $N \ge 1$. For $\pdtl_N(\alpha,\beta)$, this occurs in the representations $\repW_{N,0,\omega}$ with $\alpha = \omega+\omega^{-1} = 1$ and $\repW_{N,1,\omega=1}$, also for all $N \ge 1$. Our numerical experiments for small system sizes allow us to make a claim for the inhomogeneous transfer matrices
\be
\Tbh(u,\xi) = \ \
\psset{unit=0.9cm}
\begin{pspicture}[shift=-0.9](-0.2,-0.5)(5.2,1.0)
\facegrid{(0,0)}{(5,1)}
\psline[linewidth=0.025]{-}(0,0.15)(0.15,0.15)(0.15,0)
\rput(1,0){\psline[linewidth=0.025]{-}(0,0.15)(0.15,0.15)(0.15,0)}
\rput(2,0){\psline[linewidth=0.025]{-}(0,0.15)(0.15,0.15)(0.15,0)}
\rput(4,0){\psline[linewidth=0.025]{-}(0,0.15)(0.15,0.15)(0.15,0)}
\psline[linewidth=1.5pt,linecolor=blue,linestyle=dashed,dash=2pt 2pt]{-}(0,0.5)(-0.3,0.5)
\psline[linewidth=1.5pt,linecolor=blue,linestyle=dashed,dash=2pt 2pt]{-}(5,0.5)(5.3,0.5)
\rput(0.5,.5){\footnotesize$u\!-\!\xi_1$}
\rput(1.5,.5){\footnotesize$u\!-\!\xi_2$}
\rput(2.5,0.5){\footnotesize$u\!-\!\xi_3$}
\rput(3.5,0.5){$\ldots$}
\rput(4.5,.5){\footnotesize$u\!-\!\xi_N$}
\psline{<->}(0,-0.3)(5,-0.3)\rput(2.5,-0.55){$_N$}
\end{pspicture}\ \ ,
\qquad
\Dbh(u,\xi) = \ \
\psset{unit=0.9cm}
\begin{pspicture}[shift=-1.4](-0.4,-0.5)(5.4,2.0)
\facegrid{(0,0)}{(5,2)}
\psline[linewidth=0.025]{-}(0,0.15)(0.15,0.15)(0.15,0)
\rput(1,0){\psline[linewidth=0.025]{-}(0,0.15)(0.15,0.15)(0.15,0)}
\rput(2,0){\psline[linewidth=0.025]{-}(0,0.15)(0.15,0.15)(0.15,0)}
\rput(4,0){\psline[linewidth=0.025]{-}(0,0.15)(0.15,0.15)(0.15,0)}
\multirput(1,1)(1,0){3}{\psline[linewidth=0.025]{-}(0,0.15)(-0.15,0.15)(-0.15,0)}
\rput(5,1){\psline[linewidth=0.025]{-}(0,0.15)(-0.15,0.15)(-0.15,0)}
\psarc[linewidth=1.5pt,linecolor=blue,linestyle=dashed,dash=2pt 2pt]{-}(0,1){.5}{90}{270}
\psarc[linewidth=1.5pt,linecolor=blue,linestyle=dashed,dash=2pt 2pt]{-}(5,1){.5}{270}{90}
\rput(0.5,.5){\footnotesize$u\!-\!\xi_1$}
\rput(1.5,.5){\footnotesize$u\!-\!\xi_2$}
\rput(2.5,0.5){\footnotesize$u\!-\!\xi_3$}
\rput(3.5,0.5){$\ldots$}
\rput(4.5,.5){\footnotesize$u\!-\!\xi_N$}
\rput(0.5,1.5){\footnotesize$u\!+\!\xi_1$}
\rput(1.5,1.5){\footnotesize$u\!+\!\xi_2$}
\rput(2.5,1.5){\footnotesize$u\!+\!\xi_3$}
\rput(3.5,1.5){$\ldots$}
\rput(4.5,1.5){\footnotesize$u\!+\!\xi_N$}
\psline{<->}(0,-0.3)(5,-0.3)\rput(2.5,-0.55){$_N$}
\end{pspicture}\ \ .
\ee
We conjecture that these transfer matrices have the special eigenvalues
\begin{subequations}
\label{eq:simple.eigenvalues}
\begin{alignat}{2}
\widehat T(u,\xi) &= \frac {4^N}{3^{N/2}}\bigg[\prod_{j=1}^N  \cos\!\big(\tfrac12 (u-\xi_j+\tfrac \pi 3)\big)\sin\!\big(\tfrac12 (u-\xi_j-\tfrac \pi 3)\big)  
+\prod_{j=1}^N \cos\!\big(\tfrac12 (u-\xi_j-\tfrac \pi 3)\big)\sin\!\big(\tfrac12 (u-\xi_j+\tfrac \pi 3)\big)\bigg],\\
\widehat D(u,\xi) &= \frac {4^{2N}}{3^{N}}\bigg[\prod_{j=1}^N  \cos\!\big(\tfrac12 (u-\xi_j+\tfrac \pi 3)\big)\sin\!\big(\tfrac12 (u-\xi_j-\tfrac \pi 3)\big)\cos\!\big(\tfrac12 (u+\xi_j+\tfrac \pi 3)\big)\sin\!\big(\tfrac12 (u+\xi_j-\tfrac \pi 3)\big)  
\nonumber\\
&+\prod_{j=1}^N \cos\big(\tfrac12 (u-\xi_j-\tfrac \pi 3)\big)\sin\big(\tfrac12 (u-\xi_j+\tfrac \pi 3)\big)\cos\big(\tfrac12 (u+\xi_j-\tfrac \pi 3)\big)\sin\big(\tfrac12 (u+\xi_j+\tfrac \pi 3)\big)\bigg].
\end{alignat}
\end{subequations}
We also find that, in the homogeneous limit $\xi_j \to 0$, these are the eigenvalues of largest modulus at the isotropic point $u = \frac{3\lambda}2= \frac\pi 2$, in the corresponding standard modules.

%
\section{Continuous scaling limit and conformal partition functions}\label{sec:partition.functions}
%

This section contains the main results of this paper. We first study the large-$N$ behavior of the transfer matrix eigenvalues, and in particular the leading finite-size correction of $\log \widehat T(u)$ and $\log \widehat D(u)$, proportional to $\frac 1N$. We conjecture formulas for the traces of the transfer matrices in the continuum scaling limit, for both periodic and strip boundary conditions and in each standard module. These conjectures are the result of extensive numerical work presented in subsequent sections. 

In the present section, after formulating the conjectures, we combine them with the Markov traces defined in \cref{sec:Markov} to compute the conformal partition functions of critical site percolation on the torus and cylinder. These are expressed in terms of affine $u(1)$ characters and Kac characters of the Virasoro algebra. Remarkably, we find that these conformal partitions are identical to those that we obtained previously for the model of critical bond percolation \cite{MDKP2017}.

\subsection{Free energies, finite-size corrections and conjectures for the traces}

For the $\Atwotwo$ model with $0 < \lambda < \tfrac \pi 2$, the finite-size corrections for the leading eigenvalues $T(u)$ and $D(u)$ of the single and double row transfer matrices take the form~\cite{BCN86,Affleck86}\begin{subequations}
\label{eq:logTD.asy}
\begin{alignat}{2}
\log T(u) &= - N f_{\text{bulk}}(u) -  \tfrac{2 \pi }{N} \Big(\ir\,\eE^{-\frac{\ir\pi u}{3\lambda}}(\Delta - \tfrac c {24})  -\ir\,\eE^{\frac{\ir \pi u}{3\lambda}} (\bar \Delta -\tfrac c{24}) \Big) + o(\tfrac1N)\nonumber\\
&= - N f_{\text{bulk}}(u) -  \tfrac{2\pi}{N} \Big(\sin \tfrac{\pi u}{3\lambda}\,(\Delta+\bar\Delta - \tfrac c {12}) + \ir\cos \tfrac{\pi u}{3\lambda}\, (\Delta -\bar\Delta) \Big) + o(\tfrac1N)
\label{eq:logTD.asy.T},\\[0.1cm]
\log D(u) &= - 2N f_{\text{bulk}}(u)  -  f_\text{bdy}(u)-  \tfrac{2 \pi}{N}  \sin\tfrac{\pi u}{3\lambda}\,(\Delta - \tfrac c {24})  + o(\tfrac1N).
\label{eq:logTD.asy.D}
\end{alignat}
\end{subequations}
Here, $f_{\text{bulk}}(u)$ and $f_{\text{bdy}}(u)$ are the bulk and boundary free energies. There is no boundary free energy for the single row transfer matrix, reflecting the fact that there is no boundary. The term proportional to $\frac1N$ depends on the central charge and the conformal weights of the conformal state describing this eigenstate in the scaling limit. 

For critical site percolation corresponding to $\lambda = \frac \pi 3$, the expansions for the eigenvalues $\widehat T(u)$ and $\widehat D(u)$ of the suitably normalised transfer matrices read
\begin{subequations}
\label{eq:logTDh.asy}
\begin{alignat}{2}
\log \widehat T(u) &= - N f_{\text{bulk}}(u)- \tfrac{2 \pi }{N} \Big(\ir\,\eE^{-\ir u}\Delta  -\ir\,\eE^{\ir u} \bar \Delta \Big) + o(\tfrac1N)\nonumber\\
&= - N f_{\text{bulk}}(u)-  \tfrac{2\pi}{N} \Big(\sin u\,(\Delta+\bar\Delta) + \ir\cos u\, (\Delta -\bar\Delta) \Big) + o(\tfrac1N),
\label{eq:logTD.asy.T}\\[0.1cm]
\log \widehat D(u) &= - 2N f_{\text{bulk}}(u)  -  f_\text{bdy}(u)-  \tfrac{2 \pi}{N}  \sin (u)\, \Delta  + o(\tfrac1N),
\label{eq:logTD.asy.D}
\end{alignat}
\end{subequations}
where we set the central charge to its known value $c=0$. The bulk and boundary energies can be obtained directly from the homogeneous limit of the special eigenvalues \eqref{eq:simple.eigenvalues}:
\be
\label{eq:fbulkbdy}
f_{\text{bulk}}(u) = - \log \Big[\tfrac4{\sqrt 3} \sin\big(\tfrac12(u+\tfrac \pi 3)\big)\sin\big(\tfrac12(u+\tfrac {2\pi}3)\big)\Big], \qquad f_{\text{bdy}}(u) = 0.
\ee
For these special eigenvalues, the following corrections are exponentially small in $N$, and in particular the $\frac1N$ correction vanishes consistently with $\Delta = \bar\Delta = 0$. These bulk and boundary free energies can alternatively be computed using Baxter's inversion relation techniques~\cite{BaxInv82,OPB95,OP97}. In the present case, this involves solving the functional equations
\begin{subequations}
\begin{alignat}{2}
f_{\textrm{bulk}}(u) - f_{\textrm{bulk}}(u - \tfrac \pi 3) - f_{\textrm{bulk}}(u + \tfrac \pi 3) &= \log s(u) = \log \tfrac{2}{\sqrt{3}}\,\sin u, 
\\
f_{\textrm{bdy}}(u) - f_{\textrm{bdy}}(u - \tfrac \pi 3) - f_{\textrm{bdy}}(u + \tfrac \pi 3) &= 0.
\end{alignat}
\end{subequations}
These can be solved using Fourier transforms and assuming the correct analyticity properties and asymptotics for these functions. The above formulas \eqref{eq:fbulkbdy} for the free energies will also be recovered in \cref{sec:NLIE.FSS} using the technique of nonlinear integral equations.

In \cite{MDKP2017}, we studied the transfer matrices of the dense loop model for $\lambda = \frac \pi 3$ corresponding to critical bond percolation on the square lattice. 
For this $s\ell(2)$ model, we were able to conjecture a complete classification of the patterns of zeros, for both periodic single row and strip double row transfer matrices, in each standard module with $d$ defects. Using this data, we derived and analytically solved nonlinear integral equations for the finite-size corrections of every eigenvalue of the fundamental transfer matrices and obtained formulas for their traces in terms of {\em finitized\/} characters. In the continuum scaling limit, we obtained the conformal partition functions on the cylinder and torus in terms of Virasoro and affine $u(1)$ characters. This calculation involved skew $q$-binomials~\cite{PRdimers07,PRVO20} and a $q$-series identity (B.7) in \cite{MDKP2017} for which we gave a proof but is in fact\footnote{We thank Ole Warnaar for pointing this out to us.} the $q$-Saalsch\"utz identity, see formula (3.3.12) in \cite{AndrewsPartitions}.

Critical site percolation is intrinsically an $s\ell(3)$ model and the classification of the patterns of zeros is significantly more difficult. We are unable to obtain a complete classification of the patterns of zeros. Nevertheless, our extensive numerical results suggest the following conjectures for the scaling limit of the trace of the transfer matrices in the standard modules.
\begin{Conjecture}[Traces of single row transfer matrices] 
\label{torusConj}
The scaling limit of the traces of $\widehat \Tb(u)^M$ over the standard modules $\repW_{N,d,\omega}$ of $\pdtl_N(\alpha,\beta)$ are given by
\be
\label{eq:conj.trace.T}
\lim_{\substack{M,N \to \infty\\M/N = \delta}} \eE^{MN f_\mathrm{bulk}(u)}\,\mathrm{tr}_{\raisebox{-0.05cm}{\tiny$\repW_{N,d,\omega}$}} \widehat \Tb(u)^M =
\frac{1}{(q)_\infty (\bar q)_\infty}
\sum_{\ell=-\infty}^{\infty} q^{\Delta_{\gamma/\pi - 2\ell, d/2}} \bar q^{\Delta_{\gamma/\pi-2\ell, -d/2}}
\ee
where
\be
\omega = \eE^{\ir \gamma}, \qquad q = \exp\big(\!-\!2 \pi \ir \delta\,\eE^{-\ir u} \big), \qquad
\bar q = \exp\big(2 \pi \ir \delta\, \eE^{\ir u} \big),\qquad (q)_\infty = \prod_{k=1}^\infty (1-q^k),
\ee
and the bulk free energy $f_\mathrm{bulk}(u)$ is given in \eqref{eq:fbulkbdy}.
\end{Conjecture}
It is clear that these traces are invariant under the symmetry $\gamma \to \gamma + 2 \pi$, as this amounts to shifting $\ell$ by one in the sum.
\begin{Conjecture}[Traces of double row transfer matrices]
\label{cylConj}
The scaling limit of the traces of $\widehat \Db(u)^M$ over the standard modules $\repV_{N,d}$ of $\dtl_N(\beta)$ are given by
\be
\label{eq:conj.trace.D}
\lim_{\substack{M,N \to \infty\\M/N = \delta}}\eE^{2MN f_\mathrm{bulk}(u)+M f_\mathrm{bdy}(u)}\,\mathrm{tr}_{\raisebox{-0.05cm}{\tiny$\repV_{N,d}$}} \widehat \Db(u) = q^{\Delta_{1,d+1}} \frac{1-q^{d+1}}{(q)_\infty}, 
\ee
where  $f_\mathrm{bdy}(u)=0$ and
\be
q = \exp\big(\!-\!2 \pi \delta \sin u\big).
\ee
\end{Conjecture}
The right side of \eqref{eq:conj.trace.T} is a sesquilinear form in Verma characters whereas the right side of \eqref{eq:conj.trace.D} is precisely the Kac character $\chit_{1,d+1}(q)$.

In the conjectures, the leading term in the power-series corresponds to the groundstate, and the corresponding conformal weight coincides with \eqref{eq:Delta.periodic} and \eqref{eq:Delta.strip}. In \cref{sec:NLIE.FSS}, we give an exact derivation for these ground-state conformal weights, in each standard representation. This derivation is obtained using nonlinear integral equations, themselves obtained from the functional relations that we describe in \cref{sec:fun.rel} and in fact hold more generally for all eigenvalues. We then present numerical evidence for the above conjectures in \cref{sec:Numerics} and \cref{sec:Tabulated}. In this case, we solve numerically the logarithmic Bethe ansatz equations to high precision for large numbers of excited states of the transfer matrix. 

\subsection{Lattice partition functions}

The Markov trace allows us to compute the partition function of the $\Atwotwo$ loop model. The repeated applications of the single and double row transfer matrices produce stacks of face operators that reproduce the lattices of the loop model on sections of the torus and cylinder, respectively. These lattices are illustrated in \cref{fig:lattices.loops}. Applying $\mathcal F$ then respectively closes these sections into a proper torus and cylinder. The corresponding partition functions for the loop model with $\lambda = \frac \pi 3$, as defined in \cref{sec:perco.and.A22}, are 
\be
\widetilde Z_{\textrm{tor}} = \mathcal F\big(\Tbh(u)^M\big), 
\qquad 
\widetilde Z_{\textrm{cyl}} = \mathcal F\big(\Dbh(u)^M\big).
\ee  

\paragraph{Torus partition functions.}
For the periodic dilute Temperley-Lieb algebra, we use \eqref{eq:trace.Cdj} and \eqref{eq:trace.torus} and write
\be
\textrm{tr}_{\raisebox{-0.05cm}{\tiny$\repW_{N,d,\omega}$}} \Tbh(u)^M = \sum_{j = -M}^M  \omega^{-j}C_{d,j},
\qquad
\label{eq:Zh.loops}
\widetilde Z_{\textrm{tor}} =  \textrm{tr}_{\raisebox{-0.05cm}{\tiny$\repW_{N,0,\omega}$}} \Tbh(u)^M \big|_{\alpha \to \alpha_{1,0}} + \sum_{d=1}^N \sum_{j=-M}^M 2\, T_{d \wedge j} (\tfrac12 \alpha_{\frac j{d \wedge j},\frac{d}{d \wedge j}})\, C_{d,j}.
\ee
As argued in \cref{sec:antiperiodic}, the torus partition function for critical site percolation, defined in \eqref{eq:Z.perco.tor}, can be written in terms of partition functions in the loop model. For the four possible boundary conditions, the weights $\alpha_{i,j}$ of the non-contractible loops must be assigned according to the parities of $i$ and $j$ as
\be
\label{eq:loop.specifications}
\hspace{-0.4cm}
\begin{array}{lll}
Z_{\textrm{tor}}^{\textrm{\tiny$(0,0)$}}(\alpha):\hspace{0.2cm}
\alpha_{i,j} \to
\left\{\begin{array}{ll}
\alpha & (i,j) \equiv (1,0) \textrm{ mod } 2, \\
\alpha & (i,j) \equiv (0,1) \textrm{ mod } 2,\\
\alpha & (i,j) \equiv (1,1) \textrm{ mod } 2,
\end{array}\right.
\hspace{0.5cm}&
Z_{\textrm{tor}}^{\textrm{\tiny$(0,1)$}}(\alpha):\hspace{0.2cm}
\alpha_{i,j} \to
\left\{\begin{array}{ll}
\alpha & (i,j) \equiv (1,0) \textrm{ mod } 2, \\
0 & (i,j) \equiv (0,1) \textrm{ mod } 2,\\
0 & (i,j) \equiv (1,1) \textrm{ mod } 2,
\end{array}\right.
\\[0.85cm]
Z_{\textrm{tor}}^{\textrm{\tiny$(1,0)$}}(\alpha):\hspace{0.2cm}
\alpha_{i,j} \to
\left\{\begin{array}{ll}
0 & (i,j) \equiv (1,0) \textrm{ mod } 2, \\
\alpha & (i,j) \equiv (0,1) \textrm{ mod } 2,\\
0 & (i,j) \equiv (1,1) \textrm{ mod } 2,
\end{array}\right.
\hspace{0.35cm}&
Z_{\textrm{tor}}^{\textrm{\tiny$(1,1)$}}(\alpha):\hspace{0.2cm}
\alpha_{i,j} \to
\left\{\begin{array}{ll}
0 & (i,j) \equiv (1,0) \textrm{ mod } 2, \\
0 & (i,j) \equiv (0,1) \textrm{ mod } 2,\\
\alpha & (i,j) \equiv (1,1) \textrm{ mod } 2.
\end{array}\right.\end{array}
\ee
Moreover, for periodic-periodic boundary conditions, the number of non-contractible loops is always even, whereas it is odd in the three antiperiodic cases. It follows that
\begin{subequations}
\label{eq:ZZ}
\begin{alignat}{2}
Z_{\textrm{tor}}^{\textrm{\tiny$(0,0)$}}(\alpha)
&= \lim_{u\to \frac \pi 3}\Big[\widetilde Z_{\textrm{tor}} + \widetilde Z_{\textrm{tor}}\big|_{\alpha_{i,j} \to -\alpha_{i,j}}\Big]\bigg|_{\alpha_{i,j}\,\to \textrm{rules \eqref{eq:loop.specifications}}}\ ,\label{eq:ZZ1}
\\
Z_{\textrm{tor}}^{\textrm{\tiny$(h,v)$}}(\alpha) &= \lim_{u\to \frac \pi 3}\Big[\widetilde Z_{\textrm{tor}} - \widetilde Z_{\textrm{tor}}\big|_{\alpha_{i,j} \to -\alpha_{i,j}}\Big]\bigg|_{\alpha_{i,j}\,\to \textrm{rules \eqref{eq:loop.specifications}}}\ , \qquad \textrm{for } (h,v) = (0,1),(1,0),(1,1).\label{eq:ZZ2}
\end{alignat}
\end{subequations}
The linear combinations in \eqref{eq:ZZ1} and \eqref{eq:ZZ2} serve to keep only the configurations with even or odd numbers of non-contractible loops, respectively. Morever, we did not divide the right sides of \eqref{eq:ZZ} by~$2$, because each loop configuration must be counted twice due to their invariance under the interchange of white and purple cells. Rewriting these expressions using \eqref{eq:Zh.loops}, we find
\begin{subequations}
\label{eq:Markov.trace.general.alpha}
\allowdisplaybreaks
\begin{alignat}{2}
Z_{\textrm{tor}}^{\textrm{\tiny$(0,0)$}}(\alpha) &= 
\textrm{tr}_{\raisebox{-0.05cm}{\tiny$\repW_{N,0,\omega}$}} \widehat\Tb(u)^M
+
\textrm{tr}_{\raisebox{-0.05cm}{\tiny$\repW_{N,0,-\omega}$}} \widehat\Tb(u)^M 
+ 
 \sum_{\substack{1 \le d \le N\\ d\textrm{\,even}}} \sum_{\substack{-M\le j \le M\\ j\textrm{\,even}}} 4\, T_{d \wedge j} (\tfrac \alpha2)\, C_{d,j}, 
 \\[0.1cm]
Z_{\textrm{tor}}^{\textrm{\tiny$(0,1)$}}(\alpha) &= 
\textrm{tr}_{\raisebox{-0.05cm}{\tiny$\repW_{N,0,\omega}$}} \widehat\Tb(u)^M
-
\textrm{tr}_{\raisebox{-0.05cm}{\tiny$\repW_{N,0,-\omega}$}} \widehat\Tb(u)^M 
+ 
 \sum_{\substack{1\le d \le N\\ d\textrm{\,even}}} \sum_{\substack{-M\le j \le M\\ j\textrm{\,odd}}} 4\, T_{d \wedge j} (\tfrac \alpha2)\, C_{d,j},
\\[0.1cm]
Z_{\textrm{tor}}^{\textrm{\tiny$(1,0)$}}(\alpha) &= 
\sum_{\substack{1\le d \le N\\ d\textrm{\,odd}}} \sum_{\substack{-M\le j \le M\\ j\textrm{\,even}}} 4\, T_{d \wedge j} (\tfrac \alpha2)\, C_{d,j},
\\[0.1cm]
Z_{\textrm{tor}}^{\textrm{\tiny$(1,1)$}}(\alpha) &= 
\sum_{\substack{1\le d \le N\\ d\textrm{\,odd}}} \sum_{\substack{-M\le j \le M\\ j\textrm{\,odd}}} 4\, T_{d \wedge j} (\tfrac \alpha2)\, C_{d,j}.
\end{alignat}
\end{subequations}
A drastic simplification arises for the special value $\alpha = 2$. In this case, the above formulas reduce nicely to 
\begin{subequations}
\allowdisplaybreaks
\label{eq:Markov.trace.alpha=2}
\begin{alignat}{2}
Z_{\textrm{tor}}^{\textrm{\tiny$(0,v)$}}(\alpha = 2) &= 
\sum_{\substack{0\le d \le N\\ d\textrm{\,even}}} 
(2-\delta_{d,0})\Big(\textrm{tr}_{\raisebox{-0.05cm}{\tiny$\repW_{N,d,1}$}} \widehat\Tb(u)^M 
+(-1)^v\,
\textrm{tr}_{\raisebox{-0.05cm}{\tiny$\repW_{N,d,-1}$}} \widehat\Tb(u)^M\Big),
\\[0.1cm]
Z_{\textrm{tor}}^{\textrm{\tiny$(1,v)$}}(\alpha = 2) &= 
2\sum_{\substack{0\le d \le N\\ d\textrm{\,odd}}} 
\textrm{tr}_{\raisebox{-0.05cm}{\tiny$\repW_{N,d,1}$}} \widehat\Tb(u)^M
+(-1)^v\,
\textrm{tr}_{\raisebox{-0.05cm}{\tiny$\repW_{N,d,-1}$}} \widehat\Tb(u)^M.
\end{alignat}
\end{subequations}

\paragraph{Cylinder partition functions.}
To compute the cylinder partition function of the loop model at $\lambda = \frac \pi 3$, we use the Markov trace \eqref{eq:cylinder.Markov.trace} of the dilute Temperley-Lieb algebra and find 
\be
\label{eq:traces.with.V}
\widetilde Z_{\textrm{cyl}} = \mathcal F(\Dbh(u)^M)= \sum_{d=0}^N U_d (\tfrac \alpha 2) \textrm{tr}_{\raisebox{-0.05cm}{\tiny$\repV_{N,d}$}} \Dbh(u)^M.
\ee
For site percolation, we recall from \cref{sec:def.model.torus.cylinder} that the model with identical and different colours fixed at the ends of the cylinder respectively maps to loop configurations with even and odd numbers of non-contractible loops. As a result,
\begin{subequations}
\be
Z_{\textrm{cyl}}^{\textrm{\tiny$(0)$}}(\alpha) = \frac 12 \Big[\mathcal F(\Dbm^M) + \mathcal F(\Dbm^M)\big|_{\alpha \to -\alpha}\Big]
=\frac 1{2^{M+1}} \lim_{u \to \frac \pi 3} \widetilde Z_{\textrm{cyl}} + \widetilde Z_{\textrm{cyl}}\big|_{\alpha \to -\alpha}\ ,
\ee
\be
Z_{\textrm{cyl}}^{\textrm{\tiny$(1)$}}(\alpha) = \frac 12 \Big[\mathcal F(\Dbm^M) - \mathcal F(\Dbm^M)\big|_{\alpha \to -\alpha}\Big]
=\frac 1{2^{M+1}}\lim_{u \to \frac \pi 3} \widetilde Z_{\textrm{cyl}} - \widetilde Z_{\textrm{cyl}}\big|_{\alpha \to -\alpha}\ .
\ee
\end{subequations}
Here the factors of $\frac12$ after the first equalities are included because each loop configuration has a unique pre-image in the percolation model and must thus be counted only once. In contrast, the factor $\frac1{2^M}$ arises due to the relation $\Dbm = \frac12 \Dbh(\frac \pi 3)$, see \eqref{eq:DbmDbh}. Using \eqref{eq:traces.with.V}, we find
\be
\label{eq:ZD}
Z_{\textrm{cyl}}^{\textrm{\tiny$(0)$}}(\alpha) = \frac1{2^M}\sum_{\substack{0 \le d\le N\\[0.05cm]d \textrm{ even}}} U_d (\tfrac \alpha 2) \textrm{tr}_{\raisebox{-0.05cm}{\tiny$\repV_{N,d}$}} \widehat\Db(\tfrac \pi 3)^M,
\qquad
Z_{\textrm{cyl}}^{\textrm{\tiny$(1)$}}(\alpha) = \frac1{2^M}\sum_{\substack{0 \le d\le N\\[0.05cm]d \textrm{ odd}}} U_d (\tfrac \alpha 2) \textrm{tr}_{\raisebox{-0.05cm}{\tiny$\repV_{N,d}$}} \widehat\Db(\tfrac \pi 3)^M.
\ee
For $\alpha = 2$, this reads
\be
\label{eq:Zcyl.alpha.2}
Z_{\textrm{cyl}}^{\textrm{\tiny$(0)$}}(\alpha=2) = \frac1{2^M}\sum_{\substack{0 \le d\le N\\[0.05cm]d \textrm{ even}}} (d\!+\!1)\, \textrm{tr}_{\raisebox{-0.05cm}{\tiny$\repV_{N,d}$}} \Dbh(\tfrac \pi 3)^M,
\qquad
Z_{\textrm{cyl}}^{\textrm{\tiny$(1)$}}(\alpha=2) =\frac1{2^M}\sum_{\substack{0 \le d\le N\\[0.05cm]d \textrm{ odd}}}  (d\!+\!1)\, \textrm{tr}_{\raisebox{-0.05cm}{\tiny$\repV_{N,d}$}} \Dbh(\tfrac \pi 3)^M.
\ee

\subsection[Conformal partition functions for $\alpha = 2$]{Conformal partition functions for $\boldsymbol{\alpha = 2}$}

In this section, we consider the scaling limit of the lattice partition functions for $\alpha = 2$ and use \cref{torusConj,cylConj} to express them in terms of characters.
\paragraph{Torus partition functions.}
Using the symmetry $\Delta_{r,s+3\ell} = \Delta_{r-2\ell,s}$, the conjectured
formula \eqref{eq:conj.trace.T} for the scaling limit of the trace of the transfer matrix can be rewritten as
\be
\label{eq:ZTC}
\lim_{\substack{M,N \to \infty\\M/N = \delta}} \eE^{MN f_\mathrm{bulk}(u)}\,\mathrm{tr}_{\raisebox{-0.05cm}{\tiny$\repW_{N,d,\omega}$}} \widehat \Tb(u)^M =
\frac{1}{(q)_\infty (\bar q)_\infty}
\sum_{\ell=-\infty}^{\infty}  
q^{\Delta_{\gamma/\pi, 3\ell+d/2}} \bar q^{\Delta_{\gamma/\pi, 3\ell-d/2}}.
\ee
Setting $u = \frac \pi 3$ throughout, we define the conformal torus partition functions of the model of site percolation at $\alpha = 2$ as
\be
\mathcal Z_{\textrm{tor}}^{\textrm{\tiny$(h,v)$}} = \lim_{\substack{M,N \to \infty\\M/N = \delta}} \eE^{MN f_\text{bulk}(\frac \pi 3)} Z_{\textrm{tor}}^{\textrm{\tiny$(h,v)$}}(\alpha=2), \qquad h,v \in \{0,1\}.
\ee
Using \eqref{eq:ZTC}, these partition functions are written as infinite sesquilinear forms in Verma characters. They can be reformulated as finite sesquilinear forms in the affine $u(1)$ characters \eqref{eq:u1chars}. Let us define
\be
Z_{d,\gamma} = \frac{1}{(q)_\infty (\bar q)_\infty}
\sum_{\ell \in \mathbb Z} q^{\Delta_{\gamma/\pi, 3\ell+d/2}} \bar q^{\Delta_{\gamma/\pi, 3\ell-d/2}}, \qquad d \in \mathbb Z, \qquad \gamma \in \mathbb R.
\ee
Because $Z_{d,0} = Z_{-d,0}$ and $Z_{d,\pi} = Z_{-d,\pi}$, we can use \eqref{eq:Markov.trace.alpha=2} to write
\be
\mathcal Z_{\textrm{tor}}^{\textrm{\tiny$(0,v)$}} = 
\sum_{d \in 2 \mathbb Z} 
\big[Z_{d,0} + (-1)^v Z_{d,\pi}\big],
\qquad
\mathcal Z_{\textrm{tor}}^{\textrm{\tiny$(1,v)$}} = 
\sum_{d \in 2 \mathbb Z+1} 
\big[Z_{d,0} + (-1)^v Z_{d,\pi}\big],
\ee
We now group the functions $Z_{d,\gamma}$ in sums according to the value of $d$ mod $6$:
\be
Y_{j,\gamma} = \sum_{d\in 6 \mathbb Z+j} Z_{d,\gamma}, \qquad j = 0, 1, \dots, 5.
\ee
Using the simplification
\begingroup
\allowdisplaybreaks
\begin{align}
(q)_\infty(\qbar)_\infty\, Y_{j,\gamma}
&=\sum_{k,\ell\in{\Bbb Z}} q^{\Delta_{\gamma/\pi,3\ell+3k+j/2}}\qbar^{\Delta_{\gamma/\pi,3\ell-3k-j/2}}
=\sum_{k,\ell\in{\Bbb Z}} q^{\Delta_{\gamma/\pi,3\ell+6k+j/2}}\qbar^{\Delta_{\gamma/\pi,3\ell-j/2}}\nonumber\\
&=\sum_{k,\ell\in{\Bbb Z}} q^{\Delta_{\gamma/\pi,6\ell+6k+j/2}}\qbar^{\Delta_{\gamma/\pi,6\ell-j/2}}
+\sum_{k,\ell\in{\Bbb Z}} q^{\Delta_{\gamma/\pi,6\ell+6k+3+j/2}}\qbar^{\Delta_{\gamma/\pi,6\ell+3-j/2}}\nonumber\\
&=\sum_{k\in{\Bbb Z}} q^{\Delta_{\gamma/\pi,6k+j/2}} \sum_{\ell\in{\Bbb Z}} \qbar^{\Delta_{\gamma/\pi,6\ell-j/2}}
+\sum_{k\in{\Bbb Z}} q^{\Delta_{\gamma/\pi,6k+3+j/2}}\sum_{\ell\in{\Bbb Z}} \qbar^{\Delta_{\gamma/\pi,6\ell+3-j/2}}\\
&\hspace{-2cm}=\left\{\begin{array}{ll}
\displaystyle(q\qbar)^{-\frac{1}{24}}\Big(\sum_{k\in{\Bbb Z}} q^{\frac{(12k+j)^2}{24}}\sum_{\ell\in{\Bbb Z}} \qbar^{\frac{(12\ell+j)^2}{24}}
+\sum_{k\in{\Bbb Z}} q^{\frac{(12k+6+j)^2}{24}}\sum_{\ell\in{\Bbb Z}} \qbar^{\frac{(12\ell+6-j)^2}{24}}\Big)&\gamma=0,\\[14pt]
\displaystyle(q\qbar)^{-\frac{1}{24}}\Big(\sum_{k\in{\Bbb Z}} q^{\frac{(12k+j-3)^2}{24}}\sum_{\ell\in{\Bbb Z}} \qbar^{\frac{(12\ell+j+3)^2}{24}}
+\sum_{k\in{\Bbb Z}} q^{\frac{(12k+j+3)^2}{24}}\sum_{\ell\in{\Bbb Z}} \qbar^{\frac{(12\ell+j-3)^2}{24}}\Big)&\gamma=\pi,
\end{array}\right.\nonumber
\end{align}
\endgroup
it follows that
\be
Y_{j,0} = |\varkappa_j(q)|^2+|\varkappa_{6-j}(q)|^2, 
\qquad 
Y_{j,\pi} = \varkappa_{j+3}(q)\varkappa_{3-j}(\qbar)+\varkappa_{3-j}(q)\varkappa_{j+3}(\qbar).
\ee
With this result, we find
\begin{subequations}
\label{toroidalPartFns}
\begin{alignat}{2}
\mathcal Z_{\textrm{tor}}^{\textrm{\tiny$(0,v)$}} &= |\varkappa_0(q)|^2+ |\varkappa_6(q)|^2+ 2 |\varkappa_2(q)|^2+2 |\varkappa_4(q)|^2 
\nonumber\\[0.1cm]&\hspace{1cm} 
+2(-1)^v |\varkappa_3(q)|^2+ 2(-1)^v [\varkappa_1(q)\varkappa_5(\bar q)+ \varkappa_5(q)\varkappa_1(\bar q)],
\\[0.2cm]
\mathcal Z_{\textrm{tor}}^{\textrm{\tiny$(1,v)$}} &=2|\varkappa_1(q)|^2+2 |\varkappa_5(q)|^2+ 2 |\varkappa_3(q)|^2
\nonumber\\[0.1cm]&\hspace{1cm}
+(-1)^v[\varkappa_0(q)\varkappa_6(\bar q)  + \varkappa_6(q)\varkappa_0(\bar q)]+ 2(-1)^v [\varkappa_2(q)\varkappa_4(\bar q)+  \varkappa_4(q)\varkappa_2(\bar q)],
\end{alignat}
\end{subequations}
with $v = 0,1$. Using the identity
\be
\label{eq:K1K5relation}
\varkappa_1(q)\varkappa_5(\bar q) + \varkappa_5(q)\varkappa_1(\bar q) = |\varkappa_1(q)|^2 + |\varkappa_5(q)|^2-1,
\ee
these partition functions are expressed compactly as
\be
\begin{array}{l}
\displaystyle \mathcal Z_{\textrm{tor}}^{\textrm{\tiny$(0,v)$}} = \sum_{j=0}^6 \Big((-1)^{v j} d_j |\varkappa_j(q)|^2\Big) - 2(-1)^v,
\\[0.3cm]
\displaystyle \mathcal Z_{\textrm{tor}}^{\textrm{\tiny$(1,v)$}} = \sum_{j=0}^6 \Big((-1)^{v (j+1)} d_j \varkappa_j(q)\varkappa_{6-j}(\bar q)\Big) + 2,
\end{array}
\qquad
d_j=
\left\{\begin{array}{ll}
1&j=0, 6,\\[0.15cm]
2&j=1, \dots, 5.
\end{array}\right.
\ee
This is the final form of the conformal torus partition functions as sesquilinear forms in affine $u(1)$ characters. We now discuss properties under the action of the modular group.
Under the $T$ transformation, the $u(1)$ characters behave as
\be
\big|\varkappa_j(\eE^{2 \pi \ir (\tau+1)})\big| = \big|\varkappa_j(\eE^{2 \pi \ir \tau})\big|,
\qquad
\varkappa_j(\eE^{2 \pi \ir (\tau+1)})\varkappa_{6-j}(\eE^{-2 \pi \ir (\bar\tau+1)}) = (-1)^{j+1}
\varkappa_j(\eE^{2 \pi \ir \tau})\varkappa_{6-j}(\eE^{-2 \pi \ir \bar\tau}),
\ee
Their similar behaviour under the $S$ transformation is given in Equation (4.107) of \cite{MDKP2017}. It then results that the four partition functions behave under the $T$ and $S$ transformations as
\be
\label{eq:Z.modular.props}
\begin{array}{c}
\mathcal Z_{\textrm{tor}}^{\textrm{\tiny$(0,0)$}}(\tau+1) = \mathcal Z_{\textrm{tor}}^{\textrm{\tiny$(0,0)$}}(\tau), 
\\[0.2cm]
\mathcal Z_{\textrm{tor}}^{\textrm{\tiny$(0,1)$}}(\tau+1) = \mathcal Z_{\textrm{tor}}^{\textrm{\tiny$(0,1)$}}(\tau), 
\\[0.2cm]
\mathcal Z_{\textrm{tor}}^{\textrm{\tiny$(1,0)$}}(\tau+1) = \mathcal Z_{\textrm{tor}}^{\textrm{\tiny$(1,1)$}}(\tau),
\\[0.2cm]
\mathcal Z_{\textrm{tor}}^{\textrm{\tiny$(1,1)$}}(\tau+1) = \mathcal Z_{\textrm{tor}}^{\textrm{\tiny$(1,0)$}}(\tau),
\end{array}
\qquad
\begin{array}{c}
\mathcal Z_{\textrm{tor}}^{\textrm{\tiny$(0,0)$}}(-\tfrac1\tau) = 
\mathcal Z_{\textrm{tor}}^{\textrm{\tiny$(0,0)$}}(\tau),
\\[0.2cm]
\mathcal Z_{\textrm{tor}}^{\textrm{\tiny$(0,1)$}}(-\tfrac1\tau) = 
\mathcal Z_{\textrm{tor}}^{\textrm{\tiny$(1,0)$}}(\tau),
\\[0.2cm]
\mathcal Z_{\textrm{tor}}^{\textrm{\tiny$(1,0)$}}(-\tfrac1\tau) = 
\mathcal Z_{\textrm{tor}}^{\textrm{\tiny$(0,1)$}}(\tau),
\\[0.2cm]
\mathcal Z_{\textrm{tor}}^{\textrm{\tiny$(1,1)$}}(-\tfrac1\tau) = 
\mathcal Z_{\textrm{tor}}^{\textrm{\tiny$(1,1)$}}(\tau).
\end{array}
\ee
Therefore $\mathcal Z_{\textrm{tor}}^{\textrm{\tiny$(0,0)$}}$ is a modular invariant, whereas the other three partition functions are covariant under the modular group. We also note that these partition functions have only positive integer coefficients for $v = 0$, but have also negative ones for $v=1$.

\paragraph{Cylinder partition functions.}
Setting $u = \frac \pi 3$ throughout, we define the conformal cylinder partition functions of critical site percolation at $\alpha = 2$ as
\be
\mathcal Z_{\textrm{cyl}}^{\textrm{\tiny$(i)$}} = \lim_{\substack{M,N \to \infty\\M/N = \delta}} \eE^{M(2N-1) f_\text{bulk}(\frac \pi 3)} Z_{\textrm{cyl}}^{\textrm{\tiny$(i)$}}(\alpha=2), \qquad i \in \{0,1\},
\ee
where we recall that $M(2N-1)$ is the number of free sites for the lattice model on the cylinder. We note that this limit is well-defined because $\eE^{2MN f_\text{bulk}(\frac \pi 3)}$ cancels with the similar factor in \eqref{eq:conj.trace.D}, whereas $\eE^{-M f_\text{bulk}(\frac \pi 3)}$ cancels with the factor $2^{-M}$ in \eqref{eq:Zcyl.alpha.2}.
After some simplification, we find that these partition functions are expressed in terms of either sums of Kac characters or affine $u(1)$ characters as
\begin{subequations}
\label{CylPartFns}
\begin{alignat}{2}
\mathcal Z_{\textrm{cyl}}^{\textrm{\tiny$(0)$}} &= \frac1{(q)_\infty}
\sum_{d=0,2,4,...}^\infty\,
(d+1) q^{\Delta_{1,d+1}}(1-q^{d+1})
\nonumber\\[0.15cm]
&= \varkappa_1(q) + 3\varkappa_3(q) - \varkappa_5(q) -6\frac{\dd}{\dd z} \Big(\varkappa_1(q,z) - \varkappa_3(q,z) + \varkappa_5(q,z) \Big)\Big|_{z=1}\ ,
\\[0.15cm]
\mathcal Z_{\textrm{cyl}}^{\textrm{\tiny$(1)$}} &= \frac1{(q)_\infty}
\sum_{d=1,3,5,...}^\infty\,
(d+1) q^{\Delta_{1,d+1}}(1-q^{d+1})
\nonumber\\[0.15cm]
&= 2\varkappa_1(q)  + 4 \varkappa_5(q) +6\frac{\dd}{\dd z} \Big(\varkappa_1(q,z) - \varkappa_3(q,z) + \varkappa_5(q,z) \Big)\Big|_{z=1}\ .
\end{alignat}
\end{subequations}

\paragraph{Universal behavior of the partition functions.}
The above results are to be compared with the conformal partition functions obtained in \cite{MDKP2017} for the dense $\Aoneone$ loop model corresponding to critical bond percolation on the torus and cylinder. Let us denote these partition functions for critical bond percolation at $\alpha = 2$ as $\widehat{\mathcal Z}_{\textrm{tor}}$ and $\widehat{\mathcal Z}_{\textrm{cyl}}$. Four torus partition functions were computed corresponding to odd and even values of $M$ and $N$, and two cylinder partition functions corresponding to odd and even values of $N$.  Remarkably, we find the equalities\footnote{In comparing with our previous results \cite{MDKP2017}, we found two errors. First, there is a typo in (3.112) and (3.113): the prefactor of $\varkappa_3(q,z)$ in the parenthesis should be $-1$ not $-3$. Second, the partition function $\widehat {\mathcal Z}_{\textrm{tor}}^{\textrm{\tiny $M$\,odd,\,$N$\,odd}}$ has an incorrect overall sign. This can be traced back to incorrect signs in equation (4.93) of \cite{MDKP2017} for $M$ odd, which should instead read 
\begin{subequations}
\label{eq:bond.perco.conj}
\begin{alignat}{2}
&d \equiv 0 \textrm{ mod } 3 : \quad
&&Z_d(q,\bar q) = 
\frac{(-1)^{Md}}{(q)_\infty(\bar q)_\infty} \sum_{\ell\in \mathbb Z} \Big(q^{\Delta_{0,3\ell -d }} \bar q^{\Delta_{0,3 \ell}} + (-1)^{M} q^{\Delta_{1,3\ell - d}} \bar q^{\Delta_{1,3 \ell}} \Big),\\[0.1cm]
&d \equiv 1 \textrm{ mod } 3 : \quad
&&Z_d(q,\bar q) = 
\frac{(-1)^{Md}}{(q)_\infty(\bar q)_\infty} \sum_{\ell\in \mathbb Z} \Big(q^{\Delta_{0,3\ell -d +2}} \bar q^{\Delta_{0,3 \ell+2}} + (-1)^{M}q^{\Delta_{1,3\ell - d+2}} \bar q^{\Delta_{1,3 \ell+2}} \Big),\\[0.1cm]
&d \equiv 2 \textrm{ mod } 3  : \quad
&&Z_d(q,\bar q) = 
\frac{(-1)^{Md}}{(q)_\infty(\bar q)_\infty} \sum_{\ell\in \mathbb Z} \Big(q^{\Delta_{0,3\ell -d +1}} \bar q^{\Delta_{0,3 \ell+1}} + (-1)^{M}q^{\Delta_{1,3\ell - d+1}} \bar q^{\Delta_{1,3 \ell+1}} \Big).
\end{alignat}
\end{subequations}
}
\be
\label{eq:site.and.bond}
\begin{array}{llllll}
\mathcal Z_{\textrm{tor}}^{\textrm{\tiny$(0,0)$}} = \widehat {\mathcal Z}_{\textrm{tor}}^{\textrm{\tiny $M$\,even,\,$N$\,even}},
&\ \ &
\mathcal Z_{\textrm{tor}}^{\textrm{\tiny$(0,1)$}} = \widehat {\mathcal Z}_{\textrm{tor}}^{\textrm{\tiny $M$\,odd,\,$N$\,even}},
&\qquad&
\mathcal Z_{\textrm{cyl}}^{\textrm{\tiny$(0)$}} = \widehat {\mathcal Z}_{\textrm{cyl}}^{\textrm{\tiny $N$\,even}},
\\[0.15cm]
\mathcal Z_{\textrm{tor}}^{\textrm{\tiny$(1,0)$}} = \widehat {\mathcal Z}_{\textrm{tor}}^{\textrm{\tiny $M$\,even,\,$N$\,odd}},
&&
\mathcal Z_{\textrm{tor}}^{\textrm{\tiny$(1,1)$}} = \widehat {\mathcal Z}_{\textrm{tor}}^{\textrm{\tiny $M$\,odd,\,$N$\,odd}},
&\qquad &
\mathcal Z_{\textrm{cyl}}^{\textrm{\tiny$(1)$}} = \widehat {\mathcal Z}_{\textrm{cyl}}^{\textrm{\tiny $N$\,odd}}.
\end{array}
\ee

\begin{figure}
\begin{center}
\psset{unit=.6}
\begin{pspicture}(0,0.5)(7,6)
\facegrid{(0,0)}{(7,6)}
\rput(0,5){\loopbr}\rput(1,5){\loopar}\rput(2,5){\loopab}\rput(3,5){\loopar}\rput(4,5){\loopab}\rput(5,5){\loopar}\rput(6,5){\loopab}
\rput(0,4){\loopbb}\rput(1,4){\loopab}\rput(2,4){\loopbb}\rput(3,4){\loopbr}\rput(4,4){\loopar}\rput(5,4){\loopab}\rput(6,4){\loopar}
\rput(0,3){\loopab}\rput(1,3){\loopar}\rput(2,3){\loopab}\rput(3,3){\loopbb}\rput(4,3){\loopab}\rput(5,3){\loopbb}\rput(6,3){\loopab}
\rput(0,2){\loopar}\rput(1,2){\loopbr}\rput(2,2){\loopbb}\rput(3,2){\loopab}\rput(4,2){\loopar}\rput(5,2){\loopab}\rput(6,2){\loopbb}
\rput(0,1){\loopab}\rput(1,1){\loopar}\rput(2,1){\loopab}\rput(3,1){\loopar}\rput(4,1){\loopbr}\rput(5,1){\loopar}\rput(6,1){\loopab}
\rput(0,0){\loopbb}\rput(1,0){\loopab}\rput(2,0){\loopar}\rput(3,0){\loopbr}\rput(4,0){\loopbb}\rput(5,0){\loopbr}\rput(6,0){\loopbb}
\multiput(0,0)(0,2){4}{\multiput(0,0)(2,0){4}{\psarc[fillstyle=solid,fillcolor=black](0,0){0.1}{0}{360}}}
\multiput(0,0)(0,2){3}{\multiput(1,1)(2,0){4}{\psarc[fillstyle=solid,fillcolor=black](0,0){0.1}{0}{360}}}
\end{pspicture}
\qquad \qquad \qquad
\begin{pspicture}(-1,0.5)(8,6)
\facegrid{(0,0)}{(7,6)}
\rput(0,5){\loopab}\rput(1,5){\loopbb}\rput(2,5){\loopbr}\rput(3,5){\loopbb}\rput(4,5){\loopbr}\rput(5,5){\loopar}\rput(6,5){\loopbr}
\rput(0,4){\loopar}\rput(1,4){\loopab}\rput(2,4){\loopbb}\rput(3,4){\loopab}\rput(4,4){\loopbb}\rput(5,4){\loopbr}\rput(6,4){\loopbb}
\rput(0,3){\loopbr}\rput(1,3){\loopbb}\rput(2,3){\loopbr}\rput(3,3){\loopar}\rput(4,3){\loopbr}\rput(5,3){\loopar}\rput(6,3){\loopab}
\rput(0,2){\loopbb}\rput(1,2){\loopbr}\rput(2,2){\loopar}\rput(3,2){\loopab}\rput(4,2){\loopbb}\rput(5,2){\loopab}\rput(6,2){\loopbb}
\rput(0,1){\loopab}\rput(1,1){\loopar}\rput(2,1){\loopab}\rput(3,1){\loopbb}\rput(4,1){\loopbr}\rput(5,1){\loopbb}\rput(6,1){\loopbr}
\rput(0,0){\loopbb}\rput(1,0){\loopab}\rput(2,0){\loopbb}\rput(3,0){\loopab}\rput(4,0){\loopar}\rput(5,0){\loopbr}\rput(6,0){\loopar}
\psarc[linecolor=blue,linewidth=1.5pt](0,1){.5}{90}{270}
\psarc[linecolor=blue,linewidth=1.5pt](0,3){.5}{90}{270}
\psarc[linecolor=blue,linewidth=1.5pt](0,5){.5}{90}{270}
\psarc[linecolor=blue,linewidth=1.5pt](7,1){.5}{270}{90}
\psarc[linecolor=blue,linewidth=1.5pt](7,3){.5}{270}{90}
\psarc[linecolor=blue,linewidth=1.5pt](7,5){.5}{270}{90}
\psline[linewidth=3pt,linecolor=red](0,0)(-1,1)(0,2)(-1,3)(0,4)(-1,5)(0,6)
\rput(7,0){\psline[linewidth=3pt,linecolor=blue](0,0)(1,1)(0,2)(1,3)(0,4)(1,5)(0,6)}
\multiput(0,0)(0,2){4}{\multiput(0,0)(2,0){4}{\psarc[fillstyle=solid,fillcolor=black](0,0){0.1}{0}{360}}}
\multiput(0,0)(0,2){3}{\multiput(1,1)(2,0){4}{\psarc[fillstyle=solid,fillcolor=black](0,0){0.1}{0}{360}}}
\end{pspicture}
\end{center}
\caption{Configurations of the dense loop model on the torus and cylinder of odd width $N$, with their two dual sets of bond percolation clusters shown in blue and red. On the torus, this corresponds to antiperiodic-periodic boundary conditions for the underlying percolation configuration. On the cylinder, the odd parity of $N$ with double rows and boundary conditions made of simple arcs on the left and right edges results in wired clusters having opposite colours on the left and right boundaries.}
\label{fig:bond.perco}
\end{figure}
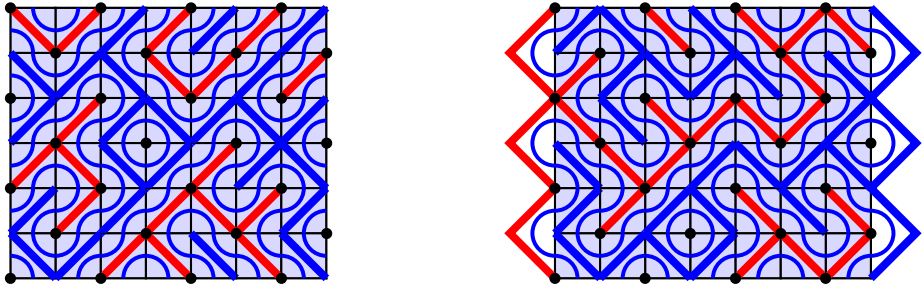

For the dense loop model, the loops draw the contours of the clusters of bonds. This is illustrated in \cref{fig:bond.perco}, where the lattice for the bond percolation model is made of the black sites and is a square lattice tilted by $45^\circ$. Its corresponding percolation clusters are drawn in red. In this figure, we also draw in blue the dual clusters, on the dual square lattice. On the torus, odd and even values of $M$ or $N$ then correspond to periodic and antiperiodic boundary conditions for the percolation clusters, respectively. This means that if we follow a cluster moving from one elementary domain to the next, a red cluster becomes a blue cluster if the length along this axis is odd, but remains a red cluster if this length is even. On the cylinder, the lattice is made of an integer number of double rows, so the percolation clusters always have periodic boundary conditions in the vertical direction (as opposed to antiperiodic). In this case, a boundary with simple half-arcs corresponds to having the clusters of one colour wired and the clusters of the other colour not wired on this edge. If $N$ is even, the two boundaries have the same colour of wired clusters, whereas this colour is different if $N$ is odd. Thus the relations \eqref{eq:site.and.bond} reveal a strong form of universality for bond and site percolation, as it shows the equality of the partition functions for precisely the same boundary conditions for the two models of percolation: periodic or antiperiodic for the torus, and wired or not wired for the cylinder.

\subsection[Torus partition functions for all $\alpha$]{Torus partition functions for all $\boldsymbol\alpha$}

In this section, we compute the torus partition functions for all values of $\alpha$. We present the computations separately for site and bond percolation, with the latter complementing our earlier work in \cite{MDKP2017}.

\paragraph{Partition functions for site percolation.}
Starting from \eqref{eq:Zh.loops} and \eqref{eq:Markov.trace.general.alpha}, we compute the coefficients $C_{d,j}$ using an integral over $\gamma$:
\be
C_{d,j} = \frac1{2\pi} \int_{0}^{2\pi} \dd\gamma\, \eE^{\ir \gamma j}\, \textrm{tr}_{\raisebox{-0.05cm}{\tiny$\repW_{N,d,\eE^{\ir \gamma}}$}} \Tbh(u)^M.
\ee
 We then define the scaled coefficients and partition functions
\be
\mathcal C_{d,j} =  \lim_{\substack{M,N \to \infty\\M/N = \delta}} \eE^{MN f_\text{bulk}(u)} C_{d,j},
\qquad
\mathcal Z_{\textrm{tor}}^{\textrm{\tiny$(h,v)$}}(\alpha) = \lim_{\substack{M,N \to \infty\\M/N = \delta}} \eE^{MN f_\text{bulk}(u)} Z_{\textrm{tor}}^{\textrm{\tiny$(h,v)$}}(\alpha).
\ee
Because $\repW_{N,0,\omega} = \repW_{N,0,\omega^{-1}}$, we know that $\mathcal C_{0,j} = \mathcal C_{0,-j}$. Then defining $C_{-d,j} = C_{d,-j}$, we write
\be
\label{eq:Zhv.alpha}
\mathcal Z_{\textrm{tor}}^{\textrm{\tiny$(h,v)$}}(\alpha) = 
\sum_{\substack{d\,\in\,\mathbb Z_{\ge 0}\\ d\, \equiv \, h\,\textrm{mod}\,2}} \hspace{0.2cm} \sum_{\substack{j \,\in\,\mathbb Z\\ j\, \equiv \, v\,\textrm{mod}\,2}} (4-2 \delta_{d,0})\, T_{d \wedge j} (\tfrac \alpha2)\, \mathcal C_{d,j}
= \sum_{\substack{d\,\in\,\mathbb Z\\ d\, \equiv \, h\,\textrm{mod}\,2}} \hspace{0.2cm} \sum_{\substack{j \,\in\,\mathbb Z\\ j\, \equiv \, v\,\textrm{mod}\,2}} 2\, T_{d \wedge j} (\tfrac \alpha2)\, \mathcal C_{d,j},
\ee
valid for $h,v \in \{0,1\}$. We then find
\begin{alignat}{2}
\mathcal C_{d,j} &= \frac1{2\pi}\frac1{(q)_\infty (\bar q)_\infty} \sum_{\ell = -\infty}^
\infty  \int_{0}^{2\pi} \dd\gamma\, \eE^{\ir \gamma j}\, q^{\Delta_{(\gamma-2\pi \ell)/\pi,d/2}} \bar q^{\Delta_{(\gamma-2\pi \ell)/\pi,-d/2}} 
\nonumber\\&=
\frac1{2\pi} \frac1{(q)_\infty (\bar q)_\infty}  \int_{-\infty}^{\infty} \dd\gamma\, \eE^{\ir \gamma j}\, q^{\Delta_{\gamma/\pi,d/2}} \bar q^{\Delta_{\gamma/\pi,-d/2}}.
\end{alignat}
Concatenating the sum of integrals into a single integral thus produces a Gaussian integral that is easily evaluated. Indeed, setting $q = \exp(2 \pi \ir \tau)$ and $\bar q = \exp(-2 \pi \ir \bar\tau)$ with $\tau = \delta \,\eE^{\ir u}$, we find
\begin{alignat}{2}
\mathcal C_{d,j} &= \frac1{2\pi} \frac{\exp\big[\tfrac{\ir \pi}{12}(\tau-\bar \tau)(d^2 -1)\big]}{(q)_\infty (\bar q)_\infty}  \int_{-\infty}^{\infty} \dd\gamma\, \exp\Big[\ir \gamma(j-\tfrac d2(\tau+\bar\tau))+\frac{3 \ir \gamma^2}{4 \pi} (\tau-\bar\tau)\Big]
\nonumber\\
& = \Big(\frac1{6 \tau_i}\Big)^{1/2} \frac{\exp(\frac{\pi \tau_i}{6})}{(q)_\infty (\bar q)_\infty} \exp\Big[-\frac \pi{6 \tau_i} \big(j^2 + d^2 (\tau_r^2 + \tau_i^2) -2 d\, j\, \tau_r\big)\Big]
= \mathcal Z_{d,j}(\tfrac16)
\label{eq:Cdj.final}
\end{alignat}
where 
\be
\mathcal Z_{m,m'}(g) = \Big(\frac g{\tau_i}\Big)^{1/2} \frac1{\eta(q) \eta(\bar q)} \exp\Big[-\frac {\pi g}{ \tau_i} \big|m \tau - m'\big|^2\Big].
\ee
Here, $\tau_r$ and $\tau_i$ are the real and imaginary parts of $\tau$, and the Dedekind function is $\eta(q) = q^{1/24} (q)_\infty$. The functions $\mathcal Z_{m,m'}(g)$ were defined in \cite{FSZ87,RS01} in the context of the Coulomb gas approach, and it is remarkable that these are reproduced here from a completely different calculation. Thus \eqref{eq:Zhv.alpha} and \eqref{eq:Cdj.final} are the final form of the partition function for critical site percolation, valid for all values of~$\alpha$. Because
\be
C_{d,j}(\tau+1) = C_{d,j-d}(\tau),
\qquad 
C_{d,j}(-\tfrac1 \tau) = C_{j,-d}(\tau),
\ee
it is clear that $\mathcal Z_{\textrm{tor}}^{\textrm{\tiny$(0,0)$}}(\alpha)$ is a modular invariant, whereas the three others partition functions are modular covariant, as in \eqref{eq:Z.modular.props}.

Following \cite{FSZ87}, we define
\be
\mathcal Z(g,f) = f \sum_{m,m' \in f \mathbb Z} \mathcal Z_{m,m'}(g),
\ee 
and note that, for $\alpha = 2$, the partition function for periodic-periodic boundary conditions can be rewritten as
\be
\mathcal Z_{\textrm{tor}}^{\textrm{\tiny$(0,0)$}}(\alpha=2) = \sum_{d,j \,\in\, 2 \mathbb Z}  2\,\mathcal C_{d,j} = \mathcal Z(\tfrac16,2) = \mathcal Z(\tfrac23,1)
\ee
where we used the identity 
\be
\mathcal Z(g,f) = \mathcal Z(g f^2,1).
\ee
The function $\mathcal Z(g,1)$ is the Coulombic partition function defined in \cite{FSZ87}:
\be
\mathcal Z_{\textrm{Coul}}(g) = \mathcal Z(g,1) = \sum_{m,m'\,\in\,\mathbb Z} {\mathcal Z}_{m,m'}(g)
= \frac{1}{\eta(q)\eta(\bar q)}\sum_{d,j\in{\Bbb Z}} q^{(d/\sqrt{g}+j\sqrt{g})^2/4}\bar q^{(d/\sqrt{g}-j\sqrt{g})^2/4}.
\ee

\paragraph{Partition functions for bond percolation.}
We now proceed with the same calculation for bond percolation. In this case, the Markov trace gives \cite{MDKP2017}
\be
\label{eq:ZTK}
\widehat Z_{\textrm{tor}}^{\textrm{\tiny $M$,\,$N$\,}}(\alpha) = 
\sum_{\substack{0 \le d \le N\\ d\,\equiv\,N\,\textrm{mod}\,2}} \sum_{\substack{-M\le j \le M\\ j \,\equiv\,M\,\textrm{mod}\,2}} (2-\delta_{d,0})\, T_{d \wedge j} (\tfrac \alpha2)\, K_{d,j},
\qquad
\textrm{tr}_{\raisebox{-0.05cm}{\tiny$\repW_{N,d,\omega}$}} \Tb(u)^M = \sum_{\substack{-M\le j \le M\\ j \,\equiv\,M\,\textrm{mod}\,2}}  \omega^{-j}K_{d,j},
\ee
where in this context, $\Tb(u)$ and $\repW_{N,d,\omega}$ refer to the transfer matrix and standard modules of the dense Temperley-Lieb algebra at $\beta = 1$. The trace conjecture in this case is
\be
\lim_{\substack{M,N \to \infty\\M/N = \delta}} \eE^{MN f_\mathrm{bulk}(u)}\,\mathrm{tr}_{\raisebox{-0.05cm}{\tiny$\repW_{N,d,\omega}$}} \widehat \Tb(u)^M =
\frac{1}{(q)_\infty (\bar q)_\infty}
\sum_{\ell=-\infty}^{\infty} (-1)^{M \ell}  q^{\Delta_{\gamma/\pi - \ell, d/2}} \bar q^{\Delta_{\gamma/\pi-\ell, -d/2}}.
\ee
One can show that this is equivalent to \eqref{eq:bond.perco.conj} for $\gamma = 0$. It is also consistent with the results given in Equation (2.69) of \cite{PS90}, stated for all $\beta$ and $\gamma$ but only for even parities of $M$. Defining the scaled coefficients $\mathcal K_{d,j}$ as 
\be
\mathcal K_{d,j} =  \lim_{\substack{M,N \to \infty\\M/N = \delta}} \eE^{MN f_\text{bulk}(u)} K_{d,j},
\ee
we compute them with an integral over $\gamma$ as before:
\begin{alignat}{2}
\mathcal K_{d,j} &= \frac1{2\pi}\frac1{(q)_\infty (\bar q)_\infty} \sum_{\ell = -\infty}^
\infty  \int_{0}^{2\pi} \dd\gamma\, \eE^{\ir \gamma j}\, \Big[q^{\Delta_{(\gamma-2\pi \ell)/\pi,d/2}} \bar q^{\Delta_{(\gamma-2\pi \ell)/\pi,-d/2}} 
\nonumber\\&\hspace{5.5cm}+ (-1)^M q^{\Delta_{(\gamma-2\pi \ell-\pi)/\pi,d/2}} \bar q^{\Delta_{(\gamma-2\pi \ell-\pi)/\pi,-d/2}}\Big] 
\nonumber\\&=
 \frac1{2\pi}\frac{1+(-1)^{M+j}}{(q)_\infty (\bar q)_\infty} \int_{-\infty}^{\infty} \dd\gamma\, \eE^{\ir \gamma j} q^{\Delta_{\gamma/\pi,d/2}} \bar q^{\Delta_{\gamma/\pi,-d/2}} = (1+(-1)^{M+j})\, \mathcal C_{d,j}.
\end{alignat}
That $\mathcal K_{d,j}$ vanishes for $j$ and $M$ with different parities is not a surprise, as the parity of the number of defects crossing the veritical line between the nodes $N$ and $1$ always has the same parity as $M$ for dense loop models. For $j$ and $M$ of the same parity, we instead have $\mathcal K_{d,j}= 2\, \mathcal C_{d,j}$. Combining this with \eqref{eq:ZTK}, we find that the partition functions for site and bond percolation are exactly equal, for all~$\alpha$. This in fact holds even more generally in the case where separate loop fugacities $\alpha_{i,j}$ are assigned to non-contractible loops depending on their windings. This thus gives even more evidence for a strong form of universality between the two percolation models.

%
\section{Functional relations and auxiliary functions}\label{sec:fun.rel}
%

In this section, we write down Bethe ansatz equations and other functional relations satisfied by the transfer matrix and its eigenvalues. We find that the structural form of these equations are independent of the choice of boundary conditions. We thus develop these functional equations in such a way that the nonlinear integral equations of \cref{sec:NLIE.FSS} are written in a unified way that covers both periodic and strip boundary conditions.

\subsection[Functional relations and Bethe ansatz equations for $\Tbh(u)$]{Functional relations and Bethe ansatz equations for $\Tbh\boldsymbol{(u)}$}\label{FuncCyl}

For the general dilute $\Atwotwo$ loop models, it was shown in \cite{MDP19} that the fundamental transfer matrix $\Tb(u)$ generates a family of fused transfer matrices $\Tb^{m,n}(u)$ where $m$ and $n$ are integers and
\be
\Tb^{1,0}(u) = \Tb(u), \qquad
\Tb^{0,1}(u) = \Tb(u+\lambda).
\ee
With the current choice of Boltzmann weights \eqref{eq:weights} without factors $\sin 2\lambda \sin 3\lambda$ in the denominators, this construction of fused transfer matrices applies to all values of $\lambda \in (0,\pi)$ and, of particular relevance here, to the value $\lambda = \frac \pi 3$. The fusion hierarchy relations derived in \cite{MDP19} also hold for this value of $\lambda$, but now with $f(u)$ as in \eqref{eq:fk}. Specializing to $\lambda = \frac \pi 3$, the derivation of the closure relations readily extends to this case and yields
\be
\label{eq:closure}
f(u+\tfrac{2\pi}{3})\Tb^{3,0}(u) = \Tb^{1,1}(u+\tfrac{2\pi}{3}) + (-1)^N f(u) f(u+\tfrac{\pi}{3}) f(u+\tfrac{2\pi}{3})\Jb,
\ee
where $\Jb$ is a central element of $\pdtl_N(\alpha,\beta)$ that does not depend on $u$. On each standard module $\repW_{N,d,\omega}$, this element is proportional to the identity matrix, with the unique eigenvalue 
\be
J=
\left\{\begin{array}{cl}
\alpha^3 - 3 \alpha & d = 0, \\[0.1cm]
(-1)^{d}(\omega^3 + \omega^{-3}) & d > 0. \\[0.1cm]
\end{array}\right.
\ee
With the convention \eqref{eq:alpha.omega} for $\alpha$, this eigenvalue is $J=(-1)^{d}(\omega^3 + \omega^{-3})$ for all $d$. The closure relation \eqref{eq:closure} can equivalently be rewritten as a cubic functional relation satisfied by $\Tbh(u)$:
\begin{alignat}{2}
\Tbh(u)\Tbh(u+\tfrac{\pi}{3})\Tbh(u+\tfrac{2\pi}{3}) &= (-1)^N f(u) \Tbh(u)^2 + f(u+\tfrac{\pi}{3}) \Tbh(u+\tfrac{\pi}{3})^2 + (-1)^Nf(u+\tfrac{2\pi}{3}) \Tbh(u+\tfrac{2\pi}{3})^2 
\nonumber\\[0.15cm]&
+ f(u) f(u+\tfrac{\pi}{3}) f(u+\tfrac{2\pi}{3}) (\Jb - 2 \Ib).\label{Tcubic}
\end{alignat}
The eigenvalues $\widehat{T}(u)$ of $\Tbh(u)$ in the module $\repW_{N,d,\omega}$ satisfy this scalar cubic equation. We emphasize that the $s\ell(3)$ cubic relation differs from the $s\ell(2)$ cubic relation, (3.10) of \cite{PearceADEfunc92}, implied by the closure relation for critical bond percolation on the square lattice~\cite{MDKP2017}.

The Bethe equations for the $\Atwotwo$ model are well known~\cite{BNW89,WBN92,ZB97}. For $\lambda = \frac \pi 3$, the eigenvalues $\widehat{T}(u)$ are expressed in terms of the auxiliary function $Q(u)$ as
\be
\widehat T(u) Q(u+\tfrac \pi 3) Q(u+\tfrac {2\pi} 3) = \sigma \nu f(u) Q(u)^2 + \omega\zeta f(u+\tfrac \pi 3) Q(u+\tfrac \pi 3)^2 + \sigma (\omega\zeta)^{-1} f(u+\tfrac {2\pi} 3) Q(u+\tfrac {2\pi} 3)^2\label{TQeqn}
\ee
where $\zeta$ is a complex parameter independent of $u$ and $\omega$, the function $Q(u)$ is a centered Laurent polynomial in $\eE^{\ir u}$ of degree $M$
\be
\label{QpolyT}
Q(u) = \prod_{j=1}^{M} \sin(u-u_j),\qquad 0\le M\le N-d,
\ee
and
\be
\sigma = (-1)^N, \qquad \nu = (-1)^d.
\ee
This is the $s\ell(3)$ analog of Baxter's famous $T$-$Q$ relation~\cite{BaxterQ}. Indeed, substituting \eqref{TQeqn} into \eqref{Tcubic}, one verifies that the cubic equation is satisfied provided
\be
(-1)^M=\sigma\nu\qquad\mbox{and}\qquad \zeta^3=1.
\ee
This implies that $N-d-M$ is an even number. The root of unity $\zeta$ is fixed by considering the braid transfer matrices
\be
\Tbh_{\pm \infty} = \lim_{u \to \pm\ir \infty} \frac{\Tbh(u)}{f(u)}.
\ee
They are central elements of $\pdtl_N(\alpha,\beta)$, each one with the unique eigenvalue 
\be
\widehat T_{\pm \infty} = \omega\, \eE^{\mp\ir \pi d/3}+1 + \omega^{-1}\eE^{\pm\ir \pi d/3}
\ee 
on $\repW_{N,d,\omega}$. By computing $\widehat T_\infty$ or $\widehat T_{-\infty}$ using \eqref{TQeqn}, we find 
\be
\zeta = \eE^{\tfrac{\ir \pi}3 (N-d-M)},
\ee 
thus fixing $\zeta$ uniquely. From numerics, we find that the roots $u_j$ for the ground state in each standard module $\repW_{N,d,\omega}$ are pure imaginary for $\omega = \pm 1$. Their number $M$ is $(N\!-\!d)$ for all $d$.

The zeros $u_j \in {\Bbb C}$ of $Q(u)$ are the Bethe roots and satisfy the Bethe ansatz equations
\be
\bigg(\frac{\sin (u_i-\frac {\pi} 3)}{\sin (u_i+\frac \pi 3)}\bigg)^N = - \frac{\nu}{\omega\zeta} \prod_{j=1}^{M} \bigg(\frac{\sin(u_i-u_j+\frac{\pi}3)}{\sin(u_i-u_j-\frac{\pi}3)}\bigg)^2, \qquad i = 1,2,\dots, M.\label{TBetheAnsatz}
\ee
Given the eigenvalue $\widehat T(u)$ and following \cite{FabriciusMcCoy01}, it is extremely useful to have linear equations determining $Q(u)$. To achieve this, we introduce a second auxiliary function $P(u)$ defined as 
\be
P(u) = \frac{f(u+\frac \pi 3)\,Q(u+\frac \pi 3)^2 + \sigma\nu\omega\zeta f(u+\frac {2\pi} 3)\, Q(u+\frac {2\pi} 3)^2}{Q(u)}.
\ee 
From the Bethe equations, it follows that any zero of the denominator is also a zero of the numerator. As a result, the function $P(u)$ has no poles and can be written as a centered Laurent polynomial in $\eE^{\ir u}$ of degree $M+N$. It is straightforward to show that the transfer matrix eigenvalue $\widehat T(u)$ and the auxiliary functions $Q(u)$ and $P(u)$ satisfy the linear relations
\begin{subequations}
\label{eq:TLinearsys}
\begin{alignat}{2}
\widehat T(u)\,Q(u+\tfrac \pi 3) &= \nu P(u+\tfrac{2\pi}3) + \sigma (\omega\zeta)^{-1}\, f(u+\tfrac{2\pi}3)\, Q(u+\tfrac{2\pi}3),
\\[0.1cm]
\widehat T(u)\,Q(u+\tfrac {2\pi} 3) &= \sigma (\omega\zeta)^{-1} P(u+\tfrac{\pi}3) + \omega\zeta f(u+\tfrac{\pi}3)\, Q(u+\tfrac{\pi}3).
\end{alignat}
\end{subequations}
The functions in these functional relations satisfy the periodicity properties
\be
\label{eq:period.fTQP}
f(u+\pi) = \sigma f(u), \qquad
\widehat T(u+\pi) = \sigma\, \widehat T(u),\qquad
Q(u+\pi) = \sigma \nu\, Q(u), \qquad
P(u+\pi) = \nu\, P(u).
\ee

\subsection[Functional relations and Bethe ansatz equations  for $\Dbh(u)$]{Functional relations and Bethe ansatz equations  for $\Dbh\boldsymbol{(u)}$}\label{FuncStrip}

Analogous functional equations hold for critical site percolation with strip boundary conditions. 
Indeed, the reduced double row transfer matrix $\Dbh(u)$, defined in \eqref{redDTM}, satisfies the cubic functional relation
\begin{align}
\Dbh(u)\Dbh(u+\tfrac{\pi}{3})\Dbh(u+\tfrac{2\pi}{3})&=f(u)\Dbh(u)^2+f(u+\tfrac{\pi}{3})\Dbh(u+\tfrac{\pi}{3})^2+f(u+\tfrac{2\pi}{3})\Dbh(u+\tfrac{2\pi}{3})^2
\nonumber\\[0.15cm]
&-4f(u)f(u+\tfrac{\pi}{3})f(u+\tfrac{2\pi}{3}) \Ib.\label{Dcubic}
\end{align}
This functional equation agrees with the functional relations derived for the dilute $\Atwotwo$ loop models for strip boundary conditions with more general roots of unity~\cite{BMDSA21}. This relation has the same structure as the cubic equation \eqref{Tcubic} satisfied by the single row transfer matrix, with $N$ even and with the identification $\Jb=-2\Ib$. This last identification can equivalently be written as $\omega^3=(-1)^{d+1}$. 

The analog of Baxter's $T$-$Q$ relation for strip boundary conditions is
\be
\widehat D(u) \widehat Q(u+\tfrac \pi 3) \widehat Q(u+\tfrac {2\pi} 3) = - f(u)\widehat Q(u)^2 + f(u+\tfrac \pi 3) \widehat Q(u+\tfrac \pi 3)^2 
+ f(u+\tfrac {2\pi} 3) \widehat Q(u+\tfrac {2\pi} 3)^2\label{DQeqn}
\ee
where $\widehat D(u)$ denotes an eigenvalue of $\Dbh(u)$ and $\widehat Q(u) = \sin u\, Q(u)$ succinctly incorporates in its definition the $O(1)$ contributions originating from the presence of the boundary. Substituting \eqref{DQeqn} into \eqref{Dcubic}, it is easily verified that the eigenvalues $\widehat D(u)$ satisfy the required cubic. 

The Bethe roots $u_j\in\Bbb C$ are the zeros of $Q(u)$ in the complex $u$-plane. Due to the crossing symmetry $\Db(u) = \Db(-u)$, the Bethe roots come in pairs $(u_j,-u_j)$, with $j = 1,2, \dots, M$. Thus $\widehat Q(u)$ is the centered Laurent polynomial of degree $2M+1$
\be
\widehat Q(u)=\sin u\,Q(u)=\prod_{j=-M}^M  \sin(u-u_j) = \sin u\prod_{j=1}^M \sin(u-u_j)\sin(u+u_j)\label{Qpoly}
\ee
with 
\be
u_0=0,\qquad u_{-j}=-u_j,\qquad j=0,1,\ldots,M.\label{BetheRootSym}
\ee
From the functional relations \eqref{DQeqn}, it follows that the Bethe roots $u_j$ satisfy the Bethe ansatz equations
\be
\bigg(\frac{\sin (u_i-\frac \pi 3)}{\sin (u_i+\frac {\pi} 3)}\bigg)^{2N} = \prod_{j=-M}^M \bigg(\frac{\sin(u_i-u_j+\frac{\pi}3)}{\sin(u_i-u_j-\frac{\pi}3)}\bigg)^2, \qquad i = -M,-M+1, \ldots, M.\label{DBetheAnsatz}
\ee
From numerics, we find that the roots $u_j$ for the ground state in each standard module $\repV_{N,d}$ are pure imaginary. Their number $2M$, excluding $u_0$, is $2(N\!-\!d)$ for $d \ge 1$ and $2(N-1)$ for $d=0$. For certain eigenstates, this number $M$ is reduced in accord with the presence of removable $3$-strings and roots at infinity.

Given $\widehat D(u)$, to obtain linear equations determining $\widehat Q(u)$, we introduce a second auxiliary function $P(u)$ defined as
\be
P(u) = \frac{f(u+\frac {\pi} 3)\,\widehat Q(u+\frac {\pi} 3)^2 - f(u+\frac {2\pi} 3)\,\widehat Q(u+\frac {2\pi} 3)^2}{\widehat Q(u)}.
\ee 
From the Bethe ansatz equations, it follows that any zero of the denominator is also a zero of the numerator. As a result, the function $P(u)$ has no poles and can be written as a centered Laurent polynomial in $\eE^{\ir u}$, with maximal degree $2M+2N+1$. If $Q(u)$ contains a $3$-string, then $P(u)$ contains the same $3$-string, and its degree $P(u)$ will be reduced accordingly. Given $\widehat D(u)$, the auxiliary functions $Q(u)$ and $P(u)$ satisfy the linear relations
\begin{subequations}
\label{eq:DLinearSys}
\begin{alignat}{2}
\widehat D(u)\,\widehat Q(u+\tfrac \pi 3) &= - P(u+\tfrac{2\pi}3) + f(u+\tfrac{2\pi}3)\,\widehat Q(u+\tfrac{2\pi}3),
\\[0.1cm]
\widehat D(u)\,\widehat Q(u+\tfrac {2\pi} 3) &=P(u+\tfrac{\pi}3) + f(u+\tfrac{\pi}3)\,\widehat Q(u+\tfrac{\pi}3).
\end{alignat}
\end{subequations}
We note that the $T$-$Q$ and linear relations given here for strip boundary conditions can all be obtained from those given in \cref{FuncCyl} for periodic boundary conditions, by setting 
\be
\label{eq:strip.constants}
\sigma = 1, \qquad
\omega = 1, \qquad
\zeta = 1, \qquad 
\nu = -1.
\ee
With these values, the functions $f(u)$, $\widehat T(u)$, $Q(u)$ and $P(u)$ satisfy the periodicity properties \eqref{eq:period.fTQP}.

\subsection{Auxiliary functions}\label{sec:aux.functions}
To set up our nonlinear integral equations, we need to introduce further auxiliary functions built from $f(u)$ and $Q(u)$. Following \cite{FK99,J08}, we define the functions
\begin{subequations}
\begin{alignat}{2}
\Lambda^1(u) &= \frac{f(u+\frac \pi 6)\, Q(u+\frac \pi 6)}{\omega\zeta\, Q(u-\frac \pi 6)}, 
\\[0.15cm]
\Lambda^2(u) &= \frac{\sigma \nu f(u+\frac \pi 2)\, Q(u+\frac \pi 2)^2}{Q(u-\frac \pi 6)\,Q(u+\frac \pi 6)},
\\[0.15cm]
\Lambda^3(u) &= \frac{\sigma \omega\zeta\, f(u-\frac \pi 6)\, Q(u-\frac \pi 6)}{Q(u+\frac \pi 6)}.
\end{alignat}
\end{subequations}
This holds for both periodic and strip boundary conditions, with the constants $\sigma$, $\omega$, $\zeta$ and $\nu$ set to \eqref{eq:strip.constants} in the latter case. The new auxiliary functions are related to the eigenvalues $\widehat T(u)$ and $\widehat D(u)$ by
\be
\Lambda^1(u) + \Lambda^2(u) + \Lambda^3(u) = 
\left\{\begin{array}{ll}
\widehat T(u+\tfrac \pi 2) &\textrm{periodic},\\[0.15cm]
\widehat D(u+\tfrac \pi 2) &\textrm{strip}.
\end{array}
\right.
\ee
In both cases, the six auxiliary functions that enter the nonlinear integral equations of \cref{sec:NLIE.FSS} are defined as
\begingroup
\allowdisplaybreaks
\begin{subequations}
\label{eq:aA.def}
\begin{alignat}{3}
\amf^1(z) &= \frac{\Lambda^1(\ir z)}{\Lambda^2(\ir z)+\Lambda^3(\ir z)}, 
\\
\amf^2(z) &= \frac{\Lambda^3(\ir z)}{\Lambda^1(\ir z)+\Lambda^2(\ir z)},
\\
\amf^3(z) &= \frac{\Lambda^1(\ir z)\Lambda^3(\ir z)}{\Lambda^2(\ir z)\big(\Lambda^1(\ir z)+\Lambda^2(\ir z)+\Lambda^3(\ir z)\big)}, 
\\
\Amf^1(z) &= 1+\amf^1(z) = \frac{\Lambda^1(\ir z)+\Lambda^2(\ir z)+\Lambda^3(\ir z)}{\Lambda^2(\ir z)+\Lambda^3(\ir z)}, 
\\
\Amf^2(z) &= 1+\amf^2(z) = \frac{\Lambda^1(\ir z)+\Lambda^2(\ir z)+\Lambda^3(\ir z)}{\Lambda^1(\ir z)+\Lambda^2(\ir z)},
\\
\Amf^3(z) &= 1+\amf^3(z) = \frac{\big(\Lambda^1(\ir z)+\Lambda^2(\ir z)\big)\big(\Lambda^2(\ir z)+\Lambda^3(\ir z)\big)}{\Lambda^2(\ir z)\big(\Lambda^1(\ir z)+\Lambda^2(\ir z)+\Lambda^3(\ir z)\big)}.
\end{alignat}
\end{subequations}
\endgroup
Setting
\be
\bmf^1(z)  = 
\left\{\begin{array}{ll}
\widehat T(\ir z+\tfrac \pi 2) &\textrm{periodic},\\[0.1cm]
\widehat D(\ir z+\tfrac \pi 2) &\textrm{strip},
\end{array}\right.
\qquad
\bmf^2(z) = 
\left\{\begin{array}{ll}
Q(\ir z + \tfrac \pi 2) &\textrm{periodic},\\[0.1cm]
\widehat Q(\ir z + \tfrac \pi 2) &\textrm{strip},
\end{array}\right.
\qquad
\begin{array}{c}
\bmf^3(z) = P(\ir z),
\\[0.15cm]
\Phi(z) = f(\ir z),
\end{array}
\ee
the six additional auxiliary functions can be expressed as
\begingroup
\allowdisplaybreaks
\begin{subequations}
\label{eq:a.U.and.b.T}
\begin{alignat}{3}
\amf^1(z) &= \frac{\Phi(z-\frac{\ir \pi}6) \bmf^2(z+\frac{\ir \pi}3)}{ \sigma  \nu  \omega  \zeta\,\bmf^3(z-\frac{\ir \pi}6)},
\qquad
&&\Amf^1(z) = \frac{\bmf^1(z) \bmf^2(z-\frac{\ir \pi}3)}{\bmf^3(z-\frac{\ir \pi}6)},
\\[0.15cm]
\amf^2(z) &= \frac{ \nu  \omega^2  \zeta^2\,\Phi(z+\frac{\ir \pi}6) \bmf^2(z-\frac{\ir \pi}3)}{\bmf^3(z+\frac{\ir \pi}6)},\qquad
&&\Amf^2(z) = \frac{\omega\zeta\,\bmf^1(z) \bmf^2(z+\frac{\ir \pi}3)}{\bmf^3(z+\frac{\ir \pi}6)},
\\[0.15cm]
\amf^3(z) &= \frac{\Phi(z+\frac{\ir \pi}6)\Phi(z-\frac{\ir \pi}6)}{\Phi(z-\frac{\ir \pi}2)}\frac{\bmf^2(z-\frac{\ir \pi}3)\bmf^2(z+\frac{\ir \pi}3)}{\bmf^1(z)\big(\bmf^2(z)\big)^2},
\qquad
&&\Amf^3(z) = \frac{\bmf^3(z-\frac{\ir \pi}6) \bmf^3(z+\frac{\ir \pi}6)}{\omega\zeta\,\Phi(z-\frac{\ir \pi}2)\bmf^1(z)\big(\bmf^2(z)\big)^2}.
\end{alignat}
\end{subequations}
\endgroup
These are identical for the two types of boundary conditions, up to the different values taken by the constants $\sigma$, $\omega$, $\zeta$ and $\mu$ which play no role in the derivation of the nonlinear integral equations in the next section.

%
\section{Nonlinear integral equations and finite-size corrections}\label{sec:NLIE.FSS}
%

In this section, we use the Yang-Baxter integrability of the model of critical site percolation on the triangular lattice to obtain the leading terms in the $\frac 1N$ expansions \eqref{eq:logTD.asy} of the eigenvalues $\widehat T(u)$ and $\widehat D(u)$. Starting with the functional equations and following \cite{KBP91,KP92,FK99,J08}, we first derive suitable NLIEs that hold for any eigenvalue of the tranfer matrix and turn out to be identical for both types of boundary conditions. We then analytically solve them for the groundstate in each standard module, separately for the periodic and strip boundary conditions, and thus compute the leading finite-size eigenvalue corrections and the corresponding conformal weight in the scaling limit.

\subsection{Derivation of the nonlinear integral equations}\label{sec:NLIEs}

To obtain the NLIEs, we take the second derivative of the logarithm of the relations \eqref{eq:a.U.and.b.T} and take their Fourier transforms. These Fourier transforms exist for the logarithm of an auxiliary function in open vertical strips where the function is analytic, non-zero and has constant asymptotics (ANZC property). The analyticity strips for the functions $\bmf^2(z)$ and $\bmf^3(z)$ are horizontal strips centered on the real line. Their widths are determined by the maximal imaginary shifts of the functions in the functional equations \eqref{eq:a.U.and.b.T}, and are respectively $\frac {2\ir \pi} 3$ and $\frac {\ir \pi} 3$. In contrast, the function $\bmf^1(z)$ always appears with its argument unshifted. It is therefore not necessary to define an analyticity strip for this function.

Depending on the eigenstate considered, the eigenvalues of $\bmf^2(z)$ and $\bmf^3(z)$ may have zeros located inside the analyticity strips. These can take the form of $1$-strings, namely real zeros lying on the real $z$ axis, or short $2$-strings, namely pairs of zeros lying symmetrically above and below this axis. There can also be long $2$-strings, namely symmetrical pairs of zeros that lie outside the strips, but these do not impact the derivation of the nonlinear integral equations. Let us denote by $\setS^2$ and $\setS^3$ the sets of zeros of $\bmf^2(z)$ and $\bmf^3(z)$ inside their corresponding analyticity strips. We define the functions
\be
\widehat \bmf^1(z) = \bmf^1(z), \qquad
\widehat \bmf^2(z) = \frac{\bmf^2(z)}{\prod_{x_2 \in \setS^2} \tanh \frac12(z-x_2)}, \qquad
\widehat \bmf^3(z) = \frac{\bmf^3(z)}{\prod_{x_3 \in \setS^3} \tanh \frac12(z-x_3)},
\ee
whose analyticity strips are free of zeros. The second derivative of $\log \widehat \bmf^n(z)$ vanishes for $z \to \pm \infty$, allowing us to define the Fourier transforms
\be
B^n(k) = \frac1{2\pi} \int_{-\infty}^\infty \dd z\, \eE^{-\ir k z}\big(\log \widehat \bmf^n(z)\big)'',
\qquad
(\log \widehat \bmf^n(z)\big)'' = \int_{-\infty}^\infty \dd k\, \eE^{\ir k z} B^n(k).
\ee

We likewise define functions $\widehat \amf^n(z)$ and $\widehat \Amf^n(z)$ in such a way that the relations \eqref{eq:a.U.and.b.T} hold unchanged with $\bmf^n, \amf^n, \Amf^n \to \widehat\bmf^n, \widehat\amf^n, \widehat\Amf^n$:
\begingroup
\allowdisplaybreaks
\begin{subequations}
\label{eq:hat.aA.defs}
\begin{alignat}{2}
\widehat \amf^1(z) &= \amf^1(z)\,\frac{\prod_{x_3 \in \setS^3} \tanh\frac12(z-x_3-\frac {\ir \pi} 6)}{\prod_{x_2 \in \setS^2} \tanh\frac12(z-x_2+\frac {\ir \pi} 3)},
\\[0.15cm]
\widehat \amf^2(z) &= \amf^2(z)\,\frac{ \prod_{x_3 \in \setS^3} \tanh\frac12(z-x_3+\frac {\ir \pi} 6)}{\prod_{x_2 \in \setS^2} \tanh\frac12(z-x_2-\frac {\ir \pi} 3)},
\\[0.15cm]
\widehat \amf^3(z) &= \amf^3(z)\,\frac{\prod_{x_2 \in \setS^2} \tanh\frac12(z-x_2)^2}{\prod_{x_2 \in \setS^2} \tanh\frac12(z-x_2-\frac {\ir \pi} 3)\tanh\frac12(z-x_2+\frac {\ir \pi} 3)},
\\[0.15cm]
\widehat \Amf^1(z) &= \Amf^1(z)\,\frac{ \prod_{x_3 \in \setS^3} \tanh\frac12(z-x_3-\frac {\ir \pi} 6)}{\prod_{x_2 \in \setS^2} \tanh\frac12(z-x_2-\frac {\ir \pi} 3)},\\[0.15cm]
\widehat \Amf^2(z) &= \Amf^2(z)\,\frac{ \prod_{x_3 \in \setS^3} \tanh\frac12(z-x_3+\frac {\ir \pi} 6)}{\prod_{x_2 \in \setS^2} \tanh\frac12(z-x_2+\frac {\ir \pi} 3)},\\[0.15cm]
\widehat \Amf^3(z) &= \Amf^3(z)\,\frac{\prod_{x_2 \in \setS^2} \tanh\frac12(z-x_2)^2}{\prod_{x_3 \in \setS^3} \tanh\frac12(z-x_3 - \frac {\ir \pi}6)\tanh\frac12(z-x_3 + \frac {\ir \pi}6)}.
\end{alignat}
\end{subequations}
\endgroup
We define the Fourier transforms of their second logarithmic derivative as
\begin{subequations}
\begin{alignat}{3}
L^n(k) &= \frac1{2\pi} \int_{-\infty+\ir \epsilon_n}^{\infty+\ir \epsilon_n} \dd z\, \eE^{-\ir k z}\big(\log \widehat \amf^n(z)\big)'', 
\qquad
&\big(\log \widehat \amf^n(z+\ir \epsilon_n)\big)'' &&= \int_{-\infty}^\infty \dd k\, \eE^{\ir k(z+\ir \epsilon_n)} L^n(k),
\\
A^n(k) &= \frac1{2\pi} \int_{-\infty+\ir \epsilon_n}^{\infty+\ir \epsilon_n} \dd z\, \eE^{-\ir k z}\big(\log \widehat \Amf^n(z)\big)'', 
\qquad
&\big(\log \widehat \Amf^n(z+\ir \epsilon_n)\big)'' &&= \int_{-\infty}^\infty \dd k\, \eE^{\ir k(z+\ir \epsilon_n)} A^n(k).
\end{alignat}
\end{subequations}
Here, the paths are moved away from the real axis by certain infinitesimal shifts $\epsilon_n$ which are chosen individually for each $n$. For reasons detailed below, we choose these parameters such that 
\be
\label{eq:epsilon.inequalities}
0< \epsilon_2  < \epsilon_3 < \epsilon_1.
\ee
These shifts are chosen small enough that the function $\bmf^n(z)$ has no zeroes in the region $\mbox{$0 < \textrm{Im}(z) < \epsilon_n$}$. 

Applying the Fourier transforms to \eqref{eq:a.U.and.b.T}, we find
\begin{subequations}
\label{eq.AB.relations}
\begin{alignat}{2}
\begin{pmatrix}
L^1(k) \\ L^2(k) \\ L^3(k)
\end{pmatrix} &=
\begin{pmatrix}
0 & \eE^{-\frac{\pi k}3} & -\eE^{\frac{\pi k}6}  \\
0 & \eE^{\frac{\pi k}3} & -\eE^{-\frac{\pi k}6} \\
-1 & \eE^{\frac{\pi k}3}+\eE^{-\frac{\pi k}3}-2 & 0 \\
\end{pmatrix}
\begin{pmatrix}
B^1(k) \\ B^2(k) \\ B^3(k)
\end{pmatrix} 
+ \frac{\mu N k}{2 \sinh(\frac{\pi k}2)}
\begin{pmatrix}
\eE^{-\frac{\pi k}3} \\ \eE^{\frac{\pi k}3} \\ \eE^{\frac{\pi k}3}+\eE^{-\frac{\pi k}3}-1
\end{pmatrix},
\\[0.15cm]
\begin{pmatrix}
A^1(k) \\ A^2(k) \\ A^3(k)
\end{pmatrix} &=
\begin{pmatrix}
1 & \eE^{\frac{\pi k}3} & -\eE^{\frac{\pi k}6}  \\
1 & \eE^{-\frac{\pi k}3} & -\eE^{-\frac{\pi k}6} \\
-1 & -2 & \eE^{\frac{\pi k}6}+\eE^{-\frac{\pi k}6} \\
\end{pmatrix}
\begin{pmatrix}
B^1(k) \\ B^2(k) \\ B^3(k)
\end{pmatrix} 
-  \frac{\mu N k}{2 \sinh(\frac{\pi k}2)}
\begin{pmatrix}
0 \\ 0 \\ 1
\end{pmatrix}.
\end{alignat}
\end{subequations}
where
\be
\mu = \left\{\begin{array}{ll}
1 &\textrm{periodic,}\\[0.1cm]
2 &\textrm{strip.}
\end{array}\right.
\ee
The rightmost terms in these equations arise from the Fourier transform of the logarithm of the function $\Phi(z)$, itself obtained from the integral
\be
\frac 1{2\pi} \int_{-\infty+\ir \epsilon}^{\infty+\ir \epsilon} \dd z\, \eE^{-\ir k z} \big(\log \sinh z \big)'' = \frac{k}{1-\eE^{-\pi k}},  
\qquad \epsilon >0.
\ee
We invert the matrix appearing in \eqref{eq.AB.relations}, and find after some simplifications
\be
\begin{pmatrix}
L^1(k) \\ L^2(k) \\ L^3(k) 
\end{pmatrix} 
= \tilde K(k) \cdot \begin{pmatrix}
A^1(k) \\ A^2(k) \\ A^3(k)
\end{pmatrix}
+ \frac{\mu N k}{\eE^{\frac{\pi k}3} + \eE^{-\frac{\pi k}3}-1}
\begin{pmatrix}
-\eE^{-\frac{\pi k}6} \\ \eE^{\frac{\pi k}6} \\ \eE^{\frac{\pi k}6}-\eE^{-\frac{\pi k}6}
\end{pmatrix},
\ee
where the kernel matrix is
\be
\label{eq:eqns.in.k.space}
\tilde K(k) = \frac1{\eE^{\frac{\pi k}3} + \eE^{-\frac{\pi k}3}-1}
\begin{pmatrix}
-2 & -\eE^{\frac{\pi k}3}+\eE^{-\frac{\pi k}3}+1 & -\eE^{\frac{\pi k}3}+\eE^{-\frac{\pi k}3}-1\\
\eE^{\frac{\pi k}3}-\eE^{-\frac{\pi k}3}+1 & -2 & \eE^{\frac{\pi k}3}-\eE^{-\frac{\pi k}3}-1\\
\eE^{\frac{\pi k}3}-\eE^{-\frac{\pi k}3}-1 & -\eE^{\frac{\pi k}3} +\eE^{-\frac{\pi k}3}-1 & \eE^{\frac{\pi k}3}+\eE^{-\frac{\pi k}3}-3
\end{pmatrix}.
\ee
This matrix satisfies the symmetry
\be
\tilde K(k)^\intercal = \tilde K(-k).
\ee
It also has non-zero asymptotic values in the limits $k \to \pm \infty$. Defining the symmetric matrix
\be
\Gamma = \begin{pmatrix}
0 & 1 & 1 \\
1 & 0 & 1 \\
1 & 1 & 1 
\end{pmatrix},
\ee
we observe that each entry of the matrix $\tilde K(k) - \Gamma$ has at most one non-zero asymptotic value, for $k \to + \infty$ and $k \to-\infty$. Applying the inverse Fourier transform to \eqref{eq:eqns.in.k.space}, we find
\be
\label{eq:prelim.NLIEs}
\log \widehat\amf^n(z + \ir \epsilon_n)'' = \mu N\fmf^n(z + \ir \epsilon_n)'' + \sum_{m=1}^3 \Gamma_{nm} \log \widehat\Amf^m(z + \ir \epsilon_n)'' + \sum_{m=1}^3 \big(K_{nm} * (\log \widehat\Amf^m)''\big) (z+\ir \epsilon_m),
\ee
where the driving terms are
\be
\fmf^n(z) =
\left\{\begin{array}{ll}
\displaystyle  \log\bigg[\frac{\tanh (\frac z2 - \frac{\ir \pi}{12})}{\tanh (\frac z2 + \frac{\ir \pi}4)}\bigg] \quad & n=1,\\[0.4cm]
\displaystyle  \log\bigg[\frac{\tanh (\frac z2 + \frac{\ir \pi}{12})}{\tanh (\frac z2 - \frac{\ir \pi}4)}\bigg] \quad & n=2,\\[0.4cm]
\displaystyle \log\bigg[\frac{\tanh (\frac z2 - \frac{\ir \pi}{12})}{\tanh (\frac z2 - \frac{5\ir \pi}{12})}\bigg] \quad & n=3.
\end{array}\right.
\ee
The convolution of two functions is
\be
(f * g)(z) = \int_{-\infty}^\infty \dd y f(z-y) g(y) =  \int_{-\infty}^\infty \dd y f(y) g(z-y),
\ee
and the kernel functions are 
\be
\label{eq:kernel.Knm}
K_{nm}(z) = \frac1{2\pi}\int_{-\infty}^{\infty} \dd k\, \eE^{\ir k (z + \ir(\epsilon_n-\epsilon_m))}\big(\tilde K(k)_{nm} - \Gamma_{nm}\big), \qquad n,m = 1,2,3.
\ee
The choice \eqref{eq:epsilon.inequalities} ensures that the integrands in \eqref{eq:kernel.Knm} are all finite at $\pm \infty$, thus leading to convergent integrals. The explicit expression for the kernel matrix is
\be
\lim_{\epsilon_1,\epsilon_2,\epsilon_3 \to 0}K(z) =  -\frac {2 \sqrt 3}{\pi \sinh 3z} 
\begin{pmatrix}
\sinh 2z & -\sinh (2z - \frac{\ir \pi}3) & \sinh (2z+ \frac{\ir \pi}3)\\[0.15cm]
-\sinh(2z + \frac{\ir \pi}3)  & \sinh 2z & \sinh (2z - \frac{\ir \pi}3)\\[0.15cm]
\sinh (2z - \frac {\ir \pi}3) & \sinh (2z + \frac {\ir \pi}3) & \sinh 2z 
\end{pmatrix}.
\ee
For simplicity, the result is given here in the limit $\epsilon_1, \epsilon_2, \epsilon_3 \to 0$. The $\epsilon$-dependence is recovered by simply reinstating $z \to z + \ir(\epsilon_n-\epsilon_m)$ for each matrix entry $K_{nm}(z)$. Moreover, we note that
\be
K(z)^\intercal = K(-z).
\ee

Using \eqref{eq:hat.aA.defs} and \eqref{eq:prelim.NLIEs}, we obtain NLIEs for the six functions $\amf^n(z)$ and $\Amf^n(z)$ with $n=1,2,3$. The dependence on the zeros in the sets $\setS^2$ and $\setS^3$ appear in two ways: some terms appear directly as algebraic functions of $z$, whereas others arise inside the convolution integrals. These integrals can be evaluated explicitly using the residue theorem, combined with the algebraic terms and then simplified. However, we prefer here a different presentation, whereby the integration path is deformed around certain zeros of the functions in such a way that there are no extra algebraic terms depending on the zeros of $\setS^2$ and $\setS^3$. In this setup, we find 
\begin{alignat}{2}
\log \amf^n(z + \ir \epsilon_n) -\phi^n &= \mu N \fmf^n(z + \ir \epsilon_n) + \sum_{m=1}^3 \Gamma_{nm} \log \Amf^m(z + \ir \epsilon_n) 
\nonumber\\&
+ \sum_{m=1}^3 \int_{\mathcal C^m}\dd y \, K_{nm}(z-y) \log \Amf^m(y+\ir \epsilon_m)\big).
\label{eq:aA.NLIEs}
\end{alignat}
To obtain this relation, we integrated twice with respect to $z$ and $\phi^n$ are the integration constants. In doing so, we assumed that the functions $\amf^n(z)$ and $\Amf^n(z)$ have finite, non-zero asymptotics in the limit $z \to \infty$, allowing us to set to zero the linear terms in $z$ obtained from the two integrations. The paths $\mathcal C^m$ are paths in the complex $y$-plane that run from $-\infty$ to $+\infty$ and have non trivial trajectories above or below the zeros and poles of the functions $\Amf^m(z + \ir \epsilon_m)$. We define the sets
\be
\label{eq:S.subsets}
\setS^{n,\uparrow} = \{x \in \setS^n \,\big|\, \textrm{Im}(x)> 0\}, \qquad
\setS^{n,\downarrow} = \{x \in \setS^n \,\big|\, \textrm{Im}(x)\le 0\}, \qquad
n = 2,3.
\ee
The path $\mathcal C^m$ then has the following properties: (i) it passes above the points $x + \frac{\ir \pi}3-\ir \epsilon_m$ and $y + \frac{\ir \pi}6-\ir \epsilon_m$, with $x \in \setS^{2,\downarrow}$ and $y \in \setS^{3,\downarrow}$, (ii) it passes below the points $x - \frac{\ir \pi}3-\ir \epsilon_m$ and $y - \frac{\ir \pi}6-\ir \epsilon_m$, with $x \in \setS^{2,\uparrow}$ and $y \in \setS^{3,\uparrow}$, and (iii) it passes above or below the poles of the kernel function $K_{nm}(z-y)$ in precisely the same way as the straight path from $-\infty$ to $+\infty$ does. 

Using the same technique, we obtain an integral equation for $\log \widehat\bmf^1(z)''$:
\be
\label{eq:hatbmf.NLIE}
\log \widehat\bmf^1(z + \ir \epsilon_0)'' =  - \mu N \Fmf(z+\ir \epsilon_0) + \sum_{n=1}^3 \int_{-\infty}^{\infty} \dd y \, K^n(z-y+\ir \epsilon_0-\ir \epsilon_n) \log \widehat\Amf^n(y+\ir \epsilon_n)'',
\ee
where the kernel functions are
\be
K^1(z) = \frac{\sqrt 3 \sinh (2z- \frac {\ir \pi} 3)}{\pi \sinh 3z},
\qquad
K^2(z) = \frac{\sqrt 3 \sinh (2z+ \frac {\ir \pi} 3)}{\pi \sinh 3z},
\qquad
K^3(z) = \frac{\sqrt 3 \sinh 2z}{\pi \sinh 3z},
\ee
the shift $\epsilon_0$ is infinitesimal and chosen such that
\be
0< \epsilon_2 < \epsilon_0 < \epsilon_3 < \epsilon_1,
\ee
and 
\be
\Fmf(z) = -\int_{-\infty}^\infty \dd k \frac{k\, \eE^{\ir k z}}{(\eE^{\frac {\pi k}3} + \eE^{-\frac {\pi k}3}-1)(\eE^{\frac {\pi k}2} - \eE^{-\frac {\pi k}2})}.\label{FF}
\ee
Using the definitions of $\widehat\bmf^1(z)$ and $\widehat\Amf^n(z)$, we rewrite \eqref{eq:hatbmf.NLIE} in terms of $\bmf^1(z)$ and $\Amf^n(z)$, so that the new formulas depend on the positions of the zeros of $\setS^2$ and $\setS^3$. The resulting  nonlinear integral equation for $\bmf^1(z)$ then contains extra terms that depend explicitly on the position of the zeros in the sets $\setS_2$ and $\setS_3$. As before, we encode this dependence entirely in the choice of integration paths, and we find after simplification
\be
\label{eq:b1}
\log \bmf^1(z+\ir \epsilon_0)'' = - \mu N \Fmf(z) + \sum_{n=1}^3 \int_{\mathcal C^n}\dd y\, K^n(z-y +\ir \epsilon_0- \ir \epsilon_n)'' \log \Amf^n(y + \ir \epsilon_n),
\ee
where the paths $\mathcal C^n$ are the same as above.

To obtain the bulk free energy $f_\textrm{bulk}(u)$, we integrate the expression \eqref{FF} for $\Fmf(z)$ twice with respect to $z$. After simplifications, we obtain the integral expression
\be
\label{eq:fb}
f_{\textrm {bulk}}(u) = - \log 2 + 2\int_{-\infty}^{\infty} \dd k\, \frac{\sinh[(\frac \pi 3-u)k]\sinh[(\frac {2\pi}3-u)k] \cosh \frac{\pi k}3}{k \sinh 2 \pi k}.
\ee
The two integration constants that arose in deriving this formula were fixed as follows: (i) by imposing that $f_\textrm{bulk}(u)$ has the correct behaviour in the braid limit, namely $\lim_{u \to \pm\ir \infty}\frac{\dd}{\dd u}f_\textrm{bulk}(u)= \mp \ir $, and (ii) from the assumption that $f_\textrm{bulk}(u)$ is independent of $\alpha$ and is known for $u=\frac \pi3, \alpha = 1$, namely $f_\textrm{bulk}(\frac \pi 3)=-\log 2$. This readily follows from the known value of the partition function at $\alpha = 1$, see \eqref{eq:power.of.2} and \eqref{eq:power.of.2.v2}. The formula \eqref{eq:fb} can then be simplified to \eqref{eq:fbulkbdy} using contour integrals techniques with the residue theorem.

\subsection{Scaling limit of the nonlinear integral equations and finite-size corrections}\label{sec:scaling.limit}

In the nonlinear integral equations \eqref{eq:aA.NLIEs}, the explicit dependence on $N$  arises only in the driving terms $\mu N \fmf^n(z)$. For $z$ of order $\pm \log N$ with $N$ large, these functions behave as exponentials:
\be
\fsf^n_\pm(z \pm \ir \epsilon_n) = \lim_{N\to \infty} \mu N \fmf^n\big(\!\pm\!(z + \log N)+\ir \epsilon_n\big) = -2\mu \sqrt 3\, \eE^{-(z\pm \ir \epsilon_n)}\times\left\{
\begin{array}{cll}
\eE^{\pm \frac{\ir\pi}3} && n = 1,\\[0.1cm]
\eE^{\mp \frac{\ir\pi}3} && n = 2,\\[0.15cm]
1 && n = 3.
\end{array}\right.
\ee
We study sequences of eigenstates on increasing values of $N$. In a given sequence, the eigenvalues of $\bmf^n(z)$ have finitely many zeros inside the analyticity strips, with the patterns of zeros inside the strips remaining stable as $N$ increases. The positions $x(N)$ of these zeros depend on $N$ and drift off to infinity as either $+\log N$ or $-\log N$. We define the sets of zeros in the scaling limit
\be 
\setS^n_\pm = \Big\{ \lim_{N\to \infty} \pm\big(x(N) - \log N\big) \, \big| \, x(N) \in \setS^n(N)\textrm{ with } x(N) \xrightarrow{N \to \infty}\pm \infty \Big\}.
\ee
We also assume the existence of the scaling functions
\be
\asf^{n}_\pm(z\pm\ir \epsilon_n) = \lim_{N\to \infty} \amf^n\big(\!\pm\!(z + \log N)+\ir \epsilon_n\big), 
\qquad
\Asf^{n}_\pm(z\pm\ir \epsilon_n) = \lim_{N\to \infty} \Amf^n\big(\!\pm\!(z + \log N)+\ir \epsilon_n\big).
\ee
These satisfy the nonlinear integral equations
\begin{alignat}{2}
\log \asf^{n}_\pm(z \pm \ir \epsilon_n) -\phi^n_\pm &= \fsf^{n}_\pm(z\pm \ir \epsilon_n) + \sum_{m=1}^3 \Gamma_{nm} \log \Asf^m_\pm(z \pm \ir \epsilon_n) 
\nonumber\\&
+ \sum_{m=1}^3 \int_{\mathcal C_\pm^m}\dd y \, K_{nm}\big(\!\pm\!(z-y)\big) \log \Asf^m(y\pm\ir \epsilon_m)\big),
\label{eq:scaling.NLIEs}
\end{alignat}
where $\phi^n_\pm$ are integration constants. The integration paths $\mathcal C_\pm^m$ are defined as
\be
\mathcal C^m_\pm = \lim_{N\to \infty} \pm \, \mathcal C^m - \log N.
\ee
They are thus defined similarly to $\mathcal C^m$, but passing over and below certain zeros of $\setS^2_\pm$ and $\setS^3_\pm$. Indeed, let us define $\setS^{n,\uparrow}_\pm$ and $\setS^{n,\downarrow}_\pm$, similarly to \eqref{eq:S.subsets}, to be the subsets of zeros of $\setS^n_\pm$ in the upper and lower half planes, with $n = 2,3$. The paths $\mathcal C^m_\pm$ then run from $-\infty$ to $+\infty$, passing above the points $x + \frac{\ir \pi}3-\ir \epsilon_m$ and $y + \frac{\ir \pi}6-\ir \epsilon_m$, with $x \in \setS_\pm^{2,\downarrow}$ and $y \in \setS_\pm^{3,\downarrow}$, and below the points $x - \frac{\ir \pi}3-\ir \epsilon_m$ and $y - \frac{\ir \pi}6-\ir \epsilon_m$, with $x \in \setS_\pm^{2,\uparrow}$ and $y \in \setS_\pm^{3,\uparrow}$. Moreover, these paths pass above and below the poles of the kernel functions in the same way as the straight path from $-\infty$ to $\infty$ does.

The functions $\asf^m_\pm(z)$ and $\Asf^m_\pm(z)$ are analytic functions of $z$ along the paths $\mathcal C_\pm^m$. In the above equations, we understand $\log \asf^m_\pm(z)$ and $\log\Asf^m_\pm(z)$ to be continuous functions as well, namely functions whose imaginary parts do not have discontinuities as their arguments have phases that exit the principal branch $(-\pi,\pi]$. We choose the branches of the functions $\log \Asf^n_\pm(z)$ such that
\be
\label{eq:branchAn}
\lim_{z \to -\infty } \textrm{Im} \big(\log \Asf^n_\pm(z)\big) = 0.
\ee
This condition is necessary for the integrals in \eqref{eq:b1.fsc} below to be well-defined. In contrast, for $\log \asf^n_\pm(z)$ there is a freedom in the choice of the logarithmic branches. The values of the constants $\phi^n_\pm$ depend on this choice, but not the final result for the leading finite-size correction. In \cref{sec:standard.groundstates.cylinder,sec:standard.groundstates.strip}, we will discuss further the choice of branches of $\log \asf^n_\pm(z)$ for the particular cases of the groundstates of the standard modules.

To obtain the finite-size correction to $\bmf^1(z)$, we compute the asymptotic expansion of the integral terms in \eqref{eq:b1} as 
\begin{alignat}{2}
\int_{\mathcal C^n}\dd y\, & K^n(z\!-\!y\! +\!\ir \epsilon_0- \ir \epsilon_n)'' \log \Amf^n(y + \ir \epsilon_n) 
=\Big[\int_{\mathcal C^n_{\textrm{Re}<0}} + \int_{\mathcal C^n_{\textrm{Re}>0}}\Big]\dd y\, K^n(z\!-\!y\! +\!\ir \epsilon_0- \ir \epsilon_n)'' \log \Amf^n(y + \ir \epsilon_n)
\nonumber\\ &
\simeq \frac 1N\int_{\mathcal C^n_-}\dd y\,  K^n_+(z+y +\ir \epsilon_0- \ir \epsilon_n)'' \log \Asf^n_-(y - \ir \epsilon_n) 
\\ &
+ \frac 1N\int_{\mathcal C^n_+}\dd y\, K^n_-(-z+y -\ir \epsilon_0+\ir \epsilon_n)'' \log \Asf^n_+(y + \ir \epsilon_n),
\nonumber
\end{alignat}
where we define
\be
K^n_{\pm}(z) = \lim_{N\to \infty} N K^n\big(\!\pm\!(z+\log N)\big)= \frac{\sqrt 3\, \eE^{-z}}{\pi} \times 
\left\{\begin{array}{cl}
\eE^{\mp\frac{\ir \pi}3} & n=1,\\[0.1cm]
\eE^{\pm\frac{\ir \pi}3} & n=2,\\[0.1cm]
1 & n=3.
\end{array}\right.
\ee
We thus find
\begin{alignat}{2}
\log \bmf^1(z + \ir \epsilon_0)'' + \mu N \Fmf(z + \ir \epsilon_0)&\simeq \frac {\sqrt 3}{\pi N} \,\eE^{-z-\ir \epsilon_0} \sum_{n=1}^3 \int_{\mathcal C^n_-} \dd y \, \eE^{-y+\ir \epsilon_n}\big(\eE^{-\frac{\ir \pi} 3},\eE^{\frac{\ir\pi} 3},1\big)_n \log \Asf^n_-(y-\ir \epsilon_n) 
\nonumber \\[0.15cm]
&+\frac {\sqrt 3}{\pi N} \, \eE^{z+\ir \epsilon_0} \sum_{n=1}^3 \int_{\mathcal C^n_+} \dd y \, \eE^{-y-\ir \epsilon_n}\big(\eE^{\frac{\ir \pi} 3},\eE^{-\frac{\ir\pi} 3},1\big)_n \log \Asf^n_+(y+\ir \epsilon_n).
\label{eq:b1.fsc}
\end{alignat}
We see from this equation that the $\frac 1N$ expansion for $\log \bmf^1(z)''$ has no constant term, for both periodic and strip boundary conditions. For periodic boundary conditions, there is no boundary and therefore no boundary free energy is expected in this expansion, see \eqref{eq:logTD.asy.D}. For strip boundary conditions, the above equation implies that the boundary free energy has the form $f_\text{bdy}(u) = \alpha_0 + \alpha_1 u$ for some constants $\alpha_0$ and $\alpha_1$. This free energy should be crossing symmetric, $f_\text{bdy}(u) = f_\text{bdy}(\pi-u)$, from which we deduce that $\alpha_1 = 0$. Thus $f_\text{bdy}(u)$ is independent of $u$. We obtain $\alpha_0$ by using the known value \eqref{eq:power.of.2.v2} of the partition function at $\alpha = 1$ on the cylinder, and its decompositions \eqref{eq:ZD} in terms of traces of over the standard modules $\repV_{N,d}$. In agreement with \eqref{eq:fbulkbdy}, we then find $f_\text{bdy}(u) = 0.$

The right-hand side of \eqref{eq:b1.fsc} is re-expressed using the dilogarithm technique. We define
\be
\label{eq:J.def}
\mathcal J_\pm = \sum_{n=1}^3 \int_{\mathcal C^n_\pm} \dd y \bigg[\big(\log \asf_\pm^n(y\pm\ir \epsilon_n)\big)' \log \Asf_\pm^n(y\pm\ir \epsilon_n) - \big(\log 
\asf_\pm^n(y\pm\ir \epsilon_n) - \phi^n_\pm\big) \big(\log \Asf_\pm^n(y\pm\ir \epsilon_n)\big)'\bigg].
\ee
These are evaluated in two ways. For the first, we use the nonlinear integral equations \eqref{eq:scaling.NLIEs}. Many simplifications occur due to the symmetry properties of the matrices $K(z)$ and $\Gamma$. We find that only integrals involving driving terms survive:
\be
\label{eq:J.computation.1}
\mathcal J_\pm = 4\mu \sqrt 3\, \sum_{n=1}^3 \int_{\mathcal C^n_\pm} \dd y\, \eE^{-(y\pm \ir \epsilon_n)} \big(\eE^{\pm \frac{\ir \pi}3},\eE^{\mp \frac{\ir \pi}3},1\big)_n \log \Asf^n_\pm(y\pm \ir \epsilon_n) .
\ee
We see that $\mathcal J_-$ and $\mathcal J_+$ are equal up to prefactors to the terms appearing on the right side of \eqref{eq:b1.fsc}:
\be
\log \bmf^1(z + \ir \epsilon_0)'' + \mu N \Fmf(z + \ir \epsilon_0)
\simeq \frac {1}{4\pi\mu N} \Big( \eE^{-z - \ir \epsilon_0} \mathcal J_- + \eE^{z + \ir \epsilon_0}\mathcal J_+\Big).
\ee
Comparing this with \eqref{eq:logTD.asy}, we find
\be
\label{eq:Delta.J}
\Delta = -\frac{\mathcal J_+}{8 \pi ^2}, \qquad \bar\Delta = -\frac{\mathcal J_-}{8 \pi ^2},
\ee
where we set the central charge of critical percolation to the known value $c=0$.

The second way of computing $\mathcal J_\pm$ is to evaluate the derivative explicitly and make a change of variables from $y$ to $\asf^n_\pm$:
\begin{alignat}{2}
\mathcal J_\pm &= \sum_{n=1}^3 \int_{\mathcal C^n_\pm} \dd y\, \frac{\dd \asf^n_\pm(y\pm \ir \epsilon_n)}{\dd y} \bigg(\frac{\log \big(1+\asf^n_{\pm}(y\pm \ir \epsilon_n)\big)}{\asf^n_{\pm}(y\pm \ir \epsilon_n)} - \frac{\log \asf^n_{\pm}(y\pm \ir \epsilon_n) - \phi_n^\pm}{1+\asf^n_{\pm}(y\pm \ir \epsilon_n)}\bigg)
\nonumber\\ 
&= \sum_{n=1}^3 \int_{{\mathcal A}^n_\pm} 
\dd \asf^n_\pm \bigg(\frac{\log \big(1+\asf^n_{\pm}\big)}{\asf^n_{\pm}} - \frac{\log \asf^n_{\pm} - \phi^n_\pm}{1+\asf^n_{\pm}}\bigg).
\label{eq:J.ints}
\end{alignat}
Here, $\mathcal A^n_\pm$ is the trajectory in the complex plane visited by the function $\asf^n_\pm(y \pm \ir \epsilon_n)$ as $y$ runs over $\mathcal C^n_\pm$. Thus \eqref{eq:J.ints} is a sum of regular integrals carried over a trajectory in the complex plane that may wrap non-trivially around the poles at $0$ and $-1$. Each eigenstate of the transfer matrix is then characterised by the homotopies of its trajectory $\mathcal A^n_\pm$ around these two points.

%
%

\subsection{Analytic solution for ground states for periodic boundary conditions}\label{sec:standard.groundstates.cylinder}

In this section, we work with periodic boundary conditions and compute the integrals $\mathcal J_+,\mathcal J_-$ and the corresponding conformal dimensions $\Delta,\bar\Delta$ for the simplest patterns of zeros, namely those corresponding to the ground state eigenvalues of the transfer matrix $\widehat \Tb(u)$ in the standard modules $\repW_{N,d,\omega}$. 

\begin{figure}
\centering
\begin{pspicture}(-4.2,-2.1)(2.8,2.1)
\rput(0,0){\includegraphics[width=.35\textwidth]{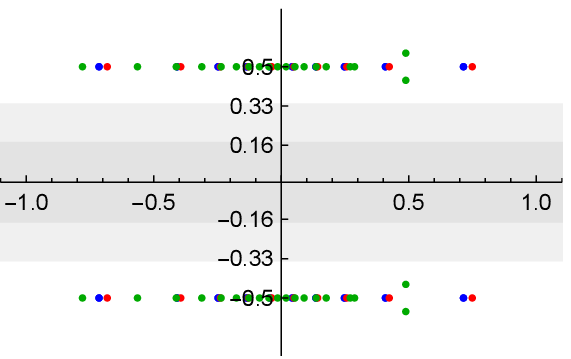}}
\rput(-3.9,0){$d=0:$}
\end{pspicture}
\qquad \qquad \qquad 
\begin{pspicture}(-4.2,-2.1)(2.8,2.1)
\rput(0,0){\includegraphics[width=.35\textwidth]{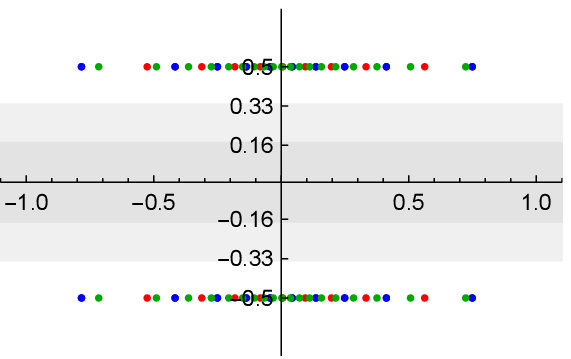}}
\rput(-3.9,0){$d=1:$}
\end{pspicture}
\begin{pspicture}(-4.2,-2.1)(2.8,2.1)
\rput(0,0){\includegraphics[width=.35\textwidth]{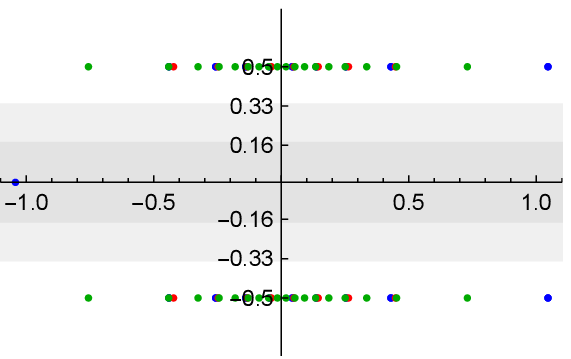}}
\rput(-3.9,0){$d=2:$}
\end{pspicture}
\qquad \qquad \qquad 
\begin{pspicture}(-4.2,-2.1)(2.8,2.1)
\rput(0,0){\includegraphics[width=.35\textwidth]{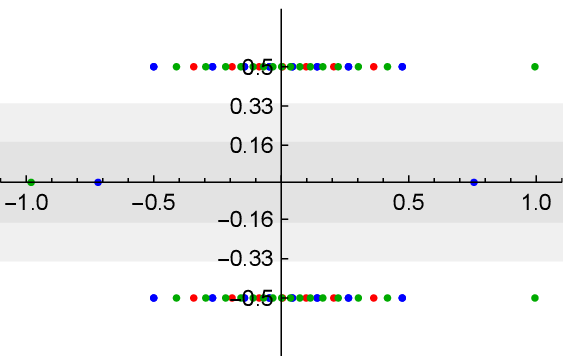}}
\rput(-3.9,0){$d=3:$}
\end{pspicture}
\begin{pspicture}(-4.2,-2.1)(2.8,2.1)
\rput(0,0){\includegraphics[width=.35\textwidth]{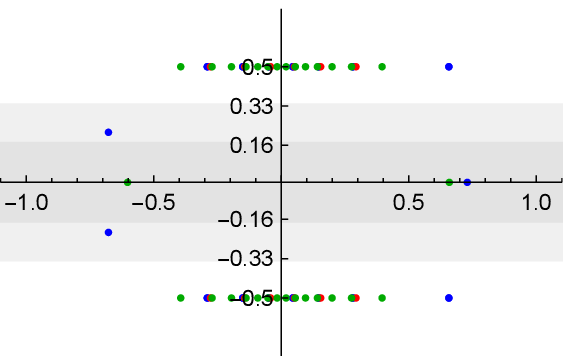}}
\rput(-3.9,0){$d=4:$}
\end{pspicture}
\qquad \qquad \qquad 
\begin{pspicture}(-4.2,-2.1)(2.8,2.1)
\rput(0,0){\includegraphics[width=.35\textwidth]{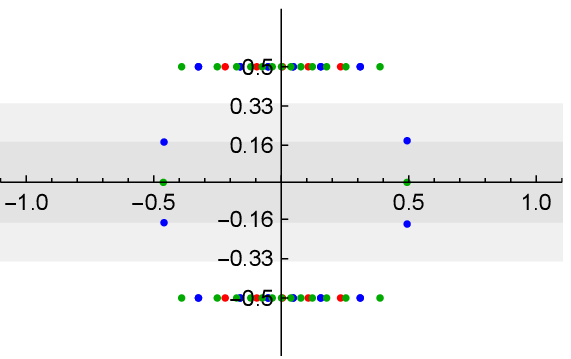}}
\rput(-3.9,0){$d=5:$}
\end{pspicture}
\begin{pspicture}(-4.2,-2.1)(2.8,2.1)
\rput(0,0){\includegraphics[width=.35\textwidth]{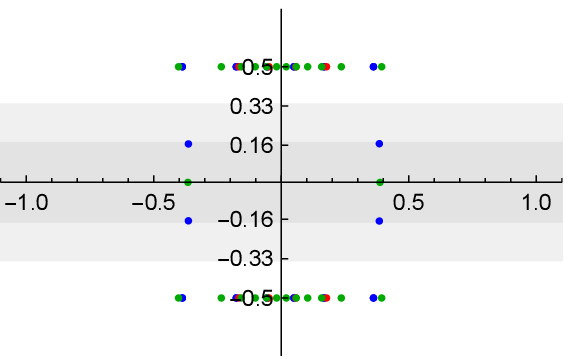}}
\rput(-3.9,0){$d=6:$}
\end{pspicture}
\qquad \qquad \qquad 
\begin{pspicture}(-4.2,-2.1)(2.8,2.1)
\rput(0,0){\includegraphics[width=.35\textwidth]{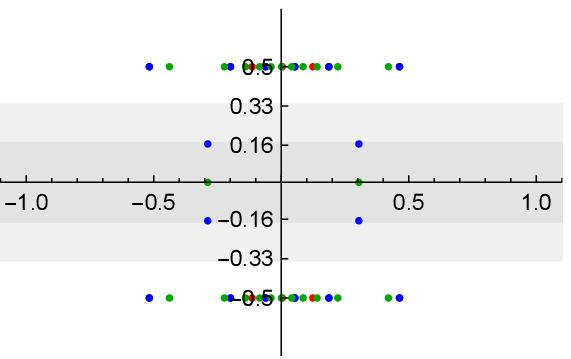}}
\rput(-3.9,0){$d=7:$}
\end{pspicture}
\begin{pspicture}(-4.2,-2.1)(2.8,2.1)
\rput(0,0){\includegraphics[width=.35\textwidth]{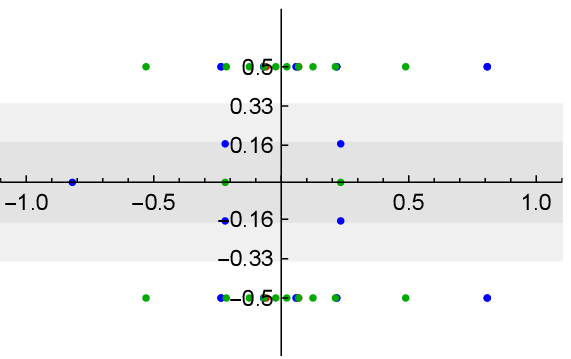}}
\rput(-3.9,0){$d=8:$}
\end{pspicture}
\qquad \qquad \qquad 
\begin{pspicture}(-4.2,-2.1)(2.8,2.1)
\rput(0,0){\includegraphics[width=.35\textwidth]{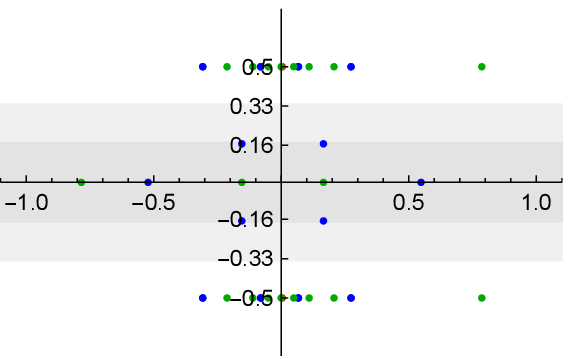}}
\rput(-3.9,0){$d=9:$}
\end{pspicture}
\caption{The zeros of the functions $\bmf^1(z)$, $\bmf^2(z)$ and $\bmf^3(z)$, coloured in blue, red and green respectively, for $N = 10$, $d = 0,1, \dots, 9$ and $\gamma = 0.15$. The analyticity strips of $\bmf^2(z)$ and $\bmf^3(z)$ are respectively coloured in lighter and darker shades of gray. The axes are in units of $\pi$.}
\label{fig:patterns.of.zeros}
\end{figure}

We set the twist parameter to $\omega = \eE^{\ir \gamma}$ and focus on the interval $0 < \gamma < \frac \pi 3$. Using our computer implementation of the transfer matrix, we computed the functions $\bmf^n(z)$ for various small values of~$N$, for these ground states. In \cref{fig:patterns.of.zeros}, we plot the patterns of zeros of these functions for  $\gamma = 0.15$, $N = 10$ and $d = 0,1,\dots, 9$. Our analysis reveals that, for the ground states, the function $\bmf^2(z)$ never has zeros inside its analyticity strip $-\frac \pi 3<\textrm{Im}(z)<\frac \pi 3$. For $d \ge 3$, the function $\bmf^3(z)$ has a finite number of zeros that lie on the real axis, but none for $d = 0,1,2$. For a fixed $d$, the number of such zeros is unchanged as $N$ is increased. Their number however grows with $d$. Moreover, for $d \ge 2$, the function $\bmf^1(z)$ also has zeros that are located either on the real axis or in the neighborhood of the lines $\textrm{Im}(z) = \pm \frac\pi6$. As discussed in the previous section, this function has an analyticity strip of width zero and its zeros do not impact the form of  the nonlinear integral equations. Our exploration on various small values of $N$ and $d$ reveals that these patterns of zeros have a simple modulo $6$ property. Indeed, knowing the patterns of zeros for a given value of $d$ inside the analyticity strips, we produce the same patterns of zeros for $d+6$ defects by adding two real zeros for the function $\bmf^3(z)$, a positive and a negative one.

The bulk behavior of the function $\asf^n_\pm(z)$, as dictated by the driving terms in its nonlinear integral equations \eqref{eq:aA.NLIEs}, vanish:
\be
\asf^n_\pm(-\infty) = 0, \qquad n = 1,2,3.
\ee
The braid limits for these functions are
\be
\label{eq:braid.limits}
\asf^n_\pm(\infty)=\psi^n\big(\eE^{-\frac{\ir \gamma}2 \pm \frac{\ir \pi d}6}\big), \qquad n = 1,2,3,
\ee
where
\be
\label{eq:psi.functions}
\psi^1(s)=\frac{s^3}{s+s^{-1}},\qquad
\psi^2(s)=\frac{s^{-3}}{s+s^{-1}},\qquad
\psi^3(s)=\frac{1}{s^2+1+s^{-2}}.
\ee
The trajectory $\mathcal A^n_\pm$ for the function $\asf^n_\pm(z)$ starts at the origin for $z \to -\infty$ and ends at its braid value for $z \to +\infty$. Computing the dilogarithm integrals \eqref{eq:J.ints} requires the knowledge of the winding of these trajectories around the points $0$ and $-1$. The logarithms of $\asf^n_\pm(z)$ and $\Asf^n_\pm(z)$ are functions whose imaginary part is understood to be continuous. The branch of $\log \Asf^n_\pm(z)$ is fixed from \eqref{eq:branchAn}, whereas the choice of branch of $\log \asf^n_\pm(z)$ is discussed below.

\begin{figure}[p] 
\centering
\begin{pspicture}(0,-3)(16,2)
\rput(0,0){$\amf^1(z)$:}
\rput(4.5,0){\includegraphics[width=7.5cm]{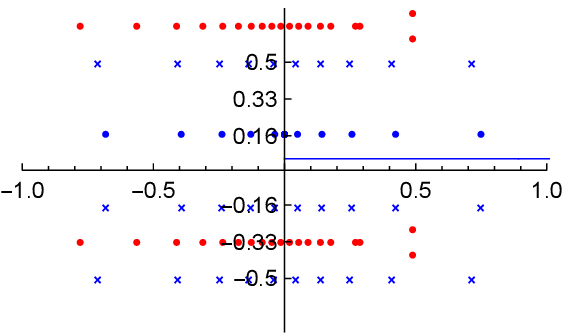}}
\rput(12.5,0){\includegraphics[width=7.5cm]{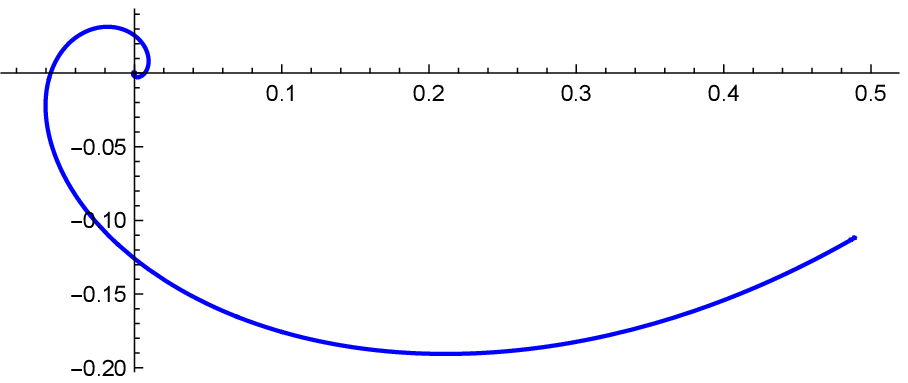}}
\end{pspicture}
\begin{pspicture}(0,-3)(16,2)
\rput(0,0){$\amf^2(z)$:}
\rput(4.5,0){\includegraphics[width=7.5cm]{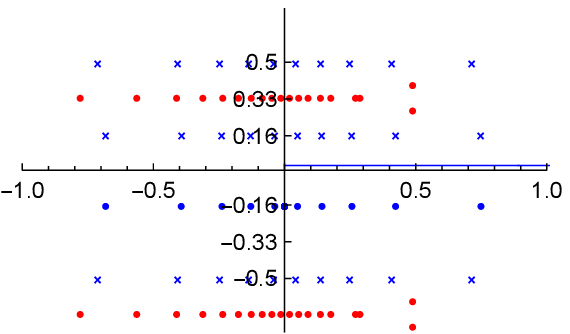}}
\rput(12.5,0){\includegraphics[width=7.5cm]{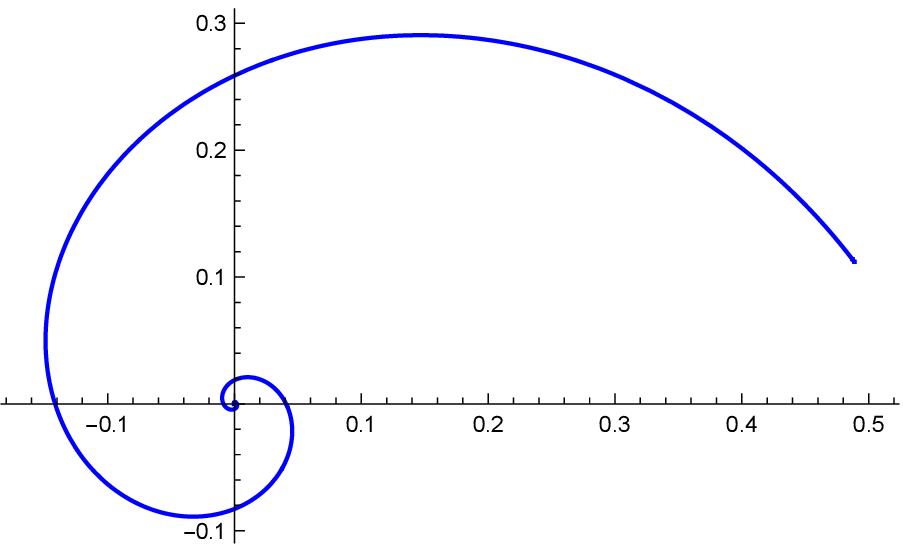}}
\end{pspicture}
\begin{pspicture}(0,-2.5)(16,2)
\rput(0,0){$\amf^3(z)$:}
\rput(4.5,0){\includegraphics[width=7.5cm]{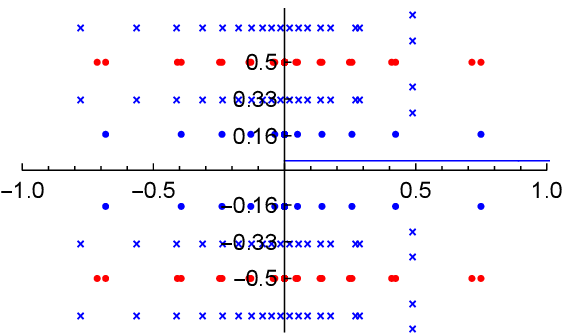}}
\rput(12.5,0){\includegraphics[width=7.5cm]{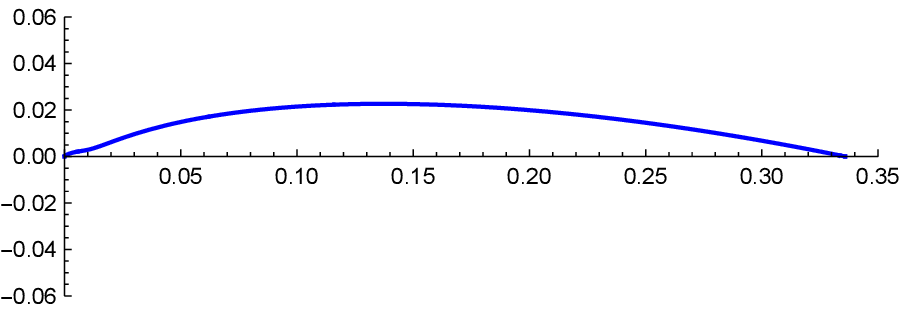}}
\end{pspicture}
\caption{Left panel: The zeros of $\amf^n(z)$, the zeros of $\Amf^n(z)$ and their joint poles, coloured with blue circles, blue crosses and red circles, respectively, for $(N,d) = (10,0)$ and $\gamma = 0.15$. Indicated by a blue line is the portion of the path $\mathcal C^n$ with a positive real part, shifted by $\ir \epsilon_n$. This path scales to $\mathcal C^n_+$ in the limit $N \to \infty$. Right panel: the corresponding trajectories $\mathcal A^n_+(10)$ for the functions $\amf^n(z)$, as we progress along this path. The functions $\amf^1(z)$ and $\amf^2(z)$ wind around the origin in the counter-clockwise and clockwise directions, respectively, whereas the function $\amf^3(z)$ has no winding. The corresponding trajectories are $\mathsf x^5$, $\mathsf x^{-5}$ and $1$.
}
\label{fig:patterns.of.zeros.0}
\end{figure}

\begin{figure}[p] 
\centering
\begin{pspicture}(0,-3.5)(16,2)
\rput(0,0){$\amf^1(z)$:}
\rput(4.5,0){\includegraphics[width=7.5cm]{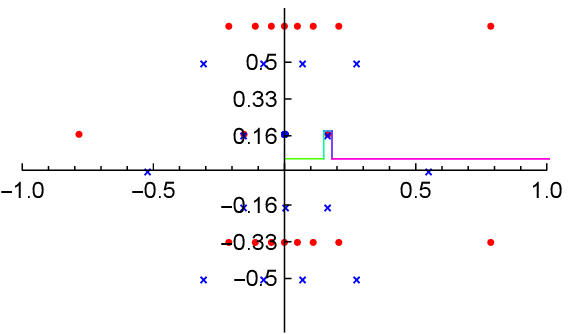}}
\rput(11.0,0){\includegraphics[width=2.4cm]{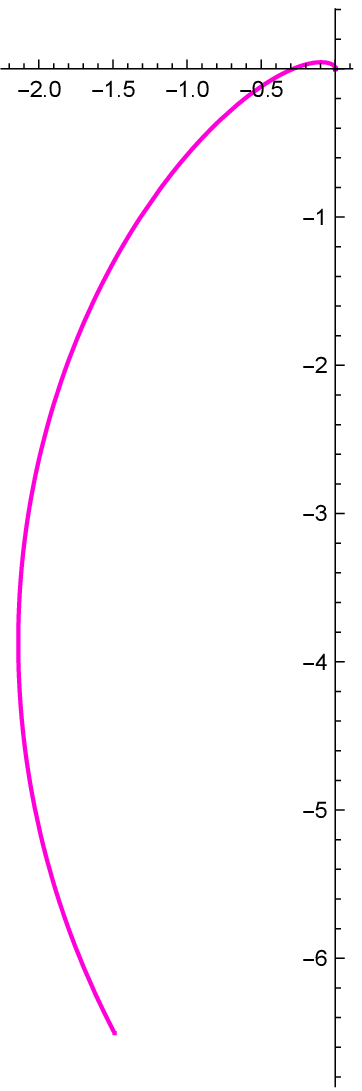}}
\end{pspicture}
\\[0.8cm]
\begin{pspicture}(0,-3.5)(16,2)
\rput(0,0){$\amf^2(z)$:}
\rput(4.5,0){\includegraphics[width=7.5cm]{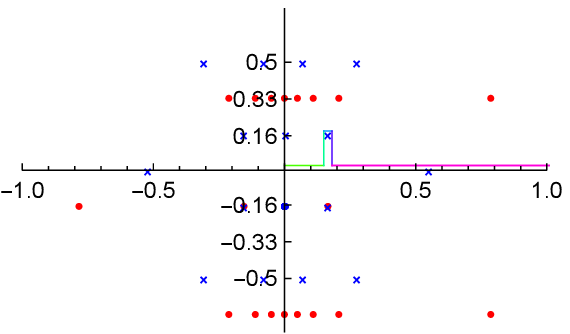}}
\rput(14.0,0){\includegraphics[width=3.0cm]{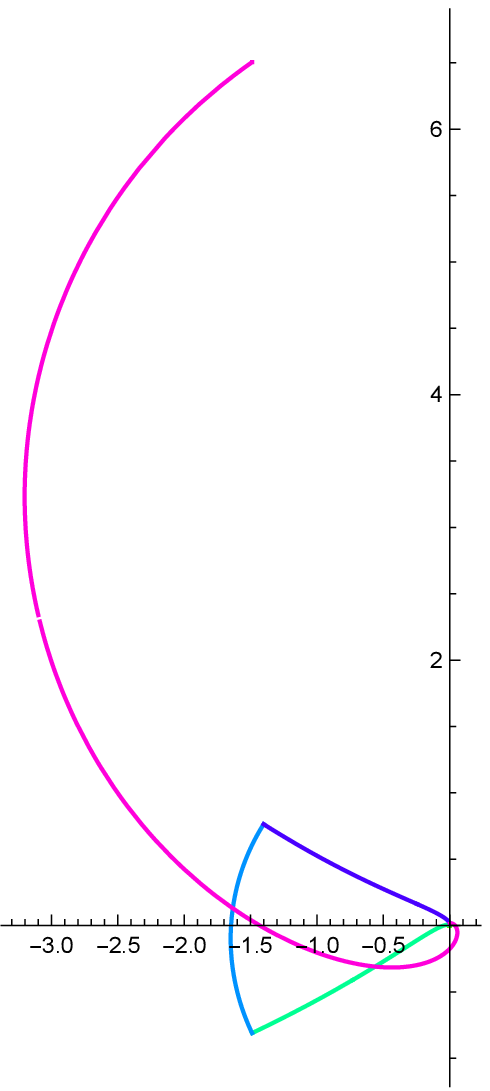}}
\end{pspicture}
\\[0.5cm]
\begin{pspicture}(0,-2.5)(16,2)
\rput(0,0){$\amf^3(z)$:}
\rput(4.5,0){\includegraphics[width=7.5cm]{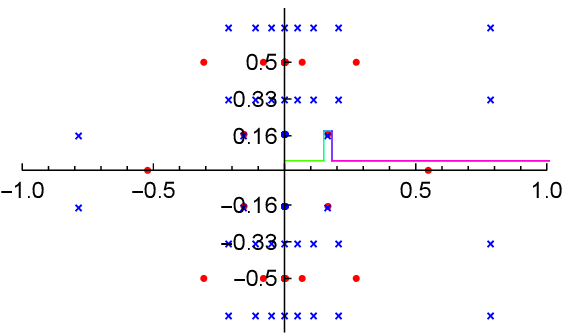}}
\rput(12.5,0){\includegraphics[width=5.cm]{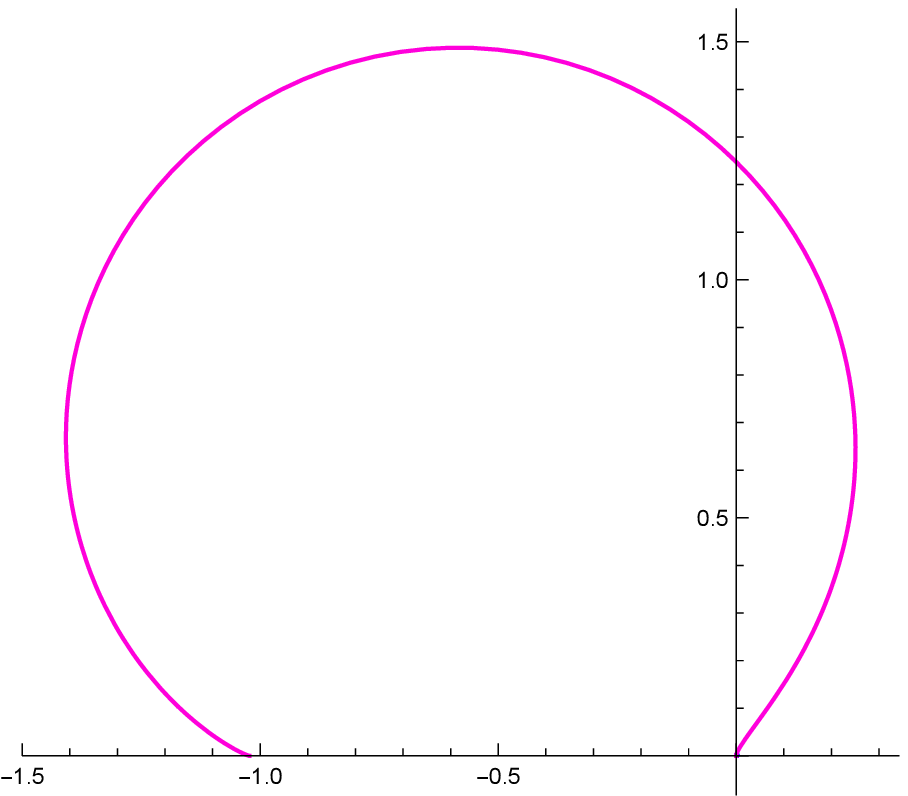}}
\end{pspicture}
\caption{Left panel: the zeros of $\amf^n(z)$, the zeros of $\Amf^n(z)$ and the poles of both functions, coloured with blue circles, blue crosses and red circles, respectively, for $(N,d) = (10,9)$ and $\gamma = 0.15$. The portion of the path $\mathcal C^n$ with a positive real part consists of different portions, which are depicted using different colours. Right panel: the corresponding trajectories $\mathcal A^n_+(10)$ for the functions $\amf^n(z)$ in the complex plane, as we move along this path. The function $\amf^2(z)$ has a non trivial homotopy around the points $-1$ and $0$. The corresponding trajectories are $\mathsf x^{3}$, $\mathsf x^{-1}\mathsf y^{-1} \mathsf x^{-1} \mathsf y^{-1}$ and $\mathsf x\,\mathsf y^{1/2}$.
}
\label{fig:patterns.of.zeros.9}
\end{figure}

With our computer program, we are able to compute the functions $\amf^n(z)$ for the ground states, for many small values of $N$. This way, we obtain approximations to the trajectories $\mathcal A^n_-$ and $\mathcal A^n_+$ at these finite values of $N$, by looking at the trajectories for $\amf^n(z)$ as $z$ visits the portions of the paths $\mathcal C^n$ that have a positive and a negative real part, respectively. We denote these finite-size trajectories as $\mathcal A^n_-(N)$ and $\mathcal A^n_+(N)$. \cref{fig:patterns.of.zeros.0,fig:patterns.of.zeros.9} give examples of the trajectories $\mathcal A^n_+(N)$ for $(N,d)=(10,0)$ and $(N,d)=(10,9)$, respectively. 

We describe the homotopy of trajectories around the points $0$ and $-1$ in terms of a word made of the non-commuting letters $\mathsf x, \mathsf y$ and their inverses $\mathsf x^{-1}$, $\mathsf y^{-1}$. For generic $\gamma$ in $(0,\frac \pi 3)$, all trajectories leave the origin at an angle that is different from $0$ and $\pi$. Then before reaching the braid value, the trajectory crosses the half-lines $(0,\infty)$ and $(-\infty,-1)$ a number of times. We assign the word $\mathsf x^{m_1}\mathsf y^{n_1}\mathsf x^{m_2}\mathsf y^{n_2} \dots$, with $m_i,n_i \in \mathbb Z$, to a trajectory that first crosses $m_1$ times the half-line $(0,\infty)$, then crosses $n_1$ times the half-line $(-\infty,-1)$, then $m_2$ times the half-line $(0,\infty)$, and so on. The integers $m_i$ and $n_i$ are positive and negative if the half-lines are crossed in the counter-clockwise and clockwise directions, respectively. For example, the word $\mathsf x^m$ is assigned to a trajectory that winds $m$ times around the origin without ever passing to the left of the point $-1$. Clearly, if a function passes to the right of the origin once in the clockwise direction, and subsequently once in the anti-clockwise direction, its trajectory can be continuously deformed in such a way that it does not wind around the origin at all, consistently with $\mathsf x^{-1}\mathsf x=1$. Likewise, a trajectory that winds to the left of the point $-1$ in one direction and subsequently in the other direction is assigned a word that can be reduced using the relation $\mathsf y^{-1}\mathsf y = 1$. The resulting dilogarithm integrals are likewise unchanged by such deformations of the trajectories.

In \cref{fig:patterns.of.zeros.0}, we see that, for $(N,d)=(10,0)$, the trajectories $\mathcal A_+^1(10)$ and $\mathcal A_+^2(10)$ cross the half-line $(0,\infty)$ a number of times, in the counter-clockwise and clockwise directions explicitly, and without ever crossing the half-line $(-\infty,-1)$. The figure does not reveal how many such windings there are. This requires a closer zoom around the origin. Our computer program is nonetheless able to measure this winding. It finds that both functions in this example wind around the origin five times, in opposite directions. In contrast, the function $\amf^3(z)$ crosses none of the two half-lines. The resulting trajectories are therefore $\mathcal A_+^1(10) = \mathsf x^5$, $\mathcal A_+^2(10) =\mathsf x^{-5}$ and $\mathcal A_+^3(10) =1$.

\cref{fig:patterns.of.zeros.9} gives a second example with $(N,d)=(10,9)$. In this case, the paths $\mathcal C^m$ are not straight lines. In the figure, we see that the paths pass above the point $z_3 + \frac {\ir \pi} 6$, where $z_3$ is a real zero in $\setS^3$. We see from \eqref{eq:a.U.and.b.T} that the functions $\amf^1(z)$ and $\Amf^1(z)$ have a pole and $\Amf^3(z)$ has a zero at this value of~$z$. There is also a zero $z_1$ of $\bmf_1(z)$ very close to $z_3 + \frac {\ir \pi} 6$. This is also a zero of $\amf^1(z)$ and $\amf^2(z)$, and a pole of $\amf^3(z)$ and $\Amf^3(z)$. In this same figure, the resulting path is chosen to pass above~$z_1$. There is indeed some freedom in the choice of this path. As discussed below \eqref{eq:S.subsets}, the integration paths are constrained in terms of the zeros in the sets $\setS^2$ and $\setS^3$. We are however free to choose whether they pass above or below the zeros of $\bmf^1(z)$. Doing so changes the windings of the functions $\amf^n(z)$ and $\Amf^n(z)$ around the points $0$ and $-1$. But one can show that this does not change the linear combination of dilogarithm integrals \eqref{eq:J.ints}. We see from the right panel of \cref{fig:patterns.of.zeros.9} that the second function crosses the half-line $(-\infty,-1)$ twice, whereas the other two functions never cross this half-line. The experimentations made with our program reveal that, for the ground states, it is always possible to use this freedom to choose the trajectories in such a way that only the second function crosses the half-line $(-\infty,-1)$. All three functions will however cross the half-line $(0,\infty)$ a number of times.

We also observe from \cref{fig:patterns.of.zeros.9} that $\asf_+^3(z)$ has a braid limit that is real, negative and slightly smaller than $-1$. Another important information for the computation of the dilogarithm integrals is whether the trajectory starting at the origin and ending at this braid value travels above or below the point~$-1$. In general, for $0 < \gamma < \frac \pi 3$, $\asf^3_+(z)$ has a negative braid limit for $d \equiv 3 \textrm{ mod } 6$ and $d \equiv 4 \textrm{ mod } 6$. Similarly, the function $\asf^3_-(z)$ has a negative braid limit for $d \equiv 2 \textrm{ mod } 6$ and $d \equiv 3 \textrm{ mod } 6$. In the cases when their endpoint is negative, we assign to the trajectories $\mathcal A_\pm^3(N) $ the symbols $\mathsf y^{1/2}$ or $\mathsf y^{-1/2}$ to indicate that the trajectory passes above or below the point $-1$, respectively. In the example of \cref{fig:patterns.of.zeros.9}, the precise details of the windings again require a zoom around the origin. Our program reveals that the trajectories are $\mathcal A_+^1(10) = \mathsf x^{3}$, $\mathcal A_+^2(10) =\mathsf x^{-1}\mathsf y^{-1} \mathsf x^{-1} \mathsf y^{-1}$ and $\mathcal A_+^3(10) =\mathsf x\,\mathsf y^{1/2}$.

As $N$ is increased, the trajectories $\mathcal A_\pm^n(N)$ remain essentially unchanged. Only the first windings $\mathsf x^{m_1}$ around the origin change for $\asf_\pm^1(z)$ and $\asf_\pm^2(z)$, with $|m_1|$ growing roughly as $\frac N2$. For example, the ground states for $d=0$ have the trajectories 
\be
d=0: \qquad \mathcal A^n_-(N) = \big(\mathsf x^{\lfloor \frac {N+1} 2 \rfloor}, \mathsf x^{-\lfloor \frac {N+1} 2 \rfloor},1\big)_n,
\qquad
\mathcal A^n_+(N) = \big(\mathsf x^{-\lfloor \frac N 2 \rfloor}, \mathsf x^{\lfloor \frac N  2\rfloor},1\big)_n.
\ee
As $N \to \infty$, these trajectories thus start by winding an infinite number of times around the origin. This initial winding however plays no role in the computation of the dilogarithm integrals. Indeed, we have
\be
\label{eq:winding.integrals}
\int_{\varepsilon \eE^{\ir \theta_1}}^{\varepsilon \eE^{\ir \theta_2}} 
\dd \asf \bigg(\frac{\log \big(1+\asf\big)}{\asf} - \frac{\log \asf - \phi}{1+\asf}\bigg) \xrightarrow{\varepsilon \to 0} 0.
\ee
This holds true when the imaginary part of the function $\log (1+\asf)$ goes to zero for $\varepsilon \to 0$. With our choice \eqref{eq:branchAn} of the branch of $\Asf^n(z)$, this condition is guaranteed to hold at the start of the trajectory. As a result, the dilogarithm integrals \eqref{eq:J.ints} can instead be performed on truncated trajectories $\widehat{\mathcal A}^n_\pm$, where the initial winding $\mathsf x^{m_1}$ around the origin is removed. With this truncation, each trajectory either starts with $y^{n_1}$ for some non-zero $n_1$, or is the trivial trajectory $1$. Moreover the truncated trajectories are independent of $N$, justifying our choice not to include $N$ as an argument for $\widehat{\mathcal A}^n_\pm$. Empirically, we find that the truncated trajectories for the ground states are given by
\be
\label{eq:An05}
\begin{array}{ll}
d = 0: \qquad
\widehat{\mathcal A}^n_- = \big(1,1,1\big)_n
\quad
&\widehat{\mathcal A}^n_+ = \big(1,1,1\big)_n 
\\[0.15cm]
d = 1: \qquad
\widehat{\mathcal A}^n_- = \big(1,1,1\big)_n
\quad
&\widehat{\mathcal A}^n_+ = \big(1,1,1\big)_n 
\\[0.15cm]
d = 2: \qquad
\widehat{\mathcal A}^n_- = \big(1,\mathsf y,\mathsf y^{-1/2}\big)_n
\quad
&\widehat{\mathcal A}^n_+ = \big(1,1,1\big)_n 
\\[0.15cm]
d = 3: \qquad
\widehat{\mathcal A}^n_- = \big(1,1,\mathsf y^{-1/2}\big)_n
\quad
&\widehat{\mathcal A}^n_+ = \big(1,\mathsf y^{-1},\mathsf y^{1/2}\big)_n
\\[0.15cm]
d = 4: \qquad
\widehat{\mathcal A}^n_- = \big(1,\mathsf y,1\big)_n
\quad
&\widehat{\mathcal A}^n_+ = \big(1,1,\mathsf y^{1/2}\big)_n 
\\[0.15cm]
d = 5: \qquad
\widehat{\mathcal A}^n_- = \big(1, \mathsf y,1\big)_n
\quad
&\widehat{\mathcal A}^n_+ = \big(1, \mathsf y^{-1},1\big)_n \ 
\end{array}
\ee
and similarly for $d \ge 6$.

\begin{figure}[t] 
\centering
\begin{pspicture}(-7.8,-6.0)(7.8,6.5)
\rput(-4.3,3.7){\includegraphics[width=.45\textwidth]{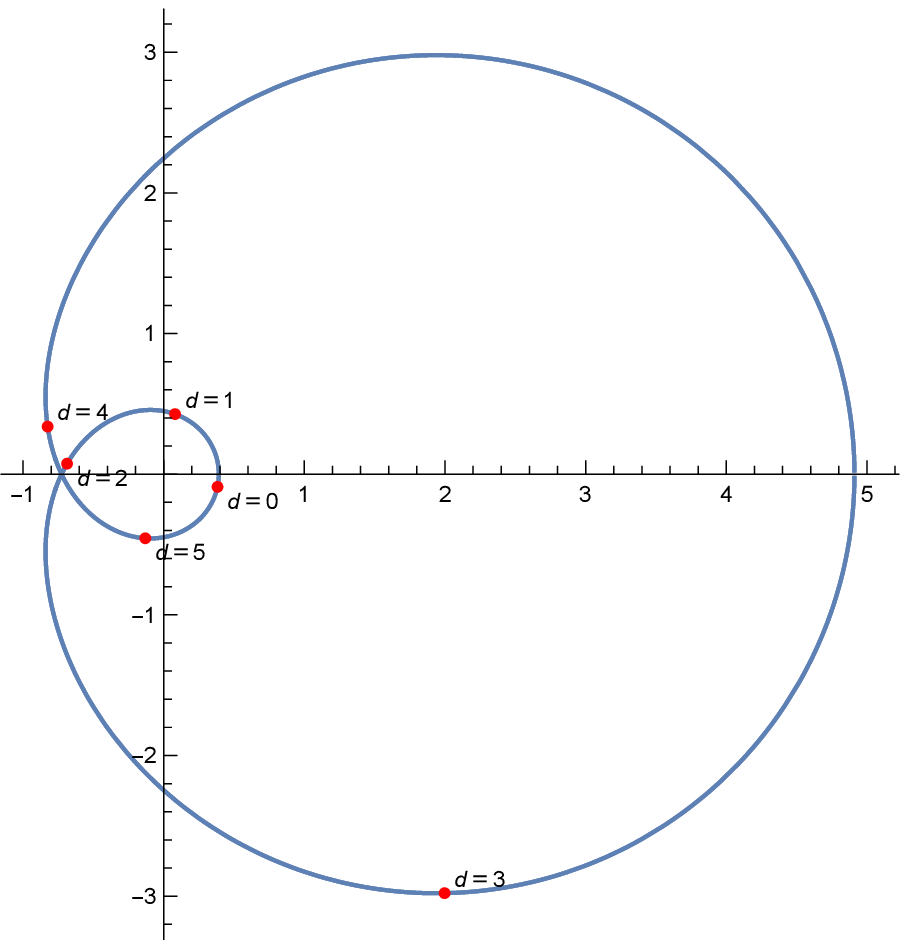}}\rput(-7.5,6.5){$\psi_1(s)$}
\rput(-1.0,-3.1){\includegraphics[width=.55\textwidth]{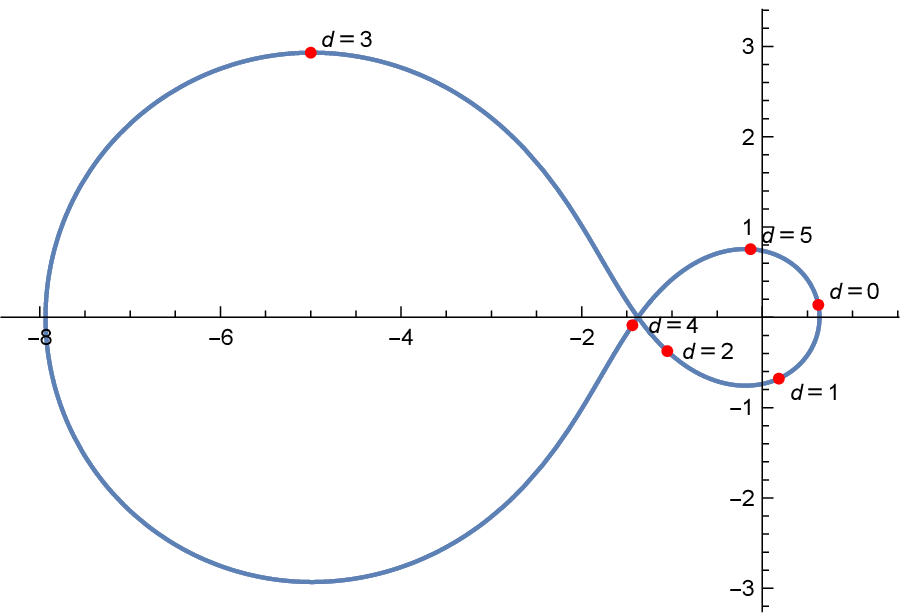}}\rput(3.8,6.5){$\psi_3(s)$}
\rput(5.9,2.7){\includegraphics[width=.28\textwidth]{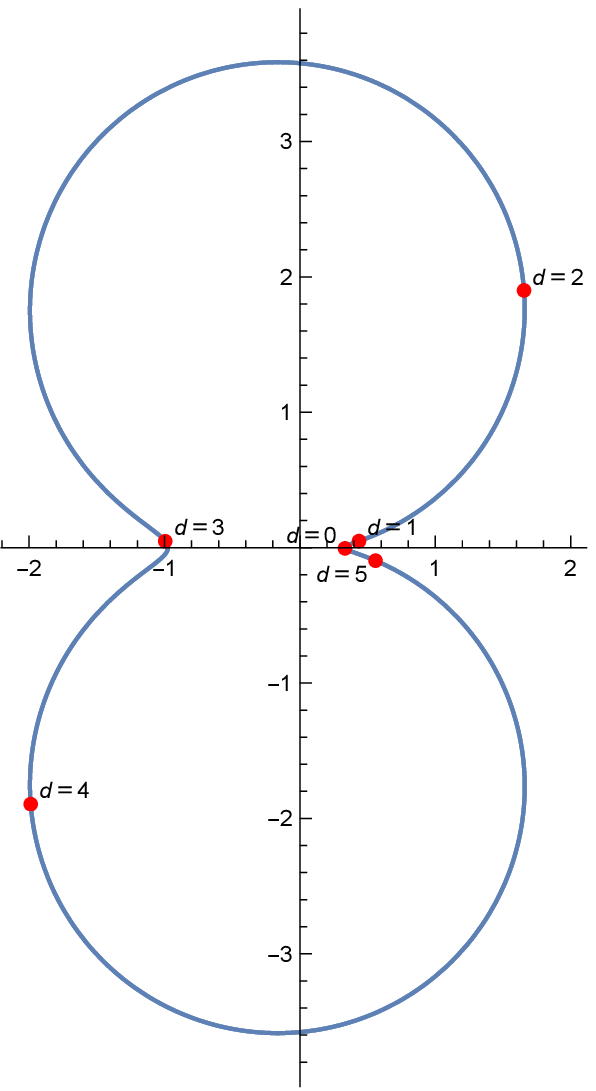}}\rput(-5,-1){$\psi_2(s)$}
\end{pspicture}
\caption{Parametric plots of the functions $\psi_1(s)$, $\psi_2(s)$ and $\psi_3(s)$ for $s = \eE^{\ir (t+\ir \epsilon)}$ with $\epsilon = 0.08$ and $t \in [0,\pi)$. The red dots are the values at $t = -\frac{\gamma}2+\frac{\pi d}3$ with $\gamma = 0.15$, which converge to $\asf^n_+(\infty)$ in the limit $\epsilon \to 0$.}
\label{fig:psi.functions}
\end{figure}

There is an alternative way to describe these truncated trajectories, using the functions $\psi^n(z)$ defined in \eqref{eq:psi.functions}, that allows us to completely characterize the truncated ground-state trajectories for all $d$. These functions incorporate the braid limits $\asf^n_\pm(\infty)$ in the different standard modules in terms of three continuous functions. In \cref{fig:psi.functions}, we plot the functions $\psi_1(s)$, $\psi_2(s)$ and $\psi_3(s)$ in the complex plane for $s=\eE^{\ir (t+\ir \epsilon)/2}$ with $t \in [0,2\pi)$ and $\epsilon$ a real infinitesimal parameter larger than zero, here set to $\epsilon=0.08$. The red dots mark the values at $t =-\frac\gamma2 + \frac{\pi d}6$ with $d \in \{0,1, \dots, 5\}$ and $\gamma = 0.15$. In the limit $\epsilon \to 0$, these values approach the braid values $\asf^n_+(\infty)$. Similarly, to get the positions of the braid limits $\asf^n_-(\infty)$ in $\repW_{N,d,\omega}$, one must select the red dot with label $d' = -d \textrm{ mod } 6$. Crucially, it turns out that the trajectories followed by the functions $\psi_n(\eE^{\ir(-\gamma \pm d\pi/3)t})$, from $t=0$ to $t = \frac12$, are precisely the same as those in \eqref{eq:An05}. These parametric curves give the correct ground-state trajectories for all $d$.

To continue with the calculation, we must fix the branches of the functions $\log \asf^n_\pm(z)$. We choose these branches to be such that its imaginary part is in the range $(-\pi,\pi)$ after the first winding $\mathsf x^{m_1}$ around the origin of the trajectory. Equivalently, due to \eqref{eq:winding.integrals}, this choice implies that the dilogarithm integrals that run over the trajectories $\widehat{\mathcal A}^n_\pm$ are computed with the imaginary part of $\log \asf^n_\pm(z)$ taken in $(-\pi,\pi)$ as $z \to -\infty$. Then the imaginary part of $\log \asf^n_\pm(\infty)$ is in $[(2k-1)\pi,(2k+1)\pi]$ for some integer~$k$, and one can compute this integer by following the corresponding curves in \cref{fig:psi.functions}.

With this information, we are able to compute $\log \asf_\pm^n(\infty)$ and $\log \big(1+\asf_\pm^n(\infty)\big)$ in the limit $z \to \infty$, including their imaginary parts. This allows us to compute the constants $\phi_\pm^n$. By taking the braid limit of \eqref{eq:aA.NLIEs}, we find
\be
\phi_\pm^n = \log \asf^n_\pm(\infty) - \sum_{m=1}^3 \tilde K_{nm}(0) \log \big(1+\asf^m_\pm(\infty)\big),
\ee
where $\tilde K(k)$ is defined in \eqref{eq:eqns.in.k.space}. Using the braid limits \eqref{eq:braid.limits} and the trajectories \eqref{eq:An05}, we find that the constants $\phi_\pm^n$ are given by the simple formulas
\be
\label{eq:phis}
\phi_1^\pm=-\phi_2^\pm=-(3\gamma\mp d\pi)\ir, \qquad \phi_3^\pm = 0.
\ee
The functions $\mathcal J_+$ and $\mathcal J_-$ thus read
\be
\mathcal J_\pm = \sum_{n=1}^3 \int_{\widehat{\mathcal A}^n_\pm} 
\dd \asf^n_\pm \bigg(\frac{\log \big(1+\asf^n_{\pm}\big)}{\asf^n_{\pm}} - \frac{\log \asf^n_{\pm} - \phi^n_\pm}{1+\asf^n_{\pm}}\bigg),
\ee
and are to be evaluated with the trajectories as described above. These integrals are computed in \cref{app:dilog.ints}. The final result reads
\be
\label{eq:J+-}
\mathcal J_\pm = \frac {\pi^2}3\bigg[ 1-\Big(\frac{3 \gamma}\pi \mp d\Big)^2\bigg],
\ee
and the resulting conformal dimensions are
\be
\label{eq:delta.bardelta}
\Delta = \frac1{24} \bigg[\displaystyle\Big(\frac{3\gamma}\pi-d\Big)^2-1\bigg] = \Delta_{\gamma/\pi,d/2}\ ,
\qquad
\bar\Delta = \frac1{24} \bigg[\displaystyle\Big(\frac{3\gamma}\pi+d\Big)^2-1\bigg] = \Delta_{\gamma/\pi,-d/2}\ .
\ee

\subsection{Analytic solution for ground states for strip boundary conditions}\label{sec:standard.groundstates.strip}

In this subsection, we focus on the strip boundary conditions and compute the finite-size corrections for the ground state eigenvalues of the standard modules $\repV_{N,d}$. With our computer program, we computed the transfer matrices $\Dbh(u)$ in these modules, their ground state eigenvalues $\widehat D(u)$, and the corresponding functions $\amf^n(z)$, $\Amf^n(z)$ and $\bmf^n(z)$, for various small values of $N$ and $d$. For all these ground states, we find that the zeros of $\widehat D(u)$ are symmetrically distributed between the two half-planes.  Moreover, we find that the ground state eigenvalue $\widehat D(u)$ is exactly the same for $d = 0$ and $d=1$. In the scaling limit, they thus share the same conformal dimensions. In the following, we therefore assume $d\ge 1$.

The bulk behavior of the functions $\asf^n_\pm(z)$, as dictated by the driving terms in the nonlinear integral equations \eqref{eq:aA.NLIEs}, vanishes:
\be
\asf^n_\pm(-\infty) = 0, \qquad n = 1,2,3.
\ee
The braid limits for these functions are 
\begin{subequations}
\label{eq:strip.braid.asy}
\begin{alignat}{2}
\asf^1_\pm(\infty) &= 
\left\{\begin{array}{cl}
\mp \frac{\ir}{\sqrt 3} & d \equiv 0 \textrm{ mod }3, \\[0.15cm]
\pm \frac{\ir}{\sqrt 3} & d \equiv 1 \textrm{ mod }3, \\[0.15cm]
\infty & d \equiv 2 \textrm{ mod }3,
\end{array}\right.
\\[0.15cm]
\asf^2_\pm(\infty) &=
\left\{\begin{array}{cl}
\pm \frac{\ir}{\sqrt 3} & d \equiv 0 \textrm{ mod }3, \\[0.15cm]
\mp \frac{\ir}{\sqrt 3} & d \equiv 1 \textrm{ mod }3, \\[0.15cm]
\infty & d \equiv 2 \textrm{ mod }3,
\end{array}\right.
\\[0.15cm]
\asf^3_\pm(\infty) &=
\left\{\begin{array}{cl}
 \frac12 & d \equiv 0,1 \textrm{ mod }3, \\[0.15cm]
-1 & d \equiv 2 \textrm{ mod }3.
\end{array}\right.
\end{alignat}
\end{subequations}

For $d=1$, the functions $\bmf^n(z)$ have no zeros in their analyticity strips. In contrast, for $d \ge 2$ and $d \ge 3$, the functions $\bmf^1(z)$ and $\bmf^3(z)$ respectively have zeros on the real line, and their numbers grow as $d$ increases. As explained under \eqref{eq:S.subsets}, the paths $\mathcal C^n$ depend on the locations of the zeros of $\bmf^3(z)$. As we move along these paths, the functions $\amf^n(z)$ follow trajectories that wind non trivially around the points $0$ and $-1$. Like in the previous section, we express the trajectories $\mathcal A_\pm^n$, in the scaling limit, by words of the form $\mathsf x^{m_1}\mathsf y^{n_1}\mathsf x^{m_2}\mathsf y^{n_2}\dots$, with $m_i$ and $n_i$ measuring the number of times the trajectory crosses the half-lines $(0,\infty)$ and $(-\infty,-1)$, respectively. The dilogarithm integrals are again computed on truncated trajectories $\widehat{\mathcal A}_\pm^n$ wherein the first windings around the origin are removed. Using our computer program, we find empirically that these truncated trajectories for $d = 1,2,3$ are 
\be
\begin{array}{ll}
d = 1: \qquad
\widehat{\mathcal A}^n_- = \big(1,1,1\big)_n 
\quad
&\widehat{\mathcal A}^n_+ = \big(1,1,1\big)_n
\\[0.15cm]
d = 2: \qquad
\widehat{\mathcal A}^n_- = \big(1,\mathsf y,\mathsf y^{-1/2}\big)_n 
\quad
&\widehat{\mathcal A}^n_+ = \big(1,\mathsf y^{-1},\mathsf y^{1/2}\big)_n
\\[0.15cm]
d = 3: \qquad
\widehat{\mathcal A}^n_- = \big(1,\mathsf y,1\big)_n
\quad
&\widehat{\mathcal A}^n_+ = \big(1,\mathsf y^{-1},1\big)_n\ 
\end{array}
\ee
and so on for $d \ge 4$. With our computer implementation, we produced these trajectories for $d = 0, \dots, 10$ on small system sizes.

For $d \equiv 2 \textrm{ mod }3$, the braid limits $\asf^1_\pm(\infty)$ and $\asf^2_\pm(\infty)$ are not finite. In contrast, the derivation of finite-size corrections in \cref{sec:scaling.limit} assumed finite asymptotics of the functions in the NLIEs. The final results of this calculation, namely the formulas \eqref{eq:J.computation.1} and \eqref{eq:J.ints} for $\mathcal J_{\pm}$ and their relation \eqref{eq:Delta.J} with the conformal weights, hold for continuous values of the twist $\gamma \in (0,\frac \pi 3)$. Crucially, these formulas also have well-defined limits for $\gamma \to 0$ and $\gamma \to \frac \pi 3$, even in the cases where the braid limits of the functions $\asf^n_\pm(\infty)$ are infinite (this depends on $d$). From this, we infer that \eqref{eq:delta.bardelta} in fact holds for $\gamma \in [0, \frac \pi 3]$. We use this to obtain the conformal weights for the strip boundary conditions, by considering limiting cases of the same calculation for periodic boundary conditions. Indeed, for $d \ge 1$, we observe that the braid limits and the trajectories for $\asf^n_+(x)$ are obtained from \eqref{eq:braid.limits} and \eqref{eq:An05} by replacing $d$ by $2d-1$ and by taking the limit $\gamma \to 0$. Likewise, for $\asf^n_-(x)$, the braid limits and trajectories are obtained from the periodic case by replacing $d$ by $2d-2$ and $\gamma \to \frac \pi 3$. This holds for all $d$. Thus, the functions $\asf^n_\pm(x)$ for the two geometries satisfy the same NLIEs and have the same trajectories and braid limits. As a result, the dilogarithm integral for the strip boundary conditions are directly obtained from \eqref{eq:J+-}, and the conformal dimensions read
\be
\Delta = \Delta_{0,d-1/2}, \qquad \bar \Delta = \Delta_{1/3,-d+1}.
\ee 
These can be equivalently written as
\be
\Delta = \bar \Delta = \frac{d(d-1)}6 = \Delta_{1,d+1}.
\ee
By the above remark, this result also holds for $d = 0$.

\section{Numerical confirmation of conformal partition functions\label{sec:Numerics}}

The Bethe ansatz equations \eqref{TBetheAnsatz} and \eqref{DBetheAnsatz} can be used to obtain numerical solutions for the transfer matrix eigenvalues for critical site percolation on the triangular lattice. But they do not constitute a solution in and of themselves. It is not easy to extract information directly from these algebraic equations. First, any numerical solution relies on a sufficiently accurate initial guess for the Bethe roots. Second, the form of the Bethe ansatz equations is not well suited for iterative numerical solutions. In this section, we solve the more stable logarithmic form of the Bethe ansatz equations to obtain the Bethe roots to high accuracy. We do this separately for the strip and periodic boundary conditions, restricting to $\omega=\pm1$ for the latter. Initial guesses for the Bethe roots of the ground state and excited states are provided by means of direct diagonalization for small system sizes $N\le 12$. Our numerical calculations are carried out to (i) find the patterns of zeros needed to determine the analyticity as input in our analytic calculations, (ii) confirm our analytic results for the ground states in \cref{sec:NLIE.FSS} and (iii) confirm more generally \cref{torusConj,cylConj} for the scaling limit of the transfer matrix traces. The numerics are performed using 100 digit precision in Mathematica~\cite{Wolfram} on a Mac Pro with 384GB of memory. 

\subsection{Logarithmic form of the Bethe ansatz equations}\label{sec:logBAEs} 
\def\arccot{\mathop{\mbox{arccot}}}

For numerical stability, the nonlinear Bethe ansatz equations are better converted to their {\em logarithmic} form. To do this, we write the Bethe roots as $u_j=\ir v_j$ and use the identity
\be
\log \frac{\sin(\ir v-\tfrac\pi 3)}{\sin(\ir v+\tfrac\pi 3)}=2\,\ir\arccot\!\Big(\frac{\tanh v}{\sqrt{3}}\Big),\qquad v\in\Bbb C,\label{logId}
\ee
where there is a branch cut along the line segment from $v=-\tfrac{\pi \ir}{3}$ to $v=\tfrac{\pi \ir}{3}$. This implies a jump discontinuity on the real $v$ line of $2\pi\ir$ at $v=0$. Numerically, we find that the zeros of $Q(u)$ occur in strings of length 1, 2 and 3 as shown in \cref{strings}.

\begin{table}[t]
\begin{center}
\begin{pspicture}(-5,-3)(5,3)
\rput(0,0){\includegraphics[width=3.5in]{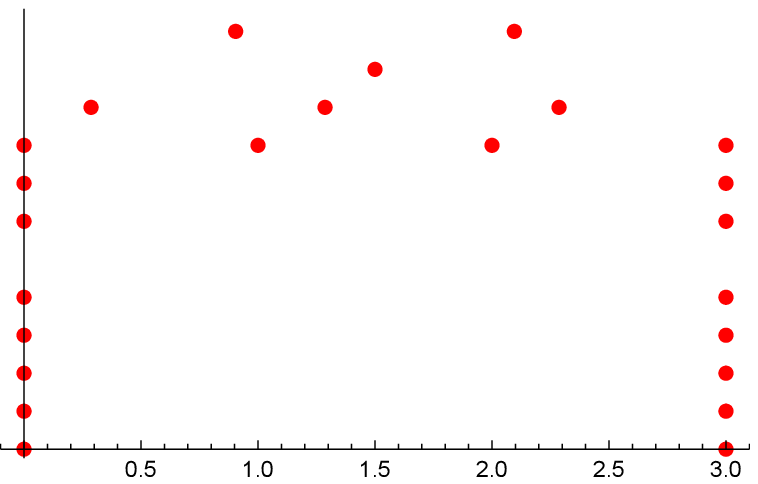}}
\rput(5.55,2.10){$1$-string}
\rput(5.55,2.50){$2$-string}
\rput(-6.3,-0.15){hole $1$-string}\psline{->}(-5,-0.15)(-4.3,-0.15)
\rput(5.4,1.4){$\left.\begin{pspicture}(0,-0.20)(0,0.20)\end{pspicture}\right\}$ $3$-strings}
\rput(5.8,-0.79){$\left.\begin{pspicture}(0,-0.91)(0,0.91)\end{pspicture}\right\}$ long $2$-strings}
\end{pspicture}
\\[0.5cm]
\begin{tabular}{l|l|c|c}
{Type of String}&{$Q(u)$ String Locations}&{Number}&{Order}\\
\hline
Long 2-strings&$\{v_j,v_j+\pi \ir\}, v_j\in\Bbb R$&$n_2$&$O(N)$\rule{0pt}{14pt}\\
Hole 1-strings&$\{\tilde v_j\}, \tilde v_j\in\Bbb R$&$m_h$&$O(1)$\rule{0pt}{12pt}\\[-2pt]
1-strings&$\{v_j+\tfrac{\pi \ir}{2}\}, v_j\in\Bbb R$&$m_1$&$O(1)$\rule{0pt}{14pt}\\[2pt]
2-strings&$\{(v_j+\epsilon+\tfrac{\pi\ir}{3},v_j-\epsilon+\tfrac{2\pi\ir}{3})\}, v_j\in\Bbb R$&$m_2$&$O(1)$\\[2pt]
3-strings&$\{(v_j,v_j+\tfrac{\pi\ir}{3},v_j+\tfrac{2\pi\ir}{3})\}, v_j\in\Bbb C$&$m_3$&$O(1)$\\[2pt]
\hline
\end{tabular}
\caption{The different 1-, 2- and 3-string types of $Q(u)$, where the Bethe roots are written as $u_j = \ir v_j$ and $\epsilon$ denotes a complex number with small modulus. 
The apparent two zeros of the ``long 2-string" are identified under $\pi\ir$ periodicity, so there is really only one zero. The location $\tilde v_j$ of the hole 1-string marks the location on the real axis where there is a hole in the distribution of long 2-strings of $Q(u)$. This hole location $\tilde v_j$ coincides with the location $v_j$ of a 2-string in $\widehat T(u)$ or $\widehat D(u)$. On the cylinder, we distinguish strings in the upper and lower half $u$-planes using the notation $m_h^{\pm},m_1^{\pm},m_2^{\pm},m_3^{\pm}$. By convention, strings on the real $u$-axis are assigned to the upper half plane. In some cases, the degree of $Q(u)$ is reduced by 1 or more, which we interpret as due to Bethe roots at infinity. The number of zeros at infinity is denoted by $m_\infty$. The top panel depicts the the various Bethe roots strings in the upper half $u$ plane, with the horizontal axis marked in units of~$\frac\pi3$.}
\label{strings}
\end{center}
\end{table}

\paragraph{Logarithmic Bethe ansatz equations for strip boundary conditions.} 
Using the identity \eqref{logId}, the Bethe ansatz equations on the strip \eqref{DBetheAnsatz} take the logarithmic form
\be
4N \arccot\!\bigg(\frac{\tanh v_j}{\sqrt{3}}\bigg)+\!\!\!\sum_{k=-M}^{M} \!\!4\arccot\!\bigg(\frac{\tanh(v_j\!-\!v_k)}{\sqrt{3}}\bigg)=2\pi I_j,\quad j=0,1,2,\ldots, M,\label{logBetheAnsatz}
\ee
where the Bethe integers $I_j\in{\Bbb N}$ are due to the jump discontinuities of the logarithm, when applied to \eqref{DBetheAnsatz}. From \eqref{BetheRootSym}, we have $v_{-j}=-v_j$ and $v_0=0$. For convenience, we count $v_0=0$, related to the boundary contribution, as a Bethe root of $\widehat Q(u)$. 
We need only consider the logarithmic Bethe ansatz equations for the $M+1$ Bethe roots    $j=0,1,2,\ldots,M$ in the fundamental domain
\be
0\le \textrm{Re}(u)\le \tfrac{\pi}{2},\qquad  0\le\textrm{Im}(u).\label{FundDomain}
\ee
The Bethe integers $I_j$ specify the different branches of the logarithm in \eqref{logId}. Let us define the reference integers
\be
I^0_j=N\!+\!j\!-\!1.
\ee
For the ground state in the standard module $\repV_{N,d}$, the number of Bethe roots is $M = N-d$ and the Bethe integers are $I_j = I^0_j$. For the excited states, we write the Bethe integers as
\be
I_j=I^0_j-E_j,\qquad j=0,1,2,\ldots,M,
\ee
where $E_j$ are integer root excitation energies, with $0\le E_j<\infty$. Thus for the ground state, we have $I_j=I_j^0$ and $E_j=0$ for all $j$. For convenience, we reverse the order of the Bethe roots and introduce the quantum integer sequences
\be
\{I^0_j\}=\{I^0_{\!M},I^0_{\!M-1},\ldots,I^0_0\}\!=\!\{N\!+\!M\!+\!1,N\!+\!M,\ldots,N\!+\!1\},\quad \{E_j\}\!=\!\{E_{M},E_{M-1},\ldots,E_0\}.
\ee
As introduced, the excitation energy of an individual Bethe root contributes a fixed amount of energy. The labelling order of these roots is arbitrary, however the actual order is immaterial since the conformal weights $\Delta$ only depend on the sum $\sum_j E_j$ which is invariant under labelling permutations. For convenience, in the tables of excitations in \cref{sec:Tabulated}, we label the Bethe roots in an order such that the sequence $ \{E_j\}$ is a non-decreasing sequence. For finite excitations, only a finite number of the excitation energies $E_j$ are positive as $N\to\infty$.

\paragraph{Logarithmic Bethe ansatz equations for periodic boundary conditions.}
\def\Re{\mathop{\mbox{Re}}}
\def\Im{\mathop{\mbox{Im}}}
On the cylinder, the patterns of zeros in the upper and lower half $u$-planes are independent and relate, through finite-size corrections, to $\Delta$ and $\bar\Delta$ respectively. 
Using the identity \eqref{logId} and setting $\zeta=\eE^{\ir {\cal P}}$, the Bethe ansatz equations on the cylinder \eqref{TBetheAnsatz} take the logarithmic form
\be
2N \arccot\!\Big(\frac{\tanh v_j}{\sqrt{3}}\Big)+\!\!\!\sum_{k=-M_-}^{M_+-1} \!\!\!4\arccot\!\Big(\frac{\tanh(v_j\!-\!v_k)}{\sqrt{3}}\Big)=-\gamma -{\cal P}+\pi I_j,
\label{TlogBetheAnsatz}
\ee
where the Bethe integers satisfy 
\be
I_j\in\left\{\begin{array}{cl}
2{\Bbb Z}+1&\mbox{$d$ even,}\\[0.1cm]
2{\Bbb Z}&\mbox{$d$ odd.}
\end{array}\right.
\ee
The Bethe roots $v_j$ are assigned the integer labels $j=-M_-,-M_-\!+\!1,\ldots, M_+\!-\!1$ in such a way that $\Re (v_j)<0$ for $j<0$ and $\Re (v_j)\ge 0$ for $j\ge 0$. The total number of roots is $M=M_-+M_+$.

For the ground state of the standard modules $\repW_{N,d,\omega}$ with $\omega = \pm 1$, we find that the number of Bethe roots is $M=N-d$ and the quantum integer sequences $\{I^0_j\}=\{I^0_{-M_-},I^0_{-M_-+1},\ldots,I^0_{M_+-1}\}$, for $N$ even, are
\be
\{I^0_j\}=
\left\{\begin{array}{cl}
\{-(M\!+\!N\!-\!3),-(M\!+\!N\!-\!5),\ldots,-(N\!-\!1),N\!+\!3,\ldots,M\!+\!N\!+\!1\}&d \textrm{ even}.\\[0.2cm]
\{-(M\!+\!N\!-\!3),-(M\!+\!N\!-\!5),\ldots,-N,N\!+\!2,\ldots,M\!+\!N\!+\!1\}&d \textrm{ odd}.
\end{array}\right.
\ee
Similar expressions hold for odd values of $N$.
For excited states, the excitation energies $\{E_j\}=\{E_{-M_-},E_{-M_-+1},\ldots,E_{M_+-1}\}$ are defined by
\be
E_j=\left\{\begin{array}{ll}
I^0_j-I_j&j\ge 0,\\[0.15cm]
I_j-I^0_j&j< 0.
\end{array}\right.\label{Tdiff}
\ee
For finite excitations, as $N\to\infty$, only a finite number of the excitation energies $E_j$ are nonzero on the left and the right extremities of the sequence $\{E_j\}$. 
Reading from the left gives the energy excitations $E_j$ relevant to the lower half $u$-plane and the conformal weight $\Delta$. 
Reading from the right gives the energy excitations $\Ebar_j$ relevant to the upper half $u$-plane and the conformal weight $\deltabar$.

\subsection{Numerics for the double row transfer matrix}\label{sec:NumericsStrip}
\nc{\mb}{\mbox{\boldmath $m$}}

The double row transfer matrices of site percolation on the triangular lattice were ``numerically diagonalized" for system sizes $N\le 13$ in the standard representations $\repV_{N,d}$ with $d=0,1,\ldots,8$. Importantly, this yields the patterns of zeros for the leading eigenvalues which are used as input in the logarithmic Bethe ansatz equations \eqref{DBetheAnsatz}. For these eigenvalues, this allows the system size $N$ to be systematically increased and the conformal energies $E$ to be obtained by extrapolation. 
More precisely, the relevant quantities were calculated in the following sequence:

\begin{enumerate} 
\item First, matrix representations $\Dbh(u)$ of the double row transfer matrices are constructed using the action of the dilute Temperley-Lieb algebra on link states in the various standard modules $\repV_{N,d}$. The generators of the dilute Temperley-Lieb algebra and the face transfer operators are implemented as sparse matrices but the transfer matrices are dense. The matrix entries of $\Dbh(u)$ are centered Laurent polynomials in $z=\eE^{\ir u}$ of degree width $2N$ with numerical coefficients. The eigenvalues of the commuting transfer matrices $\Dbh(u)$ are obtained by a process of direct ``numerical diagonalization". More specifically, the numerical eigenvectors of $\Dbh(\tfrac{\pi}{3})$ form a set of common eigenvectors of $\Dbh(u)$. In practice, as $N$ becomes larger, only a subset of the leading eigenvectors are obtained using the Arnoldi method. Acting on these eigenvectors with $\Dbh(u)$ produces the eigenvalues $\widehat D(u)$ as Laurent polynomials in $z=\eE^{\ir u}$ with numerical coefficients. Strictly speaking, referring to this process as {\em numerical diagonalization} is a misnomer since $\Dbh(\tfrac{\pi}{3})$ can exhibit nontrivial Jordan blocks. In such cases, the matrix is defective, there is no complete set of eigenvectors and the transfer matrix is {\em not diagonalizable}.
\item For each polynomial eigenvalue $\widehat D(u)$, the linear system \eqref{eq:DLinearSys} is solved numerically for the corresponding functions $\widehat Q(u)$ and $P(u)$ as Laurent polynomials in $z=\eE^{\ir u}$ with numerical coefficients. The degrees of $\widehat Q(u)$ and $P(u)$ used in this process are $2N-2d+1-2\delta_{d,0}$ and $4N-2d+1-2\delta_{d,0}$ respectively.
\item For each eigenvalue $\widehat Q(u)$, the zeros in the fundamental domain are obtained numerically and any 3-strings among these zeros are counted and removed along with any zeros at infinity. The 3-strings can be removed because the renormalized $\widehat Q(u)$, obtained by dividing out by the polynomial associated with the 3-string zeros, satisfies the same $T$-$Q$ relation as the original $\widehat Q(u)$. Due to numerical error, the ``zeros at infinity" are very large but not actually infinite. The number of remaining zeros of $\widehat Q(u)$ in the fundamental domain is
\be
M+1=N-d+1-\delta_{d,0}-m_\infty-3m_3=n_2+m_1+2m_2.\label{Mvalue}
\ee
The $M+1$ zeros of $\widehat Q(u)$ so found are substituted into the logarithmic Bethe ansatz equations to solve for the Bethe integers $I_j$. The difference between the known ground state Bethe integers 
\be
\{I^0_j\}=\{I^0_{\!M},I^0_{\!M-1},\ldots,I^0_0\}\!=\!\{N\!+\!M\!+\!1,N\!+\!M,\ldots,N\!+\!1\}
\ee
and the excited Bethe integers
\be
\{I_j\}=\{I_{\!M},I_{\!M-1},\ldots,I_0\}
\ee
give the $M$ excitation energies $E_j=I_j^0-I_j$. 
\item Considering a given eigenvalue $\widehat Q(u)$ and starting at a given system size $N=N_0$, the values of $N$ and $M$ in the logarithmic Bethe ansatz equations are incremented, one unit at a time, by adding successive long 2-strings at the edges of the analyticity strip while imposing $I_j=I_j^0-E_j$ with the excitation energies $E_j$ held fixed as $N$ is increased. The excitation energies $E_j$ of all the zeros associated with the additional long 2-strings are set to zero. This process is informally described as ``filling the Fermi sea". The logarithmic Bethe ansatz equations are solved for each incremented value of $N$ yielding the Bethe roots $v_j$ and the Laurent polynomial $\widehat Q(u)$ through \eqref{Qpoly}. As $N$ is incremented, the polynomials $\widehat D(u)$ are obtained from the Bethe ansatz equations \eqref{DQeqn} and the patterns of zeros are checked for each value of $N$. This process is typically stable out to $N=36$ or beyond. 
It is clear that the size of the truncated subset of eigenvalues obtained, at system size $N$, is limited by the number of eigenvalues of $\widehat D(u)$ kept with system size $N_0$ at the ``direct diagonalization" stage.
\item As $N$ is increased, starting with the patterns of zeros for $N=N_0$, the sequences of values $\widehat D(u_0)$, say for $u_0=\tfrac{\pi}{6}$, can be extrapolated to extract the conformal data $\{c, \Delta_{1,d+1}, E\}$ from the finite-size corrections \eqref{eq:logTD.asy.D} using Vanden Broeck~\cite{vBS} sequence acceleration (see \cref{TypicalDEigVBS}):
\be
\Delta_N=-\frac{N}{\pi}\big[\log \widehat D(\tfrac{\pi}{6})+2N f_\text{bulk}(\tfrac{\pi}{6})\big] 
\quad 
\xrightarrow{N \to \infty} \quad\Delta=\Delta_{1,d+1}+E.\label{Dextrapolant}
\ee 
We stress that some of the eigenvalues $\widehat D(u_0)$ are complex and that in this case $\Delta_N$ is complex. Indeed, the eigenvalues of $\widehat D(u)$ involving 2-strings are typically complex and appear in complex-conjugate pairs. In such cases, we find that the extrapolated value of $\Delta$ is nonetheless real. Equivalently, performing the extrapolation with the real part or modulus of $\widehat D(u_0)$ gives the same real conformal weight $\Delta$. In this sense, the imaginary parts of $\widehat D(u_0)$ are negligibly small and do not contribute in the continuum scaling limit. 
The generating functions for the leading sequences of conformal energies $E$ yield the {\em truncated\/} conformal characters
\be
\tilde \chit_{1,s}^{(L)}(q)=q^{-\frac{c}{24}+\Delta_{1,s}}\sum_E q^E+O(q^{L+1}),\qquad s=d+1,\label{chiTrunc}
\ee
where the truncation level observed from the numerics is
\be
L=N_0-d-2\delta_{d,0}-\delta_{d,1}\ge 0.
\ee
This implies that each increment in $N$ by 1 systematically gives correctly the next term in the \mbox{$q$-expansion} of the conformal character. In principle, $N_0$ can be chosen to be arbitrarily large but, in practice, it is limited by the maximum system size $N_\text{max}$ of the systems that can be handled numerically. Typically, in our calculations, $N_\text{max}\le 13$. As $N_0$ is increased, only a very small proportion of the eigenvalues ultimately contribute to the {\em truncated\/} conformal characters \eqref{chiTrunc}. 
These {\em truncated\/} conformal characters $\tilde\chit_{1,s}^{(L)}(q)$ should not be confused with {\em finitized\/} conformal characters $\chit_{1,s}^{(N)}(q)$ which, unlike the  {\em truncated\/} conformal characters, satisfy the property $\chit_{1,s=d+1}^{(N)}(1)=\dim \repV_{N,d}$.
\item From our numerics, we conjecture that the extrapolated conformal energies are given by
\be
E=E_\text{base}+\sum_j E_j
\ee
where
\begin{subequations}
\be
E_\text{base}=\left\{\begin{array}{ll}
\tfrac{1}{2} m(3m+1)& d=0,\\[0.1cm]
\tfrac{1}{2}\lceil \tfrac{m+2t-1}{2}\rceil\big(3 \lceil \tfrac{m+2t-1}{2}\rceil-(-1)^{m}\big)-\Delta_{1,d+1}& d=3t>0,\\[0.1cm]
\tfrac{1}{2}\lceil \tfrac{m+2t}{2}\rceil\big(3 \lceil \tfrac{m+2t}{2}\rceil+(-1)^{m}\big)-\Delta_{1,d+1}& d=3t+1,\\[0.1cm]
\tfrac{3}{2}(m+t)(m+t+1)-(\Delta_{1,d+1}-\tfrac{1}{3})& d=3t+2,
\end{array}\right.
\ee
\be
m=\left\{\begin{array}{ll}
m_3& d=0,\\[0.1cm]
m_\infty+2m_3& d=3t>0,\\[0.1cm]
\tfrac{1}{2} m_\infty+2m_3& d=3t+1,\\[0.1cm]
m_3& d=3t+2,
\end{array}\right.\qquad\quad
\Delta_{1,d+1}=\left\{\begin{array}{ll}
\tfrac{1}{2} t(3t-1)& d=3t>0,\\[0.1cm]
\tfrac{1}{2} t(3t+1)& d=3t+1,\\[0.1cm]
\tfrac{3}{2} t(t+1)+\tfrac{1}{3}& d=3t>0.\\[0.1cm]
\end{array}\right.
\ee
\end{subequations}
Note that although the excitation energies $E_j$ initially refer to a fixed finite system size $N=N_0$, they are held fixed as $N\to\infty$ and consequently each $E_j$ emerges as a quantum number in the continuum scaling limit.
\item The string content $(m_h,m_1,m_2)$, quantum numbers $(m_\infty,m_3,m)$, conformal eigenenergies $E$ as well as the excitation energies $E_\text{base}$ and $E_j$ for the leading eigenvalues $\widehat D(u)$ of ${\cal DLM}(2,3)$ are tabulated for the standard modules $\repV_{N,d}$ with $d=0,1,2,\ldots,8$ in the Tables~\ref{d=0Eigs}--\ref{d=8Eigs}. 
The string content $(m_h,m_1,m_2)$ for complex eigenvalues containing 2-strings is more easily discerned by looking at the patterns of zeros of the function
\be
\widetilde Q(u) = \frac12 \sum_{j=-M}^M (c_j + \bar c_j)\, \eE^{\ir j u}
\qquad \textrm{where $c_j$ is defined from}
\qquad Q(u) = \sum_{j=-M}^M c_j \eE^{\ir j u}
\ee
and $\bar c_j$ is the complex conjugate of $c_j$. In contrast to bond percolation on the square lattice~\cite{MDKP2017}, we have not been able to find a complete classification of these eigenvalues by quantum numbers and patterns of zeros for arbitrarily large system sizes $N$.
\end{enumerate}
The Vanden Broeck-Schwartz extrapolation data for a typical eigenvalue $\widehat D(u)$ is presented in \cref{TypicalDEigVBS} in \cref{sec:Typical}.

In summary, to high precision, our numerics are consistent with the central charge and conformal weights
\be
c=0,\quad\ 
\Delta_{1,d+1}=\frac{d(d-1)}{6}=0,0,\tfrac 13, 1,2,\tfrac{10}{3},5,7,\tfrac{28}{3},\ldots \quad\textrm{for}\quad\  d=0,1,\ldots,8,\ldots
\ee
and yield the following truncated conformal characters $\tilde\chit_{1,d+1}^{(L)}(q)$ for the standard representations $\repV_{N,d}$ with $d=0,1,2,\ldots,8$:
\begin{subequations}
\allowdisplaybreaks
\begin{align}
\tilde\chit_{1,1}^{(11)} (q)&=1+q^2+q^3+2q^4+2q^5+4q^6+4q^7+7q^8+8q^9+12q^{10}+14q^{11}+O(q^{12}),\\
\tilde\chit_{1,2}^{(9)}(q)&=1+q+q^2+2q^3+3q^4+4q^5+6q^6+8q^7+11q^8+15q^9+O(q^{10}),\\
\tilde\chit_{1,3}^{(8)}(q)&=q^{1/3}(1+q+2q^2+2q^3+4q^4+5q^5+8q^6+10q^7+15q^8+O(q^{9})),\\
\tilde\chit_{1,4}^{(9)}(q)&=q(1+q+2q^2+3q^3+4q^4+6q^5+9q^6+12q^7+17q^8+23q^9+O(q^{10})),\\
\tilde\chit_{1,5}^{(8)}(q)&=q^2(1+q+2q^2+3q^3+5q^4+6q^5+10q^6+13q^7+19q^8+O(q^9)),\\
\tilde\chit_{1,6}^{(8)}(q)&=q^{10/3}(1+q+2q^2+3q^3+5q^4+7q^5+10q^6+14q^7+20q^8+O(q^{9})),\\
\tilde\chit_{1,7}^{(7)}(q)&=q^5(1+q+2q^2+3q^3+5q^4+7q^5+11q^6+14q^7+O(q^8)),\\
\tilde\chit_{1,8}^{(7)}(q)&=q^7(1+q+2q^2+3q^3+5q^4+7q^5+11q^6+15q^7+O(q^8)),\\
\chit_{1,9}^{(6)}(q)&=q^{28/3}(1+q+2q^2+3q^3+5q^4+7q^5+11q^6+O(q^7)).
\end{align}
\end{subequations}
To the order indicated, these truncated conformal characters exactly reproduce the degeneracies of the ${\cal LM}(2,3)$ logarithmic characters $\chit_{1,s}(q)$ with $s = d+1$ given in \eqref{eq:rsChars}.

\subsection{Numerics for the single row transfer matrix}\label{sec:NumericsCyl}

The single row transfer matrices of site percolation on the triangular lattice were ``numerically diagonalized" for system sizes $N\le 13$ in the standard modules $\repW_{N,d,\omega}$ with $\omega=\pm 1$ and $d=0,1,\ldots,6$. Importantly, this yields the patterns of zeros for the leading eigenvalues which are used as input in the logarithmic Bethe ansatz equations \eqref{TBetheAnsatz}. For these eigenvalues, this allows the system size $N$ to be systematically increased and the conformal energies to be obtained by extrapolation. More precisely, similarly to our calculations for $\Dbh(u)$, the relevant quantities were calculated in the following sequence:

\begin{enumerate}
\item First, matrix representations $\Tbh(u)$ of the single row transfer matrices are constructed using the action of the dilute Temperley-Lieb algebra on link states in the various standard modules with $\omega= 1$ and $\omega= -1$. The generators of the dilute Temperley-Lieb algebra and the face transfer operators are implemented as sparse matrices but the transfer matrices are dense. 
The matrix entries of $\Tbh(u)$ are centered Laurent polynomials in $z=\eE^{\ir u}$ of degree $N$ with numerical coefficients. 
The eigenvalues of the commuting transfer matrices $\Tbh(u)$ are obtained by a process of direct ``numerical diagonalization". More specifically, the numerical eigenvectors of $\Tbh(\tfrac{\pi}{3})$ form a set of common eigenvectors of $\Tbh(u)$. In practice, as $N$ becomes larger, only a subset of the leading eigenvectors are obtained using the Arnoldi method. 
Acting on these eigenvectors with $\Tbh(u)$ produces the eigenvalues $\widehat T(u)$ as Laurent polynomials in $z=\eE^{\ir u}$ with numerical coefficients. Strictly speaking, referring to this process as {\em numerical diagonalization} is a misnomer since $\Tbh(\tfrac{\pi}{3})$ can exhibit nontrivial Jordan blocks. In such cases, the matrix is defective, there is no complete set of eigenvectors and the transfer matrix is {\em not diagonalizable}.
\item For each polynomial eigenvalue $\widehat T(u)$, the linear system \eqref{eq:TLinearsys} is solved numerically for the corresponding functions $Q(u)$ and $P(u)$ as Laurent polynomials in $z=\eE^{\ir u}$ with numerical coefficients. The degrees of $Q(u)$ and $P(u)$ used in this process are $N-d$ and $2N-d$ respectively.
\item For each eigenvalue $Q(u)$, the zeros in the fundamental strip $0\le \Re (u)<\pi$ are obtained numerically and any 3-strings among these zeros are counted and removed. For $\omega = \pm 1$, we do not observe any Bethe roots at infinity. The numbers of remaining zeros of $Q(u)$ in the fundamental strip are
\begin{subequations}
\begin{align}
&M_-=m_h+m_1+2m_2,\qquad M_+=\mbar_h+\mbar_1+2\mbar_2,&\\[0.15cm]
&M=M_-+M_+=N-d-3m_3-3\mbar_3,\qquad M_\pm=O(\tfrac{1}{2}N),&
\label{MvalueT}
\end{align}
\end{subequations}
where the patterns of zeros in the upper and lower half planes are not simply related by complex conjugation and are treated as independent. The $M$ zeros of $Q(u)$ so found are substituted into the logarithmic Bethe ansatz equations \eqref{TBetheAnsatz} to solve for the Bethe integers $I_j$. The differences \eqref{Tdiff}  between the known ground state Bethe integers $I^0_j$ and the excited Bethe integers $I_j$
give the $M$ excitation energies $E_j$ with $j=-M_,-M_-+1,\ldots, M_+-1$. Reading the nonzero entries from the left and right gives the finite excitation energies $\{E_j|\Ebar_j\}$.
\item Considering a given eigenvalue $Q(u)$ and starting at a given system size $N=N_0$, the values of $N$ and $M$ in the logarithmic Bethe ansatz equations are incremented, one unit at a time, by adding successive long 2-strings at the edges of the analyticity strip while holding the finite excitation energies $\{E_j|\Ebar_j\}$ fixed as $N$ is increased. 
The $M$ long 2-strings are divided suitably between the upper and lower half planes. 
The excitation energies $E_j,\Ebar_j$ of all the zeros associated with the additional long 2-strings are set to zero. This process is informally described as ``filling the Fermi sea". 
The logarithmic Bethe ansatz equations are solved for each incremented value of $N$ yielding the Bethe roots $v_j$ and the Laurent polynomial $Q(u)$ through \eqref{QpolyT}. 
As $N$ is incremented, the polynomials $\widehat T(u)$ are obtained from the Bethe ansatz equations \eqref{TQeqn} and the patterns of zeros are checked for each value of $N$. This process is typically stable out to $N=36$ or beyond. 
It is clear that the size of the truncated subset of eigenvalues obtained, at system size $N$, is limited by the number of eigenvalues of $\widehat T(u)$ kept with system size $N_0$ at the ``direct diagonalization" stage.
\item As $N$ is increased, starting with the patterns of zeros for $N=N_0$, the sequences of values $\widehat T(u_0)$, say for $u_0=\tfrac{\pi}{6}$, can be extrapolated to extract the conformal data $c, \Delta, \bar\Delta$ from the finite-size corrections \eqref{eq:logTD.asy.T} using Vanden Broeck~\cite{vBS} sequence acceleration (see \cref{TypicalEigVBS})
\begin{subequations}
\begin{alignat}{3}
X_N&=\Delta_N+\bar\Delta_N=-\tfrac{N}{\pi}\big(\Re \log \widehat T(\tfrac{\pi}{6})+Nf_\text{bulk}(\tfrac{\pi}{6})\big)&&\xrightarrow{N\to\infty}\ X = \Delta+\bar\Delta,\\
S_N&=\Delta_N-\bar\Delta_N=-\tfrac{N}{\sqrt{3}\,\pi}\big(\Im \log \widehat T(\tfrac{\pi}{6})\big)&&\xrightarrow{N\to\infty}\ S = \Delta-\bar\Delta.
\label{Dextrapolant}
\end{alignat}
\end{subequations}
In the tables of \cref{sec:Tabulated}, we summarize the results compactly in the form
\be
(\Delta,\bar\Delta)=(\Delta_{\gamma/\pi,\frac{d}{2}}+E,\Delta_{\gamma/\pi,\frac{\eta d}{2}}+\Ebar),\quad  E=E_\text{base}+\sum_j E_j,\quad \Ebar=\Ebar_\text{base}+\sum_j \Ebar_j. \\[-6pt]
\ee
where $\eta=-1$ for $\gamma=0$ ($\omega=1$) and $\eta=1$ for $\gamma=\pi$ ($\omega=-1$).
\item The string content $(\mb;\bar\mb)=(m_h,m_1,m_2;\mbar_h,\mbar_1,\mbar_2)$, $(m_3,\mbar_3)$, conformal eigenenergies $(E,\Ebar)$ as well as the excitation energies 
$(E_\text{base},\Ebar_\text{base})$ and $\{E_j|\Ebar_j\}$ for the leading eigenvalues $\widehat T(u)$ are tabulated for the standard modules $\repW_{N,d,\omega=\pm 1}$ with $d$ even/odd in the range $d=0,1,2,\ldots,6$ in {Tables~\ref{omega=1dEvenEigs}--\ref{omega=-1dOddEigs}}. 
We have been unable to classify the patterns of zeros of $\widehat T(u)$, $Q(u)$ and $P(u)$ for arbitrarily large system sizes $N$. In contrast to the eigenvalues $\widehat D(u)$, for $\widehat T(u)$ we have also not been able to find explicit expressions for $E_\text{base}$ and $\Ebar_\text{base}$ in terms of the various quantum numbers in the lower and upper half planes.
\end{enumerate}
The Vanden Broeck-Schwartz extrapolation data for a typical eigenvalue $\widehat T(u)$ is presented in \cref{TypicalEigVBS} in \cref{sec:Typical}.

To illustrate, our numerics reveal that the contributions to the modular invariant partition function $\mathcal Z_{\textrm{tor}}^{\textrm{\tiny$(0,0)$}}(q,\qbar)=\mathcal Z_+(q,\qbar)+\mathcal Z_-(q,\qbar)$ from the leading 162 eigenvalues of $\Tbh(u)$ in the standard modules with $\omega=1$ and $\omega = -1$ are
\begin{subequations}
\label{eq:Z+Z-}
\begin{alignat}{2}
\mathcal Z_+(q,\qbar)&=(q\qbar)^{-1/24}[1+(q+\qbar)+(2q^2+q\qbar+2\qbar^2)+(3q^3+2q^2\qbar+4q^{3/2}\qbar^{3/2}+2q\qbar^2+3\qbar^3)]\nonumber\\[0.1cm]
&+2(q\qbar)^{1/8}[1+(q+\qbar)+(2q^2+q\qbar+2\qbar^2)+(3q^3+q^{5/2}\qbar^{1/2}+2q^2\qbar+2q\qbar^2+q^{1/2}\qbar^{5/2}+3\qbar^3)]\nonumber\\[0.1cm]
&+2(q\qbar)^{5/8}[1+(q+\qbar)+(2q^2+q\qbar+2\qbar^2)]+\cdots\\[0.1cm]
\mathcal Z_-(q,\qbar)&=2(q\qbar)^{1/3}[1+(q+\qbar)+(2q^2+q\qbar+2\qbar^2)+(4q^3+2q^2\qbar+2q\qbar^2+4\qbar^3)]\nonumber\\[0.1cm]
&+2[(q+\qbar)+(2q^2+2q\qbar+2\qbar^2)+(3q^3+4q^2\qbar+4q\qbar^2+3\qbar^2)]+\cdots
\end{alignat}
\end{subequations}
The modular invariant partition functions for critical site percolation are thus obtained as linear combinations of the modular invariants of the associated $\Atwotwo$ vertex model. In this context, the factors of $2$ for $d \neq 0$ appearing first in \eqref{eq:Markov.trace.alpha=2} and here in \eqref{eq:Z+Z-} are accounted for by the twofold multiplicities of the equivalent magnetization sectors $S^z=\pm d$ of the vertex model. 

%
\section{Conclusion}\label{sec:conclusion}
%

In this paper, we studied the continuum scaling limit of critical site percolation on the triangular lattice, working with the associated dilute $\Atwotwo$ loop model and its single and double row commuting transfer matrices. We showed that the model is integrable in the presence of a boundary and expressed the partition functions in terms of traces of the transfer matrices. We  obtained various functional relations including cubic functional equations, Baxter $T$-$Q$ equations, Bethe ansatz equations and linear functional equations satisfied by the transfer matrix eigenvalues, the auxiliary function $Q(u)$ and an additional auxiliary function $P(u)$. Assuming the known central charge $c=0$, we derived and solved nonlinear integral equations for the ground state conformal weights. We found that the ground states in the standard modules with $d$ defects have conformal weights $\Delta_{1,d+1}$ for strip boundary conditions and $\Delta_{\gamma/\pi,\pm d/2}$ for periodic boundary conditions, where $\Delta_{r,s}=\frac1{24}\big((3r-2s)^2-1\big)$ and $\omega = \eE^{\ir \gamma}$ is the twist. For periodic boundary conditions, the ground state conformal weights we obtained analytically for $\lambda=\tfrac{\pi}{3}$ in the standard module $\repW_{N,0,\omega}$ coincide with the limit $\lambda\to \tfrac{\pi}{3}$ of the known conformal weights~\cite{BNW89,WBN92,ZB97} for $\lambda \in (0,\frac \pi 3)$.

Our investigations of the excited states relied on extensive numerics, which solved the logarithmic form of the Bethe ansatz equations to large order and high precision. Indeed, in contrast with our previous work for bond percolation on the square lattice~\cite{MDKP2017}, we have not managed to obtain a complete classification of the patterns of zeros of the transfer matrix eigenvalues and auxiliary functions for critical site percolation. Nevertheless, our numerics confirmed the analytically calculated ground-state conformal weights, and allowed us to give conjectures for the scaling limit of the traces of the transfer matrix in the continuum scaling limit. These are expressed in terms of Kac characters for $\Dbh(u)$ and as sesquilinear forms of Verma characters for $\Tbh(u)$. Using these conjectures and the Markov trace of the dilute Temperley-Lieb algebra, we obtained the two conformal cylinder partition functions and the four modular invariant torus partition functions for $\alpha=2$. In particular,
\be
\mathcal Z_{\textrm{tor}}^{\textrm{\tiny$(0,0)$}}= |\varkappa_0(q)|^2\!+ 2 |\varkappa_2(q)|^2\!+2 |\varkappa_3(q)|^2\! 
+2 |\varkappa_4(q)|^2\!+ 2[\varkappa_1(q)\varkappa_5(\bar q) \!+\! \varkappa_5(q)\varkappa_1(\bar q)] + |\varkappa_6(q)|^2,
\ee
is a modular invariant sesquilinear form in affine $u(1)$ characters whose power-series expansion in $q$ and $\bar q$ has only positive integer coefficients. We also derived expressions for the partition function at the other values of $\alpha$, for both critical site and bond percolation, and found that they involve functions $Z_{m,m'}(g)$ that appeared previously in the Coulomb gas approach \cite{FSZ87,RS01}. 

The concurrence of all this conformal data provides compelling evidence supporting a strong form of universality between bond percolation on the square lattice and site percolation on the triangular lattice as logarithmic CFTs. To push this investigation further, a next step 
would be to compare the Jordan cells of these models to see if the observed universality also extends to the Jordan cell structure~\cite{PRZ2006,MDSA2011,MDSA2013} and the structure of the indecomposable yet reducible Virasoro representations.

For strip boundary conditions, the conformal weights arising in the partition functions do not exhaust all of the conformal weights of the infinitely extended Kac tables in \cref{fig:VirKac}. This is because we only considered a limited set of integrable boundary conditions. It would be of interest to extend this study further to the second boundary Yang-Baxter solution in \eqref{eq:bdy.face.op} as well as to other types of boundary conditions~\cite{PRZ2006,JSbdy,PTC2015,MDRR2015,BPT2016}. It is expected that the missing thermal conformal weight $\Delta_h=\Delta_{\frac{1}{2},0}=\tfrac{5}{96}$ will be among the additional conformal weights to be found.

In many cases, integrable lattice models that lie in the same universality class turn out to have transfer matrices that satisfy the same set of universal functional relations and NLIEs. This occurs for instance if we consider the square lattice Ising model on different two-dimensional topologies. For bond and site percolation however, the transfer matrices satisfy different cubic functional relations, different $T$-$Q$ relations, and ultimately different NLIEs. We stress that, for site percolation, we avoided resorting to the $Y$-system of the $\Atwotwo$ model to derive the NLIEs. Indeed, the $Y$-system that we obtained in \cite{MDP19} involves auxiliary functions that do not have the correct analyticity properties to allow for a computation of the conformal eigenenergies. The generalised set of ``snake'' $Y$-systems, similar to those found by Kuniba, Sakai and Suzuki for the six-vertex model \cite{KSS98}, are not yet known for the $\Atwotwo$ model. After writing this paper, we realised that the auxiliary functions $\amf^1(z)$, $\amf^2(z)$ and $\amf^3(z)$ defined in \cref{sec:aux.functions} can be understood as three of five functions satisfying a $Y$-system studied by Gliozzi and Tateo \cite{GT95}, see \cref{app:Y.system}. It remains to be seen whether this approach also allows for the calculation of the finite-size corrections to the transfer matrix eigenvalues. In the end, although the NLIEs of this paper appear somewhat more elaborate, one advantage of the current approach is that they can be used for arbitrary values of $\lambda \in (0, \frac \pi 3]$, not only for roots of unity. Moreover, it applies to all values of the twist parameter, can be applied to arbitrary excitations, and thus allows for a systematic approach for $s\ell(3)$ models.

It would also be desirable to extend the calculation of the finite-size spectra and of the modular invariant partition functions to other $\Atwotwo$ loop models. In particular, the special value $\lambda = \frac{5 \pi}{16}$ corresponds to the model ${\cal DLM}(3,4)$ with central charge $c = \frac 12$. Moreover, it would be interesting to see if the technique of nonlinear integral equations can be applied favourably in Regime~III of the $\Atwotwo$ loop model to understand the conformal properties of this model whose ground state energy is known to become infinitely degenerate in the scaling limit \cite{VJS2014}.

\subsection*{Acknowledgments}

We thank Gyuri Feh\'er, Sasha Garbali, Junji Suzuki and Ole Warnaar for discussions. We also thank an anonymous referee for useful comments. AMD is supported by the EOS Research Project, project number 30889451. He also acknowledges support and hospitality from the Max-Planck-Institut f\"ur Mathematik in Bonn and the Leibniz Universit\"at Hannover, where parts of this work were done. AK was supported by DFG through the program FOG 2316. This work was completed while PAP was visiting the APCTP as an ICTP Visiting Scholar.

\appendix

\section{Dilogarithm integrals}\label{app:dilog.ints}

We want to calculate the sum of three dilogarithmic integrals
\begin{alignat}{2}
J &= \sum_{n=1}^3 \int_{\widehat {\mathcal A}^n} 
\dd \asf\bigg(\frac{\log \big(1+\asf\big)}{\asf} - \frac{\log \asf}{1+\asf}\bigg),
\label{eq:J.intsApp}
\end{alignat}
with the truncated trajectories $\widehat {\mathcal A}^n$ as discussed in \cref{sec:standard.groundstates.cylinder}. For each
state of the transfer matrix, we have a specific set of functions satisfying
the NLIEs where the indices $\pm$ refer to the contributions of left- and
right-movers (in a field-theoretic language) which are identified
in the scaling limit.

For each state and in general differently for the left- and right- scaling limits
($\pm$), the functions $\asf^n_{\pm}(z)$ are considered on paths $\mathcal
C^n_\pm$ that take $z$ from $-\infty$ to $+\infty$ along the real axis
with certain occasional indentations, as described in \cref{sec:standard.groundstates.cylinder}. For the evaluation of the above dilogarithmic integral, we only need to work with
the trajectories $\widehat {\mathcal A}^n$ that are taken by the functions $\asf^n_{\pm}(z)$
evaluated on $\mathcal C^n_\pm$, namely by the concatenation of
$\asf^n_{\pm}(z)$ with $\mathcal C^n_\pm$ yielding the above $\widehat {\mathcal A}^n$.
For all the eigenstates, the initial point of $\widehat {\mathcal A}^n$ is always $0$, for all
$n=1, 2, 3$. In contrast, the end points are given by the braid limits \eqref{eq:braid.limits}
which are written for our purposes in terms of the functions defined in \eqref{eq:psi.functions} as $\asf^n_\pm(\infty) = \psi_n(s)$ with $s=\eE^{\pm\ir\pi d/6}\omega^{-1/2}$ and $\omega = \eE^{\ir \gamma}$.

We note the important property of the function
\be
J(s) = \sum_{n=1}^3 \int_0^{\psi_n(s)}
\dd \asf\bigg(\frac{\log \big(1+\asf\big)}{\asf} - \frac{\log \asf}{1+\asf}\bigg)
\label{eq:J.intsApp2}
\ee
to be independent of $s$, where for the moment we assume that $s$ takes
positive real values. The independence of $s$ is proved by taking the
derivative and using elementary identities for the standard log-function to
show that all terms cancel each other. The value of $J(s)$ can be evaluated
for instance by inserting $s=\infty$, such that
\be
J(s)=J(\infty)=\int_0^\infty
\dd \asf\bigg(\frac{\log \big(1+\asf\big)}{\asf} - \frac{\log \asf}{1+\asf}\bigg)
=\frac{\pi^2}3.\label{eq2:J.intsApp2}
\ee
Moreover, because the logarithms in these integrals are understood to be continuous functions, the function $J(s)$ is by definition also continuous.

Next, we address the question of the value of $J(s)$ in the case where $s$ leaves the positive
real semi-axis. Of course, the combination of integrals that defines $J(s)$ stays
constant as long as we move~$s$ in the complex plane along a trajectory for which
all values of $\psi_n(s)$ avoid the singularities $0$ and $-1$ of the
integrand \eqref{eq:J.intsApp} and the logarithms of the integrand are kept
continuous along the
integration trajectories. The smoothness of all involved logarithm functions is naturally incorporated in the method of nonlinear integral equations. Here, the
smoothness of the logarithms allows us to extend the vanishing of the derivative
of $J(s)$ with respect to $s$ for positive real values to any complex value
under the before mentioned conditions.

For the ground state in $\repW_{N,0,\omega=\eE^{\ir \gamma}}$, the paths $\mathcal C^n_\pm$
follow the real axis. The evaluation of the functions
$\asf^n_{\pm}(z)$ on these paths yields trajectories $\widehat {\mathcal A}^n$ that
connect the initial points with the end points in a practically direct
manner, namely trajectories that stay close to the real axis and do not wind around the singularities $0$ and $-1$. We can find a simple replacement trajectory
yielding the same integral and allowing for a straight evaluation.  For each~$n$, we define this trajectory by two pieces, the first from $0$ to $\psi_n(1)$
along the real axis, the second (usually short) piece parameterized by
$\psi_n(\eE^{-\ir\gamma t})$ with real values of $t$ starting at $t=0$
(evaluating to $\psi_n(1)$) and ending at $t=1/2$ (evaluating to
$\psi_n(\eE^{-\ir\gamma/2}$). The trajectory obtained by the concatenation of these
two pieces and the true trajectory ${\mathcal A}^n$ lie in the same homotopy
class, namely they can be deformed continuously into each other without crossing
the points $0$ and $-1$. Of course, they share the same initial point, namely
0, but avoid this strictly in the later course. The integral of type
\eqref{eq:J.intsApp} on the replacement trajectories evaluates of course to
$J(1)+0=\pi^2/3$.

For the ground state in $\repW_{N,d,\omega}$ with $d>0$, the paths $\mathcal
C^n_\pm$ leave the real axis and the evaluation of the functions
$\asf^n_{\pm}(z)$ on these paths yields trajectories ${\mathcal A}^n$ that
connect the initial points with the end points in a more involved manner than
in the case $d=0$. This is described in \cref{sec:standard.groundstates.cylinder}. As involved
as these trajectories are, we may again find a simple replacement trajectory yielding the
same integral and allowing for a straight evaluation.  Again, for each $n$, we
define this trajectory by two pieces. The first piece consists of the trajectory from $0$ to $\psi_n(1)$ along
the real axis. The second (now longer) piece is parameterized by
$\psi_n(\eE^{\ir(-\gamma \pm d\pi/3)t})$ with values of $t$ starting at $t=0$
(evaluating to $\psi_n(1)$), then taking values slightly above (below) the
real axis for the case $+$ ($-$) and ending at $t=1/2$ (evaluating to
$\psi_n(\eE^{-\ir \gamma/2}\eE^{\pm\ir\pi d/6})$). The trajectory obtained by the
concatenation of these two pieces and the true trajectory $\widehat {\mathcal A}^n$
lie in the same homotopy class: they can be deformed continuously into
each other without crossing the points $0$ and $-1$. In other words, the second piece of these trajectories can be read off directly from \cref{fig:psi.functions}, by starting at the value for $d=0$ and $\gamma = 0$ where $s=1$ and moving along the parametric curve up to $\eE^{-\ir \gamma/2}\eE^{\pm\ir\pi d/6}$. The integral of type \eqref{eq:J.intsApp} on the replacement trajectories again evaluates
to $J(1)+0=\pi^2/3$. 

We have shown that in all cases so far, the dilogarithmic integrals evaluate
to the same value. The full finite size amplitude in \eqref{eq:J.ints}
contains the additional term
\be
\sum_{n=1}^3 \phi_n^\pm\int_{\widehat{\mathcal A}^n} 
\frac{\dd \asf}{1+\asf}=\phi_1^\pm \log\frac{1+\psi_1(s)}{1+\psi_2(s)}
=-\frac13(3\gamma\mp d\pi)^2,
\label{eq:J.intsApp3}
\ee
because of $\phi_1^\pm=-\phi_2^\pm=-(3\gamma\mp d\pi)\ir$, $\phi_3^\pm=0$ and
$s=\eE^{\pm\ir\pi d/6}\omega^{-1/2}$.
Finally we obtain
\be
\mathcal J_\pm=J-\frac13(3\gamma\mp d\pi)^2=\frac{\pi^2}3\Bigg[1-\left(\frac{3\gamma}\pi\mp d\right)^2\Bigg],
\ee
ending the proof of \eqref{eq:J+-}.

\section{Tabulated numerical results}\label{sec:Tabulated}

\subsection{Typical Vanden Broeck-Schwartz extrapolations}
\label{sec:Typical}

\begin{table}[H]
\begin{center}
\includegraphics[width=3.5in]{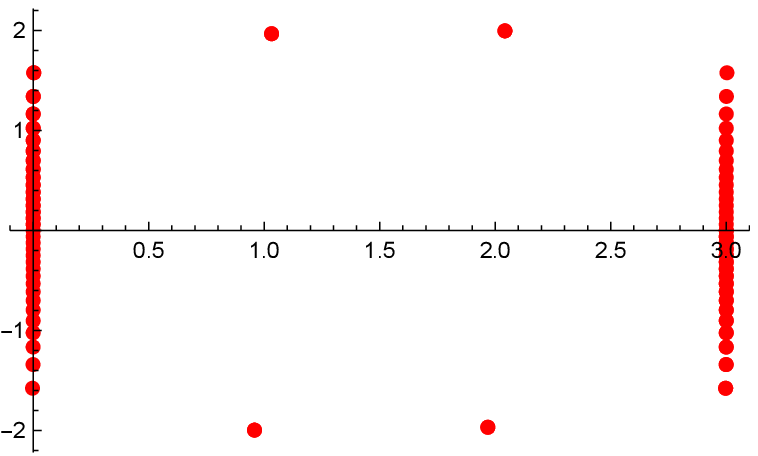}\\[16pt]
\footnotesize
\begin{tabular}{ccccc}
\hline\\[-8pt]
\multicolumn{5}{c}{\normalsize Conformal Weight Sequence $\mathop{\mbox{Re}} \Delta_N$ Extrapolation}\\[2pt]
\hline
6.06901630386251542	&&&&\\
6.05777824560024541	&6.01948611928504238	&&&\\
6.04909002547372132	&6.01652675142341243	&6.00007947399388592	&&\\
6.04223168255463500	&6.01419802777344113	&6.00005895801027374	&6.00001168315100050	&\\
6.03672140998680299	&6.01233171964422558	&6.00004466537468217	&6.00000885458703860	&6.00000005446810154\\
6.03222662709910136	&6.01081243522604241	&6.00003445838039663	&6.00000683477535594	&6.00000003289996419\\
6.02851161544488649	&6.00955880659702628	&6.00002701046403082	&6.00000536053373507	&\\
6.02540545419116287	&6.00851207971313484	&6.00002147214424538	&&\\
6.02278171602283199	&6.00762894576065053	&&&\\
6.02054523061579928	&&&&\\
\hline\\[-8pt]
\multicolumn{5}{c}{\normalsize Conformal Weight Sequence $|\Delta_N|$ Extrapolation}\\[2pt]
\hline
6.06901630386251407	&&&&\\
6.05777824560024503	&6.01948611928503589	&&&\\
6.04909002547372120	&6.01652675142341042	&6.00007947399381232	&&\\
6.04223168255463496	&6.01419802777344045	&6.00005895801024914	&6.00001168315088705	&\\
6.03672140998680297	&6.01233171964422532	&6.00004466537467324	&6.00000885458699844	&6.00000005446782494\\
6.03222662709910136	&6.01081243522604231	&6.00003445838039314	&6.00000683477534060	&6.00000003289986152\\
6.02851161544488648	&6.00955880659702624	&6.00002701046402938	&6.00000536053372882	&\\
6.02540545419116287	&6.00851207971313482	&6.00002147214424474	&&\\
6.02278171602283199	&6.00762894576065052	&&&\\
6.02054523061579928	&&&&\\
\hline
\end{tabular}
\caption{\label{TypicalDEigVBS}
Typical Vanden Broeck-Schwartz extrapolation~\cite{vBS} for $\widehat D(u)$ of the real part and modulus of the conformal weight sequence $\Delta_N$ in \eqref{Dextrapolant}. The tabulated system sizes range from $N=22$ to $N=40$ in increments of 2. This data relates to the 
eigenvalue with $\Delta=6$ and labelled by $k=9$ in the standard module $\repV_{N,d=0}$ tabulated in Table~\ref{d=0Eigs}. The initial sequences for Re\,$\Delta_N$ and $|\Delta_N|$ are presented in the first column. Sequences of extrapolated values are given in each subsequent column added to the right. The convergence to the $N \to \infty$ value improves with each additional column. The pattern of zeros of the associated $Q(u)$, after a 3-string has been removed, is shown in the upper panel for $N=22$. 
The string content for this eigenvalue is $(m_h,\!m_1,\!m_2)=(0,0,1)$ with $m_3=1$. To the extent that the finite-size extrapolations of the real part and modulus of $\log\widehat D(\tfrac{\pi}{6})$ have converged, they agree and yield the conformal weight $\Delta_N\to \Delta=6$. 
In principle, the accuracy of the extrapolation could be improved by increasing the system size further.}
\end{center}
\end{table}

\begin{table}[H]
\begin{center}
\includegraphics[width=3.5in]{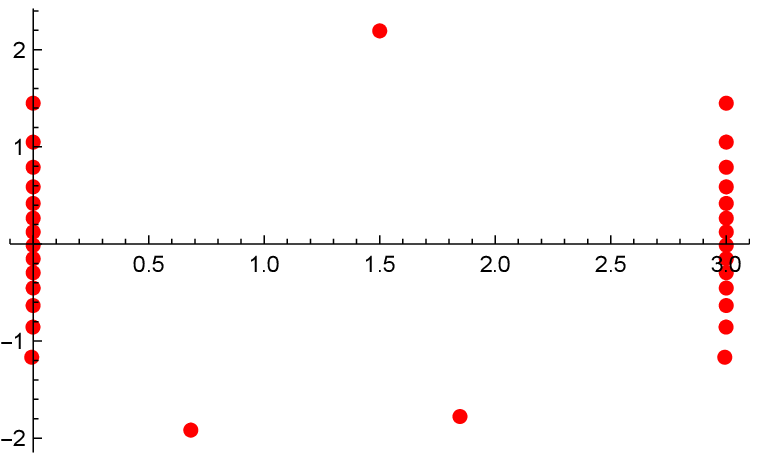}\\[16pt]
\footnotesize
\begin{tabular}{ccccc}
\hline\\[-8pt]
\multicolumn{5}{c}{\normalsize Scaling Dimension Sequence $X_N$ Extrapolation}\\[2pt]
\hline
3.72653094565	&&&&\\
3.71563081287	   	&3.68344486835	&&&\\
3.70748825057	  	&3.68059854046	&3.66672744801	&&\\
3.70123826663 	&3.67842751199	&3.66671983518	&3.66691941781	&\\
3.69633244327		&3.67673220654 	&3.66671192580	&3.66663137806	&3.66665105591\\
3.69240870772		&3.67538209874	&3.66670472881	&3.66666000526	&3.66666675603\\
3.68921983840		&3.67428879092	&3.66669853386	&3.66666622790	&\\
3.68659216854		&3.67339062325	&3.66669333926	&&\\
3.68440069522		&3.67264348526	&&&\\
3.68255352360	&&&&\\
\hline\\[-8pt]
\multicolumn{5}{c}{\normalsize Spin Sequence $S_N$ Extrapolation}\\[2pt]
\hline
1.00057890750	\\
1.00039082315		&1.00007921152\\
1.00027353329		&1.00005531841	&1.00000007406\\
1.00019724705		&1.00003982344	&1.00000004553	&1.00000000624\\
1.00014586174		&1.00002940992	&1.00000002901	&1.00000000411	&0.999999999392\\
1.00011020864		&1.00002219761	&1.00000001908	&1.00000000272	&0.999999999919\\
1.00008483454		&1.00001707211	&1.00000001290	&1.00000000184\\
1.00006637336		&1.00001334743	&1.00000000894\\
1.00005267968		&1.00001058736\\
1.00004234736\\
 \hline
\end{tabular}
\caption{\label{TypicalEigVBS}
Typical Vanden Broeck-Schwartz extrapolation~\cite{vBS} for $\widehat T(u)$ of the scaling dimension and spin sequences $X_N$, $S_N$. The tabulated system sizes range from $N=20$ to $N=38$ in increments of~2. 
This data relates to the $k=25$ eigenvalue in the module $\repW_{N,d=0,\omega=-1}$ in \cref{omega=-1dEvenEigs}, with $(\Delta,\bar\Delta)=(\tfrac{7}{3},\tfrac{4}{3})$. Sequences of extrapolated values are given in each subsequent column added to the right. The convergence to the $N \to \infty$ value improves with each additional column.
The pattern of zeros of the associated $Q(u)$, after a 3-string has been removed, is shown in the upper panel for $N=20$. The string content for this eigenvalue is $(m_h,\!m_1,\!m_2;\mbar_h,\!\mbar_1,\!\mbar_2)=(0,0,1;0,1,0)$ with $(m_3,\!\mbar_3)=(0,1)$. 
For this eigenvalue, we have $M=\tfrac{1}{2}N-3$, $M_-=\tfrac{1}{2}N-1$ and $M_+=\tfrac{1}{2}N-2$. The finite-size extrapolation of the real and imaginary parts of $\log \widehat T(\tfrac{\pi}{6})$ yields the scaling dimension and spin $X_N\to X=\Delta+\bar\Delta=\tfrac{11}{3}$, $S_N\to S=\Delta-\bar\Delta=1$. The value of ${\cal P}$ appears in the Bethe ansatz equations  and is a multiple of $\tfrac{2\pi}{3}$. In this case, it is zero. 
In principle, the accuracy of the extrapolation could be improved by increasing the system size further.}
\end{center}
\end{table}
\clearpage

\subsection{Tabulated numerical results for strip boundary conditions}

In the following pages, we tabulate our numerical results for strip boundary conditions, in the standard modules $\repV_{N,d}$ with $d=0,1,\ldots,8$. 
The label $k$ is the decreasing ordering of the eigenvalues in absolute value, at $u = \frac\pi6$, for the chosen system size $N_0$ used in the numerics.
\begin{table}[h!]
\begin{center}
\scriptsize
\begin{tabular}{|c||c|c||c||c|c|c|}
\multicolumn{7}{c}{\bf \normalsize Standard module $\boldsymbol{\repV_{N,d=0}}$}\\[4pt]
\hline
\hline
{Label}&\textbf{$(m_h,\!m_1,\!m_2)$}&$m_3$&\textbf{$E$}&$E_\text{base}$&$\sum_j E_j$&\textbf{$\{E_j\}$}\\
\hline
$k=1$&$(0,0,0)$&0&$0$&0&0&$\{0,0,0,0,0,0,0,0,0,\ldots\}$\rule{0pt}{8pt}\\[2pt]
\hline
$k=2$&$(1,0,0)$&1&$2$&2&0&$\{0,0,0,0,0,0,0,0,0,\ldots\}$\rule{0pt}{8pt}\\[2pt]
\hline
$k=3$&$(1,0,0)$&1&$3$&2&1&$\{1,0,0,0,0,0,0,0,0,\ldots\}$\\[2pt]
\hline
$k=4$&$(0,1,0)$&1&$4$&2&2&$\{2,0,0,0,0,0,0,0,0,\ldots\}$\\[2pt]
$k=5$&$(1,0,0)$&1&$4$&2&2&$\{1,1,0,0,0,0,0,0,0,\ldots\}$\\[2pt]
\hline
$k=6$&$(0,1,0)$&1&$5$&2&3&$\{3,0,0,0,0,0,0,0,0,\ldots\}$\\[2pt]
$k=7$&$(1,0,0)$&1&$5$&2&3&$\{1,1,1,0,0,0,0,0,0,\ldots\}$\\[2pt]
\hline
$k=8$&$(0,1,0)$&1&$6$&2&4&$\{4,0,0,0,0,0,0,0,0,\ldots\}$\\[2pt]
$k=9,10$&$(0,0,1)$&1&$6,6$&2&4&$\{1,3,0,0,0,0,0,0,0,\ldots\}$\\[2pt]
$k=11$&$(1,0,0)$&1&$6$&2&4&$\{1,1,1,1,0,0,0,0,0,\ldots\}$\\[2pt]
\hline
$k=12$&$(0,1,0)$&1&$7$&2&5&$\{5,0,0,0,0,0,0,0,0,\ldots\}$\\[2pt]
$k=13,14$&$(2,0,0)$&2&$7,7$&7&0&$\{0,0,0,0,0,0,0,0,0,\ldots\}$\\[2pt]
$k=15$&$(1,0,0)$&1&$7$&2&5&$\{1,1,1,1,1,0,0,0,0,\ldots\}$\\[2pt]
\hline
$k=16$&$(0,1,0)$&1&$8$&2&6&$\{6,0,0,0,0,0,0,0,0,\ldots\}$\\[2pt]
$k=17$&$(0,2,0)$&1&$8$&2&6&$\{2,4,0,0,0,0,0,0,0,\ldots\}$\\[2pt]
$k=18,19$&$(0,0,1)$&1&$8,8$&2&6&$\{1,2,3,0,0,0,0,0,0,\ldots\}$\\[2pt]
$k=20,21$&$(2,0,0)$&2&$8,8$&7&1&$\{1,0,0,0,0,0,0,0,0,\ldots\}$\\[2pt]
$k=22$&$(1,0,0)$&1&$8$&2&6&$\{1,1,1,1,1,1,0,0,0,\ldots\}$\\[2pt]
\hline
$k=23$&$(0,1,0)$&1&$9$&2&7&$\{7,0,0,0,0,0,0,0,0,\ldots\}$\\[2pt]
$k=24,25$&$(0,1,1)$&1&$9,9$&2&7&$\{2,2,3,0,0,0,0,0,0,\ldots\}$\\[2pt]
$k=26,27$&$(2,0,0)$&2&$9,9$&7&2&$\{2,0,0,0,0,0,0,0,0,\ldots\}$\\[2pt]
$k=28,29$&$(2,0,0)$&2&$9,9$&7&2&$\{1,1,0,0,0,0,0,0,0,\ldots\}$\\[2pt]
$k=30$&$(1,0,0)$&1&$9$&2&7&$\{1,1,1,1,1,1,1,0,0,\ldots\}$\\[2pt]
\hline
$k=31$&$(0,1,0)$&1&$10$&2&8&$\{8,0,0,0,0,0,0,0,0,\ldots\}$\\[2pt]
$k=32,33$&$(0,1,1)$&1&$10,10$&2&8&$\{2,2,4,0,0,0,0,0,0,\ldots\}$\\[2pt]
$k=34,35$&$(1,1,0)$&2&$10,10$&7&3&$\{3,0,0,0,0,0,0,0,0,\ldots\}$\\[2pt]
$k=36,37$&$(0,0,1)$&1&$10,10$&2&8&$\{1,1,3,3,0,0,0,0,0,\ldots\}$\\[2pt]
$k=38,39$&$(2,0,0)$&2&$10,10$&7&3&$\{1,2,0,0,0,0,0,0,0,\ldots\}$\\[2pt]
$k=40,41$&$(2,0,0)$&2&$10,10$&7&3&$\{1,1,1,0,0,0,0,0,0,\ldots\}$\\[2pt]
$k=53$&$(1,0,0)$&1&$10$&2&8&$\{1,1,1,1,1,1,1,1,0,\ldots\}$\\[2pt]
\hline
$k=42$&$(0,1,0)$&1&$11$&2&9&$\{9,0,0,0,0,0,0,0,0,\ldots\}$\\[2pt]
$k=43,44$&$(0,1,1)$&1&$11,11$&2&9&$\{1,2,2,4,0,0,0,0,0,\ldots\}$\\[2pt]
$k=45,46$&$(1,1,0)$&2&$11,11$&7&4&$\{4,0,0,0,0,0,0,0,0,\ldots\}$\\[2pt]
$k=47,48$&$(1,1,0)$&2&$11,11$&7&4&$\{1,3,0,0,0,0,0,0,0,\ldots\}$\\[2pt]
$k=49,50$&$(2,0,0)$&2&$11,11$&7&4&$\{2,2,0,0,0,0,0,0,0,\ldots\}$\\[2pt]
$k=51,52$&$(2,0,0)$&2&$11,11$&7&4&$\{1,1,2,0,0,0,0,0,0,\ldots\}$\\[2pt]
$k=55,56$&$(2,0,0)$&2&$11,11$&7&4&$\{1,1,1,1,0,0,0,0,0,\ldots\}$\\[2pt]
$k=113$&$(1,0,0)$&1&$11$&2&9&$\{1,1,1,1,1,1,1,1,1,\ldots\}$\\[2pt]
\hline
\end{tabular}
\caption{\small String content $(m_h,m_1,m_2)$, conformal eigenenergies and quantum integers for the leading 56 eigenvalues $\widehat D(u)$ in the standard module $\repV_{N,0}$. The conformal energies are given by $E=E_\text{base}+\sum_j E_j$ where the pentagonal numbers are $E_\text{base}=\tfrac12 m_3(3m_3+1)=0,2,7,15,26,\ldots$ for $m_3=0,1,2,3,4,\ldots$. The associated conformal character is $\chit_{1,1}(q)=1+q^2+q^3+2q^4+2q^5+4q^6+4q^7+7q^8+8q^9+12q^{10}+14q^{11}+O(q^{12})$. 
For a finite system, $N-3m_3=n_2+m_1+2m_2=M$ is the total number of zeros of $Q(u)$ in the fundamental domain that enter the Bethe ansatz equations.
}\label{d=0Eigs}
\end{center}
\end{table}

\newpage
\begin{table}[H]
\begin{center}
\scriptsize
\begin{tabular}{|c||c|c|c|c||c||c|c|c|}
\multicolumn{9}{c}{\bf\normalsize Standard module $\boldsymbol{\repV_{N,d=1}}$}\\[4pt]
\hline
\hline
{Label}&\textbf{$(m_h,\!m_1,\!m_2)$}&$\!\!m_\infty\!\!$&$\!m_3\!$&$m$&\rule{0pt}{8pt}\textbf{$E$}&$E_\text{base}$&$\sum_j E_j$&\textbf{$\{E_j\}$}\\
\hline
$k=1$&$(0,0,0)$&0&0&0&$0$&0&0&$\{0,0,0,0,0,0,0,\ldots\}$\rule{0pt}{8pt}\\[2pt]
\hline
$k=2$\rule{0pt}{9pt}&$(1,0,0)$&2&0&1&$1$&1&0&$\{0,0,0,0,0,0,0,\ldots\}$\\[2pt]
\hline
$k=3$&$(1,0,0)$&0&1&2&$2$&2&0&$\{0,0,0,0,0,0,0,\ldots\}$\\[2pt]
\hline
$k=4$&$(0,1,0)$&2&0&1&$3$&1&2&$\{2,0,0,0,0,0,0,\ldots\}$\\[2pt]
$k=5$&$(1,0,0)$&0&1&2&$3$&2&1&$\{1,0,0,0,0,0,0,\ldots\}$\\[2pt]
\hline
$k=6$&$(0,1,0)$&0&1&2&$4$&2&2&$\{2,0,0,0,0,0,0,\ldots\}$\\[2pt]
$k=7$&$(0,0,1)$&2&0&1&$4$&1&3&$\{1,2,0,0,0,0,0,\ldots\}$\\[2pt]
$k=8$&$(1,0,0)$&0&1&2&$4$&2&2&$\{1,1,0,0,0,0,0,\ldots\}$\\[2pt]
\hline
$k=9$&$(0,1,0)$&0&1&2&$5$&2&3&$\{3,0,0,0,0,0,0,\ldots\}$\\[2pt]
$k=10,11$&$(2,0,0)$&2&1&3&$5,5$&5&0&$\{0,0,0,0,0,0,0,\ldots\}$\\[2pt]
$k=12$&$(1,0,0)$&0&1&2&$5$&2&3&$\{1,1,1,0,0,0,0,\ldots\}$\\[2pt]
\hline
$k=13$&$(0,1,0)$&0&1&2&$6$&2&4&$\{4,0,0,0,0,0,0,\ldots\}$\\[2pt]
$k=14,15$&$(0,0,1)$&0&1&2&$6,6$&2&4&$\{2,2,0,0,0,0,0,\ldots\}$\\[2pt]
$k=16,17$&$(2,0,0)$&2&1&3&$6,6$&5&1&$\{1,0,0,0,0,0,0,\ldots\}$\\[2pt]
$k=18$&$(1,0,0)$&0&1&2&$6$&2&4&$\{1,1,1,1,0,0,0,\ldots\}$\\[2pt]
\hline
$k=19$&$(0,1,0)$&0&1&2&$7$&2&5&$\{5,0,0,0,0,0,0,\ldots\}$\\[2pt]
$k=20,21$&$(0,1,1)$&2&0&1&$7,7$&1&6&$\{1,2,3,0,0,0,0,\ldots\}$\\[2pt]
$k=22,23$&$(2,0,0)$&0&2&4&$7,7$&7&0&$\{0,0,0,0,0,0,0,\ldots\}$\\[2pt]
$k=24,25$&$(2,0,0)$&2&1&3&$7,7$&5&2&$\{1,1,0,0,0,0,0,\ldots\}$\\[2pt]
$k=26$&$(1,0,0)$&0&1&2&$7$&2&5&$\{1,1,1,1,1,0,0,\ldots\}$\\[2pt]
\hline
$k=27$&$(0,1,0)$&0&1&2&$8$&2&6&$\{6,0,0,0,0,0,0,\ldots\}$\\[2pt]
$k=28$&$(0,2,0)$&0&1&2&$8$&2&6&$\{2,4,0,0,0,0,0,\ldots\}$\\[2pt]
$k=29,30$&$(1,1,0)$&2&1&3&$8,8$&5&3&$\{3,0,0,0,0,0,0,\ldots\}$\\[2pt]
$k=31,32$&$(0,0,1)$&0&1&2&$8,8$&2&6&$\{3,3,0,0,0,0,0,\ldots\}$\\[2pt]
$k=33,34$&$(2,0,0)$&0&2&4&$8,8$&7&1&$\{1,0,0,0,0,0,0,\ldots\}$\\[2pt]
$k=35,36$&$(2,0,0)$&2&1&3&$8,8$&5&3&$\{1,1,1,0,0,0,0,\ldots\}$\\[2pt]
$k=42$&$(1,0,0)$&0&1&2&$8$&2&6&$\{1,1,1,1,1,1,0,\ldots\}$\\[2pt]
\hline
$k=37$&$(0,1,0)$&0&1&2&$9$&2&7&$\{7,0,0,0,0,0,0,\ldots\}$\\[2pt]
$k=38,39$&$(1,1,0)$&0&2&4&$9,9$&7&2&$\{2,0,0,0,0,0,0,\ldots\}$\\[2pt]
$k=40,41$&$(0,1,1)$&0&1&2&$9,9$&2&7&$\{2,2,3,0,0,0,0,\ldots\}$\\[2pt]
$k=43$&$(1,0,2)$&2&0&1&$9$&1&8&$\{1,2,2,3,0,0,0,\ldots\}$\\[2pt]
$k=44,45$&$(1,1,0)$&2&1&3&$9,9$&5&4&$\{2,2,0,0,0,0,0,\ldots\}$\\[2pt]
$k=46,47$&$(2,0,0)$&0&2&4&$9,9$&7&2&$\{2,0,0,0,0,0,0,\ldots\}$\\[2pt]
$k=48,49$&$(2,0,0)$&0&2&4&$9,9$&7&2&$\{1,1,0,0,0,0,0,\ldots\}$\\[2pt]
$k=53,54$&$(2,0,0)$&2&1&3&$9,9$&5&4&$\{1,1,1,1,0,0,0,\ldots\}$\\[2pt]
$k=87$&$(1,0,0)$&0&1&2&$9$&2&7&$\{1,1,1,1,1,1,1,\ldots\}$\\[2pt]
\hline
\end{tabular}
\caption{\small String content $(m_h,m_1,m_2)$, conformal eigenenergies and quantum integers for the leading 52 eigenvalues $\widehat D(u)$ in the standard module $\repV_{N,1}$. The conformal energies are given by $E=E_\text{base}+\sum_j E_j$ where the generalized pentagonal numbers are $E_\text{base}=\Delta_{1,3\lfloor\frac{m+2}{2}\rfloor-(-1)^m}=\tfrac{1}{2}\lceil \tfrac{m}{2}\rceil(3\lceil \tfrac{m}{2}\rceil+(-1)^m)=0,1,2,5,7,12,\ldots$ with $m=0,1,2,3,4,5,\ldots$. 
The quantum number $m$ is given by $m=\tfrac{1}{2} m_{\infty}+2m_3$ where $m_{\infty}=0,2$ is the number of zeros at infinity. The associated conformal character is $\chit_{1,2}(q)=1+q+q^2+2q^3+3q^4+4q^5+6q^6+8q^7+11q^8+15q^9+O(q^{10})$.
For a finite system, $N-d+1-m_\infty-3m_3=n_2+m_1+2m_2=M$ is the total number of zeros of $Q(u)$ in the fundamental domain that enter the Bethe ansatz equations.
}\label{d=1Eigs}
\end{center}
\end{table}

\newpage
\begin{table}[H]
\begin{center}
\scriptsize
\begin{tabular}{|c||c|c||c||c|c|c|}
\multicolumn{7}{c}{\bf \normalsize Standard module $\boldsymbol{\repV_{N,d=2}}$}\\[4pt]
\hline
\hline
{Label}&\textbf{$(m_h,\!m_1,\!m_2)$}&$m_3$&\textbf{$E$}&$E_\text{base}$&$\sum_j E_j$&\textbf{$\{E_j\}$}\\
\hline
$k=1$&$(0,0,0)$&0&$0$&$0$&$0$&$\{0,0,0,0,0,0,0,\ldots\}$\rule{0pt}{8pt}\\[2pt]
\hline
$k=2$&$(0,1,0)$&0&$1$&$0$&$1$&$\{1,0,0,0,0,0,0,\ldots\}$\rule{0pt}{8pt}\\[2pt]
\hline
$k=3,4$&$(0,0,1)$&0&$2,2$&$0$&$2$&$\{1,1,0,0,0,0,0,\ldots\}$\rule{0pt}{8pt}\\[2pt]
\hline
$k=5,6$&$(1,0,0)$&1&$3,3$&$3$&$0$&$\{0,0,0,0,0,0,0,\ldots\}$\rule{0pt}{8pt}\\[2pt]
\hline
$k=7,8$&$(0,1,1)$&0&$4,4$&$0$&$4$&$\{1,1,2,0,0,0,0,\ldots\}$\rule{0pt}{8pt}\\[2pt]
$k=9,10$&$(1,0,0)$&1&$4,4$&$3$&$1$&$\{1,0,0,0,0,0,0,\ldots\}$\\[2pt]
\hline
$k=11$&$(0,2,0)$&0&$5$&$0$&$5$&$\{1,1,1,2,0,0,0,\ldots\}$\rule{0pt}{8pt}\\[2pt]
$k=12,13$&$(1,1,0)$&1&$5,5$&$3$&$2$&$\{2,0,0,0,0,0,0,\ldots\}$\\[2pt]
$k=14,15$&$(1,0,0)$&1&$5,5$&$3$&$2$&$\{1,1,0,0,0,0,0,\ldots\}$\\[2pt]
\hline
$k=16,17$&$(0,1,0)$&1&$6,6$&$3$&$3$&$\{3,0,0,0,0,0,0,\ldots\}$\rule{0pt}{8pt}\\[2pt]
$k=18,19$&$(0,0,2)$&0&$6,6$&$0$&$6$&$\{1,1,2,2,0,0,0,\ldots\}$\\[2pt]
$k=20,21$&$(1,1,0)$&1&$6,6$&$3$&$3$&$\{1,2,0,0,0,0,0,\ldots\}$\\[2pt]
$k=22,23$&$(1,0,0)$&1&$6,6$&$3$&$3$&$\{1,1,1,0,0,0,0,\ldots\}$\\[2pt]
\hline
$k=24,25$&$(0,1,0)$&1&$7,7$&$3$&$4$&$\{4,0,0,0,0,0,0,\ldots\}$\rule{0pt}{8pt}\\[2pt]
$k=26$--$29$&$(1,0,1)$&1&$7,7,7,7$&$3$&$4$&$\{2,2,0,0,0,0,0,\ldots\}$\\[2pt]
$k=30,31$&$(1,1,0)$&1&$7,7$&$3$&$4$&$\{1,1,2,0,0,0,0,\ldots\}$\\[2pt]
$k=32,33$&$(1,0,0)$&1&$7,7$&$3$&$4$&$\{1,1,1,1,0,0,0,\ldots\}$\\[2pt]
\hline
$k=34,35$&$(0,1,0)$&1&$8,8$&$3$&$5$&$\{5,0,0,0,0,0,0,\ldots\}$\rule{0pt}{8pt}\\[2pt]
$k=36$&$(0,3,0)$&0&$8$&$0$&$8$&$\{1,1,2,4,0,0,0,\ldots\}$\\[2pt]
$k=37$--$40$&$(1,0,1)$&1&$8,8,8,8$&$3$&$5$&$\{2,3,0,0,0,0,0,\ldots\}$\\[2pt]
$k=41$--$44$&$(1,0,1)$&1&$8,8,8,8$&$3$&$5$&$\{1,2,2,0,0,0,0,\ldots\}$\\[2pt]
$k=45,46$&$(1,0,1)$&1&$8,8$&$3$&$5$&$\{1,1,1,2,0,0,0,\ldots\}$\\[2pt]
$k=47,48$&$(1,0,0)$&1&$8,8$&$3$&$5$&$\{1,1,1,1,1,0,0,\ldots\}$\\[2pt]
\hline
\end{tabular}
\caption{\small String content $(m_h,m_1,m_2)$, conformal eigenenergies and quantum integers for the leading 48 eigenvalues $\widehat D(u)$ in the standard module $\repV_{N,2}$. The conformal energies are given by $E=E_\text{base}+\sum_j E_j$ where the triangular matchstick numbers are $E_\text{base}=\Delta_{1,3(m+1)-\frac{1}{3}}=\tfrac{3}{2}m_3(m_3+1)=0,3,9,18,30,\ldots$ with $m_3=0,1,2,3,4,\ldots$. 
The associated conformal character is $\chit_{1,3}(q)=q^{1/3} \sum_E q^E=q^{1/3}(1+q+2q^2+2q^3+4q^4+5q^5+8q^6+10q^7+15q^8+O(q^{9}))$. For a finite system, $N-d+1-3m_3=n_2+m_1+2m_2=M$ is the total number of zeros of $Q(u)$ in the fundamental domain that enter the Bethe ansatz equations.
}\label{d=2Eigs}
\end{center}
\end{table}

\newpage
\begin{table}[H]
\begin{center}
\scriptsize
\mbox{}\vspace{-6pt}
\begin{tabular}{|c||c|c|c|c||c||c|c|c|}
\multicolumn{9}{c}{\bf\normalsize Standard module $\boldsymbol{\repV_{N,d=3}}$}\\[4pt]
\hline
\hline
{Label}&\textbf{$(m_h,\!m_1,\!m_2)$}&$\!\!m_\infty\!\!$&$\!m_3\!$&$m$&\textbf{$E$}&$E_\text{base}$&$\sum_j E_j$&\textbf{$\{E_j\}$}\\
\hline
$k=1$&$(1,0,0)$&0&0&0&$0$&0&0&$\{0,0,0,0,0,0,0,0,\ldots\}$\rule{0pt}{8pt}\\[2pt]
\hline
$k=2$&$(1,0,0)$&1&0&1&$1$&1&0&$\{0,0,0,0,0,0,0,0,\ldots\}$\rule{0pt}{8pt}\\[2pt]
\hline
$k=3$&$(0,1,0)$&0&0&0&$2$&0&2&$\{2,0,0,0,0,0,0,0,\ldots\}$\rule{0pt}{8pt}\\[2pt]
$k=4$&$(1,0,0)$&1&0&1&$2$&1&1&$\{1,0,0,0,0,0,0,0,\ldots\}$\\[2pt]
\hline
$k=5$&$(0,1,0)$&1&0&1&$3$&1&2&$\{2,0,0,0,0,0,0,0,\ldots\}$\rule{0pt}{8pt}\\[2pt]
$k=6$&$(1,0,1)$&0&0&0&$3$&0&3&$\{1,2,0,0,0,0,0,0,\ldots\}$\\[2pt]
$k=7$&$(1,0,0)$&1&0&1&$3$&1&2&$\{1,1,0,0,0,0,0,0,\ldots\}$\\[2pt]
\hline
$k=8$&$(0,1,0)$&1&0&1&$4$&1&3&$\{3,0,0,0,0,0,0,0,\ldots\}$\rule{0pt}{8pt}\\[2pt]
$k=9,10$&$(2,0,0)$&0&1&2&$4,4$&4&0&$\{0,0,0,0,0,0,0,0,\ldots\}$\\[2pt]
$k=11$&$(1,0,0)$&1&0&1&$4$&1&3&$\{1,1,1,0,0,0,0,0,\ldots\}$\\[2pt]
\hline
$k=12$&$(0,1,0)$&1&0&1&$5$&1&4&$\{4,0,0,0,0,0,0,0,\ldots\}$\rule{0pt}{8pt}\\[2pt]
$k=13,14$&$(1,0,1)$&1&0&1&$5,5$&1&4&$\{2,2,0,0,0,0,0,0,\ldots\}$\\[2pt]
$k=15,16$&$(2,0,0)$&0&1&2&$5,5$&4&1&$\{1,0,0,0,0,0,0,0,\ldots\}$\\[2pt]
$k=17$&$(1,0,0)$&1&0&1&$5$&1&4&$\{1,1,1,1,0,0,0,0,\ldots\}$\\[2pt]
\hline
$k=18$&$(0,1,0)$&1&0&1&$6$&1&5&$\{5,0,0,0,0,0,0,0,\ldots\}$\rule{0pt}{8pt}\\[2pt]
$k=19,20$&$(0,1,1)$&0&0&0&$6,6$&0&6&$\{1,2,3,0,0,0,0,0,\ldots\}$\\[2pt]
$k=21$--$23$&$(2,0,0)$&1&1&3&$6,6,6$&6&0&$\{0,0,0,0,0,0,0,0,\ldots\}$\\[2pt]
$k=24,25$&$(2,0,0)$&0&1&2&$6,6$&4&2&$\{1,1,0,0,0,0,0,0,\ldots\}$\\[2pt]
$k=26$&$(1,0,0)$&1&0&1&$6$&1&5&$\{1,1,1,1,1,0,0,0,\ldots\}$\\[2pt]
\hline
$k=27$&$(0,1,0)$&1&0&1&$7$&1&6&$\{6,0,0,0,0,0,0,0,\ldots\}$\rule{0pt}{8pt}\\[2pt]
$k=28$&$(0,2,0)$&1&0&1&$7$&1&6&$\{2,4,0,0,0,0,0,0,\ldots\}$\\[2pt]
$k=29,30$&$(0,1,0)$&0&1&2&$7,7$&4&3&$\{3,0,0,0,0,0,0,0,\ldots\}$\\[2pt]
$k=31,32$&$(0,1,1)$&1&0&1&$7,7$&1&6&$\{1,2,3,0,0,0,0,0,\ldots\}$\\[2pt]
$k=33$--$35$&$(2,0,0)$&0&1&2&$7,7,7$&4&3&$\{1,2,0,0,0,0,0,0,\ldots\}$\\[2pt]
$k=36,37$&$(2,0,0)$&0&1&2&$7,7$&4&3&$\{1,1,1,0,0,0,0,0,\ldots\}$\\[2pt]
$k=38$&$(1,0,0)$&1&0&1&$7$&1&6&$\{1,1,1,1,1,1,0,0,\ldots\}$\\[2pt]
\hline
$k=39$&$(0,1,0)$&1&0&1&$8$&1&7&$\{7,0,0,0,0,0,0,0,\ldots\}$\rule{0pt}{8pt}\\[2pt]
$k=40,41$&$(1,1,0)$&0&1&2&$8,8$&4&4&$\{4,0,0,0,0,0,0,0,\ldots\}$\\[2pt]
$k=42,43$&$(0,1,1)$&1&0&1&$8,8$&1&7&$\{2,2,3,0,0,0,0,0,\ldots\}$\\[2pt]
$k=44$&$(1,0,2)$&0&0&0&$8$&0&8&$\{1,2,2,3,0,0,0,0,\ldots\}$\\[2pt]
$k=45,46$&$(1,1,0)$&0&1&2&$8,8$&4&4&$\{1,3,0,0,0,0,0,0,\ldots\}$\\[2pt]
$k=47$--$49$&$(2,0,0)$&1&1&3&$8,8,8$&6&2&$\{2,0,0,0,0,0,0,0,\ldots\}$\\[2pt]
$k=50$--$52$&$(2,0,0)$&1&1&3&$8,8,8$&6&2&$\{1,1,0,0,0,0,0,0,\ldots\}$\\[2pt]
$k=53,54$&$(2,0,0)$&0&1&2&$8,8$&4&4&$\{1,1,1,1,0,0,0,0,\ldots\}$\\[2pt]
\hline
$k=55$&$(0,1,0)$&1&0&1&$9$&1&8&$\{8,0,0,0,0,0,0,0,\ldots\}$\rule{0pt}{8pt}\\[2pt]
$k=56,57$&$(1,1,0)$&0&1&2&$9,9$&4&5&$\{5,0,0,0,0,0,0,0,\ldots\}$\\[2pt]
$k=58,59$&$(0,1,1)$&1&0&1&$9,9$&1&8&$\{2,2,4,0,0,0,0,0,\ldots\}$\\[2pt]
$k=60$--$62$&$(1,1,0)$&1&1&3&$9,9,9$&6&3&$\{3,0,0,0,0,0,0,0,\ldots\}$\\[2pt]
$k=63,64$&$(2,0,1)$&0&1&2&$9,9$&4&5&$\{2,3,0,0,0,0,0,0,\ldots\}$\\[2pt]
$k=65,66$&$(0,0,1)$&1&0&1&$9,9$&1&8&$\{1,1,3,3,0,0,0,0,\ldots\}$\\[2pt]
$k=67,68$&$(1,1,0)$&0&1&2&$9,9$&4&5&$\{1,1,3,0,0,0,0,0,\ldots\}$\\[2pt]
$k=69$--$71$&$(2,0,0)$&1&1&3&$9,9,9$&6&3&$\{1,2,0,0,0,0,0,0,\ldots\}$\\[2pt]
$k=72$&$(1,0,0)$&1&0&1&$8$&1&7&$\{1,1,1,1,1,1,1,0,\ldots\}$\\[2pt]
$k=73$--$75$&$(2,0,0)$&1&1&3&$9,9,9$&6&3&$\{1,1,1,0,0,0,0,0,\ldots\}$\\[2pt]
$k=93,94$&$(2,0,0)$&0&1&2&$9,9$&4&5&$\{1,1,1,1,1,0,0,0,\ldots\}$\\[2pt]
$k=138$&$(1,0,0)$&1&0&1&$9$&1&8&$\{1,1,1,1,1,1,1,1,\ldots\}$\\[2pt]
\hline
\end{tabular}
\caption{\small String content $(m_h,m_1,m_2)$, conformal eigenenergies and quantum integers for the leading 78 eigenvalues $\widehat D(u)$ in the standard module $\repV_{N,3}$. The conformal energies are given by $E=E_\text{base}+\sum_j E_j$ where the shifted generalized pentagonal numbers are $E_\text{base}=\tfrac{1}{2}\lceil \tfrac{m+1}{2}\rceil(3\lceil \tfrac{m+1}{2}\rceil-(-1)^m)-1=0,1,4,6,11,14,\ldots$ with $m=0,1,2,3,4,5,\ldots$.  
The quantum number $m$ is given by $m=m_{\infty}+2m_3$ where $m_{\infty}=0,1$ is the number of zeros at infinity. The associated conformal character is $\chit_{1,4}(q)=q(1+q+2q^2+3q^3+4q^4+6q^5+9q^6+12q^7+17q^8+23q^9+O(q^{10}))$.
For a finite system, $N-d+1-m_\infty-3m_3=n_2+m_1+2m_2=M$ is the total number of zeros of $Q(u)$ in the fundamental domain that enter the Bethe ansatz equations.
}\label{d=3Eigs}
\end{center}
\end{table}

\newpage
\begin{table}[H]
\begin{center}
\scriptsize
\begin{tabular}{|c||c|c|c|c||c||c|c|c|}
\multicolumn{9}{c}{\bf\normalsize Standard module $\boldsymbol{\repV_{N,d=4}}$}\\[4pt]
\hline
\hline
{Label}&\textbf{$(m_h,\!m_1,\!m_2)$}&$\!\!m_\infty\!\!$&$\!m_3\!$&$m$&\textbf{$E$}&$E_\text{base}$&$\sum_j E_j$&\textbf{$\{E_j\}$}\\
\hline
$k=1$&$(1,0,0)$&0&0&0&$0$&0&0&$\{0,0,0,0,0,0,0,0,\ldots\}$\rule{0pt}{8pt}\\[2pt]
\hline
$k=2$&$(1,0,0)$&0&0&0&$1$&0&1&$\{1,0,0,0,0,0,0,0,\ldots\}$\rule{0pt}{8pt}\\[2pt]
\hline
$k=3$&$(0,1,0)$&0&0&0&$2$&0&2&$\{2,0,0,0,0,0,0,0,\ldots\}$\rule{0pt}{8pt}\\[2pt]
$k=4$&$(1,0,0)$&0&0&0&$2$&0&2&$\{1,1,0,0,0,0,0,0,\ldots\}$\\[2pt]
\hline
$k=5$&$(0,1,0)$&0&0&0&$3$&0&3&$\{3,0,0,0,0,0,0,0,\ldots\}$\rule{0pt}{8pt}\\[2pt]
$k=6$&$(2,0,0)$&2&0&1&$3$&3&0&$\{0,0,0,0,0,0,0,0,\ldots\}$\\[2pt]
$k=7$&$(1,0,0)$&0&0&0&$3$&0&3&$\{1,1,1,0,0,0,0,0,\ldots\}$\\[2pt]
\hline
$k=8$&$(0,1,0)$&0&0&0&$4$&0&4&$\{4,0,0,0,0,0,0,0,\ldots\}$\rule{0pt}{8pt}\\[2pt]
$k=9,10$&$(0,0,1)$&0&0&0&$4,4$&0&2&$\{2,2,0,0,0,0,0,0,\ldots\}$\\[2pt]
$k=11$&$(2,0,0)$&2&0&1&$4$&3&1&$\{1,0,0,0,0,0,0,0,\ldots\}$\\[2pt]
$k=12$&$(1,0,0)$&0&0&0&$4$&0&4&$\{1,1,1,1,0,0,0,0,\ldots\}$\\[2pt]
\hline
$k=13$&$(0,1,0)$&0&0&0&$5$&0&5&$\{5,0,0,0,0,0,0,0,\ldots\}$\rule{0pt}{8pt}\\[2pt]
$k=14$--$16$&$(2,0,0)$&0&1&2&$5,5,5$&5&0&$\{0,0,0,0,0,0,0,0,\ldots\}$\\[2pt]
$k=17$&$(2,0,0)$&2&0&1&$5$&3&2&$\{1,1,0,0,0,0,0,0,\ldots\}$\\[2pt]
$k=18$&$(1,0,0)$&0&0&0&$5$&0&5&$\{1,1,1,1,1,0,0,0,\ldots\}$\\[2pt]
\hline
$k=19$&$(0,1,0)$&0&0&0&$6$&0&6&$\{6,0,0,0,0,0,0,0,\ldots\}$\rule{0pt}{8pt}\\[2pt]
$k=20$&$(0,2,0)$&0&0&0&$6$&0&6&$\{2,4,0,0,0,0,0,0,\ldots\}$\\[2pt]
$k=21$&$(1,1,0)$&2&0&1&$6$&3&3&$\{3,0,0,0,0,0,0,0,\ldots\}$\\[2pt]
$k=22,23$&$(0,0,1)$&0&0&0&$6,6$&0&6&$\{1,2,3,0,0,0,0,0,\ldots\}$\\[2pt]
$k=24$--$26$&$(2,0,0)$&0&1&2&$6,6,6$&5&1&$\{1,0,0,0,0,0,0,0,\ldots\}$\\[2pt]
$k=27$&$(2,0,0)$&2&0&1&$6$&3&3&$\{1,1,1,0,0,0,0,0,\ldots\}$\\[2pt]
$k=28$&$(1,0,0)$&0&0&0&$6$&0&6&$\{1,1,1,1,1,1,0,0,\ldots\}$\\[2pt]
\hline
$k=29$&$(0,1,0)$&0&0&0&$7$&0&7&$\{7,0,0,0,0,0,0,0,\ldots\}$\rule{0pt}{8pt}\\[2pt]
$k=30$&$(1,1,0)$&2&0&1&$7$&3&4&$\{4,0,0,0,0,0,0,0,\ldots\}$\\[2pt]
$k=31,32$&$(0,1,1)$&0&0&0&$7,7$&0&7&$\{2,2,3,0,0,0,0,0,\ldots\}$\\[2pt]
$k=33$&$(1,1,0)$&2&0&1&$7$&3&4&$\{2,2,0,0,0,0,0,0,\ldots\}$\\[2pt]
$k=34$--$36$&$(2,0,0)$&0&1&2&$7,7,7$&5&2&$\{2,0,0,0,0,0,0,0,\ldots\}$\\[2pt]
$k=37$--$39$&$(2,0,0)$&0&1&2&$7,7,7$&5&2&$\{1,1,0,0,0,0,0,0,\ldots\}$\\[2pt]
$k=40$&$(2,0,0)$&2&0&1&$7$&3&4&$\{1,1,1,1,0,0,0,0,\ldots\}$\\[2pt]
\hline
$k=41$&$(0,1,0)$&0&0&0&$8$&0&8&$\{8,0,0,0,0,0,0,0,\ldots\}$\rule{0pt}{8pt}\\[2pt]
$k=42$&$(1,1,0)$&2&0&1&$8$&3&5&$\{5,0,0,0,0,0,0,0,\ldots\}$\\[2pt]
$k=43,44$&$(0,1,1)$&0&0&0&$8,8$&0&8&$\{2,2,4,0,0,0,0,0,\ldots\}$\\[2pt]
$k=45$--$47$&$(1,1,0)$&0&1&2&$8,8,8$&5&3&$\{3,0,0,0,0,0,0,0,\ldots\}$\\[2pt]
$k=48$&$(2,0,1)$&2&0&1&$8$&3&5&$\{2,3,0,0,0,0,0,0,\ldots\}$\\[2pt]
$k=49,50$&$(0,0,1)$&0&0&0&$8,8$&0&8&$\{1,1,3,3,0,0,0,0,\ldots\}$\\[2pt]
$k=51$&$(1,1,0)$&2&0&1&$8$&3&5&$\{1,2,2,0,0,0,0,0,\ldots\}$\\[2pt]
$k=52$--$54$&$(2,0,0)$&0&1&2&$8,8,8$&5&3&$\{1,2,0,0,0,0,0,0,\ldots\}$\\[2pt]
$k=55$&$(1,0,0)$&0&0&0&$7$&0&7&$\{1,1,1,1,1,1,1,0,\ldots\}$\\[2pt]
$k=56$--$58$&$(2,0,0)$&0&1&2&$8,8,8$&5&3&$\{1,1,1,0,0,0,0,0,\ldots\}$\\[2pt]
$k=75$&$(2,0,0)$&2&0&1&$8$&3&5&$\{1,1,1,1,1,0,0,0,\ldots\}$\\[2pt]
$k=109$&$(1,0,0)$&0&0&0&$8$&0&8&$\{1,1,1,1,1,1,1,1,\ldots\}$\\[2pt]
\hline
\end{tabular}
\caption{\small String content $(m_h,m_1,m_2)$, conformal eigenenergies and quantum integers for the leading 60 eigenvalues $\widehat D(u)$ in the standard module $\repV_{N,4}$. The conformal energies are given by $E=E_\text{base}+\sum_j E_j$ where the shifted generalized pentagonal numbers are $E_\text{base}=\tfrac{1}{2}\lceil \tfrac{m+2}{2}\rceil(3\lceil \tfrac{m+2}{2}\rceil+(-1)^m)-2=0,3,5,10,13,20,\ldots$ with $m=0,1,2,3,4,5,\ldots$. 
The quantum number $m$ is given by $m=\tfrac{1}{2} m_{\infty}+2m_3$ where $m_{\infty}=0,2$ is the number of zeros at infinity. The associated conformal character is $\chit_{1,5}(q)=q^2(1+q+2q^2+3q^3+5q^4+6q^5+10q^6+13q^7+19q^8+O(q^9))$.
For a finite system, $N-d+1-m_\infty-3m_3=n_2+m_1+2m_2=M$ is the total number of zeros of $Q(u)$ in the fundamental domain that enter the Bethe ansatz equations.
}\label{d=4Eigs}
\end{center}
\end{table}

\newpage
\begin{table}[H]
\begin{center}
\scriptsize
\begin{tabular}{|c||c|c||c||c|c|c|}
\multicolumn{7}{c}{\bf \normalsize Standard module $\boldsymbol{\repV_{N,d=5}}$}\\[4pt]
\hline
\hline
{Label}&\textbf{$(m_h,\!m_1,\!m_2)$}&$m_3$&\textbf{$E$}&$E_\text{base}$&$\sum_j E_j$&\textbf{$\{E_j\}$}\\
\hline
$k=1$&$(1,0,0)$&0&$0$&$0$&$0$&$\{0,0,0,0,0,0,0,0,\ldots\}$\rule{0pt}{8pt}\\[2pt]
\hline
$k=2$&$(1,0,0)$&0&$1$&$0$&$1$&$\{1,0,0,0,0,0,0,0,\ldots\}$\rule{0pt}{8pt}\\[2pt]
\hline
$k=3$&$(1,1,0)$&0&$2$&$0$&$2$&$\{2,0,0,0,0,0,0,0,\ldots\}$\rule{0pt}{8pt}\\[2pt]
$k=4$&$(1,0,0)$&0&$2$&$0$&$3$&$\{1,1,0,0,0,0,0,0,\ldots\}$\\[2pt]
\hline
$k=5$&$(0,1,0)$&0&$3$&$0$&$3$&$\{3,0,0,0,0,0,0,0,\ldots\}$\rule{0pt}{8pt}\\[2pt]
$k=6$&$(1,1,0)$&0&$3$&$0$&$3$&$\{3,0,0,0,0,0,0,0,\ldots\}$\\[2pt]
$k=7$&$(1,0,0)$&0&$3$&$0$&$3$&$\{1,1,1,0,0,0,0,0,\ldots\}$\\[2pt]
\hline
$k=8$&$(0,1,0)$&0&$4$&$0$&$4$&$\{4,0,0,0,0,0,0,0,\ldots\}$\rule{0pt}{8pt}\\[2pt]
$k=9,10$&$(0,0,1)$&0&$4,4$&$0$&$4$&$\{2,2,0,0,0,0,0,0,\ldots\}$\\[2pt]
$k=11$&$(0,1,0)$&0&$4$&$0$&$4$&$\{1,1,2,0,0,0,0,0,\ldots\}$\\[2pt]
$k=12$&$(1,0,0)$&0&$4$&$0$&$4$&$\{1,1,1,1,0,0,0,0,\ldots\}$\\[2pt]
\hline
$k=13$&$(0,1,0)$&0&$5$&$0$&$5$&$\{5,0,0,0,0,0,0,0,\ldots\}$\rule{0pt}{8pt}\\[2pt]
$k=14,15$&$(0,0,1)$&0&$5,5$&$0$&$5$&$\{2,3,0,0,0,0,0,0,\ldots\}$\\[2pt]
$k=16,17$&$(1,0,1)$&0&$5,5$&$0$&$5$&$\{1,2,2,0,0,0,0,0,\ldots\}$\\[2pt]
$k=18$&$(0,1,0)$&0&$5$&$0$&$5$&$\{1,1,1,2,0,0,0,0,\ldots\}$\\[2pt]
$k=19$&$(1,0,0)$&0&$5$&$0$&$5$&$\{1,1,1,1,1,0,0,0,\ldots\}$\\[2pt]
\hline
$k=20$&$(0,1,0)$&0&$6$&$0$&$6$&$\{6,0,0,0,0,0,0,0,\ldots\}$\rule{0pt}{8pt}\\[2pt]
$k=21$&$(0,2,0)$&0&$6$&$0$&$6$&$\{2,4,0,0,0,0,0,0,\ldots\}$\\[2pt]
$k=22$--$25$&$(2,0,0)$&1&$6,6,6,6$&$6$&$0$&$\{0,0,0,0,0,0,0,0,\ldots\}$\\[2pt]
$k=26,27$&$(1,0,1)$&0&$6,6$&$0$&$6$&$\{1,1,2,2,0,0,0,0,\ldots\}$\\[2pt]
$k=28$&$(1,1,0)$&0&$6$&$0$&$6$&$\{1,1,1,1,2,0,0,0,\ldots\}$\\[2pt]
$k=29$&$(1,0,0)$&0&$6$&$0$&$6$&$\{1,1,1,1,1,1,0,0,\ldots\}$\\[2pt]
\hline
$k=30$&$(0,1,0)$&0&$7$&$0$&$7$&$\{7,0,0,0,0,0,0,0,\ldots\}$\rule{0pt}{8pt}\\[2pt]
$k=31$&$(0,2,0)$&0&$7$&$0$&$7$&$\{2,5,0,0,0,0,0,0,\ldots\}$\\[2pt]
$k=32,33$&$(1,0,1)$&0&$7,7$&$0$&$7$&$\{2,2,3,0,0,0,0,0,\ldots\}$\\[2pt]
$k=34,35$&$(0,0,1)$&0&$7,7$&$0$&$7$&$\{1,3,3,0,0,0,0,0,\ldots\}$\\[2pt]
$k=36$--$39$&$(2,0,0)$&1&$7,7,7,7$&$6$&$1$&$\{1,0,0,0,0,0,0,0,\ldots\}$\\[2pt]
$k=40,41$&$(1,0,1)$&0&$7,7$&$0$&$7$&$\{1,1,1,2,2,0,0,0,\ldots\}$\\[2pt]
$k=42$&$(1,1,0)$&0&$7$&$0$&$7$&$\{1,1,1,1,1,2,0,0,\ldots\}$\\[2pt]
$k=59$&$(1,0,0)$&0&$7$&$0$&$7$&$\{1,1,1,1,1,1,1,0,\ldots\}$\\[2pt]
\hline
$k=43$&$(0,1,0)$&0&$8$&$0$&$8$&$\{8,0,0,0,0,0,0,0,\ldots\}$\rule{0pt}{8pt}\\[2pt]
$k=44$&$(0,2,0)$&0&$8$&$0$&$8$&$\{2,6,0,0,0,0,0,0,\ldots\}$\\[2pt]
$k=45,46$&$(0,1,1)$&0&$8,8$&$0$&$8$&$\{2,2,4,0,0,0,0,0,\ldots\}$\\[2pt]
$k=47,48$&$(0,1,1)$&0&$8,8$&$0$&$8$&$\{2,3,3,0,0,0,0,0,\ldots\}$\\[2pt]
$k=49,50$&$(0,1,1)$&0&$8,8$&$0$&$8$&$\{2,2,2,2,0,0,0,0,\ldots\}$\\[2pt]
$k=51$--$54$&$(2,0,0)$&1&$8,8,8,8$&$6$&$2$&$\{2,0,0,0,0,0,0,0,\ldots\}$\\[2pt]
$k=55$--$58$&$(2,0,0)$&1&$8,8,8,8$&$6$&$2$&$\{1,1,0,0,0,0,0,0,\ldots\}$\\[2pt]
$k=61,62$&$(1,0,1)$&0&$8,8$&$0$&$8$&$\{1,1,1,1,2,2,0,0,\ldots\}$\\[2pt]
$k=82$&$(1,1,0)$&0&$8$&$0$&$8$&$\{1,1,1,1,1,1,2,0,\ldots\}$\\[2pt]
$k=107$&$(1,0,0)$&0&$8$&$0$&$8$&$\{1,1,1,1,1,1,1,1,\ldots\}$\\[2pt]
\hline
\end{tabular}
\caption{\small String content $(m_h,m_1,m_2)$, conformal eigenenergies and quantum integers for the leading 63 eigenvalues $\widehat D(u)$ in the standard module $\repV_{N,5}$. The conformal energies are given by $E=E_\text{base}+\sum_j E_j$ where the shifted triangular matchstick numbers are 
$E_\text{base}=\tfrac{3}{2}(m_3\!+\!1)(m_3\!+\!2)-3=0,6,15,27,42,\ldots$ with $m_3=0,1,2,3,4,\ldots$. 
The associated conformal character is $\chit_{1,6}(q)=q^{10/3} \sum_E q^E=q^{10/3}(1+q+2q^2+3q^3+5q^4+7q^5+10q^6+14q^7+20q^8+O(q^{9}))$. 
For a finite system, $N-d+1-3m_3=n_2+m_1+2m_2=M$ is the total number of zeros of $Q(u)$ in the fundamental domain that enter the Bethe ansatz equations.
}\label{d=5Eigs}
\end{center}
\end{table}

\newpage
\begin{table}[H]
\begin{center}
\scriptsize
\begin{tabular}{|c||c|c|c|c||c||c|c|c|}
\multicolumn{9}{c}{\bf\normalsize Standard module $\boldsymbol{\repV_{N,d=6}}$}\\[4pt]
\hline
\hline
{Label}&\textbf{$(m_h,\!m_1,\!m_2)$}&$\!\!m_\infty\!\!$&$\!m_3\!$&$m$&\textbf{$E$}&$E_\text{base}$&$\sum_j E_j$&\textbf{$\{E_j\}$}\\
\hline
$k=1$&$(2,0,0)$&0&0&0&$0$&0&0&$\{0,0,0,0,0,0,0,0,\ldots\}$\rule{0pt}{8pt}\\[2pt]
\hline
$k=2$&$(2,0,0)$&0&0&0&$1$&0&1&$\{1,0,0,0,0,0,0,0,\ldots\}$\rule{0pt}{8pt}\\[2pt]
\hline
$k=3$&$(2,0,0)$&1&0&1&$2$&2&0&$\{0,0,0,0,0,0,0,0,\ldots\}$\rule{0pt}{8pt}\\[2pt]
$k=4$&$(2,0,0)$&0&0&0&$2$&0&2&$\{1,1,0,0,0,0,0,0,\ldots\}$\\[2pt]
\hline
$k=5$&$(1,1,0)$&0&0&0&$3$&0&3&$\{3,0,0,0,0,0,0,0,\ldots\}$\rule{0pt}{8pt}\\[2pt]
$k=6$&$(2,0,0)$&1&0&1&$3$&2&1&$\{1,0,0,0,0,0,0,0,\ldots\}$\\[2pt]
$k=7$&$(2,0,0)$&0&0&0&$3$&0&3&$\{1,1,1,0,0,0,0,0,\ldots\}$\\[2pt]
\hline
$k=8$&$(1,1,0)$&0&0&0&$4$&0&4&$\{4,0,0,0,0,0,0,0,\ldots\}$\rule{0pt}{8pt}\\[2pt]
$k=9$&$(1,1,0)$&0&0&0&$4$&0&2&$\{1,3,0,0,0,0,0,0,\ldots\}$\rule{0pt}{8pt}\\[2pt]
$k=10$&$(2,0,0)$&1&0&1&$4$&2&2&$\{2,0,0,0,0,0,0,0,\ldots\}$\\[2pt]
$k=11$&$(2,0,0)$&1&0&1&$4$&2&2&$\{1,1,0,0,0,0,0,0,\ldots\}$\\[2pt]
$k=12$&$(2,0,0)$&0&0&0&$4$&0&4&$\{1,1,1,1,0,0,0,0,\ldots\}$\\[2pt]
\hline
$k=13$&$(1,1,0)$&0&0&0&$5$&0&5&$\{5,0,0,0,0,0,0,0,\ldots\}$\rule{0pt}{8pt}\\[2pt]
$k=14$&$(1,1,0)$&1&0&1&$5$&2&3&$\{3,0,0,0,0,0,0,0,\ldots\}$\\[2pt]
$k=15$&$(2,0,1)$&0&0&0&$5$&0&5&$\{2,3,0,0,0,0,0,0,\ldots\}$\\[2pt]
$k=16$&$(1,1,0)$&0&0&0&$5$&0&5&$\{1,1,2,0,0,0,0,0,\ldots\}$\\[2pt]
$k=17$&$(2,0,0)$&1&0&1&$5$&2&3&$\{1,2,0,0,0,0,0,0,\ldots\}$\\[2pt]
$k=18$&$(2,0,0)$&1&0&1&$5$&2&3&$\{1,1,1,0,0,0,0,0,\ldots\}$\\[2pt]
$k=19$&$(2,0,0)$&0&0&0&$5$&0&5&$\{1,1,1,1,1,0,0,0,\ldots\}$\\[2pt]
\hline
$k=20$&$(1,1,0)$&0&0&0&$6$&0&6&$\{6,0,0,0,0,0,0,0,\ldots\}$\rule{0pt}{8pt}\\[2pt]
$k=21$&$(1,1,0)$&1&0&1&$6$&2&6&$\{4,0,0,0,0,0,0,0,\ldots\}$\\[2pt]
$k=22,23$&$(1,0,1)$&0&0&0&$6,6$&0&6&$\{3,3,0,0,0,0,0,0,\ldots\}$\\[2pt]
$k=24$&$(1,1,0)$&1&0&1&$6$&2&4&$\{1,3,0,0,0,0,0,0,\ldots\}$\\[2pt]
$k=25$&$(2,0,1)$&0&0&0&$6$&0&6&$\{1,2,3,0,0,0,0,0,\ldots\}$\\[2pt]
$k=26$&$(2,0,0)$&1&0&1&$6$&2&4&$\{2,2,0,0,0,0,0,0,\ldots\}$\\[2pt]
$k=27$&$(1,1,0)$&0&0&0&$6$&0&6&$\{1,1,2,2,0,0,0,0,\ldots\}$\\[2pt]
$k=28$&$(2,0,0)$&1&0&1&$6$&2&4&$\{1,1,2,0,0,0,0,0,\ldots\}$\\[2pt]
$k=29$&$(2,0,0)$&1&0&1&$6$&2&4&$\{1,1,1,1,0,0,0,0,\ldots\}$\\[2pt]
\hline
$k=30$&$(1,1,0)$&0&0&0&$7$&0&7&$\{7,0,0,0,0,0,0,0,\ldots\}$\rule{0pt}{8pt}\\[2pt]
$k=31$&$(1,1,0)$&1&0&1&$7$&2&5&$\{5,0,0,0,0,0,0,0,\ldots\}$\\[2pt]
$k=32$&$(1,1,0)$&1&0&1&$7$&2&5&$\{2,3,0,0,0,0,0,0,\ldots\}$\\[2pt]
$k=33$--$36$&$(3,0,0)$&0&1&2&$7,7,7,7$&7&0&$\{0,0,0,0,0,0,0,0,\ldots\}$\\[2pt]
$k=37$&$(1,1,0)$&1&0&1&$7$&2&5&$\{1,2,2,0,0,0,0,0,\ldots\}$\\[2pt]
$k=38$&$(2,0,1)$&0&0&0&$7$&0&7&$\{1,1,2,3,0,0,0,0,\ldots\}$\\[2pt]
$k=39$&$(2,0,0)$&1&0&1&$7$&2&5&$\{1,2,2,0,0,0,0,0,\ldots\}$\\[2pt]
$k=40$&$(2,0,0)$&0&0&0&$6$&0&6&$\{1,1,1,1,1,1,0,0,\ldots\}$\\[2pt]
$k=41$&$(1,1,0)$&0&0&0&$7$&0&7&$\{1,1,1,1,3,0,0,0,\ldots\}$\\[2pt]
$k=42$&$(2,0,0)$&1&0&1&$7$&2&5&$\{1,1,1,2,0,0,0,0,\ldots\}$\\[2pt]
$k=43$&$(2,0,0)$&1&0&1&$7$&2&5&$\{1,1,1,1,1,0,0,0,\ldots\}$\\[2pt]
$k=44$&$(2,0,0)$&0&0&0&$7$&0&7&$\{1,1,1,1,1,1,1,0,\ldots\}$\\[2pt]
\hline
\end{tabular}
\caption{\small String content $(m_h,m_1,m_2)$, conformal eigenenergies and quantum integers for the leading 44 eigenvalues $\widehat D(u)$ in the standard module $\repV_{N,6}$. The conformal energies are given by $E=E_\text{base}+\sum_j E_j$ where the shifted generalized pentagonal numbers $E_\text{base}=\tfrac{1}{2}\lceil \tfrac{m+3}{2}\rceil(3\lceil \tfrac{m+3}{2}\rceil-(-1)^m)=0,2,7,10,17,21,\ldots$ with $m=0,1,2,3,4,5,\ldots$. 
The quantum number $m$ is given by $m=m_{\infty}+2m_3$ where $m_{\infty}=0,1$ is the number of zeros at infinity. The associated conformal character is $\chit_{1,7}(q)=q^5(1+q+2q^2+3q^3+5q^4+7q^5+11q^6+14q^7+O(q^8))$.
For a finite system, $N-d+1-m_\infty-3m_3=n_2+m_1+2m_2=M$ is the total number of zeros of $Q(u)$ in the fundamental domain that enter the Bethe ansatz equations.
}\label{d=6Eigs}
\end{center}
\end{table}

\newpage
\begin{table}[H]
\begin{center}
\scriptsize
\begin{tabular}{|c||c|c|c|c||c||c|c|c|}
\multicolumn{9}{c}{\bf\normalsize Standard module $\boldsymbol{\repV_{N,d=7}}$}\\[4pt]
\hline
\hline
{Label}&\textbf{$(m_h,m_1,m_2)$}&$\!\!m_\infty\!\!$&$\!m_3\!$&$m$&\textbf{$E$}&$E_\text{base}$&$\sum_j E_j$&\textbf{$\{E_j\}$}\\
\hline
$k=1$&$(2,0,0)$&0&0&0&$0$&0&0&$\{0,0,0,0,0,0,0,0,\ldots\}$\rule{0pt}{8pt}\\[2pt]
\hline
$k=2$&$(2,0,0)$&0&0&0&$1$&0&1&$\{1,0,0,0,0,0,0,0,\ldots\}$\rule{0pt}{8pt}\\[2pt]
\hline
$k=3$&$(2,0,0)$&0&0&0&$2$&0&2&$\{2,0,0,0,0,0,0,0,\ldots\}$\rule{0pt}{8pt}\\[2pt]
$k=4$&$(2,0,0)$&0&0&0&$2$&0&2&$\{1,1,0,0,0,0,0,0,\ldots\}$\\[2pt]
\hline
$k=5$&$(0,1,0)$&0&0&0&$3$&0&3&$\{3,0,0,0,0,0,0,0,\ldots\}$\rule{0pt}{8pt}\\[2pt]
$k=6$&$(2,0,0)$&0&0&0&$3$&0&3&$\{2,1,0,0,0,0,0,0,\ldots\}$\\[2pt]
$k=7$&$(2,0,0)$&0&0&0&$3$&0&3&$\{1,1,1,0,0,0,0,0,\ldots\}$\\[2pt]
\hline
$k=8$&$(0,1,0)$&0&0&0&$4$&0&4&$\{4,0,0,0,0,0,0,0,\ldots\}$\rule{0pt}{8pt}\\[2pt]
$k=9$&$(1,1,0)$&0&0&0&$4$&0&2&$\{2,2,0,0,0,0,0,0,\ldots\}$\\[2pt]
$k=10$&$(2,0,0)$&0&0&0&$4$&0&4&$\{2,2,0,0,0,0,0,0,\ldots\}$\\[2pt]
$k=11$&$(2,0,0)$&0&0&0&$4$&0&4&$\{1,1,2,0,0,0,0,0,\ldots\}$\\[2pt]
$k=12$&$(2,0,0)$&0&0&0&$4$&0&4&$\{1,1,1,1,0,0,0,0,\ldots\}$\\[2pt]
\hline
$k=13$&$(0,1,0)$&0&0&0&$5$&0&5&$\{5,0,0,0,0,0,0,0,\ldots\}$\rule{0pt}{8pt}\\[2pt]
$k=14$&$(0,1,0)$&0&0&0&$5$&0&5&$\{1,4,0,0,0,0,0,0,\ldots\}$\\[2pt]
$k=15$&$(3,0,0)$&2&0&1&$5$&5&0&$\{0,0,0,0,0,0,0,0,\ldots\}$\\[2pt]
$k=16$&$(1,1,0)$&0&0&0&$5$&0&5&$\{1,1,3,0,0,0,0,0,\ldots\}$\\[2pt]
$k=17$&$(2,0,0)$&0&0&0&$5$&0&5&$\{1,2,2,0,0,0,0,0,\ldots\}$\\[2pt]
$k=18$&$(2,0,0)$&0&0&0&$5$&0&5&$\{1,1,1,2,0,0,0,0,\ldots\}$\\[2pt]
$k=19$&$(2,0,0)$&0&0&0&$5$&0&5&$\{1,1,1,1,1,0,0,0,\ldots\}$\\[2pt]
\hline
$k=20$&$(0,1,0)$&0&0&0&$6$&0&6&$\{6,0,0,0,0,0,0,0,\ldots\}$\rule{0pt}{8pt}\\[2pt]
$k=21$&$(0,1,0)$&0&0&0&$6$&0&6&$\{1,5,0,0,0,0,0,0,\ldots\}$\\[2pt]
$k=22,23$&$(0,0,1)$&0&0&0&$6,6$&0&6&$\{3,3,0,0,0,0,0,0,\ldots\}$\\[2pt]
$k=24$&$(1,1,0)$&0&0&0&$6$&0&6&$\{1,1,4,0,0,0,0,0,\ldots\}$\\[2pt]
$k=25$&$(3,0,0)$&2&0&1&$6$&5&1&$\{1,0,0,0,0,0,0,0,\ldots\}$\\[2pt]
$k=26$&$(2,0,0)$&0&0&0&$6$&0&6&$\{2,2,2,0,0,0,0,0,\ldots\}$\\[2pt]
$k=27$&$(1,1,0)$&0&0&0&$6$&0&6&$\{1,1,1,3,0,0,0,0,\ldots\}$\\[2pt]
$k=28$&$(2,0,0)$&0&0&0&$6$&0&6&$\{1,1,2,2,0,0,0,0,\ldots\}$\\[2pt]
$k=29$&$(2,0,0)$&0&0&0&$6$&0&6&$\{1,1,1,1,2,0,0,0,\ldots\}$\\[2pt]
$k=44$&$(2,0,0)$&0&0&0&$6$&0&6&$\{1,1,1,1,1,1,0,0,\ldots\}$\\[2pt]
\hline
$k=30$&$(1,1,0)$&0&0&0&$7$&0&7&$\{7,0,0,0,0,0,0,0,\ldots\}$\rule{0pt}{8pt}\\[2pt]
$k=31$&$(1,1,0)$&0&0&0&$7$&0&7&$\{1,6,0,0,0,0,0,0,\ldots\}$\\[2pt]
$k=32,33$&$(1,0,1)$&0&0&0&$7,7$&0&7&$\{3,4,0,0,0,0,0,0,\ldots\}$\\[2pt]
$k=34$&$(1,1,0)$&0&0&0&$7$&0&7&$\{1,1,5,0,0,0,0,0,\ldots\}$\\[2pt]
$k=35,36$&$(1,0,1)$&0&0&0&$7,7$&0&7&$\{1,3,3,0,0,0,0,0,\ldots\}$\\[2pt]
$k=37$&$(3,0,0)$&2&0&1&$7$&5&2&$\{2,0,0,0,0,0,0,0,\ldots\}$\\[2pt]
$k=38$&$(1,1,0)$&0&0&0&$7$&0&7&$\{1,1,1,4,0,0,0,0,\ldots\}$\\[2pt]
$k=39$&$(3,0,0)$&2&0&1&$7$&5&2&$\{1,1,0,0,0,0,0,0,\ldots\}$\\[2pt]
$k=40$&$(2,0,0)$&0&0&0&$7$&0&7&$\{1,2,2,2,0,0,0,0,\ldots\}$\\[2pt]
$k=42$&$(1,1,0)$&0&0&0&$7$&0&7&$\{1,1,1,1,3,0,0,0,\ldots\}$\\[2pt]
$k=45$&$(2,0,0)$&0&0&0&$7$&0&7&$\{1,1,1,2,2,0,0,0,\ldots\}$\\[2pt]
$k=58$&$(2,0,0)$&0&0&0&$7$&0&7&$\{1,1,1,1,1,2,0,0,\ldots\}$\\[2pt]
$k=73$&$(2,0,0)$&0&0&0&$7$&0&7&$\{1,1,1,1,1,1,1,0,\ldots\}$\\[2pt]
\hline
\end{tabular}
\caption{\small String content $(m_h,m_1,m_2)$, conformal eigenenergies and quantum integers for the leading 45 eigenvalues $\widehat D(u)$ in the standard module $\repV_{N,7}$. The conformal energies are given by $E=E_\text{base}+\sum_j E_j$ where the shifted generalized pentagonal numbers $E_\text{base}=\tfrac{1}{2}\lceil \tfrac{m+4}{2}\rceil(3\lceil \tfrac{m+4}{2}\rceil+(-1)^m)-7=0,5,8,15,19,28\ldots$ with $m=0,1,2,3,4,5,\ldots$. 
The quantum number $m$ is given by $m=\tfrac{1}{2} m_{\infty}+2m_3$ where $m_{\infty}=0,2$ is the number of zeros at infinity. The associated conformal character is $\chit_{1,8}(q)=q^7(1+q+2q^2+3q^3+5q^4+7q^5+11q^6+15q^7+O(q^8))$.
For a finite system, $N-d+1-m_\infty-3m_3=n_2+m_1+2m_2=M$ is the total number of zeros of $Q(u)$ in the fundamental domain that enter the Bethe ansatz equations.
}\label{d=7Eigs}
\end{center}
\end{table}

\newpage
\begin{table}[H]
\begin{center}
\scriptsize
\begin{tabular}{|c||c|c||c||c|c|c|}
\multicolumn{7}{c}{\bf \normalsize Standard module $\boldsymbol{\repV_{N,d=8}}$}\\[4pt]
\hline\hline
{Label}&\textbf{$(m_h,m_1,m_2)$}&$m_3$&\textbf{$E$}&$E_\text{base}$&$\sum_j E_j$&\textbf{$\{E_j\}$}\\
\hline
$k=1$&$(2,0,0)$&0&$0$&$0$&$0$&$\{0,0,0,0,0,0,0,0,\ldots\}$\rule{0pt}{8pt}\\[2pt]
\hline
$k=2$&$(2,0,0)$&0&$1$&$0$&$1$&$\{1,0,0,0,0,0,0,0,\ldots\}$\\[2pt]
\hline
$k=3$&$(2,0,0)$&0&$2$&$0$&$2$&$\{2,0,0,0,0,0,0,0,\ldots\}$\\[2pt]
$k=4$&$(2,0,0)$&0&$2$&$0$&$2$&$\{1,1,0,0,0,0,0,0,\ldots\}$\\[2pt]
\hline
$k=5$&$(2,1,0)$&0&$3$&$0$&$3$&$\{3,0,0,0,0,0,0,0,\ldots\}$\\[2pt]
$k=6$&$(2,0,0)$&0&$3$&$0$&$3$&$\{1,2,0,0,0,0,0,0,\ldots\}$\\[2pt]
$k=7$&$(2,0,0)$&0&$3$&$0$&$3$&$\{1,1,1,0,0,0,0,0,\ldots\}$\\[2pt]
\hline
$k=8$&$(1,1,0)$&0&$4$&$0$&$4$&$\{4,0,0,0,0,0,0,0,\ldots\}$\\[2pt]
$k=9$&$(2,1,0)$&0&$4$&$0$&$3$&$\{1,3,0,0,0,0,0,0,\ldots\}$\\[2pt]
$k=10$&$(2,0,0)$&0&$4$&$0$&$4$&$\{2,2,0,0,0,0,0,0,\ldots\}$\\[2pt]
$k=11$&$(2,0,0)$&0&$4$&$0$&$4$&$\{1,1,2,0,0,0,0,0,\ldots\}$\\[2pt]
$k=12$&$(2,0,0)$&0&$4$&$0$&$4$&$\{1,1,1,1,0,0,0,0,\ldots\}$\\[2pt]
\hline
$k=13$&$(1,1,0)$&0&$5$&$0$&$5$&$\{5,0,0,0,0,0,0,0,\ldots\}$\\[2pt]
$k=14$&$(1,1,0)$&0&$5$&$0$&$5$&$\{1,4,0,0,0,0,0,0,\ldots\}$\\[2pt]
$k=15$&$(2,1,0)$&0&$5$&$0$&$5$&$\{2,3,0,0,0,0,0,0,\ldots\}$\\[2pt]
$k=16$&$(2,1,0)$&0&$5$&$0$&$5$&$\{1,1,3,0,0,0,0,0,\ldots\}$\\[2pt]
$k=17$&$(2,0,0)$&0&$5$&$0$&$5$&$\{1,2,2,0,0,0,0,0,\ldots\}$\\[2pt]
$k=18$&$(2,0,0)$&0&$5$&$0$&$5$&$\{1,1,1,2,0,0,0,0,\ldots\}$\\[2pt]
$k=27$&$(2,0,0)$&0&$5$&$0$&$5$&$\{1,1,1,1,1,0,0,0,\ldots\}$\\[2pt]
\hline
$k=19$&$(1,1,0)$&0&$6$&$0$&$6$&$\{6,0,0,0,0,0,0,0,\ldots\}$\\[2pt]
$k=20$&$(1,1,0)$&0&$6$&$0$&$6$&$\{1,5,0,0,0,0,0,0,\ldots\}$\\[2pt]
$k=21,22$&$(0,0,1)$&0&$6,6$&$0$&$6$&$\{3,3,0,0,0,0,0,0,\ldots\}$\\[2pt]
$k=23$&$(1,1,0)$&0&$6$&$0$&$6$&$\{1,1,4,0,0,0,0,0,\ldots\}$\\[2pt]
$k=24$&$(2,1,0)$&0&$6$&$0$&$6$&$\{1,2,3,0,0,0,0,0,\ldots\}$\\[2pt]
$k=25$&$(2,0,0)$&0&$6$&$0$&$6$&$\{2,2,2,0,0,0,0,0,\ldots\}$\\[2pt]
$k=28$&$(2,1,0)$&0&$6$&$0$&$6$&$\{1,1,1,3,0,0,0,0,\ldots\}$\\[2pt]
$k=29$&$(2,0,0)$&0&$6$&$0$&$6$&$\{1,1,2,2,0,0,0,0,\ldots\}$\\[2pt]
$k=38$&$(2,0,0)$&0&$6$&$0$&$6$&$\{1,1,1,1,2,0,0,0,\ldots\}$\\[2pt]
$k=57$&$(2,0,0)$&0&$6$&$0$&$6$&$\{1,1,1,1,1,1,0,0,\ldots\}$\\[2pt]
\hline
\end{tabular}
\caption{\small String content $(m_h,m_1,m_2)$, conformal eigenenergies and quantum integers for the leading 30 eigenvalues $\widehat D(u)$ in the standard module $\repV_{N,8}$. The conformal energies are given by $E=E_\text{base}+\sum_j E_j$ where the shifted triangular matchstick numbers are 
$E_\text{base}=\tfrac{3}{2}(m_3\!+\!2)(m_3\!+\!3)-9=0,9,21,36,54,\ldots$ with $m_3=0,1,2,3,4,\ldots$. 
The associated conformal character is $\chit_{1,9}(q)=q^{28/3} \sum_E q^E=q^{28/3}(1+q+2q^2+3q^3+5q^4+7q^5+11q^6+O(q^{7}))$. 
For a finite system, $N-d+1-3m_3=n_2+m_1+2m_2=M$ is the total number of zeros of $Q(u)$ in the fundamental domain that enter the Bethe ansatz equations.}
\label{d=8Eigs}
\end{center}
\end{table}


\subsection{Tabulated numerical results for periodic boundary conditions}

In the following pages, we separately tabulate our numerical results for periodic boundary conditions in the standard modules $\repW_{N,d,\omega}$ with $\omega=\pm 1$ and $d$ even/odd in the range $d=0,1,\ldots,6$. In the tables, we use the term {\it multiplicity} to mean the number of times the corresponding eigenvalue is counted in the torus partition functions, whereas we use the term {\it degeneracy} to mean the number of occurrences of this eigenvalue in the given standard module.
%
\begin{table}[h!]
\begin{center}
\scriptsize
\begin{tabular}{|c|c||c|c||c|c||c|c||c|c|c|}
\multicolumn{11}{c}{\bf \normalsize Standard modules $\boldsymbol{\repW_{N,d,\omega = 1}}$ with $\boldsymbol d$ even}\\[4pt]
\hline
\hline
{Label}&{\!\!\!Defect \!\#\!\!\!}&{\!\!\!Mult\!\!\!}&{\!\!\!Deg\!\!\!}&\textbf{$\!\!\!(\mb;\bar\mb)\!\!\!$}&$\!\!\!(m_3,\!\mbar_3)\!\!\!$&\textbf{$\!\!\!(\Delta_{0,\frac{d}{2}},\Delta_{0,-\frac{d}{2}})\!\!\!$}&$(E,\bar E)$&$\!\!E_\text{base}\!\!$&\textbf{$\!\!\!\sum_j (E,\Ebar)\!\!\!$}&\textbf{$\{E_j|\Ebar_j\}$}\rule{0pt}{8pt}\\
\hline
$k=1$&$d=0$&1&1&$(0,0,0;0,0,0)$&$(0,0)$&$(-\tfrac{1}{24},-\tfrac{1}{24})$&$(0, 0)$&$(0, 0)$&$(0, 0)$&$\{0,0,0,0,\!..|0,0,0,0,\!..\}$\rule{0pt}{8pt}\\[2pt]
\hline
$k=2$&$d=0$&1&1&$(0,0,1;0,0,0)$&$(0,0)$&$(-\tfrac{1}{24},-\tfrac{1}{24})$&$(1, 0)$&$(0, 0)$&$(1, 0)$&$\{1,0,0,0,\!..|0,0,0,0,\!..\}$\rule{0pt}{8pt}\\[2pt]
$k=3$&$d=0$&1&1&$(0,0,0;0,0,1)$&$(0,0)$&$(-\tfrac{1}{24},-\tfrac{1}{24})$&$(0, 1)$&$(0, 0)$&$(0, 1)$&$\{0,0,0,0,\!..|1,0,0,0,\!..\}$\\[2pt]
\hline
$k=4,5$&$d=0$&1&2&$(0,0,1;0,0,0)$&$(0,0)$&$(-\tfrac{1}{24},-\tfrac{1}{24})$&$(2, 0)$&$(0, 0)$&$(2, 0)$&$\{1,1,0,0,\!..|0,0,0,0,\!..\}$\rule{0pt}{8pt}\\[2pt]
$k=6,7$&$d=0$&1&2&$(0,0,0;0,0,1)$&$(0,0)$&$(-\tfrac{1}{24},-\tfrac{1}{24})$&$(0, 2)$&$(0, 0)$&$(0, 2)$&$\{0,0,0,0,\!..|1,1,0,0,\!..\}$\\[2pt]
$k=8$&$d=0$&1&1&$(0,0,1;0,0,1)$&$(0,0)$&$(-\tfrac{1}{24},-\tfrac{1}{24})$&$(1, 1)$&$(0, 0)$&$(1, 1)$&$\{1,0,0,0,\!..|1,0,0,0,\!..\}$\\[2pt]
\hline
$k=9$&$d=0$&1&1&$(0,2,0;0,0,0)$&$(0,0)$&$(-\tfrac{1}{24},-\tfrac{1}{24})$&$(3, 0)$&$(0, 0)$&$(3, 0)$&\!\!\!$\{1,3,1,-2,\!..|0,0,0,0,\!..\}$\!\!\!\rule{0pt}{8pt}\\[2pt]
$k=10$&$d=0$&1&1&$(0,0,0;0,2,0)$&$(0,0)$&$(-\tfrac{1}{24},-\tfrac{1}{24})$&$(0, 3)$&$(0, 0)$&$(0, 3)$&\!\!\!$\{0,0,0,0,\!..|1,3,1,-2,\!..\}$\!\!\!\\[2pt]
$k=11,12$&$d=0$&1&2&$(0,0,1;0,0,0)$&$(0,0)$&$(-\tfrac{1}{24},-\tfrac{1}{24})$&$(3, 0)$&$(0, 0)$&$(3, 0)$&$\{1,2,0,0,\!..|0,0,0,0,\!..\}$\\[2pt]
$k=13,14$&$d=0$&1&2&$(0,0,0;0,0,1)$&$(0,0)$&$(-\tfrac{1}{24},-\tfrac{1}{24})$&$(0, 3)$&$(0, 0)$&$(0, 3)$&$\{0,0,0,0,\!..|1,2,0,0,\!..\}$\\[2pt]
$k=15,16$&$d=0$&1&2&$(0,0,1;0,0,1)$&$(0,0)$&$(-\tfrac{1}{24},-\tfrac{1}{24})$&$(2, 1)$&$(0, 0)$&$(2, 1)$&$\{1,1,0,0,\!..|1,0,0,0,\!..\}$\\[2pt]
$k=17,18$&$d=0$&1&2&$(0,0,1;0,0,1)$&$(0,0)$&$(-\tfrac{1}{24},-\tfrac{1}{24})$&$(1, 2)$&$(0, 0)$&$(1, 2)$&$\{1,0,0,0,\!..|1,1,0,0,\!..\}$\\[2pt]
$k=19,20$&$d=0$&1&2&$(0,0,0;0,0,0)$&$(1,1)$&$(-\tfrac{1}{24},-\tfrac{1}{24})$&$\!\!\!(\tfrac{3}{2},\tfrac{3}{2})\!\!\!$&$\!\!\!(\tfrac{3}{2},\tfrac{3}{2})\!\!\!$&$(0, 0)$&$\{0,0,0,0,\!..|0,0,0,0,\!..\}$\\[2pt]
\hline
$k=1,2$&$d=6$&2&1&$(0,0,0;0,0,0)$&$(1,1)$&$(\tfrac{35}{24},\tfrac{35}{24})$&$(0, 0)$&$(0, 0)$&$(0,0)$&$\{0,0,0,0,\!..|0,0,0,0,\!..\}$\rule{0pt}{8pt}\\[2pt]
\hline\hline
$k=1$&$d=2$&2&1&$(0,0,0;0,0,0)$&$(0,0)$&$(\tfrac{1}{8},\tfrac{1}{8})$&$(0, 0)$&$(0, 0)$&$(0,0)$&$\{0,0,0,0,\!..|0,0,0,0,\!..\}$\rule{0pt}{8pt}\\[2pt]
\hline
$k=2$&$d=2$&2&1&$(0,1,0;0,0,0)$&$(0,0)$&$(\tfrac{1}{8},\tfrac{1}{8})$&$(1,0)$&$(0, 0)$&$(1,0)$&$\{1,0,0,0,\!..|0,0,0,0,\!..\}$\rule{0pt}{8pt}\\[2pt]
$k=3$&$d=2$&2&1&$(0,0,0;0,1,0)$&$(0,0)$&$(\tfrac{1}{8},\tfrac{1}{8})$&$(0,1)$&$(0, 0)$&$(0,1)$&$\{0,0,0,0,\!..|1,0,0,0,\!..\}$\\[2pt]
\hline
$k=4,5$&$d=2$&2&2&$(0,0,1;0,0,0)$&$(0,0)$&$(\tfrac{1}{8},\tfrac{1}{8})$&$(2,0)$&$(0, 0)$&$(2,0)$&$\{1,1,0,0,\!..|0,0,0,0,\!..\}$\rule{0pt}{8pt}\\[2pt]
$k=6,7$&$d=2$&2&2&$(0,0,0;0,0,1)$&$(0,0)$&$(\tfrac{1}{8},\tfrac{1}{8})$&$(0,2)$&$(0, 0)$&$(0,2)$&$\{0,0,0,0,\!..|1,1,0,0,\!..\}$\\[2pt]
$k=8$&$d=2$&2&1&$(0,1,0;0,1,0)$&$(0,0)$&$(\tfrac{1}{8},\tfrac{1}{8})$&$(1,1)$&$(0, 0)$&$(1,1)$&$\{1,0,0,0,\!..|1,0,0,0,\!..\}$\\[2pt]
\hline
$k=9$&$d=2$&2&1&$(0,1,1;0,0,0)$&$(0,0)$&$(\tfrac{1}{8},\tfrac{1}{8})$&$(3,0)$&$(0, 0)$&$(3,0)$&$\{1,0,2,0,\!..|0,0,0,0,\!..\}$\rule{0pt}{8pt}\\[2pt]
$k=10$&$d=2$&2&1&$(0,0,0;0,1,1)$&$(0,0)$&$(\tfrac{1}{8},\tfrac{1}{8})$&$(0,3)$&$(0, 0)$&$(0,3)$&$\{0,0,0,0,\!..|1,0,2,0,\!..\}$\\[2pt]
$k=11,12$&$d=2$&2&2&$(0,0,1;0,0,0)$&$(0,0)$&$(\tfrac{1}{8},\tfrac{1}{8})$&$(3,0)$&$(0, 0)$&$(3,0)$&$\{1,2,0,0,\!..|0,0,0,0,\!..\}$\\[2pt]
$k=13,14$&$d=2$&2&2&$(0,0,0;0,0,1)$&$(0,0)$&$(\tfrac{1}{8},\tfrac{1}{8})$&$(0,3)$&$(0, 0)$&$(0,3)$&$\{0,0,0,0,\!..|1,2,0,0,\!..\}$\\[2pt]
$k=15^*$&$d=2$&2&1&$(0,0,0;0,0,0)$&$(1,0)$&$(\tfrac{1}{8},\tfrac{1}{8})$&$\!\!\!(\tfrac{5}{2},\tfrac{1}{2})\!\!\!$&$\!\!\!(\tfrac{5}{2},\tfrac{1}{2})\!\!\!$&$(0, 0)$&$\{0,0,0,0,\!..|0,0,0,0,\!..\}$\\[2pt]
$k=16^*$&$d=2$&2&1&$(0,0,0;0,0,0)$&$(0,1)$&$(\tfrac{1}{8},\tfrac{1}{8})$&$\!\!\!(\tfrac{1}{2},\tfrac{5}{2})\!\!\!$&$\!\!\!(\tfrac{1}{2},\tfrac{5}{2})\!\!\!$&$(0, 0)$&$\{0,0,0,0,\!..|0,0,0,0,\!..\}$\\[2pt]
$k=17,18$&$d=2$&2&2&$(0,0,1;0,1,0)$&$(0,0)$&$(\tfrac{1}{8},\tfrac{1}{8})$&$(2,1)$&$(0, 0)$&$(2,1)$&$\{1,1,0,0,\!..|1,0,0,0,\!..\}$\\[2pt]
$k=19,20$&$d=2$&2&2&$(0,1,0;0,0,1)$&$(0,0)$&$(\tfrac{1}{8},\tfrac{1}{8})$&$(1,2)$&$(0, 0)$&$(1,2)$&$\{1,0,0,0,\!..|1,1,0,0,\!..\}$\\[2pt]
\hline\hline
$k=1$&$d=4$&2&1&$(0,0,0;0,0,0)$&$(0,0)$&$(\tfrac{5}{8},\tfrac{5}{8})$&$(0, 0)$&$(0, 0)$&$(0,0)$&$\{0,0,0,0,\!..|0,0,0,0,\!..\}$\rule{0pt}{8pt}\\[2pt]
\hline
$k=2$&$d=4$&2&1&$(0,1,0;0,0,0)$&$(0,0)$&$(\tfrac{5}{8},\tfrac{5}{8})$&$(1,0)$&$(0, 0)$&$(1,0)$&$\{1,0,0,0,\!..|0,0,0,0,\!..\}$\rule{0pt}{8pt}\\[2pt]
$k=3$&$d=4$&2&1&$(0,0,0;0,1,0)$&$(0,0)$&$(\tfrac{5}{8},\tfrac{5}{8})$&$(0,1)$&$(0, 0)$&$(0,1)$&$\{0,0,0,0,\!..|1,0,0,0,\!..\}$\\[2pt]
\hline
$k=4$&$d=4$&2&1&$(0,1,0;0,0,0)$&$(0,0)$&$(\tfrac{5}{8},\tfrac{5}{8})$&$(2,0)$&$(0, 0)$&$(2,0)$&$\{2,0,0,0,\!..|0,0,0,0,\!..\}$\rule{0pt}{8pt}\\[2pt]
$k=5$&$d=4$&2&1&$(0,0,0;0,1,0)$&$(0,0)$&$(\tfrac{5}{8},\tfrac{5}{8})$&$(0,2)$&$(0, 0)$&$(0,2)$&$\{0,0,0,0,\!..|2,0,0,0,\!..\}$\\[2pt]
$k=6$&$d=4$&2&1&$(0,0,0;0,0,0)$&$(1,0)$&$(\tfrac{5}{8},\tfrac{5}{8})$&$(2,0)$&$(2, 0)$&$(0,0)$&$\{0,0,0,0,\!..|0,0,0,0,\!..\}$\rule{0pt}{8pt}\\[2pt]
$k=7$&$d=4$&2&1&$(0,0,0;0,0,0)$&$(0,1)$&$(\tfrac{5}{8},\tfrac{5}{8})$&$(0,2)$&$(0, 2)$&$(0,0)$&$\{0,0,0,0,\!..|0,0,0,0,\!..\}$\\[2pt]
$k=8$&$d=4$&2&1&$(0,1,0;0,1,0)$&$(0,0)$&$(\tfrac{5}{8},\tfrac{5}{8})$&$(1,1)$&$(0, 0)$&$(1,1)$&$\{1,0,0,0,\!..|1,0,0,0,\!..\}$\\[2pt]
\hline\hline
\end{tabular}
\caption{\small String content $(\mb;\bar\mb)=(m_h,m_1,m_2;\mbar_h,\mbar_1,\mbar_2)$, $(m_3,\mbar_3)$, reference conformal weights $(\Delta_{0,d/2},\Delta_{0,-d/2})$, eigenenergies $(E,\Ebar)$ and degeneracies for the leading 78 eigenvalues $\widehat T(u)$ in the standard modules $\repW_{N,d,1}$ with $d$ even. The multiplicities are 1 for $d=0$ and 2 otherwise. Allowing for multiplicities and degeneracies, there are 22 eigenvalues in the $d=0$ mod $6$ modules, 40 eigenvalues in the $d=2$ mod $6$ modules and 16 eigenvalues in the $d=4$ mod $6$ modules. 
The total conformal weights for each eigenvalue are $(\Delta,\bar\Delta)=(\Delta_{0,\frac{d}{2}}+E,\Delta_{0,-\frac{d}{2}}+\Ebar)$ where $E=E_\text{base}+\sum_j E_j$, $\Ebar=\Ebar_\text{base}+\sum_j \Ebar_j$, where $E=E_\text{base}+\sum_j E_j$, $\Ebar=\Ebar_\text{base}+\sum_j \Ebar_j$.  In these modules, $J=2$ and $Q(u)=\widehat T(u)$. In general, each 2-string of $\widehat T(u)$ is accompanied by a 2-string of $Q(u)$ or it is indicative of a 1-hole in $Q(u)$. Since $Q(u)=\widehat T(u)$, it follows that there are no 1-holes and $m_h=\mbar_h=0$ for all eigenvalues in these modules. However, unlike in other modules, in these $\omega=1$ modules $\widehat T(u)$ does admit 3-strings. An asterisk in the label column indicates that, for these eigenvalues, ${\cal P}$ is not zero.
}\label{omega=1dEvenEigs}
\end{center}
\end{table}

\begin{table}[h!]
\begin{center}
\scriptsize
\begin{tabular}{|c|c||c|c||c|c||c|c||c|c|c|}
\multicolumn{11}{c}{\bf \normalsize Standard modules $\boldsymbol{\repW_{N,d,\omega = 1}}$ with $\boldsymbol d$ odd}
\\[4pt]
\hline
\hline
{Label}&{\!\!\!Defect \!\#\!\!\!}&{\!\!\!Mult\!\!\!}&{\!\!\!Deg\!\!\!}&\textbf{$\!\!\!(\mb;\bar\mb)\!\!\!$}&$\!\!\!(m_3,\!\mbar_3)\!\!\!$&\textbf{$\!\!\!(\Delta_{0,\frac{d}{2}},\Delta_{0,-\frac{d}{2}})\!\!\!$}&$(E,\bar E)$&$\!\!E_\text{base}\!\!$&\textbf{$\!\!\!\sum_j (E,\Ebar)\!\!\!$}&\textbf{$\{E_j|\Ebar_j\}$}\rule{0pt}{8pt}\\
\hline
$k=1$&$d=1$&1&1&$(0,0,0;0,0,0)$&$(0,0)$&$(0,0)$&$(0, 0)$&$(0, 0)$&$(0, 0)$&$\{0,0,0,0,\!..|0,0,0,0,\!..\}$\rule{0pt}{8pt}\\[2pt]
\hline
$k=2^*$&$d=1$&1&1&$(1,0,0;0,0,0)$&$(0,0)$&$(0,0)$&$(1, 0)$&$(1, 0)$&$(0, 0)$&$\{0,0,0,0,\!..|0,0,0,0,\!..\}$\rule{0pt}{8pt}\\[2pt]
$k=3^*$&$d=1$&1&1&$(0,0,0;1,0,0)$&$(0,0)$&$(0,0)$&$(0, 1)$&$(0, 1)$&$(0, 0)$&$\{0,0,0,0,\!..|0,0,0,0,\!..\}$\\[2pt]
\hline
$k=4,5$&$d=1$&1&2&$(1,0,0;0,0,0)$&$(1,0)$&$(0,0)$&$(2, 0)$&$(2, 0)$&$(0, 0)$&$\{0,0,0,0,\!..|0,0,0,0,\!..\}$\rule{0pt}{8pt}\\[2pt]
$k=6,7$&$d=1$&1&2&$(0,0,0;1,0,0)$&$(0,1)$&$(0,0)$&$(0, 2)$&$(0, 2)$&$(0, 0)$&$\{0,0,0,0,\!..|0,0,0,0,\!..\}$\\[2pt]
$k=8$&$d=1$&1&1&$(1,0,0;1,0,0)$&$(0,0)$&$(0,0)$&$(1, 1)$&$(1,1)$&$(0,0)$&$\{0,0,0,0,\!..|0,0,0,0,\!..\}$\\[2pt]
\hline
$k=9^*$&$d=1$&1&1&$(0,0,1;0,0,0)$&$(0,0)$&$(0,0)$&$(3, 0)$&$(1, 0)$&$(2, 0)$&\!\!\!$\{2,0,0,0,\!..|0,0,0,0,\!..\}$\!\!\!\rule{0pt}{8pt}\\[2pt]
$k=10^*$&$d=1$&1&1&$(0,0,0;0,0,1)$&$(0,0)$&$(0,0)$&$(0, 3)$&$(0, 1)$&$(0, 2)$&\!\!\!$\{0,0,0,0,\!..|2,0,0,0,\!..\}$\!\!\!\\[2pt]
$k=11,12$&$d=1$&1&2&$(1,0,0;0,0,0)$&$(1,0)$&$(0,0)$&$(3, 0)$&$(0, 0)$&$(3, 0)$&$\{3,0,0,0,\!..|0,0,0,0,\!..\}$\\[2pt]
$k=13,14$&$d=1$&1&2&$(0,0,0;1,0,0)$&$(0,1)$&$(0,0)$&$(0, 3)$&$(0, 0)$&$(0, 3)$&$\{0,0,0,0,\!..|3,0,0,0,\!..\}$\\[2pt]
$\!\!k=15$--$17^*\!\!\!$&$d=1$&1&3&$(1,0,0;1,0,0)$&$(1,0)$&$(0,0)$&$(2, 1)$&$(2,1)$&$(0,0)$&$\{0,0,0,0,\!..|0,0,0,0,\!..\}$\\[2pt]
$\!\!k=18$--$20^*\!\!\!$&$d=1$&1&3&$(1,0,0;1,0,0)$&$(0,1)$&$(0,0)$&$(1, 2)$&$(1,2)$&$(0,0)$&$\{0,0,0,0,\!..|0,0,0,0,\!..\}$\\[2pt]
\hline\hline
$k=1$&$d=3$&2&1&$(0,0,0;0,0,0)$&$(0,0)$&$(\tfrac{1}{3},\tfrac{1}{3})$&$(0, 0)$&$(0, 0)$&$(0,0)$&$\{0,0,0,0,\!..|0,0,0,0,\!..\}$\rule{0pt}{8pt}\\[2pt]
\hline
$k=2$&$d=3$&2&1&$(0,1,0;0,0,0)$&$(0,0)$&$(\tfrac{1}{3},\tfrac{1}{3})$&$(1,0)$&$(0, 0)$&$(1,0)$&$\{1,0,0,0,\!..|0,0,0,0,\!..\}$\rule{0pt}{8pt}\\[2pt]
$k=3$&$d=3$&2&1&$(0,0,0;0,1,0)$&$(0,0)$&$(\tfrac{1}{3},\tfrac{1}{3})$&$(0,1)$&$(0, 0)$&$(0,1)$&$\{0,0,0,0,\!..|1,0,0,0,\!..\}$\\[2pt]
\hline
$k=4,5$&$d=3$&2&2&$(0,0,1;0,0,0)$&$(0,0)$&$(\tfrac{1}{3},\tfrac{1}{3})$&$(2,0)$&$(0, 0)$&$(2,0)$&$\{1,1,0,0,\!..|0,0,0,0,\!..\}$\rule{0pt}{8pt}\\[2pt]
$k=6,7$&$d=3$&2&2&$(0,0,0;0,0,1)$&$(0,0)$&$(\tfrac{1}{3},\tfrac{1}{3})$&$(0,2)$&$(0, 0)$&$(0,2)$&$\{0,0,0,0,\!..|1,1,0,0,\!..\}$\\[2pt]
$k=8$&$d=3$&2&1&$(0,1,0;0,1,0)$&$(0,0)$&$(\tfrac{1}{3},\tfrac{1}{3})$&$(1,1)$&$(0, 0)$&$(1,1)$&$\{1,0,0,0,\!..|1,0,0,0,\!..\}$\\[2pt]
\hline
$k=9$--$12$&$d=3$&2&4&$(1,0,0;0,0,0)$&$(1,0)$&$(\tfrac{1}{3},\tfrac{1}{3})$&$(3,0)$&$(3,0)$&$(0,0)$&$\{0,0,0,0,\!..|0,0,0,0,\!..\}$\rule{0pt}{8pt}\\[2pt]
$k=13$--$16$&$d=3$&2&4&$(0,0,0;1,0,0)$&$(0,1)$&$(\tfrac{1}{3},\tfrac{1}{3})$&$(0,3)$&$(0,3)$&$(0,0)$&$\{0,0,0,0,\!..|0,0,0,0,\!..\}$\\[2pt]
$k=17,18$&$d=3$&2&2&$(0,0,1;0,1,0)$&$(0,0)$&$(\tfrac{1}{3},\tfrac{1}{3})$&$(2,1)$&$(0, 0)$&$(2,1)$&$\{1,1,0,0,\!..|1,0,0,0,\!..\}$\\[2pt]
$k=19,20$&$d=3$&2&2&$(0,1,0;0,0,1)$&$(0,0)$&$(\tfrac{1}{3},\tfrac{1}{3})$&$(1,2)$&$(0, 0)$&$(1,2)$&$\{1,0,0,0,\!..|1,1,0,0,\!..\}$\\[2pt]
\hline\hline
$k=1$&$d=5$&2&1&$(0,0,0;0,0,0)$&$(0,0)$&$(1,1)$&$(0, 0)$&$(0, 0)$&$(0,0)$&$\{0,0,0,0,\!..|0,0,0,0,\!..\}$\rule{0pt}{8pt}\\[2pt]
\hline
$k=2^*$&$d=5$&2&1&$(1,0,0;1,0,0)$&$(0,0)$&$(1,1)$&$(1,0)$&$(1, 0)$&$(0,0)$&$\{0,0,0,0,\!..|0,0,0,0,\!..\}$\rule{0pt}{8pt}\\[2pt]
$k=3^*$&$d=5$&2&1&$(1,0,0;1,0,0)$&$(0,0)$&$(1,1)$&$(0,1)$&$(0, 1)$&$(0,0)$&$\{0,0,0,0,\!..|0,0,0,0,\!..\}$\\[2pt]
\hline
$k=4$&$d=5$&2&1&$(0,0,0;1,0,0)$&$(0,0)$&$(1,1)$&$(2,0)$&$(0, 0)$&$(2,0)$&$\{2,0,0,0,\!..|0,0,0,0,\!..\}$\rule{0pt}{8pt}\\[2pt]
$k=5$&$d=5$&2&1&$(1,0,0;0,0,0)$&$(0,0)$&$(1,1)$&$(0,2)$&$(0, 0)$&$(0,2)$&$\{0,0,0,0,\!..|2,0,0,0,\!..\}$\\[2pt]
$k=6^*$&$d=5$&2&1&$(1,0,0;1,0,0)$&$(0,0)$&$(1,1)$&$(2,0)$&$(1,0)$&$(1,0)$&$\{1,0,0,0,\!..|0,0,0,0,\!..\}$\\[2pt]
$k=7^*$&$d=5$&2&1&$(1,0,0;1,0,0)$&$(0,0)$&$(1,1)$&$(0,2)$&$(0,1)$&$(0,1)$&$\{0,0,0,0,\!..|1,0,0,0,\!..\}$\\[2pt]
$k=8$&$d=5$&2&1&$(1,0,0;1,0,0)$&$(0,0)$&$(1,1)$&$(1,1)$&$(1,1)$&$(0,0)$&$\{0,0,0,0,\!..|0,0,0,0,\!..\}$\\[2pt]
\hline\hline
\end{tabular}
\caption{\small String content $(\mb;\bar\mb)=(m_h,m_1,m_2;\mbar_h,\mbar_1,\mbar_2)$, $(m_3,\mbar_3)$, reference conformal weights $(\Delta_{0,d/2},\Delta_{0,-d/2})$, eigenenergies $(E,\Ebar)$ and degeneracies for the leading 76 eigenvalues $\widehat T(u)$ in the standard modules $\repW_{N,d,1}$ with $d$ odd. The multiplicities are 1 for $d=1$ and 2 otherwise.
Allowing for multiplicities and degeneracies, there are 20 eigenvalues in the $d=1$ mod 6 modules, 40 eigenvalues in the $d=3$ mod 6 modules and 16 eigenvalues in the $d=5$ mod 6 modules. The total conformal weights for each eigenvalue are $(\Delta,\bar\Delta)=(\Delta_{0,\frac{d}{2}}+E,\Delta_{0,-\frac{d}{2}}+\Ebar)$ where $E=E_\text{base}+\sum_j E_j$, $\Ebar=\Ebar_\text{base}+\sum_j \Ebar_j$. In these modules, $J=-2$ and $Q(u)\neq \widehat T(u)$. An asterisk in the label column indicates that, for these eigenvalues, ${\cal P}$ is not zero.
}\label{omega=1dOddEigs}
\end{center}
\end{table}

\begin{table}[h!]
\begin{center}
\scriptsize
\begin{tabular}{|c|c||c|c||c|c||c|c||c|c|c|}
\multicolumn{11}{c}{\bf \normalsize Standard modules $\boldsymbol{\repW_{N,d,\omega = -1}}$ with $\boldsymbol d$ even}\\[4pt]
\hline
\hline
{Label}&{\!\!\!Defect \!\#\!\!\!}&{\!\!Mult\!\!}&{\!\!Deg\!\!}&\textbf{$\!\!\!(\mb;\bar\mb)\!\!\!$}&$\!\!\!(m_3,\!\mbar_3)\!\!\!$&\textbf{$\!\!\!(\Delta_{1,\frac{d}{2}},\Delta_{1,\frac{d}{2}})\!\!\!$}
&$\!\!\!(E,\Ebar)\!\!\!$&$\textrm{$\!\!\!\!E_\text{base}\!\!\!\!$}$&\textbf{$\!\!\!\sum_j(E_j,\Ebar_j)\!\!\!\!$}&\textbf{$\{E_j|\Ebar_j\}$}\rule{0pt}{8pt}\\
\hline
$k=1,2$&$d=0$&1&2&$(0,0,0;0,0,0)$&$(1,0)$&$(\tfrac{1}{3},\tfrac{1}{3})$&$(0,0)$&$(0,0)$&$(0,0)$&$\{0,0,0,0,\!..|0,0,0,0,\!..\}$\rule{0pt}{8pt}\\[2pt]
\hline
$k=3,4$&$d=0$&1&2&$(0,1,0;0,0,0)$&$(0,1)$&$(\tfrac{1}{3},\tfrac{1}{3})$&$(1,0)$&$(0,0)$&$(1,0)$&$\{1,0,0,0,\!..|0,0,0,0,\!..\}$\rule{0pt}{8pt}\\[2pt]
$k=5,6$&$d=0$&1&2&$(0,0,0;0,1,0)$&$(1,0)$&$(\tfrac{1}{3},\tfrac{1}{3})$&$(0,1)$&$(0,0)$&$(0,1)$&$\{0,0,0,0,\!..|1,0,0,0,\!..\}$\\[2pt]
\hline
$k=7$--$10$&$d=0$&1&4&$(0,0,1;0,0,0)$&$(0,1)$&$(\tfrac{1}{3},\tfrac{1}{3})$&$(2,0)$&$(0,0)$&$(2,0)$&$\{1,1,0,0,\!..|0,0,0,0,\!..\}$\rule{0pt}{8pt}\\[2pt]
$k=11$--$14$&$d=0$&1&4&$(0,0,0;0,0,1)$&$(1,0)$&$(\tfrac{1}{3},\tfrac{1}{3})$&$(0,2)$&$(0,0)$&$(0,2)$&$\{0,0,0,0,\!..|1,1,0,0,\!..\}$\\[2pt]
$k=15,16$&$d=0$&1&2&$(0,0,1;0,0,1)$&$(1,0)$&$(\tfrac{1}{3},\tfrac{1}{3})$&$(1,1)$&$(0,0)$&$(1,1)$&$\{1,0,0,0,\!..|1,0,0,0,\!..\}$\\[2pt]
\hline
$k=17$--$20$&$d=0$&1&4&$(1,0,0;0,0,0)$&$(1,1)$&$(\tfrac{1}{3},\tfrac{1}{3})$&$(3,0)$&$(3,0)$&$(0,0)$&$\{0,0,0,0,\!..|0,0,0,0,\!..\}$\rule{0pt}{8pt}\\[2pt]
$k=21$--$24$&$d=0$&1&4&$(0,0,0;1,0,0)$&$(1,1)$&$(\tfrac{1}{3},\tfrac{1}{3})$&$(0,3)$&$(0,3)$&$(0,0)$&$\{0,0,0,0,\!..|0,0,0,0,\!..\}$\\[2pt]
$k=25$--$28$&$d=0$&1&4&$(0,0,1;0,1,0)$&$(0,1)$&$(\tfrac{1}{3},\tfrac{1}{3})$&$(2,1)$&$(0,0)$&$(2,1)$&$\{1,1,0,0,\!..|1,0,0,0,\!..\}$\\[2pt]
$k=29$--$32$&$d=0$&1&4&$(0,1,0;0,0,1)$&$(1,0)$&$(\tfrac{1}{3},\tfrac{1}{3})$&$(1,2)$&$(0,0)$&$(1,2)$&$\{1,0,0,0,\!..|1,1,0,0,\!..\}$\\[2pt]
\hline
$k=1,2$&$d=6$&2&2&$(1,0,0;0,0,0)$&$(0,0)$&$(\tfrac{1}{3},\tfrac{1}{3})$&$(3,0)$&$(3,0)$&$(0,0)$&$\{0,0,0,0,\!..|0,0,0,0,\!..\}$\rule{0pt}{8pt}\\[2pt]
$k=3,4$&$d=6$&2&2&$(0,0,0;1,0,0)$&$(0,0)$&$(\tfrac{1}{3},\tfrac{1}{3})$&$(0,3)$&$(0,3)$&$(0,0)$&$\{0,0,0,0,\!..|0,0,0,0,\!..\}$\\[2pt]
\hline\hline
$k=1^*$&$d=2$&2&1&$(1,0,0;0,0,0)$&$(0,0)$&$(0,0)$&$(1,0)$&$(1,0)$&$(0,0)$&$\{0,0,0,0,\!..|0,0,0,0,\!..\}$\rule{0pt}{8pt}\\[2pt]
$k=2^*$&$d=2$&2&1&$(0,0,0;1,0,0)$&$(0,0)$&$(0,0)$&$(0,1)$&$(0,1)$&$(0,0)$&$\{0,0,0,0,\!..|0,0,0,0,\!..\}$\\[2pt]
\hline
$k=3$&$d=2$&2&1&$(1,0,0;0,0,0)$&$(0,0)$&$(0,0)$&$(2,0)$&$(2,0)$&$(0,0)$&$\{0,0,0,0,\!..|0,0,0,0,\!..\}$\rule{0pt}{8pt}\\[2pt]
$k=4$&$d=2$&2&1&$(0,0,0;1,0,0)$&$(0,0)$&$(0,0)$&$(0,2)$&$(0,2)$&$(0,0)$&$\{0,0,0,0,\!..|0,0,0,0,\!..\}$\\[2pt]
$k=5,6$&$d=2$&2&2&$(1,0,0;1,0,0)$&$(1,0)$&$(0,0)$&$(1,1)$&$(1,1)$&$(0,0)$&$\{0,0,0,0,\!..|0,0,0,0,\!..\}$\\[2pt]
\hline
$k=7^*$&$d=2$&2&1&$(0,1,0;0,0,0)$&$(0,0)$&$(0,0)$&$(3,0)$&$(1,0)$&$(2,0)$&$\{2,0,0,0,\!..|0,0,0,0,\!..\}$\rule{0pt}{8pt}\\[2pt]
$k=8^*$&$d=2$&2&1&$(0,0,0;0,1,0)$&$(0,0)$&$(0,0)$&$(0,3)$&$(0,1)$&$(0,2)$&$\{0,0,0,0,\!..|2,0,0,0,\!..\}$\\[2pt]
$k=9$&$d=2$&2&1&$(1,0,0;0,0,0)$&$(0,0)$&$(0,0)$&$(3,0)$&$(2,0)$&$(1,0)$&$\{1,0,0,0,\!..|0,0,0,0,\!..\}$\\[2pt]
$k=10$&$d=2$&2&1&$(0,0,0;1,0,0)$&$(0,0)$&$(0,0)$&$(0,3)$&$(0,2)$&$(0,1)$&$\{0,0,0,0,\!..|1,0,0,0,\!..\}$\\[2pt]
$\!\!k=11$--$12^*\!\!\!$&$d=2$&2&2&$(1,0,0;1,0,0)$&$(0,1)$&$(0,0)$&$(2,1)$&$(2,1)$&$(0,0)$&$\{0,0,0,0,\!..|0,0,0,0,\!..\}$\\[2pt]
$\!\!k=13$--$14^*\!\!\!$&$d=2$&2&2&$(1,0,0;1,0,0)$&$(1,0)$&$(0,0)$&$(1,2)$&$(1,2)$&$(0,0)$&$\{0,0,0,0,\!..|0,0,0,0,\!..\}$\\[2pt]
\hline\hline
$k=1$&$d=4$&2&1&$(1,0,0;0,0,0)$&$(0,0)$&$(0,0)$&$(2,0)$&$(2,0)$&$(0,0)$&$\{0,0,0,0,\!..|0,0,0,0,\!..\}$\rule{0pt}{8pt}\\[2pt]
$k=2$&$d=4$&2&1&$(0,0,0;1,0,0)$&$(0,0)$&$(0,0)$&$(0,2)$&$(0,2)$&$(0,0)$&$\{0,0,0,0,\!..|0,0,0,0,\!..\}$\\[2pt]
\hline
$k=3$&$d=4$&2&1&$(1,0,0;0,0,0)$&$(0,0)$&$(0,0)$&$(3,0)$&$(2,0)$&$(1,0)$&$\{1,0,0,0,\!..|0,0,0,0,\!..\}$\rule{0pt}{8pt}\\[2pt]
$k=4$&$d=4$&2&1&$(0,0,0;1,0,0)$&$(0,0)$&$(0,0)$&$(0,3)$&$(0,2)$&$(0,1)$&$\{0,0,0,0,\!..|1,0,0,0,\!..\}$\\[2pt]
$k=5^*$&$d=4$&2&1&$(1,0,0;1,0,0)$&$(0,0)$&$(0,0)$&$(2,1)$&$(2,1)$&$(0,0)$&$\{0,0,0,0,\!..|0,0,0,0,\!..\}$\\[2pt]
$k=6^*$&$d=4$&2&1&$(1,0,0;1,0,0)$&$(0,0)$&$(0,0)$&$(1,2)$&$(1,2)$&$(0,0)$&$\{0,0,0,0,\!..|0,0,0,0,\!..\}$\\[2pt]
\hline\hline
\end{tabular}
\caption{\small String content $(m_h,m_1,m_2;\mbar_h,\mbar_1,\mbar_2)$, $(m_3,\mbar_3)$, reference conformal weights $(\Delta_{1,d/2},\Delta_{1,d/2})$ eigenenergies $(E,\Ebar)$ and degeneracies for the leading 80 eigenvalues $\widehat T(u)$ in the standard modules $\repW_{N,d,-1}$ with $d$ even. 
The multiplicities are 1 for $d=0$ and 2 otherwise. 
Allowing for multiplicities and degeneracies, there are 40 eigenvalues in the $d=0$ mod 6 modules, 28 eigenvalues in the $d=2$ mod 6 modules and 12 eigenvalues in the $d=4$ mod 6 modules. The total conformal weights for each eigenvalue are 
$(\Delta,\bar\Delta)=(\Delta_{1,\frac{d}{2}}+E,\Delta_{1,\frac{d}{2}}+\Ebar)$. In these modules, $J=-2$ and $Q(u)\ne \widehat T(u)$. 
An asterisk in the label column indicates that, for these eigenvalues, ${\cal P}$ is not zero.
}\label{omega=-1dEvenEigs}
\end{center}
\end{table}

\begin{table}[h!]
\mbox{}\vspace{-.95in}\mbox{}
\begin{center}
\scriptsize
\begin{tabular}{|c|c||c|c||c|c||c|c||c|c|c|}
\multicolumn{11}{c}{\bf \normalsize Standard modules $\boldsymbol{\repW_{N,d,\omega = -1}}$ with $\boldsymbol d$ odd}\\[4pt]
\hline
\hline
{Label}&{\!\!\!Defect \!\#\!\!\!}&{\!\!Mult\!\!}&{\!\!Deg\!\!}&\textbf{$\!\!(\mb;\bar\mb)\!\!$}&$\!\!\!(m_3,\!\mbar_3)\!\!\!$&\textbf{$\!\!\!\!(\Delta_{1,\frac{d}{2}},\Delta_{1,\frac{d}{2}})\!\!\!\!$}
&$\!\!\!(E,\Ebar)\!\!\!$&\textrm{$\!\!E_\text{base}\!\!$}&\textbf{$\!\!\!\sum_j(E_j,\Ebar_j)\!\!\!$}&\textbf{$\{E_j|\Ebar_j\}$}\rule{0pt}{8pt}\\
\hline
$k=1^*$&$d=1$&2&1&$(0,0,0;0,0,0)$&$(0,0)$&$(\tfrac{1}{8},\tfrac{1}{8})$&$(0,\tfrac{1}{2})$&$(0,\tfrac{1}{2})$&$(0, 0)$&$\{0,0,0,0,\!..|0,0,0,0,\!..\}$\rule{0pt}{8pt}\\[2pt]
$k=2^*$&$d=1$&2&1&$(0,0,0;0,0,0)$&$(0,0)$&$(\tfrac{1}{8},\tfrac{1}{8})$&$(\tfrac{1}{2},0)$&$(\tfrac{1}{2},0)$&$(0, 0)$&$\{0,0,0,0,\!..|0,0,0,0,\!..\}$\\[2pt]
\hline
$k=3^*$&$d=1$&2&1&$(0,1,0;0,0,0)$&$(0,0)$&$(\tfrac{1}{8},\tfrac{1}{8})$&$(\tfrac{3}{2},0)$&$(\tfrac{1}{2},0)$&$(1,0)$&$\{1,0,0,0,\!..|0,0,0,0,\!..\}$\rule{0pt}{8pt}\\[2pt]
$k=4^*$&$d=1$&2&1&$(0,0,0;0,1,0)$&$(0,0)$&$(\tfrac{1}{8},\tfrac{1}{8})$&$(0,\tfrac{3}{2})$&$(0,\tfrac{1}{2})$&$(0,1)$&$\{0,0,0,0,\!..|1,0,0,0,\!..\}$\\[2pt]
$k=5^*$&$d=1$&2&1&$(0,1,0;0,0,0)$&$(0,0)$&$(\tfrac{1}{8},\tfrac{1}{8})$&$(1,\tfrac{1}{2})$&$(0,\tfrac{1}{2})$&$(1,0)$&$\{1,0,0,0,\!..|0,0,0,0,\!..\}$\\[2pt]
$k=6^*$&$d=1$&2&1&$(0,0,0;0,1,0)$&$(0,0)$&$(\tfrac{1}{8},\tfrac{1}{8})$&$(\tfrac{1}{2},1)$&$(\tfrac{1}{2},0)$&$(0,1)$&$\{0,0,0,0,\!..|1,0,0,0,\!..\}$\\[2pt]
\hline
$k=7^*$&$d=1$&2&1&$(0,1,0;0,0,0)$&$(0,0)$&$(\tfrac{1}{8},\tfrac{1}{8})$&$(\tfrac{5}{2},0)$&$(\tfrac{1}{2},0)$&$(2,0)$&$\{2,0,0,0,\!..|0,0,0,0,\!..\}$\rule{0pt}{8pt}\\[2pt]
$k=8^*$&$d=1$&2&1&$(0,0,0;0,1,0)$&$(0,0)$&$(\tfrac{1}{8},\tfrac{1}{8})$&$(0,\tfrac{5}{2})$&$(0,\tfrac{1}{2})$&$(0,2)$&$\{0,0,0,0,\!..|2,0,0,0,\!..\}$\\[2pt]
$k=9$&$d=1$&2&1&$(0,0,0;0,0,0)$&$(1,0)$&$(\tfrac{1}{8},\tfrac{1}{8})$&$(\tfrac{5}{2},0)$&$(\tfrac{5}{2},0)$&$(0,0)$&$\{0,0,0,0,\!..|0,0,0,0,\!..\}$\\[2pt]
$k=10$&$d=1$&2&1&$(0,0,0;0,0,0)$&$(0,1)$&$(\tfrac{1}{8},\tfrac{1}{8})$&$(0,\tfrac{5}{2})$&$(0,\tfrac{5}{2})$&$(0,0)$&$\{0,0,0,0,\!..|0,0,0,0,\!..\}$\\[2pt]
$\!\!\!k=11$--$12^*\!\!\!\!$&$d=1$&2&2&$(0,0,1;0,0,0)$&$(0,0)$&$(\tfrac{1}{8},\tfrac{1}{8})$&$(2,\tfrac{1}{2})$&$(0,\tfrac{1}{2})$&$(2,0)$&$\{1,1,0,0,\!..|0,0,0,0,\!..\}$\\[2pt]
$\!\!\!k=13$--$14^*\!\!\!\!$&$d=1$&2&2&$(0,0,0;0,0,1)$&$(0,0)$&$(\tfrac{1}{8},\tfrac{1}{8})$&$(\tfrac{1}{2},2)$&$(\tfrac{1}{2},0)$&$(0,2)$&$\{0,0,0,0,\!..|1,1,0,0,\!..\}$\\[2pt]
$k=15^*$&$d=1$&2&1&$(0,1,0;0,1,0)$&$(0,0)$&$(\tfrac{1}{8},\tfrac{1}{8})$&$(\tfrac{3}{2},1)$&$(\tfrac{1}{2},0)$&$(1,1)$&$\{1,0,0,0,\!..|1,0,0,0,\!..\}$\\[2pt]
$k=16^*$&$d=1$&2&1&$(0,1,0;0,1,0)$&$(0,0)$&$(\tfrac{1}{8},\tfrac{1}{8})$&$(1,\tfrac{3}{2})$&$(0,\tfrac{1}{2})$&$(1,1)$&$\{1,0,0,0,\!..|1,0,0,0,\!..\}$\\[2pt]
\hline
$\!\!\!k=17$--$18^*\!\!\!\!$&$d=1$&2&2&$(0,0,1;0,0,0)$&$(0,0)$&$(\tfrac{1}{8},\tfrac{1}{8})$&$(\tfrac{7}{2},0)$&$(\tfrac{1}{2},0)$&$(3,0)$&$\{2,1,0,0,\!..|0,0,0,0,\!..\}$\rule{0pt}{8pt}\\[2pt]
$\!\!\!k=19$--$20^*\!\!\!\!$&$d=1$&2&2&$(0,0,0;0,0,1)$&$(0,0)$&$(\tfrac{1}{8},\tfrac{1}{8})$&$(0,\tfrac{7}{2})$&$(0,\tfrac{1}{2})$&$(0,3)$&$\{0,0,0,0,\!..|2,1,0,0,\!..\}$\\[2pt]
$k=21^*$&$d=1$&2&1&$(0,1,1;0,0,0)$&$(0,0)$&$(\tfrac{1}{8},\tfrac{1}{8})$&$(3,\tfrac{1}{2})$&$(0,\tfrac{1}{2})$&$(3,0)$&$\{2,1,0,0,\!..|0,0,0,0,\!..\}$\\[2pt]
$k=22^*$&$d=1$&2&1&$(0,0,0;0,1,1)$&$(0,0)$&$(\tfrac{1}{8},\tfrac{1}{8})$&$(\tfrac{1}{2},3)$&$(\tfrac{1}{2},0)$&$(0,3)$&$\{0,0,0,0,\!..|2,1,0,0,\!..\}$\\[2pt]
$k=23^*$&$d=1$&2&1&$(0,1,0;0,1,0)$&$(0,0)$&$(\tfrac{1}{8},\tfrac{1}{8})$&$(\tfrac{5}{2},1)$&$(\tfrac{1}{2},0)$&$(2,1)$&$\{2,0,0,0,\!..|1,0,0,0,\!..\}$\\[2pt]
$k=24^*$&$d=1$&2&1&$(0,1,0;0,1,0)$&$(0,0)$&$(\tfrac{1}{8},\tfrac{1}{8})$&$(1,\tfrac{5}{2})$&$(0,\tfrac{1}{2})$&$(1,2)$&$\{1,0,0,0,\!..|2,0,0,0,\!..\}$\\[2pt]
$\!\!\!k=25$--$26^*\!\!\!\!$&$d=1$&2&2&$(0,0,1;0,0,0)$&$(0,0)$&$(\tfrac{1}{8},\tfrac{1}{8})$&$(3,\tfrac{1}{2})$&$(0,\tfrac{1}{2})$&$(3,0)$&$\{2,1,0,0,\!..|0,0,0,0,\!..\}$\\[2pt]
$\!\!\!k=27$--$28^*\!\!\!\!$&$d=1$&2&2&$(0,0,0;0,0,1)$&$(0,0)$&$(\tfrac{1}{8},\tfrac{1}{8})$&$(\tfrac{1}{2},3)$&$(\tfrac{1}{2},0)$&$(0,3)$&$\{0,0,0,0,\!..|2,1,0,0,\!..\}$\\[2pt]
$k=29$&$d=1$&2&1&$(0,0,0;0,0,0)$&$(1,0)$&$(\tfrac{1}{8},\tfrac{1}{8})$&$(\tfrac{7}{2},0)$&$(\tfrac{5}{2},0)$&$(1,0)$&$\{1,0,0,0,\!..|0,0,0,0,\!..\}$\\[2pt]
$k=30$&$d=1$&2&1&$(0,0,0;0,0,0)$&$(0,1)$&$(\tfrac{1}{8},\tfrac{1}{8})$&$(0,\tfrac{7}{2})$&$(0,\tfrac{5}{2})$&$(0,1)$&$\{0,0,0,0,\!..|1,0,0,0,\!..\}$\\[2pt]
$k=31$&$d=1$&2&1&$(0,1,0;0,0,0)$&$(0,1)$&$(\tfrac{1}{8},\tfrac{1}{8})$&$(\tfrac{5}{2},1)$&$(\tfrac{3}{2},1)$&$(1,0)$&$\{1,0,0,0,\!..|0,0,0,0,\!..\}$\rule{0pt}{8pt}\\[2pt]
$k=32$&$d=1$&2&1&$(0,0,0;0,1,0)$&$(1,0)$&$(\tfrac{1}{8},\tfrac{1}{8})$&$(1,\tfrac{5}{2})$&$(1,\tfrac{3}{2})$&$(0,1)$&$\{0,0,0,0,\!..|1,0,0,0,\!..\}$\\[2pt]
$\!\!\!k=33$--$34^*\!\!\!\!$&$d=1$&2&2&$(0,0,1;0,1,0)$&$(0,0)$&$(\tfrac{1}{8},\tfrac{1}{8})$&$(2,\tfrac{3}{2})$&$(0,\tfrac{1}{2})$&$(2,1)$&$\{1,1,0,0,\!..|1,0,0,0,\!..\}$\\[2pt]
$\!\!\!k=35$--$36^*\!\!\!\!$&$d=1$&2&2&$(0,1,0;0,0,1)$&$(0,0)$&$(\tfrac{1}{8},\tfrac{1}{8})$&$(\tfrac{3}{2},2)$&$(\tfrac{1}{2},0)$&$(1,2)$&$\{1,0,0,0,\!..|1,1,0,0,\!..\}$\\[2pt]
\hline\hline
$k=1$&$d=3$&2&1&$(0,0,0;0,0,0)$&$(1,0)$&$(-\tfrac{1}{24},-\tfrac{1}{24})$&$(\tfrac{3}{2},0)$&$(\tfrac{3}{2},0)$&$(0,0)$&$\{0,0,0,0,\!..|0,0,0,0,\!..\}$\rule{0pt}{8pt}\\[2pt]
$k=2$&$d=3$&2&1&$(0,0,0;0,0,0)$&$(0,1)$&$(-\tfrac{1}{24},-\tfrac{1}{24})$&$(0,\tfrac{3}{2})$&$(0,\tfrac{3}{2})$&$(0,0)$&$\{0,0,0,0,\!..|0,0,0,0,\!..\}$\\[2pt]
\hline
$k=3$&$d=3$&2&1&$(0,0,0;0,0,0)$&$(1,0)$&$(-\tfrac{1}{24},-\tfrac{1}{24})$&$(\tfrac{5}{2},0)$&$(\tfrac{5}{2},0)$&$(0,0)$&$\{0,0,0,0,\!..|0,0,0,0,\!..\}$\rule{0pt}{8pt}\\[2pt]
$k=4$&$d=3$&2&1&$(0,0,0;0,0,0)$&$(0,1)$&$(-\tfrac{1}{24},-\tfrac{1}{24})$&$(0,\tfrac{5}{2})$&$(0,\tfrac{5}{2})$&$(0,0)$&$\{0,0,0,0,\!..|0,0,0,0,\!..\}$\\[2pt]
$k=5$&$d=3$&2&1&$(0,0,1;0,0,0)$&$(1,0)$&$(-\tfrac{1}{24},-\tfrac{1}{24})$&$(\tfrac{3}{2},1)$&$(\tfrac{3}{2},0)$&$(0,1)$&$\{0,0,0,0,\!..|1,0,0,0,\!..\}$\\[2pt]
$k=6$&$d=3$&2&1&$(0,0,0;0,0,1)$&$(0,1)$&$(-\tfrac{1}{24},-\tfrac{1}{24})$&$(1,\tfrac{3}{2})$&$(0,\tfrac{3}{2})$&$(1,0)$&$\{1,0,0,0,\!..|0,0,0,0,\!..\}$\\[2pt]
\hline
$k=7$&$d=3$&2&1&$(0,1,0;0,0,0)$&$(1,0)$&$(-\tfrac{1}{24},-\tfrac{1}{24})$&$(\tfrac{7}{2},0)$&$(\tfrac{5}{2},0)$&$(1,0)$&$\{1,0,0,0,\!..|0,0,0,0,\!..\}$\rule{0pt}{8pt}\\[2pt]
$k=8$&$d=3$&2&1&$(0,0,0;0,1,0)$&$(0,1)$&$(-\tfrac{1}{24},-\tfrac{1}{24})$&$(0,\tfrac{7}{2})$&$(0,\tfrac{5}{2})$&$(0,1)$&$\{0,0,0,0,\!..|1,0,0,0,\!..\}$\\[2pt]
$k=9$&$d=3$&2&1&$(0,0,1;0,0,0)$&$(1,0)$&$(-\tfrac{1}{24},-\tfrac{1}{24})$&$(\tfrac{5}{2},1)$&$(\tfrac{3}{2},0)$&$(1,1)$&$\{1,0,0,0,\!..|1,0,0,0,\!..\}$\\[2pt]
$k=10$&$d=3$&2&1&$(0,0,0;0,0,1)$&$(0,1)$&$(-\tfrac{1}{24},-\tfrac{1}{24})$&$(1,\tfrac{5}{2})$&$(0,\tfrac{3}{2})$&$(1,1)$&$\{1,0,0,0,\!..|1,0,0,0,\!..\}$\\[2pt]
$k=11$&$d=3$&2&1&$(0,0,0;0,0,0)$&$(1,0)$&$(-\tfrac{1}{24},-\tfrac{1}{24})$&$(\tfrac{7}{2},0)$&$(\tfrac{3}{2},0)$&$(2,0)$&$\{1,1,0,0,\!..|0,0,0,0,\!..\}$\\[2pt]
$k=12$&$d=3$&2&1&$(0,0,0;0,0,0)$&$(0,1)$&$(-\tfrac{1}{24},-\tfrac{1}{24})$&$(0,\tfrac{7}{2})$&$(0,\tfrac{3}{2})$&$(0,2)$&$\{0,0,0,0,\!..|1,1,0,0,\!..\}$\\[2pt]
$k=13$--$14$&$d=3$&2&2&$(0,0,1;0,0,0)$&$(0,1)$&$(-\tfrac{1}{24},-\tfrac{1}{24})$&$(2,\tfrac{3}{2})$&$(0,\tfrac{3}{2})$&$(2,0)$&$\{1,1,0,0,\!..|0,0,0,0,\!..\}$\\[2pt]
$k=15$--$16$&$d=3$&2&2&$(0,0,0;0,0,1)$&$(1,0)$&$(-\tfrac{1}{24},-\tfrac{1}{24})$&$(\tfrac{3}{2},2)$&$(\tfrac{3}{2},0)$&$(0,2)$&$\{0,0,0,0,\!..|1,1,0,0,\!..\}$\\[2pt]
\hline\hline
$k=1$&$d=5$&2&1&$(0,0,0;0,0,0)$&$(0,1)$&$(\tfrac{1}{8},\tfrac{1}{8})$&$(0,\tfrac{5}{2})$&$(0,\tfrac{5}{2})$&$(0,0)$&$\{0,0,0,0,\!..|0,0,0,0,\!..\}$\rule{0pt}{8pt}\\[2pt]
$k=2$&$d=5$&2&1&$(0,0,0;0,0,0)$&$(1,0)$&$(\tfrac{1}{8},\tfrac{1}{8})$&$(\tfrac{5}{2},0)$&$(\tfrac{5}{2},0)$&$(0,0)$&$\{0,0,0,0,\!..|0,0,0,0,\!..\}$\\[2pt]
\hline
$k=3$&$d=5$&2&1&$(0,0,0;0,0,0)$&$(1,0)$&$(\tfrac{1}{8},\tfrac{1}{8})$&$(\tfrac{7}{2},0)$&$(\tfrac{5}{2},0)$&$(1,0)$&$\{1,0,0,0,\!..|0,0,0,0,\!..\}$\rule{0pt}{8pt}\\[2pt]
$k=4$&$d=5$&2&1&$(0,0,0;0,0,0)$&$(0,1)$&$(\tfrac{1}{8},\tfrac{1}{8})$&$(0,\tfrac{7}{2})$&$(0,\tfrac{5}{2})$&$(0,1)$&$\{0,0,0,0,\!..|1,0,0,0,\!..\}$\\[2pt]
$k=5$&$d=5$&2&1&$(0,1,0;0,0,0)$&$(0,1)$&$(\tfrac{1}{8},\tfrac{1}{8})$&$(1,\tfrac{5}{2})$&$(0,\tfrac{5}{2})$&$(1,0)$&$\{1,0,0,0,\!..|0,0,0,0,\!..\}$\\[2pt]
$k=6$&$d=5$&2&1&$(0,0,0;0,1,0)$&$(1,0)$&$(\tfrac{1}{8},\tfrac{1}{8})$&$(\tfrac{5}{2},1)$&$(\tfrac{5}{2},0)$&$(0,1)$&$\{0,0,0,0,\!..|1,0,0,0,\!..\}$\\[2pt]
\hline\hline
\end{tabular}
\caption{\small String content $(\mb;\bar\mb)=(m_h,m_1,m_2;\mbar_h,\mbar_1,\mbar_2)$, $(m_3,\mbar_3)$, reference conformal weights $(\Delta_{1,d/2},\Delta_{1,d/2})$, eigenenergies $(E,\Ebar)$ and degeneracies for the leading 116 eigenvalues $\widehat T(u)$ in the standard modules $\repW_{N,d,-1}$ with $d$ odd. The multiplicities are 1 for $d=0$ and 2 otherwise. 
Allowing for multiplicities and degeneracies, there are 72 eigenvalues in the $d=1$ mod $6$ modules, 32 eigenvalues in the $d=3$ mod $6$ modules and 12 eigenvalues in the $d=4$ mod $6$ modules. 
The total conformal weights for each eigenvalue are 
$(\Delta,\bar\Delta)=(\Delta_{1,\frac{d}{2}}+E,\Delta_{1,\frac{d}{2}}+\Ebar)$.
In these modules, $J=2$ and $Q(u)=\widehat T(u)$ and $Q(u)$ has no 1-holes. 
An asterisk in the label column indicates that, for these eigenvalues, ${\cal P}$ is not zero.
}\label{omega=-1dOddEigs}
\end{center}
\end{table}

\clearpage

\section{The $\boldsymbol Y$-system of Gliozzi and Tateo}\label{app:Y.system}

Let us define the five functions
\begin{subequations}
\begin{alignat}{2} 
Y^1(z) &= \frac{\Lambda^2(\ir z)+\Lambda^3(\ir z)}{\Lambda^1(\ir z)}, 
\qquad
&&Y^2(z) = \frac{\Lambda^2(\ir z)\big(\Lambda^1(\ir z)+\Lambda^2(\ir z)+\Lambda^3(\ir z)\big)}{\Lambda^1(\ir z)\Lambda^3(\ir z)},
\\
Y^3(z) &= \frac{\Lambda^1(\ir z)+\Lambda^2(\ir z)}{\Lambda^3(\ir z)},
\qquad
&&Y^4(z) = \frac{\Lambda^1(\ir z)}{\Lambda^2(\ir z)},
\qquad
Y^5(z) = \frac{\Lambda^3(\ir z)}{\Lambda^2(\ir z)}.
\end{alignat}
Comparing with \eqref{eq:aA.def}, we see that $Y^1(z) = \amf^1(z)^{-1}$, $Y^2(z) = \amf^3(z)^{-1}$, $Y^3(z) = \amf^2(z)^{-1}$. Remarkably, these five functions satisfy the $Y$-system relations of Gliozzi and Tateo \cite{GT95}
\be
Y^{n-1}(z)Y^{n+1}(z) = 1+Y^{n}(z), \qquad n = 1, 2, \dots, 5,
\ee
\end{subequations}
with the periodicity condition $Y^{n+5}(z) = Y^{n}(z)$. 
Since there are no shifts in the spectral parameter, this $Y$-system relating
the auxiliary functions is a set of algebraic equations rather than the usual
functional equations. It implies that only two of the five $Y^n(z)$ are
independent functions. 

Importantly, we may express $\log Y^n(z)$ linearly in terms of $\log(1+Y^n(z))$ for $n=1,2,\dots,5$ with a symmetric coefficient matrix. This $Y$-system also holds in the scaling limit for the corresponding scaling functions 
\be
Y^{n}_\pm(z) = \lim_{N\to \infty} Y^n\big(\!\pm\!(z + \log N)\big).
\ee
Due to the symmetry of the matrix, we find
\be
\sum_{n=1}^5 L_+({Y^n_\pm (\infty)})-\sum_{n=1}^5 L_+({Y^n_\pm (-\infty}))=0,
\qquad
L_+(x)=L(\tfrac{x}{1+x}),
\label{C7a}
\ee
where $L(x)$ is the Rogers dilogarithm. From (\ref{eq:psi.functions}), the braid limits of the five $Y$-functions are
\be
\{ Y^1_\pm(\infty), Y^2_\pm(\infty), Y^3_\pm(\infty), Y^4_\pm(\infty), Y^5_\pm(\infty)\}=\{\tfrac{1+t}{t^2},t\!+\!1\!+\!\tfrac{1}{t},t(1\!+\!t),t,\tfrac{1}{t}\},\qquad t=s^2=\eE^{\pm \ir\pi d/3}\omega,
\ee
and the bulk limits at $z \to -\infty$ are simply
\be
\{ Y^1_\pm(-\infty), Y^2_\pm(-\infty), Y^3_\pm(-\infty), Y^4_\pm(-\infty), Y^5_\pm(-\infty)\}=\{\infty,\infty,\infty,0,0\}.
\ee
Hence we obtain
\be
\sum_{n=1}^5 L_+({Y^n_\pm(\infty)})= \sum_{n=1}^5 L_+({Y^n_\pm(-\infty)})= \frac{\pi^2}2,\qquad t\in {\Bbb C},
\label{C7b}
\ee
which is actually a one-parameter specialization of Abel's two-parameter, five-term identity.
Now, the three-term identity
\be
\sum_{n=1}^3 L_+(Y^n_\pm(\infty))=\frac{\pi^2}{3},\qquad t\in {\Bbb C}\label{C7}
\ee
follows from Euler's two-term identity $L_+(x)+L_+(\tfrac{1}{x})=\tfrac{\pi^2}{6}$.

%


\begin{thebibliography}{99}
%

\bib{BroadHamm57} 
	S.~Broadbent, J.~Hammersley, 
	{\it Percolation processes I. Crystals and mazes}, 
	Proc.~Camb.~Phil. Soc.~{\bf 53} (1957) 629.

\bib{KestonPerc82}
	H.~Kesten, 
	{\em Percolation theory for mathematicians}, 
	Springer Science (1982).

\bib{Stauffer92}
	D.~Stauffer, A.~Aharony, 
	{\it Introduction to Percolation Theory}, 
	Taylor and Francis (1992).

\bib{Grimmet97}
	G.~Grimmet, 
	{\em Percolation and disordered systems}, 
	Lectures on Probability Theory and Statistics, Springer (1997).

\bib{PhaseTransitions}
	C.~Domb, M.S.~Green, J.L.~Lebowitz (editors), 
	{\em Phase Transitions and Critical Phenomena}, 
	Vols 1--20, 
	Academic Press, (1972)--(2001).

\bib{MDKP2017}
	A.~Morin-Duchesne, A.~Kl\"umper, P.A.~Pearce, 
	{\em Conformal partition functions of critical percolation from $D_3$ thermodynamic Bethe Ansatz equations}, 
	J.~Stat.~Mech.~(2017) 083101,
	\arxiv{1701.08167}{[cond-mat.stat-mech]}.	

\bib{BaxBook}
	R.J.~Baxter, 
	{\it Exactly Solved Models in Statistical Mechanics},
	Academic Press (1982).

\bib{SykesEssam64}
	M.F.~Sykes, J.W.~Essam, 
	{\em Exact critical percolation probabilities for site and bond problems in two dimensions}, 
	J.~Math.~Phys.~{\bf 5} (1964) 1117--1127.

\bib{TL71}
	H.~Temperley, E.~Lieb,
	{\em Relations between the ``percolation'' and ``colouring'' problem and other graph-theoretical problems associated with regular planar lattices: Some exact results for the ``percolation'' problem},
	Proc.~Roy.~Soc.~London Ser.~{\bf A322} (1971) 251--280.
	
\bibitem{Jones}
	V.F.R.~Jones, 
	{\em Planar algebras I}, 
	\arxiv{math/9909027}{[math.QA]}.

\bib{GP93}
	U.~Grimm, P.A.~Pearce,
	{\em Multi-colour braid monoid algebras}, 
	J.~Phys.~A~{\bf 26} (1993) 7435--7459,
	\arxiv{hep-th/9303161}{\!\!}.

\bib{P94}
	P.A.~Pearce,
	{\em Recent progress in solving $A$-$D$-$E$ lattice models}, 
	Physica~A~{\bf 205} (1994) 15--30.

\bib{Grimm96}
	U.~Grimm,
	{\em Dilute algebras and solvable lattice models},
	Statistical Models, Yang-Baxter Equation and Related Topics, Proc. of the satellite meeting of STATPHYS-19, 
	World Scientific (1996) 110--117, 
	\arxiv{q-alg/9511020}{\!\!}.	
	
\bib{BSA14}
	J.~Bellet\^ete, Y.~Saint-Aubin,
	{\em The principal indecomposable modules of the dilute Temperley-Lieb algebra,}
	J.~Math.~Phys.~{\bf 55} (2014) 111706,
	\arxiv{1310.4791}{[math-ph]}.

\bibitem{FMS}
	P.~Di Francesco, P.~Mathieu, D.~S\'en\'echal, 
	{\em Conformal Field Theory},
	Springer (1997).

\bib{Saluer87}
	H.~Saleur, 
	{\em Conformal invariance for polymers and percolation}, 
	J.~Phys.~A~{\bf 20} (1987) 455--470.

\bib{SaleurDup87}
	H.~Saleur, B.~Duplantier, 
	{\em Exact determination of the percolation hull exponent in two dimensions}, 
	Phys.~Rev.~Lett.~{\bf 58} (1987) 2325--2328.

\bib{Saleur92}
	H.~Saleur,
	{\em Polymers and percolation in two dimensions and twisted $N=2$ supersymmetry}, 
	Nucl. Phys.~{\bf B382} (1992) 486--531.

\bib{LanglandsEtAl92}
	R.P.~Langlands, C.~Pichet, P.~Pouliot, Y.~Saint-Aubin, 
	{On the universality of crossing probabilities in two-dimensional percolation}, 
	J.~Stat.~Phys.~{\bf 67} (1992) 553--574.

\bib{Cardy92}
	J.L.~Cardy, 
	{\em Critical percolation in finite geometries}, 
	J.~Phys.~A {\bf 25} (1992) L201--L206.

\bib{Watts96}
	G.M.T.~Watts, 
	{\em A crossing probability for critical percolation in two dimensions}, 
	J.~Phys.~A {\bf 29} (1996) L363--L368.

\bibitem{AAMRH}
	J.~Asikainen, A.~Aharony, B.B.~Mandelbrot, E.M.~Rauch, J.-P.~Hovi, 
	{\em Fractal geometry of critical Potts clusters}, 
	Euro.~Phys.~J.~{\bf B34} (2003) 479--487,
	\arxiv{cond-mat/0212216}{\!\!}.

\bibitem{JankeSchakel05d} 
	W.~Janke, A.M.J.~Schakel, 
	{\em Fractal structure of spin clusters and domain walls in the two-dimensional Ising model}, 
	Phys.~Rev.~{\bf E71} (2005) 036703, 
	\arxiv{cond-mat/0410364}{\!\!}.

\bibitem{JankeSchakel06b} 
	W.~Janke, A.M.J.~Schakel, 
	{\em Two-dimensional critical Potts and its tricritical shadow}, 
	Braz. J.~Phys.~{\bf 36} (2006) 708--716,
	\arxiv{cond-mat/0612650}{\!\!}.

\bib{SAPR2009}
	Y.~Saint-Aubin, P.A.~Pearce, J.~Rasmussen,
	{\em Geometric exponents, SLE and logarithmic minimal models},
	J.~Stat.~Mech. (2009) P02028,
	\arxiv{0809.4806}{[cond-mat.stat-mech]}.

\bib{DengEtAl}
	Y.~Deng, W.~Zhang, T.M.~Garoni, A.D.~Sokal, A.~Sportiello,
	{\em Some geometric critical exponents for percolation and the random cluster model},
	Phys.~Rev.~E~{\bf 81} (2010) 020102,
	\arxiv{0904.3448}{[cond-mat.stat-mech]}.

\bib{SAPR2012}
	G.~Provencher, Y.~Saint-Aubin, P.A.~Pearce, J.~Rasmussen, 
	{\em Geometric exponents of dilute loop models}, 
	J.~Stat.~Phys.~{\bf 147} (2012) 315--350,
	\arxiv{1109.0653}{[cond-mat.stat-mech]}.

\bib{Smirnov} 
	S.~Smirnov, 
	{\em Critical percolation in the plane: conformal invariance, Cardy's formula, scaling limits}, 
	Comptes Rendus de l'Acad\'emie des Sciences, Series I - Mathematics, 
	{\bf 333} (2001) 239--244.

\bib{KSmirnov}
	M.~Khristoforov, S.~Smirnov, 
	{\em Percolation and $O(1)$ loop model}, 
	\arxiv{2111.15612}{[math.PR]}.
 
\bib{SmirnovW01}
	S.~Smirnov, W.~Werner, 
	{\em Critical exponents for two-dimensional percolation}, 
	Math.~Res.~Lett.~{\bf 8} (2001) 729--744,
	\arxiv{math/0109120}{[math.PR]}.
 
 \bib{KNienhuis}
	W.~Kager, B.~Nienhuis,
	{\em Guide to stochastic L\"owner evolution and its applications},
	J.~Stat. Phys.~{\bf 115} (2004) 1149--1229,
	\arxiv{math-ph/0312056}{\!\!}.
 
 \bib{CNewman}
	F.~Camia, C.M.~Newman,
	{\em $SLE_6$ and $CLE_6$ from critical percolation},
	Probability, Geometry and Integrable Systems, MSRI Publications {\bf 55} (2007)
	\arxiv{math/0611116}{[math.PR]}.
  
\bib{DC98}
	G.~Delfino, J.L.~Cardy,
	{\em Universal amplitude ratios in the two-dimensional q-state Potts model and percolation from quantum field theory},
	Nucl.~Phys.~{\bf B519} (1998) 551--578,
	\arxiv{hep-th/9712111}{\!\!}.	  
  
 \bib{CardyRGScaling}
	J.~Cardy,
	{\em Scaling and Renormalization in Statistical Physics}, 
	Cambridge Lecture Notes Phys., 
	Cambridge University Press (1996).

\bib{MooreSeiberg}
	G.~Moore, N.~Seiberg, 
	{\em Classical and quantum conformal field theory}, 
	Comm.~Math.~Phys.~{\bf 123} (1989) 177--254.

\bib{BPZ} 
	A.A.~Belavin, A.M.~Polyakov, A.B.~Zamolodchikov,
	{\em Infinite conformal symmetry in two-dimensional quantum field theory},
	Nucl.~Phys.~{\bf B241} (1984) 333--380.

\bib{ABF} 
	G.E.~Andrews, R.J.~Baxter, P.J.~Forrester, 
	{\em Eight-vertex SOS model and generalized Rogers-Ramanujan-type identities}, 
	J.~Stat.~Phys.~{\bf 35} (1984) 193--266.

\bib{FB} 
	P.J.~Forrester, R.J.~Baxter, 
	{\em Further exact solutions of the eight-vertex SOS model and generalizations of the Rogers-Ramanujan identities}, 
	J.~Stat.~Phys.~{\bf 38} (1985) 435--472.
 
\bibitem{Pasquier87}
	V.~Pasquier, 
	{\em Two-dimensional critical systems labelled by Dynkin diagrams}, 
	Nucl.~Phys.~{\bf B285} (1987) 162--172.

\bibitem{CIZ87}
	A.~Cappelli, C.~Itzykson, J.B.~Zuber.
	{\em The ${A}$-${D}$-${E}$ classification of minimal and ${A}\sp {(1)}\sb 1$ conformal invariant theories}, 
	Comm.~Math.~Phys.~{\bf 113} (1987) 1--26.
 
\bib{GuoBlote}
	W.~Guo, H.W.J.~Bl\"ote, 
	{\em Finite-size analysis of the hard-square lattice gas}, 
	Phys.~Rev.~E~{\bf 66} (2002) 046140.
 
 \bib{KP91} 
	A.~Kl\"umper, P.A.~Pearce, 
	{\em Analytic calculation of scaling dimensions: Tricritical hard squares and critical hard hexagons}, 
	J.~Stat.~Phys.~{\bf 64} (1991) 13--76.
 
 \bib{Pearce93}
	P.A.~Pearce, 
	{\em The 3-state Potts model, hard hexagons and universality}, 
	J.~Korean Phys.~Soc.~{\bf 26} (1993) S349--S354.
 
	\bib{Gurarie} 
	V.~Gurarie, 
	{\em Logarithmic operators in conformal field theory},
	Nucl.~Phys.~{\bf B410} (1993) 535--549,
	\arxiv{hep-th/9303160}{\!\!}.

\bib{PRZ2006}
	P.A.~Pearce, J.~Rasmussen, J.B.~Zuber,
	{\em Logarithmic minimal models}, 
	J.~Stat.~Mech.~(2006) P11017, \arxiv{hep-th/0607232}{\!\!}.
	
\bib{RS07}
	N.~Read, H.~Saleur,
	{\em Associative-algebraic approach to logarithmic conformal field theories}, 
	Nucl.~Phys.~{\bf B777} (2007) 316--351.

\bib{SpecialIssue} 
	A.~Gainutdinov, D.~Ridout, I.~Runkel (editors),
	{\em Special issue on logarithmic conformal field theory}, 
	J.~Phys.~A: Math.~Theor.~{\bf 46(49)} (2013).

\bib{IK1981} 
	A.G.~Izergin, V.E.~Korepin, 
	{\em The inverse scattering method approach to the quantum Shabat-Mikhailov model},
	Comm.~Math.~Phys.~{\bf 79} (1981) 303--316.
	
\bib{AMN1995}
	S.~Artz, L.~Mezincescu, R.I.~Nepomechie,
	{\em Analytical Bethe ansatz for $A_{2n-1}^{(2)}$, $B_n^{(1)}$, $C_n^{(1)}$, $D_n^{(1)}$ quantum-algebra-invariant open spin chains},
	J.~Phys.~A: Math.~Gen.~{\bf 28} (1996) 5131--5142,
	\arxiv{hep-th/9504085}{\!\!}.

\bib{Fan97}
	H.~Fan,
	{\em Bethe ansatz for the Izergin-Korepin model}, 
	Nucl.~Phys.~{\bf B488} (1997) 409--425.

\bib{Utiel2003}
	W.~Utiel,
	{\em Algebraic Bethe ansatz for 19-vertex models with reflection conditions}, 
	J.~Phys.~A: Math.~Gen.~{\bf 36} (2003) 9425--9447.

\bib{Kuniba1991}
	A.~Kuniba, 
	{\em Exact solution of solid-on-solid model for twisted affine Lie agebras $A_{2n}^{(2)}$ and $A_{2n-1}^{(2)}$}, 
	Nucl.~Phys.~{\bf B355} (1991) 801--821.

\bib{WNS1992}
	S.O.~Warnaar, B.~Nienhuis, K.A.~Seaton,
	{\em New construction of solvable lattice models including an Ising model in a magnetic field}, 
	Phys.~Rev.~Lett.~{\bf 69} (1992) 710--712.

\bib{Roche92}
	P.~Roche,
	{\em On the construction of integrable dilute ADE lattice models,} 
	Phys.~Lett.~{\bf B285} (1992) 49--53,
	\arxiv{hep-th/9204036}{\!\!}.

\bib{W93}	
	S.O.~Warnaar, 
	{\em Solvable loop models and graphs}, 
	Ph.D.~thesis, Amsterdam (1993).

\bib{WPSN1994}
	S.O.~Warnaar, P.A.~Pearce, K.A.~Seaton, B.~Nienhuis, 
	{\em Order parameters of the dilute $A$ models}, 
	J.~Stat.~Phys.~{\bf 74} (1994) 469--531,
	\arxiv{hep-th/9305134}{\!\!}.

\bib{BNW94}
	V.V~Bazhanov, B.~Nienhuis, S.O.~Warnaar, 
	{\em Lattice Ising model in a magnetic field: $E_8$ scattering theory}, 
	Phys.~Lett.~{\bf B322} (1994) 198--206,
	\arxiv{hep-th/9312169}{\!\!}.
	
\bib{ZPG1995}
	Y.K.~Zhou, P.A.~Pearce, U.~Grimm,
	{\em Fusion of dilute $A_L$ lattice models},
	Physica~A~{\bf 222} (1995) 261--306,
	\arxiv{hep-th/9506108}{\!\!}.

\bib{Suzuki1998}
	J.~Suzuki, 
	{\em Quantum Jacobi-Trudi formula and $E_8$ structure in the Ising model in a field}, 
	Nucl.~Phys.~{\bf B528} (1998) 638--700,
	\arxiv{cond-mat/9805241}{[cond-mat.stat-mech]}.

\bib{Nienhuis90}
	B.~Nienhuis,
	{\em Critical and multicritical $O(n)$ models}, 
	Physica~A~{\bf 163} (1990) 152--157.

\bib{DJS2010}
	J.~Dubail, J.L.~Jacobsen, H.~Saleur, 
	{\em Conformal boundary conditions in the critical $O(n)$ models and dilute loop models}, 
	Nucl.~Phys.~{\bf B827} (2010) 457--502,
	\arxiv{0905.1382}{[math-ph]}.

	\bib{Garboli12}
	A.~Garbali,
	{\em Dilute $O(1)$ loop model on a strip and the qKZ equations},	
	Master Thesis, University of Amsterdam (2012).

\bib{FeherNien2015}
	G.Z.~Feh\'er, B.~Nienhuis,
	{\em Currents in the dilute $O(n\!=\!1)$ loop model}, 
	\arxiv{1510.02721}{[math-ph]}.

\bib{GarbNien2017a}
	A.~Garbali, B.~Nienhuis,
	{\em The dilute Temperley-Lieb $O(n\!=\!1)$ loop model on a semi infinite strip: the ground state}, 
	J.~Stat.~Mech.~(2017) 043108, 
	\arxiv{1411.7020}{[math-ph]}.

\bib{GarbNien2017b}
	A.~Garbali, B.~Nienhuis,
	{\em The dilute Temperley-Lieb $O(n\!=\!1)$ loop model on a semi infinite strip: the sum rule}, 
	J.~Stat.~Mech.~(2017) 053102,
	\arxiv{1411.7160}{[math-ph]}.

\bib{MDP19}
	A.~Morin-Duchesne, P.A.~Pearce, 
	{\em Fusion hierarchies, $T$-systems and $Y$-systems for the dilute $\Atwotwo$ loop models}, 
	J.~Stat.~Mech.~(2019) 094007,
	\arxiv{1905.07973}{[math-ph]}.

\bib{BMDSA21}
	F.~Boileau. A.~Morin-Duchesne, Y.~Saint-Aubin,
	{\em Fusion hierarchies, $T$-systems and $Y$-systems for the dilute $\Atwotwo$ loop models on a strip}, 
	\arxiv{2211.09017}{[math-ph]}.

\bib{BNW89}
	M.T.~Batchelor, B.~Nienhuis, S.O.~Warnaar, 
	{\em Bethe-Ansatz results for a solvable $O(n)$ model on the square lattice},	
	Phys.~Rev.~Lett.~{\bf 62} (1989) 2425.
	
\bib{WBN92}
	S.O.~Warnaar, M.T.~Batchelor, B.~Nienhuis, 
	{\em Critical properties of the Izergin-Korepin and solvable $O(n)$ models and their related quantum spin chains},
	J.~Phys.~A {\bf 25} (1992) 3077.
	
\bib{ZB97}
	Y.K.~Zhou, M.T.~Batchelor,
	{\em Critical behaviour of the dilute $O(n)$, Izergin-Korepin and dilute $A_L$ face models: Bulk properties},
	Nucl.~Phys.~{\bf B485} (1997) 646--664,
	\arxiv{cond-mat/9611156}{\!\!}.

\bib{YB95}
	C.M.~Yung, M.T.~Batchelor,
	{\em Integrable vertex and loop models on the square lattice with open boundaries via reflection matrices}
	J.~Phys.~A: Math.~Gen. {\bf 28} (1995) L421,
	\arxiv{hep-th/9410042}{\!\!}.
		
\bib{DJS10}
	J.~Dubail, J.L.~Jacobsen, H.~Saleur,
	{\em Conformal boundary conditions in the critical $O(n)$ model and dilute loop models}
	Nucl.~Phys.~{\bf B827} (2010) 457--502,
	\arxiv{0905.1382}{[math-ph]}.
	
\bib{Jones83}
	V.~Jones,
	{\em Index for subfactors},
	Invent.~Math.~{\bf 72} (1983) 1--25.

\bib{RJ06}	
	J.F.~Richard, J.L.~Jacobsen,
	{\em Character decomposition of Potts model partition functions. I. Cyclic geometry},
	Nucl.~Phys.~{\bf B750} (2006) 250--264,
	\arxiv{math-ph/0605016}{[math-ph]}.

\bib{DJS09}
	J.~Dubail, J.L.~Jacobsen, H.~Saleur,
	{\em Conformal two-boundary loop model on the annulus},
	Nucl.~Phys.~{\bf B813} (2009) 430--459,
	\arxiv{0812.2746}{[math-ph]}.

\bib{MDPR13}
	A.~Morin-Duchesne, P.A.~Pearce,~J. Rasmussen, 
	{\em Modular invariant partition function of critical dense polymers}, 
	Nucl.~Phys.~{\bf B874} (2013) 312--357, 
	\arxiv{1303.4895}{[hep-th]}.
	
\bib{FSZ87}
	P.~di Francesco, H.~Saleur, J.B.~Zuber,
	{\em Modular invariance in non-minimal two-dimensional conformal theories},
	Nucl.~Phys.~{\bf B285} (1987) 454--480;
	{\em Relations between the Coulomb gas picture and conformal invariance of two-dimensional critical models},
	J.~Stat.~Phys.~{\bf 49} (1987) 57--79.

\bibitem{RS01}
	N.~Read, H.~Saleur,
	{\em Exact spectra of conformal supersymmetric nonlinear sigma models in two dimensions},	
	Nucl.~Phys.~{\bf B613} (2001) 409--444,
	\arxiv{hep-th/0106124}{\!\!}.

\bib{FabriciusMcCoy01}
	K.~Fabricius, B.M.~McCoy,
	{\em Bethe's equation is incomplete for the XXZ model at roots of unity}, 
	J.~Stat.~Phys.~{\bf 103} (2001) 647--678,
	\arxiv{cond-mat/0009279}{[cond-mat.stat-mech]}.

\bib{KBP91}
	A.~Kl\"umper, M.T.~Batchelor, P.A.~Pearce, 
	{\em Central charges of the 6- and 19-vertex models with twisted boundary conditions}, 
	J.~Phys.~A~{\bf 24} (1991) 3111--3133.

\bib{KP92}
	A.~Kl\"umper, P.A.~Pearce, 
	{\em Conformal weights of RSOS lattice models and their fusion hierarchies}, 
	Physica~{\bf A183} (1992) 304--350.

\bib{FK99}	
	A.~Fujii, A.~Kl\"umper, 
	{\em Anti-symmetrically fused model and non-linear integral equations in the three-state Uimin-Sutherland model}, 
	Nucl.~Phys.~{\bf B546} (1999) 751--764,
	\arxiv{cond-mat/9811234}{\!\!}.

\bib{J08}
	J.~Damerau,
	{\em Nonlinear integral equations for the thermodynamics of integrable quantum chains},
	Ph.D.~thesis, Wuppertal (2008).

\bib{PR2011}
	P.A.~Pearce, J.~Rasmussen, 
	{\em Coset graphs in bulk and boundary logarithmic minimal models}, 
	Nucl.~Phys.~B {\bf 846} (2011) 616--649,
	\arxiv{1010.5328}{[hep-th]}.	

\bib{IMD21}
	Y.~Ikhlef, A.~Morin-Duchesne,
	{\em Fusion in the periodic Temperley-Lieb algebra and connectivity operators of loop models},
	SciPost Phys.~{\bf 12} (2022) 030,
	\arxiv{2105.10240}{[math-ph]}.

\bib{TrinomMotzkin}
	T.~Motzkin, 
	{\em Relations between hypersurface cross ratios, and combinatorial formula for partitions of a polygon, for permanent preponderance, 
 and for nonassociative products},
	Bull.~Amer.~Math.~Soc.~{\bf 54} (1948) 352--360; 
	The On-Line Encyclopedia of Integer Sequences (2022), http://oeis.org/A027907 and A064189.

\bibitem{RJ07} 
	J.F.~Richard, J.L.~Jacobsen, 
	{\em Eigenvalue amplitudes of the Potts model on a torus},
	Nucl. Phys.~{\bf B769} (2007) 256--274, 
	\arxiv{math-ph/0608055}{[math-ph]}.

\bib{AK10}
	B.~Aufgebauer, A.~Kl\"umper,
	{\em Quantum spin chains of Temperley-Lieb type: periodic boundary conditions, spectral multiplicities and finite temperature},
	J.~Stat.~Mech.~(2010) P05018, 
	\arxiv{1003.1932}{[cond-mat.stat-mech]}.	

\bib{BPO96}
	R.E.~Behrend, P.A.~Pearce, D.L.~O'Brien,
	{\em Interaction-round-a-face models with fixed boundary conditions: the ABF fusion hierarchy},
	J.~Stat.~Phys.~{\bf 84} (1996) 1--48,
	\arxiv{hep-th/9507118}{\!\!}.
 
\bib{PRVO20}
	P.A.~Pearce, J.~Rasmussen, A.~Vittorini-Orgeas, 
	{\em Yang-Baxter integrable dimers on a strip}, 
	J.~Stat.~Mech.~(2020) 013107,
	\arxiv{1907.07610}{[math-ph]}.

\bib{PRGN2002} 
	P.A.~Pearce, V.~Rittenberg, J.~de Gier, B.~Nienhuis, 
	{\em Temperley-Lieb stochastic processes}, 
	J.~Phys.~{\bf A35} (2002) L661--L668, 
	\arxiv{math-ph/0209017}{\!\!}.
 
\bib{BGN2001}
	M.T.~Batchelor, J.~de~Gier, B.~Nienhuis, 
	{\em The quantum symmetric XXZ chain at $\Delta=-\tfrac{1}{2}$, alternating-sign matrices and plane partitions}, 
	J.~Phys.~A {\bf 34} (2001) L265--L271,
	\arxiv{cond-mat/0101385}{\!\!}.

\bib{RS01a}
	A.V.~Razumov, Y.G.~Stroganov, 
	{\em Spin chains and combinatorics}, 
	J.~Phys.~A: Math.~Gen. {\bf 34} (2001) 3185--3190,
	\arxiv{cond-mat/0012141}{\!\!}.

\bib{RS01b}
	A.V.~Razumov, Y.G.~Stroganov, 
	{\em Spin chains and combinatorics: twisted boundary conditions}, 
	J.~Phys.~A: Math.~Gen. {\bf 34} (2001) 5335--5340,
	\arxiv{cond-mat/0102247}{\!\!}.

\bibitem{BCN86} 
	H.W.J.~Bl\"ote, J.L.~Cardy, M.P.~Nightingale, 
	{\em Conformal invariance, the central charge, and universal finite-size amplitudes at criticality}, 
	Phys.~Rev.~Lett.~{\bf 56} (1986) 742--745.

\bibitem{Affleck86} 
	I.~Affleck, 
	{\em Universal term in the free energy at a critical point and the conformal anomaly}, 
	Phys.~Rev.~Lett.~{\bf 56} (1986) 746--748.

\bib{BaxInv82} 
	R.J.~Baxter, 
	{\em The inversion relation method for some two-dimensional exactly solved models in lattice statistics}, 
	J.~Stat.~Phys.~{\bf 28} (1982) 1--41.

\bib{OPB95} 
	D.L.~O'Brien, P.A.~Pearce, R.E.~Behrend, 
	{\em Surface free energies and surface critical behavior of the ABF models with fixed boundaries}, 
	Statistical Models, Yang-Baxter Equation and Related Topics, Tianjin, China (1996),
	\arxiv{cond-mat/9511081}{\!\!}. 

\bib{OP97}
	D.L.~O'Brien, P.A.~Pearce, 
	{\em Surface free energies, interfacial tensions and correlation lengths of the ABF models}, 
	J.~Phys.~A: Math.~Gen.~{\bf 30} (1997) 2353--2366,
	\arxiv{cond-mat/9607033}{\!\!}.

\bib{PRdimers07}
	P.A.~Pearce, J.~Rasmussen, 
	{\em Solvable critical dense polymers}, 
	J.~Stat.~Mech.~(2007) P02015,
	\arxiv{hep-th/0610273}{\!\!}.

\bib{AndrewsPartitions}
	G.E.~Andrews,
	{\em The Theory of Partitions},
	Addison-Wesley, Reading, Massachusetts (1976). 
	
\bibitem{PS90} 
	V.~Pasquier, H.~Saleur, 
	{\em Common structures between finite systems and conformal field theories through quantum groups}, 
	Nucl.~Phys. {\bf B330} (1990) 523--556.


\bib{PearceADEfunc92}
	P.A.~Pearce, 
	{\em Row transfer functional equations for $A$-$D$-$E$ lattice models}, 
	Int.~J.~Mod.~Phys.~A {\bf 7}, Suppl. 1B (1992) 791--804.

\bibitem{BaxterQ}
	R.J.~Baxter, 
	{\em Partition function of the eight-vertex lattice model}, 
	Ann.~Phys.~{\bf 70} (1972) 193--228.

\bib{Wolfram} 
	Wolfram Research, 
	{\em Mathematica Edition: Version 13}, 
	Wolfram Research Inc., Champaign, Illinois (2020).

\bib{vBS} 
	J.-M.~Vanden Broeck, L.W.~Schwartz, 
	{\em A one-parameter family of sequence transformations}, 
	SIAM J.~Math.~Anal.~{\bf 10} (1979) 658--666; 
	C.J.~Hamer, M.N.~Barber, 
	{\em Finite-lattice extrapolations for ${\Bbb Z}_3$ and ${\Bbb Z}_5$ models},
	J.~Phys.~A: Math.~Gen. {\bf 14} (1981) 2009--2025.
 
 
\bib{MDSA2011}
	A.~Morin-Duchesne, Y.~Saint-Aubin, 
	{\em The Jordan loop structure of two-dimensional loop models}, 
	J.~Stat.~Mech.~(2011) P04007,
	\arxiv{1101.2885}{[math-ph]}.

\bib{MDSA2013}
	A.~Morin-Duchesne, Y.~Saint-Aubin, 
	{\em Jordan cells of periodic loop models}, 
	J.~Phys.~A~{\bf 49} (2013) 494013,
	\arxiv{1302.5483}{[math-ph]}.
	 
 
\bib{JSbdy}
	J.L.~Jacobsen, H.~Saleur, 
	{\em Conformal boundary loop models},
	Nucl.~Phys.~{\bf B788} (2008) 137--166.
	\arxiv{math-ph/0611078}{\!\!}.

\bib{PTC2015}
	P.A.~Pearce, E.~Tartaglia, R.~Couvreur,
	{\em Kac boundary conditions of the logarithmic minimal models}, 
	J.~Stat.~Mech.~(2015) P01018,
	\arxiv{1410.0103}{[hep-th]}.

\bib{MDRR2015}
	A.~Morin-Duchesne, J.~Rasmussen, D.~Ridout,
	{\em Boundary algebras and Kac modules for logarithmic minimal models}, 
	Nucl.~Phys.~{\bf B899} (2015) 677--769,
	\arxiv{1503.07584}{[hep-ph]}.

\bib{BPT2016}
	J-E.~Bourgine, P.A.~Pearce, E.~Tartaglia,
	{\em Logarithmic minimal models with Robin boundary conditions}, 
	J.~Stat.~Mech.~(2016) 063104,
	\arxiv{1601.04760}{[hep-th]}.
 
\bib{KSS98}
	A.~Kuniba, K.~Sakai, J.~Suzuki,
	{\em Continued fraction TBA and functional relations in XXZ model at root of unity},
	Nucl.~Phys.~{\bf B525} (1998) 597--626,
	\arxiv{math/9803056}{[math.QA]}.

\bib{GT95}
	F.~Gliozzi, R.~Tateo,
	{\em ADE functional dilogarithm identities and integrable models},
	Phys. Lett.~{\bf B348} (1995) 84--88,
	\arxiv{arXiv:hep-th/9411203}.

\bib{VJS2014}
	E.~Vernier, J.L.~Jacobsen, H.~Saleur, 
	{\em Non compact conformal field theory and the $a_2^{(2)}$ (Izergin-Korepin) model in regime III}, 
	J.~Phys.~A~{\bf 47} (2014) 285202, 
	\arxiv{1404.4497}{[math-ph]}.
		
\end{thebibliography}
\end{document}